\documentclass[proof]{pasj00}

\usepackage{mathrsfs}

\def\amati{$E_{\rm p}$--$E_{\rm iso}$ }
\def\yonetoku{$E_{\rm p}$--$L_{\rm p}$ }

\def\vector#1{\mbox{\boldmath $#1$}}

\SetRunningHead{R. Tsutsui and T. Shigeyama}{Hidden subclasses in {\it Swift} long gamma-ray bursts}

\begin{document}

\title{On the Subclasses in {\it Swift} Long Gamma-Ray Bursts: \\A Clue to Different Central Engines}
\author{Ryo \textsc{Tsutsui}\altaffilmark{1}\email{tsutsui@resceu.s.u-tokyo.ac.jp}, \& Toshikazu \textsc{Shigeyama}\altaffilmark{1}}
\altaffiltext{1}{Research Center for the Early Universe, School of Science, University of Tokyo, Bunkyo-ku, Tokyo 113-0033, Japan}
%%% end:list of authors

%%% Please use the following style in case that sorting by
%%% affilation is impossible.
%
% \author{%
%   D-Firstname \textsc{D-Familyname}\altaffilmark{1}
%   E-Firstname \textsc{E-Familyname}\altaffilmark{1,2}
%   and
%   F-Firstname \textsc{F-Familyname}\altaffilmark{2}}
% \altaffiltext{1}{Address of Institute}
% \email{ddddd@xxx.xxx.xx.xx}
% \email{eeeee@xxx.xxx.xx.xx}
% \altaffiltext{2}{Address of Institute}

%% `\KeyWords{}' always has to be placed before `\maketitle'.
\KeyWords{gamma rays bursts : general ---  methods: observational ---
methods: statistical} %Do NOT move this preamble from here!
\maketitle

\begin{abstract}
Analyzing light curves of a complete sample of bright  {\it Swift}  long gamma-ray bursts (LGRBs) of which the peak photon fluxes constructed with the bin width of 1 second in the {\it Swift} 15-350 keV energy band exceed 2.6 photons cm$^{-2}$s$^{-1}$, 
we confirm that there do exist the third class in GRBs in addition to short and long GRBs. 
Being different from previous works based on the duration, fluence,  etc.  our classification method is based on two properties both quantified with light curve shapes of the prompt emission: {\it the Absolute Deviation from the Constant Luminosity 
of their cumulative light curve  $ADCL$} , and {\it the ratio of the mean counts to the maximum counts}  $\bar C/C_{\rm max}$.  These are  independent of the distance and the jet opening angle.
A cluster analysis via the Gaussian mixture model detects three subclasses: 
one consisting of LGRBs with small $ADCL$ and large $\bar C/C_{\rm max}$ values referred to as Type I, one with large $ADCL$ and 
large $\bar C/C_{\rm max}$ referred to as Type II, and one with intermediate $ADCL$ and small $\bar C/C_{\rm max}$,  which is composed of contaminating short GRBs with the extended emission. 
This result is reinforced by different temporal and spectral indices of their X-ray afterglows. 
The difference is prominent in the temporal index of the steep decay phase in particular: the indices for Type I LGRBs distribute between $-6$ and $-3$ 
while those for Type II LGRBs  between $-3$ and $-2$.
From these properties, we propose a possible scenario with different central engines: an accreting black hole and a magnetar.
 \end{abstract}
%Section heading

\section{Introduction}
The existence of the third class of gamma-ray bursts (GRBs) has been studied by some authors.  Nevertheless, it is not widely accepted unlike  short and long GRBs (SGRBs and LGRBs).
\citet{Horvath:1998} proposed the intermediate class of GRBs, 
because  three log-normal functions are needed to reproduce the distribution of durations $T_{90}$ based on the BATSE data.  This feature of the distribution was confirmed by the {\it Swift} data \citep{Horvath:2008} { and {\it BeppoSAX} data \citep{Horvath:2009}}.
Some authors have searched better methods to classify the intermediate class more significantly with  multi-dimentional analyses of  burst properties such as the duration, 
spectral hardness, and fluence { \citep{Mukherjee:1998,Hakkila:2003,Horvath:2006,Horvath:2010}}, and confirmed the  "intermediate" class.  

\citet{Hakkila:2000}, however, insisted that the fluence and the duration of some faint long bursts are underestimated due to the background of detectors and that 
this bias could be responsible for such apparent characteristics of  the intermediate class. Thus the existence of the third class has not been settled.

The purpose of this paper is to confirm the third class of GRBs with a completely different method from the previous works \citep{Horvath:1998,Mukherjee:1998,Hakkila:2000,Horvath:2010}. 
In these works, the duration, hardness and/or fluence were used to classify GRBs, but these observed values include some obstacles to identifying the difference between subclasses of GRBs as follows:
\begin{itemize}
\item fluence - duration bias
\item different intrinsic criteria for the truncation of GRBs at different distances and redshifts with the same detector limit 
\item different jet opening angles
\end{itemize} 

Recently, \citet{Kocevski:2012} quantitively estimated the effect of the bias on bursts exhibiting a single Fast Rise Exponential Decay (FRED) shape pulse by performing Monte Carlo simulations
and concluded that a burst with the signal-to-noise ratio less than $\sim 25$ significantly suffers from the bias. 
As a consequence, it can conceal the expected time dilation due to the cosmological expansion.
Besides these detector limit problems, the dependence of the duration and fluence on nuisance parameters, e.g., the distance, jet opening angle of a burst, makes the situation more complex. 
To establish the third class of GRBs, a new method free from these obstacles is inevitable. 

There seem to be  two possible ways to achieve our purpose here.
The first is to make proper 
corrections to obtain only intrinsic properties of GRBs \citep{Frail:2001,Bloom:2003,Ghirlanda:2004a}.
This approach needs information on the afterglow emission and the host galaxy in addition to the prompt emission and thus the number of sample with sufficient information significantly decreases. Furthermore, this approach needs to estimate the value of the jet opening angle  by detecting an achromatic break in the afterglow emission predicted by the standard fireball model \citep{Rhoads:1999}. Therefore the discovery of chromatic breaks in the afterglow emission has brought this approach into a crisis \citep{Panaitescu:2006}. 

Here we propose an alternative way to extract properties independent of these nuisance parameters from observed light curves. 
In addition, we avoid the fluence - duration bias by using  bright LGRBs.

We use two properties to detect subgroups in our sample:  the absolute deviation from the constant luminosity  of their cumulative light curve $ADCL$ 
and the ratio of the mean photon counts to the maximum photon counts $\bar C/C_{\rm max}$. 
The first parameter is developed by \citet{Tsutsui:2013a} to show that there are two tight fundamental planes, rather than a single wide plane\footnote{The fundamental plane is a correlation between the spectral peak energy ($E_{\rm p}$), peak luminosity ($L_{\rm p}$) and luminosity time ($T_{\rm L}$), the total energy ($E_{\rm iso}$) divided by the peak luminosity.  Well known \amati \citep{Amati:2002} and \yonetoku \citep{Yonetoku:2004} correlations are projections of the fundamental plane.} \citep{Tsutsui:2009b}. 
The cumulative light curve of each event is constructed from the photon counts normalized with the total counts as a function of time normalized with $T_{\rm 90}$ 
to eliminate the dependence on nuisance parameters. The second one is a new parameter introduced to classify LGRBs into subgroups  more 
effectively and to avoid contamination of SGRBs with extended emission (SGRBwEE).  

Here we show  that a cluster analysis via the Gaussian mixture model detects two subclasses: Type I LGRB and Type II LGRB. Furthermore, this analysis identifies  SGRBwEE creeping into the sample. We also investigate X-ray afterglow emission of LGRBs and find different properties in each type. 
Although most of them share a common canonical light curve in their X-ray afterglows, the slope of the steep decay phase for Type I LGRBs is steeper  than that of Type II LGRBs.
Furthermore, the spectra of Type I LGRBs are slightly softer than  those of Type II LGRBs. 
The existence of the third class of LGRBs therefore are confirmed from  properties of not only the prompt emission but also the X-ray afterglow emission. 

Taking into account our results, we propose that the two subclasses have different central engines: an accreting black hole engine for Type I LGRBs and a magnetar engine for Type II LGRBs.
The slower decay in the steep decay phase of Type II LGRBs might come from the temporal change of the magnetic dipole radiation due to the spin down of a magnetar.
Furthermore the common shallow to normal phase might indicate that these two phases come from a common physical process independent of energy injection from central engines such as interactions with the circumstellar matter (Shigeyama \& Tsutsui in prep).  

In the next section, we explain how we select the sample.  We classify LGRBs using the selected data of prompt emission in section \ref{sec:prompt} and X-ray afterglow emission in section \ref{sec:xray}. 
We discuss a possible connection of different properties  found in the prompt emission and the X-ray afterglow emission of different classes of GRBs with the central engines in section \ref{sec:engines}. 
We make concluding remarks in section \ref{sec:summary}.

\section{Selection criteria}
{\it Swift} has monitored the sky since November 2004 and detected 769 GRBs up to 28 May 2013.
It has three instruments to detect the prompt emission of GRBs and observe their afterglow emission in gamma-ray, X-ray, and ultraviolet/optical bands.  
After the Burst Alert Telescope BAT \citep{Barthelmy:2005} detects the prompt emission of a GRB, {\it Swift} automatically slews in the direction of the GRB 
and starts the follow up observation with the X-Ray Telescope XRT \citep{Burrows:2005} and the Ultraviolet/Optical Telescope UVOT \citep{Roming:2005}.
Although there have been some other missions that detected a larger number of GRBs than {\it Swift}, these missions did not cover most GRBs in X-ray and optical bands or measure the redshifts. These pieces of information would help us to investigate the nature of GRBs from various viewpoints. Thus we analyze exclusively the GRBs detected by {\it Swift}  in this paper. 

From the {\it Swift} GRB table on the Web\footnote{http://swift.gsfc.nasa.gov/docs/swift/archive/grb\_table.html/}, we obtain the list of LGRBs including the trigger number,  the duration $T_{90}^{\rm Swift}$ for which  90 \% of the total photons of the event in the {\it Swift} 15-350 keV band  are detected, 
the peak photon flux $P$ in the {\it Swift} 15-150 keV band, the time $T_{\rm start}^{\rm XRT}$ elapsed from the BAT trigger  to the first XRT observation,  the  initial temporal decay index $\alpha_{\rm init}$,  and  spectral index $\beta_{\rm XRT}$  obtained by XRT observations. The table contains 672 LGRBs up to 28 May 2013. 
From these 672 LGRBs, we excluded bursts not detected by onboard analyses and those with only the lower limit of $T_{90}^{\rm Swift}$, and then the number of the sample decreases to 648.
Because it is difficult to determine the correct duration and fluence of GRBs for dim events with the signal-to-noise ratio less than $\sim25$ due to the fluence - duration bias \citep{Kocevski:2012}, 
we select  events with the peak flux $P$  in excess of $2.6$ photons cm$^{-2}$ s$^{-1}$. This criterion reduces the number of the appropriate events from 648 to 168.
According to \citet{Salvaterra:2012}, this threshold flux corresponds to an instrument $\sim$6 times less sensitive than {\it Swift}.
Although the burst trigger threshold of the BAT detector is not simple, it is ranging from 4 to 11 $\sigma$ above the background noise level with a typical value of 8 
$\sigma$\footnote{http://heasarc.gsfc.nasa.gov/docs/swift/about\_swift/bat\_desc.html}.
From these facts,  we consider that our criteria are sufficient to correctly determine the duration.

For these 168 events, we derive the durations  $T_{90}^{\rm ours}$'s by ourselves via the bayesian block algorithm from light curves in the {\it Swift} 15-350 keV band constructed with the bin size of  64 ms in the observer frame. To construct light curves, we use the {\tt batgrbproduct} and {\tt batbinevt} tools in the heasoft version 6.13 according to the instruction on the Web\footnote{http://grbworkshop.wikidot.com/s9-10-swift-bat}. 
The correlation between $T_{90}^{\rm Swift}$ and $T_{90}^{\rm ours}$ is shown in Figure 1.  For most of events, $T_{90}^{\rm ours}$s are consistent with $T_{90}^{\rm Swift}$, which indicates our estimation of $T_{90}^{\rm ours}$ is reasonable.
There are some outliers in Figure 1 probably because of the updated energy calibration of the BAT detector.
{ We derive  $T_{90}$ by ourselves because we need to decide the starting time of  $T_{90}$, which is not contained in the {\it Swift} GRB table, to calculate the two important parameters introduced in the next paragraph. We use the distributions of GRBs with respect to these parameters to investigate the subclasses.}

\begin{figure}[h]
   \begin{center}
      \FigureFile(80mm,80mm){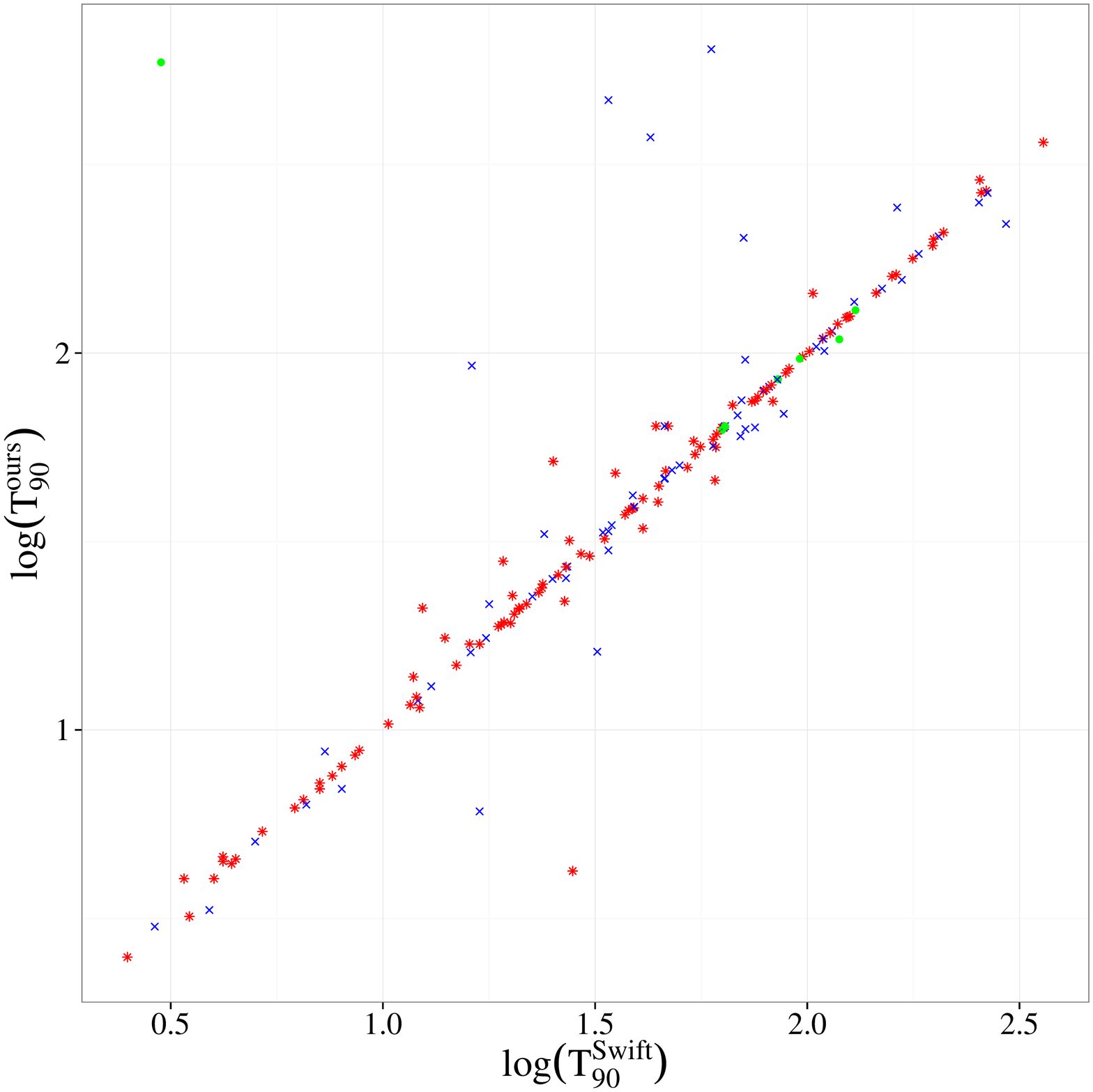}
   \end{center}
   \caption{The relationship between $T_{90}^{\rm Swift}$ and $T_{90}^{\rm ours}$. Red points indicate Type I, Blue points Type II, and green points SGRBwEE. 
   For the classification of GRBs, see section 3. }\label{fig:T90}
\end{figure}

We calculate two parameters from these light curves.
The first is the absolute deviation from the constant luminosity of their cumulative light curves $ADCL$ introduced in the previous section.
To define $ADCL$, we first calculate the normalized time $t_{i}^{\rm norm} \equiv (t_{i}-t_{\rm start}^{T_{90}})/T_{90}$ and normalized cumulative counts $C_{i}^{\rm cum}\equiv (\sum_{k=1}^{i}C_k)/C_{\rm total}$
, where $t_{i}$ ($C_i$), $t_{\rm start}^{T_{90}}$ and $C_{\rm total}$ are the time (counts) of the $i$-th bin, the starting time of $T_{90}$, and the total counts, respectively.
Both the normalized time and the normalized cumulative counts run from 0 to 1.
From these normalized cumulative light curves, the definition of $ADCL$ is given by
\begin{equation}
ADCL = \sum_{i=1}^{N_{\rm bin}} \frac{|C_{i}^{\rm cum} -t_i^{\rm
norm}|}{N_{\rm bin}},
\end{equation}
where the number $N_{\rm bin}$  of bins are different from burst to burst. 

The other parameter $\bar C/C_{\rm max}$  is the ratio of the mean counts during the period $T_{\rm 90}^{\rm ours}$, $\bar C=\sum C_i/N_{\rm bin}$, to the maximum counts $C_{\rm max}$.
$(\bar C/C_{\rm max})^{-1}$ can be regard as an index sensitive to a kind of variability, although it is much simpler than the original definition \citep{Fenimore:2000,Reichart:2001}.

We summarized the data in table~\ref{tb:data}.

\begin{longtable}{l|rrrrrrrrrl}
   \caption{List of LGRBs.}
   \label{tb:data}
\multicolumn{3}{l}{\hbox to 0pt{\parbox{180mm}{ \footnotesize
   * in the unit of seconds \\ 
   $\dagger$ in the unit of photons cm$^{-2}$ s$^{-1}$ }}}
   \endlastfoot
  \hline
 & Trigger & $T_{90}^{\rm Swift}$* & $T_{90}^{\rm ours}$* &$P\dagger$ & $T_{\rm start}^{\rm XRT}$* & $\alpha_{\rm init}$ & $\beta_{\rm XRT}$ & $ADCL$ & $\log({\bar C}/C_{\rm max})$ &  Type \\ 
  \hline
  \hline
130528A & 556870 & 59.40 & 640.00 & 3.00 & 64.87 & -1.45 & 1.91 & 0.42 & -1.62 & Type II \\ 
  130527A & 556753 & 44.00 & 64.00 & 20.10 & 105.24 & -1.50 & 1.67 & 0.21 & -1.10 & Type I \\ 
  130514A & 555821 & 204.00 & 203.90 & 2.80 & 88.83 & -2.02 & 2.06 & 0.20 & -0.83 & Type II \\ 
  130505A & 555163 & 88.00 & 68.99 & 30.00 & 96.37 & 0.64 & 1.92 & 0.36 & -1.26 & Type II \\ 
  130502A & 554996 & 3.00 & 590.72 & 2.90 & 91.41 & -0.68 & 2.21 & 0.03 & -2.41 & SGRBwEE \\ 
  130427B & 554635 & 27.00 & 25.28 & 3.00 & 77.37 & -1.71 & 1.81 & 0.30 & -0.83 & Type II \\ 
  130427A & 554620 & 162.83 & 243.26 & 331.00 & 140.19 & -2.79 & 1.79 & 0.36 & -1.43 & Type II \\ 
  130420A & 553977 & 123.50 & 124.16 & 3.40 & 735.33 & -0.75 & 2.30 & 0.14 & -0.68 & Type I \\ 
  130408A & 553132 & 28.00 & 4.22 & 4.90 & 149.89 & -0.47 & 2.03 & 0.06 & -0.49 & Type I \\ 
  130216A & 548927 & 6.50 & 6.53 & 6.50 & 317700.00 &  &  & 0.05 & -0.26 & Type I \\ 
  121209A & 540964 & 42.70 & 373.63 & 3.40 & 91.98 & 0.61 & 2.30 & 0.48 & -1.77 & Type II \\ 
  121128A & 539866 & 23.30 & 23.17 & 12.90 & 77.17 & -2.37 & 1.99 & 0.12 & -0.68 & Type I \\ 
  121125A & 539563 & 52.20 & 49.73 & 2.90 & 66.80 & -2.69 & 2.15 & 0.05 & -0.55 & Type I \\ 
  121017A & 536172 & 4.20 & 4.61 & 3.30 & 95.25 & -0.96 & 2.14 & 0.08 & -0.48 & Type I \\ 
  120918A & 534015 & 25.10 & 25.15 & 4.50 &  &  &  & 0.19 & -0.66 & Type II \\ 
  120913B & 533613 & 126.00 & 125.38 & 3.20 &  &  &  & 0.05 & -0.61 & Type I \\ 
  120911A & 533268 & 17.80 & 21.57 & 2.70 & 3333.20 & -1.24 & 2.02 & 0.23 & -0.85 & Type II \\ 
  120907A & 532871 & 16.90 & 6.08 & 2.90 & 82.02 & -0.45 & 1.82 & 0.17 & -0.51 & Type II \\ 
  120811C & 530689 & 26.80 & 21.95 & 4.10 & 68.67 & -4.62 & 1.92 & 0.09 & -0.45 & Type I \\ 
  120802A & 529486 & 50.00 & 50.37 & 3.00 & 84.78 & -2.79 & 2.04 & 0.25 & -0.93 & Type II \\ 
  120729A & 529095 & 71.50 & 62.85 & 2.90 & 68.12 & -1.12 & 1.88 & 0.27 & -0.89 & Type II \\ 
  120703A & 525671 & 25.20 & 51.58 & 10.50 & 86.73 & -0.67 & 2.00 & 0.19 & -1.19 & Type I \\ 
  120327A & 518731 & 62.90 & 63.42 & 3.90 & 75.61 & -2.96 & 1.76 & 0.10 & -0.87 & Type I \\ 
  120326A & 518626 & 69.60 & 60.16 & 4.60 & 59.54 & -3.61 & 1.86 & 0.30 & -0.93 & Type II \\ 
  120324A & 518507 & 118.00 & 119.42 & 5.90 & 75.13 & 1.50 & 2.06 & 0.10 & -0.97 & Type I \\ 
  120311A & 517469 & 3.50 & 3.20 & 3.20 & 3161.63 & -1.02 & 2.00 & 0.11 & -0.53 & Type I \\ 
  120308A & 517234 & 60.60 & 45.95 & 6.00 & 92.63 & -3.62 & 1.65 & 0.04 & -0.63 & Type I \\ 
  120218A & 515277 & 27.50 & 31.81 & 9.10 &  &  &  & 0.10 & -0.82 & Type I \\ 
  120119A & 512035 & 253.80 & 250.88 & 10.30 & 53.29 & -2.57 & 1.61 & 0.40 & -1.19 & Type II \\ 
  120116A & 511866 & 41.00 & 34.24 & 4.10 & 74.40 & -3.99 & 2.09 & 0.10 & -0.48 & Type I \\ 
  120102A & 510922 & 38.70 & 38.59 & 10.30 & 112.94 & -3.05 & 2.02 & 0.21 & -0.99 & Type I \\ 
  111228A & 510649 & 101.20 & 101.06 & 12.40 & 145.07 & -5.45 & 2.04 & 0.11 & -1.09 & Type I \\ 
  111121A & 508161 & 119.00 & 108.74 & 7.10 &  & -1.85 & 1.90 & 0.24 & -1.88 & SGRBwEE \\ 
  111103B & 506903 & 167.00 & 156.42 & 7.20 &  & -3.07 & 1.87 & 0.30 & -1.17 & Type II \\ 
  111103A & 506902 & 11.60 & 11.65 & 3.10 &  &  &  & 0.05 & -0.63 & Type I \\ 
  111008A & 505054 & 63.46 & 63.30 & 6.40 &  & -3.27 & 1.94 & 0.22 & -0.99 & Type II \\ 
  110915A & 503219 & 78.76 & 79.17 & 3.30 &  & -6.09 & 2.25 & 0.08 & -0.72 & Type I \\ 
  110731A & 458448 & 38.80 & 41.92 & 11.00 &  & 1.29 & 1.85 & 0.38 & -1.01 & Type II \\ 
  110715A & 457330 & 13.00 & 13.06 & 53.90 &  & -0.57 & 1.85 & 0.30 & -0.79 & Type II \\ 
  110709A & 456939 & 44.70 & 44.35 & 6.20 &  & -1.41 & 2.06 & 0.04 & -0.52 & Type I \\ 
  110625A & 456073 & 44.50 & 40.26 & 49.50 & 140.35 & -1.12 & 1.37 & 0.14 & -0.93 & Type I \\ 
  110610A & 455155 & 46.40 & 48.64 & 4.20 & 71.85 & -2.84 & 2.29 & 0.06 & -0.66 & Type I \\ 
  110519A & 453628 & 27.20 & 27.14 & 4.60 &  &  &  & 0.18 & -0.46 & Type II \\ 
  110422A & 451901 & 25.90 & 25.79 & 30.70 & 814.50 & -1.02 & 1.89 & 0.05 & -0.34 & Type I \\ 
  110420A & 451757 & 11.80 & 13.82 & 14.00 & 87.60 & -3.98 & 2.05 & 0.07 & -0.42 & Type I \\ 
  110402A & 450545 & 60.90 & 56.26 & 4.10 & 544.32 & -0.38 & 2.17 & 0.13 & -1.41 & Type I \\ 
  110318A & 449542 & 16.00 & 16.90 & 8.00 &  &  &  & 0.10 & -0.48 & Type I \\ 
  110205A & 444643 & 257.00 & 266.24 & 3.60 & 155.40 & -7.99 & 1.94 & 0.07 & -0.77 & Type I \\ 
  110102A & 441454 & 264.00 & 269.50 & 8.40 & 148.55 & -2.73 & 2.12 & 0.17 & -1.15 & Type I \\ 
  101117B & 438675 & 5.20 & 5.38 & 4.50 & 77.00 & -0.73 & 2.17 & 0.14 & -0.52 & Type I \\ 
  101024A & 437016 & 18.70 & 18.82 & 5.50 & 77.02 & -0.11 & 1.82 & 0.08 & -0.95 & Type I \\ 
  101017A & 436429 & 70.00 & 74.94 & 9.40 & 80.98 & -2.27 & 1.90 & 0.24 & -0.65 & Type II \\ 
  100906A & 433509 & 114.40 & 114.37 & 10.10 & 80.24 & -3.37 & 2.03 & 0.23 & -0.95 & Type II \\ 
  100816A & 431764 & 2.90 & 3.01 & 10.90 & 82.85 & -1.51 & 1.91 & 0.15 & -0.29 & Type II \\ 
  100728B & 430172 & 12.10 & 11.97 & 3.50 & 97.05 & -1.03 & 2.04 & 0.17 & -0.58 & Type II \\ 
  100728A & 430151 & 198.50 & 200.64 & 5.10 & 76.72 & -0.80 & 1.90 & 0.08 & -0.54 & Type I \\ 
  100704A & 426722 & 197.50 & 192.83 & 4.30 &  & -3.12 & 2.12 & 0.16 & -1.14 & Type I \\ 
  100621A & 425151 & 63.60 & 63.62 & 12.80 & 76.03 & -3.02 & 2.30 & 0.07 & -0.50 & Type I \\ 
  100619A & 424998 & 97.50 & 97.86 & 4.80 & 76.44 & -4.85 & 2.29 & 0.20 & -0.99 & Type I \\ 
  100615A & 424733 & 39.00 & 38.85 & 5.40 & 62.40 & -4.29 & 2.38 & 0.09 & -0.58 & Type I \\ 
  100522A & 422783 & 35.30 & 48.00 & 7.10 & 65.37 & -4.75 & 2.28 & 0.13 & -1.24 & Type I \\ 
  100119A & 383063 & 23.80 & 24.38 & 7.70 &  &  &  & 0.02 & -0.35 & Type I \\ 
  091221 & 380311 & 68.50 & 68.35 & 3.00 & 72.37 & -1.22 & 1.71 & 0.25 & -0.66 & Type II \\ 
  091208B & 378559 & 14.90 & 14.85 & 15.20 & 115.14 & -0.32 & 2.03 & 0.10 & -0.97 & Type I \\ 
  091127 & 377179 & 7.10 & 6.98 & 46.50 & 3214.62 & -1.08 & 1.80 & 0.16 & -0.63 & Type I \\ 
  091020 & 373458 & 34.60 & 34.94 & 4.20 & 81.50 & -2.98 & 2.09 & 0.20 & -0.65 & Type II \\ 
  091018 & 373172 & 4.40 & 4.42 & 10.30 & 61.49 & -0.41 & 1.98 & 0.13 & -0.31 & Type I \\ 
  090929B & 371050 & 360.00 & 362.30 & 3.30 & 84.31 & -1.02 & 1.99 & 0.15 & -1.42 & Type I \\ 
  090926B & 370791 & 109.70 & 101.31 & 3.20 & 88.76 & -1.01 & 1.56 & 0.21 & -0.81 & Type II \\ 
  090904B & 361831 & 47.00 & 64.00 & 5.30 & 134.26 & 0.47 & 1.81 & 0.04 & -0.62 & Type I \\ 
  090813 & 359884 & 7.10 & 7.23 & 8.50 & 78.69 & -0.17 & 1.90 & 0.17 & -0.84 & Type I \\ 
  090812 & 359711 & 66.70 & 72.70 & 3.60 & 76.82 & -2.32 & 1.89 & 0.15 & -0.79 & Type I \\ 
  090715B & 357512 & 266.00 & 266.05 & 3.80 & 46.25 & -1.13 & 2.03 & 0.32 & -1.26 & Type II \\ 
  090715A & 357498 & 63.00 & 62.46 & 3.90 &  &  &  & 0.20 & -1.79 & SGRBwEE \\ 
  090709A & 356890 & 89.00 & 88.58 & 7.80 & 67.81 & -1.87 & 2.01 & 0.07 & -0.50 & Type I \\ 
  090618 & 355083 & 113.20 & 113.09 & 38.90 & 120.90 & -5.86 & 1.92 & 0.10 & -0.54 & Type I \\ 
  090424 & 350311 & 48.00 & 48.90 & 71.00 & 84.46 & -1.29 & 1.96 & 0.41 & -1.19 & Type II \\ 
  090401B & 348152 & 183.00 & 183.30 & 23.10 & 73.22 & -1.13 & 1.93 & 0.39 & -1.69 & Type II \\ 
  090301A & 344582 & 41.00 & 41.09 & 18.70 &  &  &  & 0.07 & -0.60 & Type I \\ 
  090201 & 341749 & 83.00 & 74.43 & 14.70 & 2835.85 & -0.66 & 2.26 & 0.07 & -0.68 & Type I \\ 
  090129 & 341504 & 17.50 & 17.54 & 3.70 &  &  &  & 0.18 & -0.49 & Type II \\ 
  090102 & 338895 & 27.00 & 27.07 & 5.50 & 387.21 & -0.50 & 1.77 & 0.08 & -0.60 & Type I \\ 
  081222 & 337914 & 24.00 & 33.09 & 7.70 & 51.75 & -2.10 & 1.99 & 0.31 & -0.69 & Type II \\ 
  081221 & 337889 & 34.00 & 468.93 & 18.20 & 68.40 & -1.40 & 2.49 & 0.44 & -1.56 & Type II \\ 
  081203A & 336489 & 294.00 & 220.03 & 2.90 & 83.10 & -2.13 & 2.04 & 0.27 & -1.04 & Type II \\ 
  081126 & 335647 & 54.00 & 58.37 & 3.70 & 65.72 & -3.01 & 2.05 & 0.07 & -0.87 & Type I \\ 
  081121 & 335105 & 14.00 & 17.54 & 4.40 & 2813.20 & -1.45 & 1.90 & 0.02 & -0.56 & Type I \\ 
  080916A & 324895 & 60.00 & 56.70 & 2.70 & 70.21 & -1.32 & 1.88 & 0.20 & -0.56 & Type II \\ 
  080915B & 324805 & 3.90 & 3.33 & 8.50 &  &  &  & 0.15 & -0.37 & Type II \\ 
  080804 & 319016 & 34.00 & 33.66 & 3.10 & 99.04 & -1.10 & 1.82 & 0.21 & -0.68 & Type II \\ 
  080727B & 318101 & 8.60 & 8.58 & 7.60 & 101.23 & -2.50 & 1.33 & 0.09 & -0.48 & Type I \\ 
  080721 & 317508 & 16.20 & 92.61 & 20.90 & 108.03 & -0.63 & 1.94 & 0.36 & -1.29 & Type II \\ 
  080714 & 316910 & 33.00 & 33.41 & 4.20 & 79.71 & -1.13 & 1.69 & 0.26 & -0.80 & Type II \\ 
  080613B & 313954 & 105.00 & 103.94 & 2.70 & 69.46 & 1.50 & 2.07 & 0.26 & -0.94 & Type II \\ 
  080607 & 313417 & 79.00 & 79.55 & 23.10 & 82.13 & -2.35 & 2.03 & 0.29 & -0.99 & Type II \\ 
  080605 & 313299 & 20.00 & 19.20 & 19.90 & 90.39 & -0.60 & 1.75 & 0.07 & -0.54 & Type I \\ 
  080603B & 313087 & 60.00 & 59.01 & 3.50 & 61.77 & -3.46 & 1.83 & 0.16 & -0.96 & Type I \\ 
  080602 & 312958 & 74.00 & 74.30 & 2.90 & 111.59 & -4.71 & 1.90 & 0.13 & -1.02 & Type I \\ 
  080515 & 311658 & 21.00 & 20.93 & 3.90 &  & -0.91 & 1.77 & 0.06 & -0.63 & Type I \\ 
  080413B & 309111 & 8.00 & 6.98 & 18.70 & 131.25 & -0.57 & 1.93 & 0.25 & -0.56 & Type II \\ 
  080413A & 309096 & 46.00 & 46.40 & 5.60 & 60.67 & -2.77 & 2.15 & 0.23 & -0.88 & Type II \\ 
  080411 & 309010 & 56.00 & 56.38 & 43.20 & 70.15 & -0.99 & 1.98 & 0.06 & -0.88 & Type I \\ 
  080409 & 308812 & 20.20 & 22.72 & 3.70 & 84.03 & -0.71 & 2.07 & 0.17 & -1.16 & Type I \\ 
  080328 & 307931 & 90.60 & 90.88 & 5.50 & 99.44 & -7.70 & 1.95 & 0.13 & -0.80 & Type I \\ 
  080319C & 306778 & 34.00 & 29.95 & 5.20 & 223.69 & 1.50 & 1.61 & 0.28 & -0.72 & Type II \\ 
  080229A & 304379 & 64.00 & 64.00 & 5.70 & 90.34 & -3.71 & 1.81 & 0.21 & -0.93 & Type II \\ 
  080218B & 303631 & 6.20 & 6.21 & 3.10 & 930.98 & -1.04 & 2.30 & 0.11 & -0.67 & Type I \\ 
  071117 & 296805 & 6.60 & 6.34 & 11.30 & 2848.00 & -0.89 & 2.05 & 0.24 & -0.52 & Type II \\ 
  071020 & 294835 & 4.20 & 4.48 & 8.40 & 61.24 & -0.66 & 1.60 & 0.10 & -0.31 & Type I \\ 
  071003 & 292934 & 150.00 & 148.29 & 6.30 &  & -0.51 & 1.91 & 0.34 & -1.22 & Type II \\ 
  070917 & 291292 & 7.30 & 8.77 & 8.50 &  & -0.90 & 1.12 & 0.26 & -0.58 & Type II \\ 
  070911 & 290624 & 162.00 & 161.54 & 3.90 &  & -1.57 & 2.03 & 0.06 & -0.73 & Type I \\ 
  070714B & 284856 & 64.00 & 63.94 & 2.70 & 61.37 & -1.63 & 2.07 & 0.26 & -1.75 & SGRBwEE \\ 
  070628 & 283320 & 39.10 & 39.10 & 5.10 & 110.82 & -0.68 & 2.04 & 0.25 & -0.74 & Type II \\ 
  070521 & 279935 & 37.90 & 38.27 & 6.53 & 76.89 & -0.13 & 2.00 & 0.12 & -0.62 & Type I \\ 
  070508 & 278854 & 20.90 & 21.06 & 24.10 & 75.92 & -0.26 & 1.83 & 0.06 & -0.48 & Type I \\ 
  070420 & 276321 & 76.50 & 76.48 & 7.12 & 100.68 & -4.45 & 1.98 & 0.13 & -0.70 & Type I \\ 
  070328 & 272773 & 75.30 & 63.49 & 4.22 & 88.29 & -1.34 & 2.14 & 0.23 & -0.54 & Type II \\ 
  070306 & 263361 & 209.50 & 208.96 & 4.07 & 153.20 & -3.21 & 1.94 & 0.21 & -1.18 & Type I \\ 
  070220 & 261299 & 129.00 & 136.70 & 5.88 & 78.79 & -1.27 & 1.55 & 0.28 & -0.92 & Type II \\ 
  061222A & 252588 & 71.40 & 96.00 & 8.53 & 101.02 & -4.29 & 1.93 & 0.25 & -1.18 & Type II \\ 
  061210 & 243690 & 85.30 & 85.25 & 5.31 &  & -1.71 & 2.86 & 0.09 & -2.41 & SGRBwEE \\ 
  061126 & 240766 & 70.80 & 202.18 & 9.76 & 1599.69 & -1.34 & 1.92 & 0.38 & -1.49 & Type II \\ 
  061121 & 239899 & 81.30 & 81.41 & 21.10 & 55.40 & -3.88 & 1.90 & 0.29 & -1.14 & Type II \\ 
  061021 & 234905 & 46.20 & 46.53 & 6.11 & 72.79 & -2.01 & 1.99 & 0.31 & -1.02 & Type II \\ 
  061007 & 232683 & 75.30 & 74.69 & 14.60 & 80.45 & -1.95 & 2.00 & 0.09 & -0.52 & Type I \\ 
  061006 & 232585 & 129.90 & 129.92 & 5.24 & 156.58 & -0.78 & 1.87 & 0.23 & -1.98 & SGRBwEE \\ 
  060927 & 231362 & 22.50 & 22.59 & 2.70 & 64.72 & -0.73 & 1.95 & 0.18 & -0.72 & Type II \\ 
  060912A & 229185 & 5.00 & 5.06 & 8.58 & 108.88 & -1.07 & 1.89 & 0.20 & -0.47 & Type II \\ 
  060908 & 228581 & 19.30 & 19.33 & 3.03 & 71.68 & -0.53 & 2.13 & 0.09 & -0.47 & Type I \\ 
  060904A & 227996 & 80.10 & 80.13 & 4.87 & 65.96 & -3.56 & 1.25 & 0.12 & -0.74 & Type I \\ 
  060825 & 226382 & 8.00 & 8.00 & 2.66 & 65.59 & -1.04 & 1.76 & 0.03 & -0.30 & Type I \\ 
  060814 & 224552 & 145.30 & 144.32 & 7.27 & 71.54 & -1.96 & 2.12 & 0.18 & -0.82 & Type I \\ 
  060813 & 224364 & 16.10 & 16.06 & 8.84 & 76.24 & -0.28 & 1.92 & 0.22 & -0.46 & Type II \\ 
  060614 & 214805 & 108.70 & 109.25 & 11.50 & 91.40 & 1.50 & 1.90 & 0.10 & -0.92 & Type I \\ 
  060510A & 209351 & 20.40 & 20.29 & 14.70 & 146.15 & -2.29 & 1.89 & 0.12 & -0.68 & Type I \\ 
  060421 & 206257 & 12.20 & 11.46 & 2.94 & 87.56 & -1.10 & 1.55 & 0.09 & -0.48 & Type I \\ 
  060418 & 205851 & 103.10 & 144.00 & 6.52 & 77.97 & -1.54 & 1.94 & 0.15 & -1.11 & Type I \\ 
  060306 & 200638 & 61.20 & 60.99 & 5.97 & 87.52 & -3.92 & 2.29 & 0.17 & -1.32 & Type I \\ 
  060223B & 192152 & 10.30 & 10.37 & 2.87 & 68745.00 &  & 2.00 & 0.06 & -0.39 & Type I \\ 
  060210 & 180977 & 255.00 & 288.00 & 2.72 & 94.95 & -1.26 & 2.08 & 0.16 & -1.22 & Type I \\ 
  060206 & 180455 & 7.60 & 7.55 & 2.79 & 58.35 & -0.81 & 2.20 & 0.13 & -0.38 & Type I \\ 
  060117 & 177666 & 16.90 & 16.90 & 48.30 &  &  &  & 0.08 & -0.67 & Type I \\ 
  060105 & 175942 & 54.40 & 53.82 & 7.44 & 86.89 & -1.13 & 2.05 & 0.13 & -0.52 & Type I \\ 
  051111 & 163438 & 46.10 & 64.00 & 2.66 &  & -1.60 & 2.23 & 0.19 & -0.75 & Type II \\ 
  051109A & 163136 & 37.20 & 37.25 & 3.94 & 119.66 & -3.13 & 2.07 & 0.20 & -0.99 & Type I \\ 
  050922C & 156467 & 4.50 & 4.54 & 7.26 &  & -1.05 & 2.21 & 0.08 & -0.41 & Type I \\ 
  050820B & 151334 & 12.00 & 12.22 & 3.95 &  & -1.77 & 1.98 & 0.11 & -0.40 & Type I \\ 
  050802 & 148646 & 19.00 & 19.01 & 2.75 &  & 1.09 & 1.86 & 0.13 & -0.58 & Type I \\ 
  050724 & 147478 & 96.00 & 96.58 & 3.26 &  & -0.01 & 1.80 & 0.10 & -1.83 & SGRBwEE \\ 
  050717 & 146372 & 85.00 & 85.06 & 6.23 &  & -1.95 & 1.79 & 0.26 & -1.05 & Type II \\ 
  050713A & 145675 & 124.70 & 124.67 & 4.67 &  & -5.65 & 2.20 & 0.17 & -1.09 & Type I \\ 
  050701 & 143708 & 21.80 & 21.57 & 2.74 &  & -1.19 & 2.33 & 0.11 & -0.72 & Type I \\ 
  050603 & 131560 & 12.40 & 21.06 & 21.50 &  & -1.71 & 2.02 & 0.15 & -1.19 & Type I \\ 
  050525A & 130088 & 8.80 & 8.83 & 41.70 &  & -0.68 & 2.09 & 0.05 & -0.42 & Type I \\ 
  050418 & 114893 & 82.30 & 82.30 & 3.68 &  &  &  & 0.11 & -0.85 & Type I \\ 
  050416B & 114797 & 3.40 & 4.03 & 5.93 & 310326.00 &  &  & 0.08 & -0.51 & Type I \\ 
  050416A & 114753 & 2.50 & 2.50 & 4.88 &  & -0.70 & 2.06 & 0.12 & -0.51 & Type I \\ 
  050401 & 113120 & 33.30 & 32.13 & 10.70 &  & -0.58 & 1.79 & 0.12 & -0.83 & Type I \\ 
  050326 & 112453 & 29.30 & 29.31 & 12.20 &  & -1.69 & 2.04 & 0.06 & -0.70 & Type I \\ 
  050318 & 111529 & 32.00 & 16.13 & 3.16 &  & -1.12 & 1.98 & 0.33 & -1.35 & Type II \\ 
  050306 & 107547 & 158.30 & 159.81 & 3.58 & 127308.00 &  &  & 0.07 & -0.83 & Type I \\ 
  050219B & 106442 & 30.70 & 28.93 & 24.80 &  & -1.23 & 2.18 & 0.14 & -0.76 & Type I \\ 
  050219A & 106415 & 23.70 & 23.81 & 3.53 &  & -2.96 & 1.58 & 0.05 & -0.42 & Type I \\ 
  050128 & 103906 & 19.20 & 28.03 & 7.42 &  & -0.95 & 1.99 & 0.18 & -0.84 & Type I \\ 
  050124 & 103647 & 4.00 & 4.03 & 5.46 &  & -1.63 & 1.97 & 0.06 & -0.43 & Type I \\ 
  041224 & 100703 & 177.20 & 178.11 & 2.94 &  &  &  & 0.11 & -0.85 & Type I \\ 
  041223 & 100585 & 109.10 & 109.12 & 7.35 &  & -1.88 & 2.13 & 0.19 & -0.76 & Type II \\ 
   \hline
   \hline
  % \label{tb:data}
\end{longtable}

\section{Prompt Emission Properties}
\label{sec:prompt}
In this section, we search subgroups in LGRBs by applying a cluster analysis based on the Gaussian mixture model to the dataset of LGRBs each of which is composed of a pair of the two quantities $ADCL$ and $\bar C/C_{\rm max}$ introduced in the previous section.

The {\tt Mclust} package \citep{FraleyRaftery2002,Fraley:2012} in the {\tt R} language is used to select the optimal model via the EM algorithm. 
First, we apply  the method to the one dimensional distributions of LGRBs with respect to $ADCL$ and $\bar C/C_{\rm max}$  separately and find how many components are needed to describe each distribution. Then we apply the same method to  the two dimensional distribution of LGRBs.
The number of components is optimized according to the Bayesian information criterion (BIC).
Details of the Gaussian mixture model, the EM algorithm, and BIC are described in Appendix 1.

\subsection{One-dimensional cluster analysis}
\begin{figure*}[htb]
\begin{tabular}{c}
\begin{minipage}{0.50\hsize}
\begin{center}
    \FigureFile(80mm,80mm){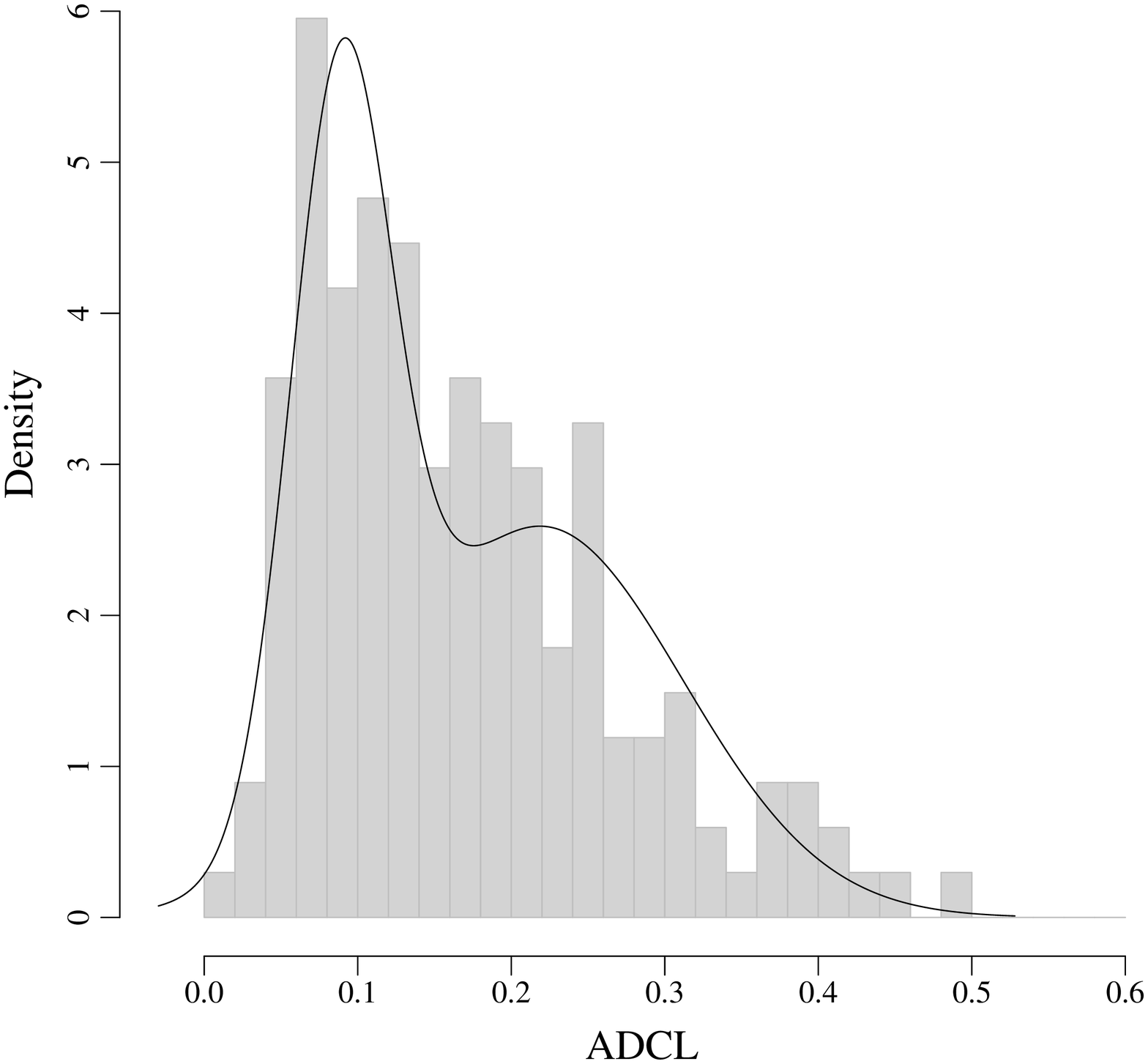}
\end{center}
\end{minipage}
\begin{minipage}{0.50\hsize}
\begin{center}
    \FigureFile(80mm,80mm){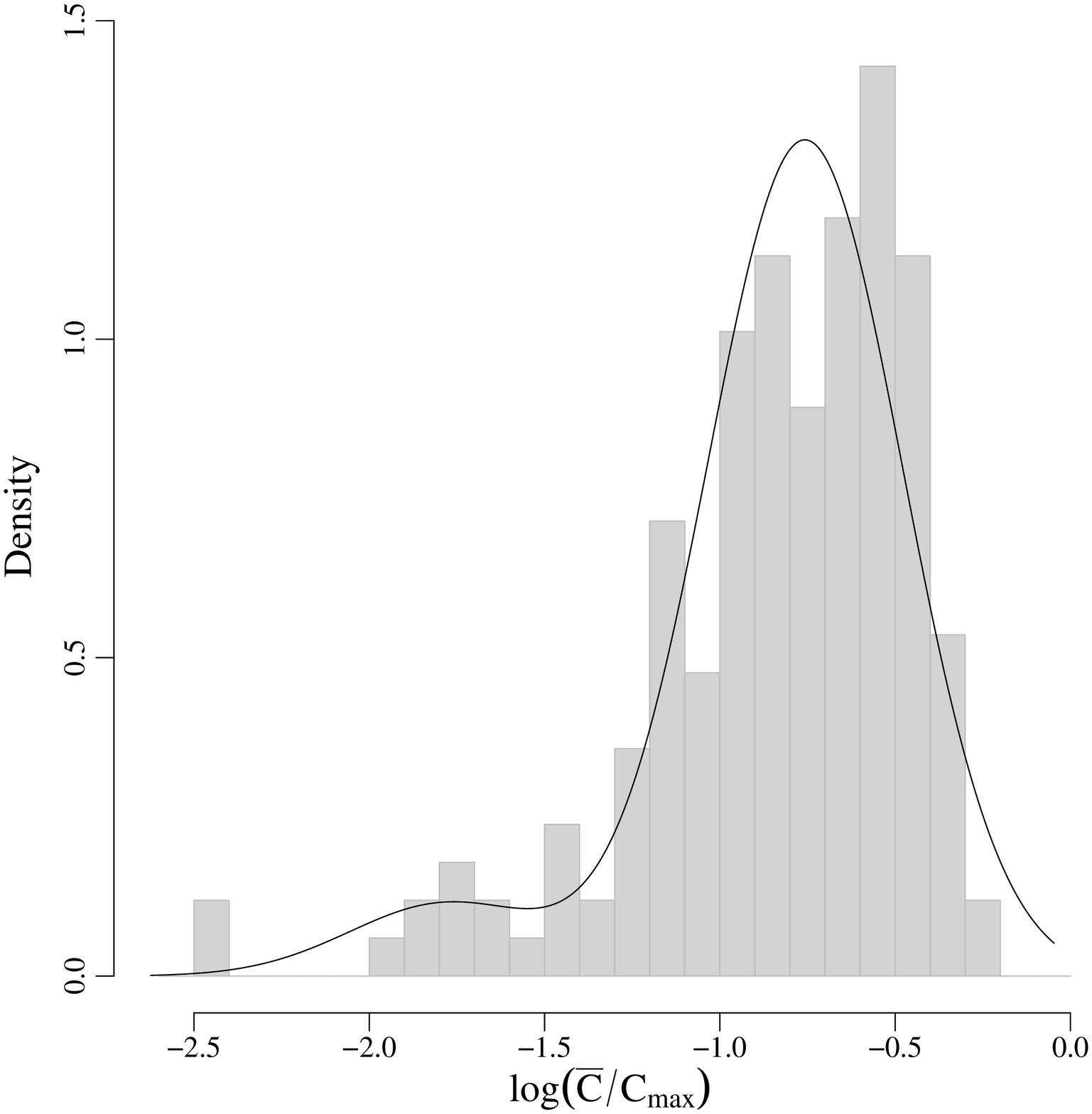}
 \end{center}
\end{minipage}
\end{tabular}
   \caption{The histograms of $ADCL$ (left) and $\log (\bar C/C_{\rm max})$ (right), respectively. The vertical axises show not frequency but frequency density. 
   The solid lines indicates the optimal models with two component Gaussian mixture model.}\label{fig:1DCluster}
   \end{figure*}

The left and right panels of Figure \ref{fig:1DCluster}  show the histograms of $ADCL$ and $\log(\bar C/C_{\rm max})$, respectively. 
The ordinate of each panel represents the frequency density with respect to the parameter. Note that the total area of the shaded region is equal to unity. 
From these $ADCL$ and $\log(\bar C/C_{\rm max})$ distributions, we estimate the optimal probability densities based on the Gaussian mixture model (see Appendix 1 for details). 
In Tables \ref{tb:ADCL} and \ref{tb:MeanFlux}, we list the number of components, the logarithm of the likelihood function, and the BIC for the optimal models. The tables indicate that two component models are best fitted to both of the $ADCL$ and $\log(\bar C/C_{\rm max})$ distributions.
The solid lines in Figures \ref{fig:1DCluster} indicate the optimal densify profiles of the two-component Gaussian mixture model.

\begin{table}[htbp]
\caption{Number of components \#, the logarithm of the likelihood function, and BIC for optimal Gaussian mixture model on the $ADCL$ distribution. }
\begin{center}
\begin{tabular}{c|cc}
    \# & log(likelihood)  & BIC \\ \hline
    1 & 151.2  & 292.2 \\ 
    2 &  169.9 &  314.1 \\
    3 & 174.0 &  307.0 \\
  \end{tabular}
  \label{tb:ADCL}
\end{center}
\end{table}

\begin{table}[htbp]
\caption{Number of components \#, logarithm of likelihood function, and BIC of Gaussian mixture model on $\log(\bar C/C_{\rm max})$ distribution.  }
\begin{center}
  \begin{tabular}{c|cc}
    \#  & log(likelihood)  & BIC \\ \hline
    1 & -81.8&  -173.8\\
    2 &  -62.9 &  -146.3\\
    3 &  -53.4  &  -147.8\\
  \end{tabular}
\end{center}
\label{tb:MeanFlux}
\end{table}

\subsection{Two-dimensional cluster analysis}
Next we apply the cluster analysis to the two-dimensional distribution in the $ADCL$ -- $\log(\bar C/C_{\rm max})$ diagram.
In Table 4, we list the number of components, the logarithm of the likelihood function, and BIC for the optimal models. 
Figure 3 shows the distribution of LGRBs in the two-dimensional space of $ADCL$ and  $\log(\bar C/C_{\rm max})$.
The result of the cluster analysis with three components is shown as coloring of symbols. The red asterisks denote Type I LGRBs and the blue crosses Type II LGRBs, 
following the definition by \citet{Tsutsui:2013a}. The green circles denote contaminating SGRBwEE. 
The solid ellipses in Figure 3 indicate the optimal density profiles of each Gaussian component. 
The optimal mixing probability $\pi$, mean vector $\vector{\mu}$, and variance matrix $\vector{\sigma}$ for each component are listed in Table 5.

Table 4 indicates that  the four-component model which divides Type I LGRBs into two subclasses fits the data best, 
but the difference of BIC between four-component model and three-component model is quite small.
Furthermore,  GRBs belonging to the two subclasses in Type I LGRBs exhibit similar light curves of the X-ray afterglow, which are clearly distinct from those of Type II LGRBs
 (See  Appendix~\ref{sec:appendix2} for the results of four-component model analysis).  Then as a tentative model, we adopt three-component model in the following to simplify the argument. To determine the true number of groups, we need a larger number of sample with more accurate estimation of parameters. 
Therefore we will leave it as a subject in future works.

We show all of light curves classified into each subclass in Figures~\ref{fig:A1-1} -~\ref{fig:A3}. 
{ Each light curve is constructed with 100 bins to reduce noise, although we use 64 ms bin light curves to calculate $ADCL$ and $\log(\bar C/C_{\rm max})$).}
These figures show that the pulses in a Type II LGRB progressively decline, while each pulse in a Type I LGRB emits a similar amount of energy. 
Because the pulse is thought to be caused by the collision of shells,  the pulse shape may be used to probe the formation mechanism of shells and thus the energy injection from the central engine.
Therefore the decline of pulse heights in a Type II LGRB indicates that the energy injected by central engine decreases. On the other hand, the central engine of a Type I LGRB seems to maintain the energy injection rate during the prompt emission.
We will discuss the origin of the difference after looking at properties of X-ray afterglow emission of these two types of LGRBs.

\begin{table}[htbp]
\caption{Number of components \#, logarithm of likelihood function, and BIC of Gaussian mixture model on $ADCL$ and logarithm  $\bar C/C_{\rm max}$ distribution. }
\begin{center}
  \begin{tabular}{c|cc}
    \# & log(likelihood)  & BIC \\ \hline
    1 & 91.6 & 157.5\\
    2 &  133.7 &  221.3 \\
    3 & 154.4 &  242.1\\
    4 & 173.8  &  245.2\\
    5 & 171.5 & 235.4 \\ \hline
  
  \end{tabular}
\end{center}
\end{table}

\begin{table*}[htbp]
\caption{The mixing probability $\pi$, mean vector $\vector{\mu}$, and variance matrix $\vector{\sigma}$ for each component of the three-components Gaussian mixture model.}
\begin{center}
\renewcommand{\arraystretch}{1.5}
  \begin{tabular}{c|ccc}
    & Type I LGRB& Type II LGRB  & SGRBwEE \\ \hline
    $\pi$ &0.61 & 0.35 & 0.042\\
     $\vector{ \mu}$ &  $(0.11, \,-0.72) $&  $(0.27, \,-0.90)$& $(0.17,\, -1.99)$ \\
   ${\vector \sigma}$ &
  $\left(\begin{array}{cc}
 0.0027 &-0.0097\\
 -0.0097 & 0.090
   \end{array}  \right)$& 
  $\left(\begin{array}{cc}
  0.0052 &-0.018\\
 -0.018  & 0.088
   \end{array}\right)$  & 
$\left( \begin{array}{cc}
0.0071 & 0.022\\
0.022 & 0.086   
\end{array}\right)$  \\
  \end{tabular}
  \renewcommand{\arraystretch}{1}
\end{center}
\end{table*}

\begin{figure}[h]
   \begin{center}
      \FigureFile(80mm,80mm){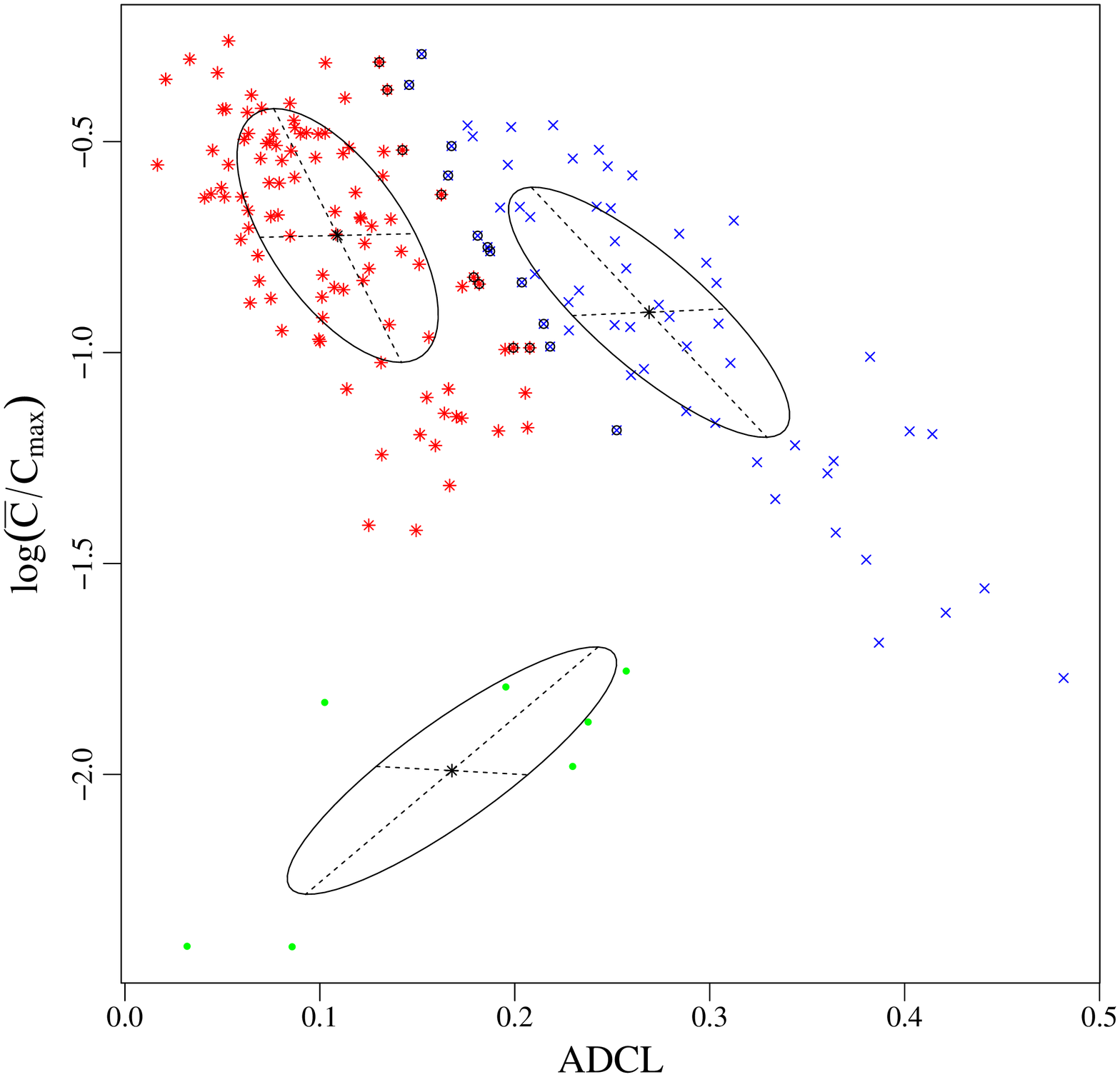}
   \end{center}
   \caption{The relationship between $ADCL$ and $\log(\bar C/C_{\rm max})$ with the result of the clustering analysis based on three-components Gaussian mixture model.
  The result of cluster analysis is shown as the coloring of symbols. The red asterisks are referred as Type I LGRBs, the blue crosses as Type II LGRBs, and the green circles as SGRBwEE.
  The optimal Gaussian distributions are indicated by ellipses. 
  The symbols marked with circles show the events whose uncertainty of classification are larger than $0.2$ and removed from following afterglow analysis (See section~\ref{sec:xray}.).
   }
   \label{fig:2DCluster}
\end{figure}

\section{X-ray Afterglow Properties}
\label{sec:xray}
Contrary to expectations before the launch of {\it Swift}, the majority of early X-ray afterglows observed by {\it Swift} show a complex time profile called the 'canonical' light curve 
which consists of three segments with power-law decays : the first steep decay phase with the decay index $\alpha \sim -3$ -- $- 5$,  
the second shallow phase with $\alpha \sim -0.5$, and the third normal phase with $\alpha \sim -1.3$ \citep{Nousek:2006,Zhang:2006}.
They sometimes  have the fourth phase called the post jet break phase with $\alpha \sim -2$.

In addition to the complexity, what makes the variation of X-ray afterglow light curves is an issue of wide interest in the GRB study.
According to \citet{Evans:2009},  4 \% of GRBs have no break in their X-ray afterglows,  30 \% one break (with flattening or steepening), 
and the rest at least two breaks.
The last group was divided into 'canonical' (42 \%) and 'oddball' (24 \%) :
if an X-ray afterglow light curve contains a flattening break with $\Delta \alpha \ge 0.5$ followed by a steepening break with $\Delta\alpha \le -0.5$, it is regarded as canonical, 
and otherwise oddball.

 \citet{Willingale:2007} first showed that X-ray afterglow light curves  can be fitted using two components with an early exponential decay phase followed by a power-law decay. Other authors \citep{Ghisellini:2009,Yamazaki:2009} showed that the two components are sufficient to explain the complexity and variation in the afterglow emission. On the other hand, what makes the second component and determines the ratio of energies injected to these components remains unresolved.
In the previous section, we have developed a new classification scheme of LGRBs based on  the light curve properties exclusively of the prompt emission.
The different subclasses thus introduced are expected to have different progenitors or central engines. 
Therefore it is natural to think that there are also some differences in  properties of their afterglow emission.
 
Figure 3 shows that there are some LGRBs around the border line on which the conditional probability of Type I is equal to that of Type II.
The probability of classification errors is not negligible around the line, and misclassifications can disturb the following analysis.  
We thereby remove events whose uncertainties of classification are greater than 0.2 from the following analysis.
In Figure 3, we marked such 19 events with circles. 
In addition, we found  7 Type II LGRBs with weak precursors (120802A, 120326A, 091221, 080229A, 070628, 061222A, 061121).
Although our classification method does not distinguish them from the other Type II events,  there is a clear difference in the shapes of their cumulative light curves. Cumulative light curves of these 7 events exhibit shapes of convex functions of time while the other Type II events have the cumulative light curves with  concave shapes (see Figs. ~\ref{fig:A2-1} -~\ref{fig:A2-2}). Therefore we exclude this small number of events in the following analyses.
Furthermore, we do not consider SGRBwEE because of the small number of the sample. 

In Figures 4 and 5, we show the histograms of $\alpha_ {\rm init}$ and $\beta_{\rm XRT}$, respectively.
The figures obviously indicate that Type I and Type II LGRBs have different XRT initial temporal and spectral indices in their X-ray afterglow emission.
Two-sample Kolmogorov-Smirnov (KS) test shows $p=0.082$ for $\alpha_ {\rm init}$, and $p=0.070$ for $\beta_{\rm XRT}$, respectively. 
These large values might be due to a feature of  KS test, i.e., the p-value is more sensitive to the median values of the distributions than the long tails that we would like to focus.
If we use only the LGRBs whose $\alpha_{\rm init}$ less than -2, the vertical line in Figure 4, 
KS test shows $p= 0.0039$. This $p$-value indicates the difference is significant at nearly 3-$\sigma$ confidence level.
To examine the temporal index in the steep decay phase, we need to know the starting time of XRT observations because the steep decay phase might be missed if the XRT starts observation too late.
Figure 6 shows the relationship between $\log(T_{\rm start}^{\rm XRT}/T_{90})$ and $\alpha_{\rm init}$ for Type I (left) and Type II (right) LGRBs, respectively.
If LGRBs do not have steep decay or XRT observations started after steep decay phase, these points distribute in a cluster whose $\alpha_{\rm init}$ is in excess -2. 
We therefore roughly estimate that the indices for Type I LGRBs distribute between -6 and -3 while those for Type II LGRBs between -3 and -2.
We will make a more thorough discussion after the analysis of XRT light curve shape  (Tsutsui \& Shigeyama in prep).

\begin{figure}[h]
   \begin{center}
      \FigureFile(80mm,80mm){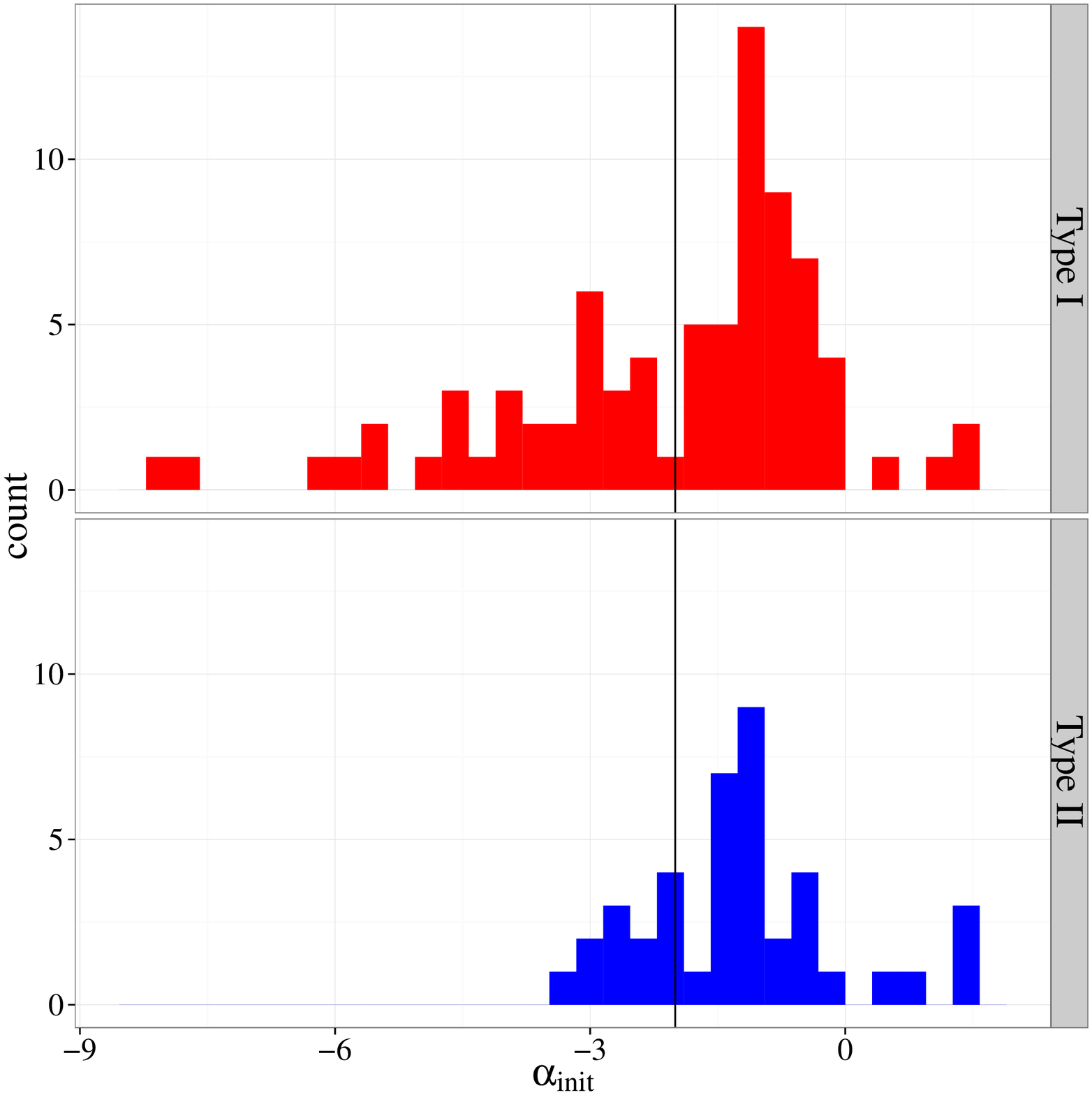}
   \end{center}
   \caption{The histograms of $\alpha_{\rm init}$ for Type I (top) and Type II (bottom) LGRBs, respectively. }\label{fig:alpha}
\end{figure}
\begin{figure}[h]
   \begin{center}
      \FigureFile(80mm,80mm){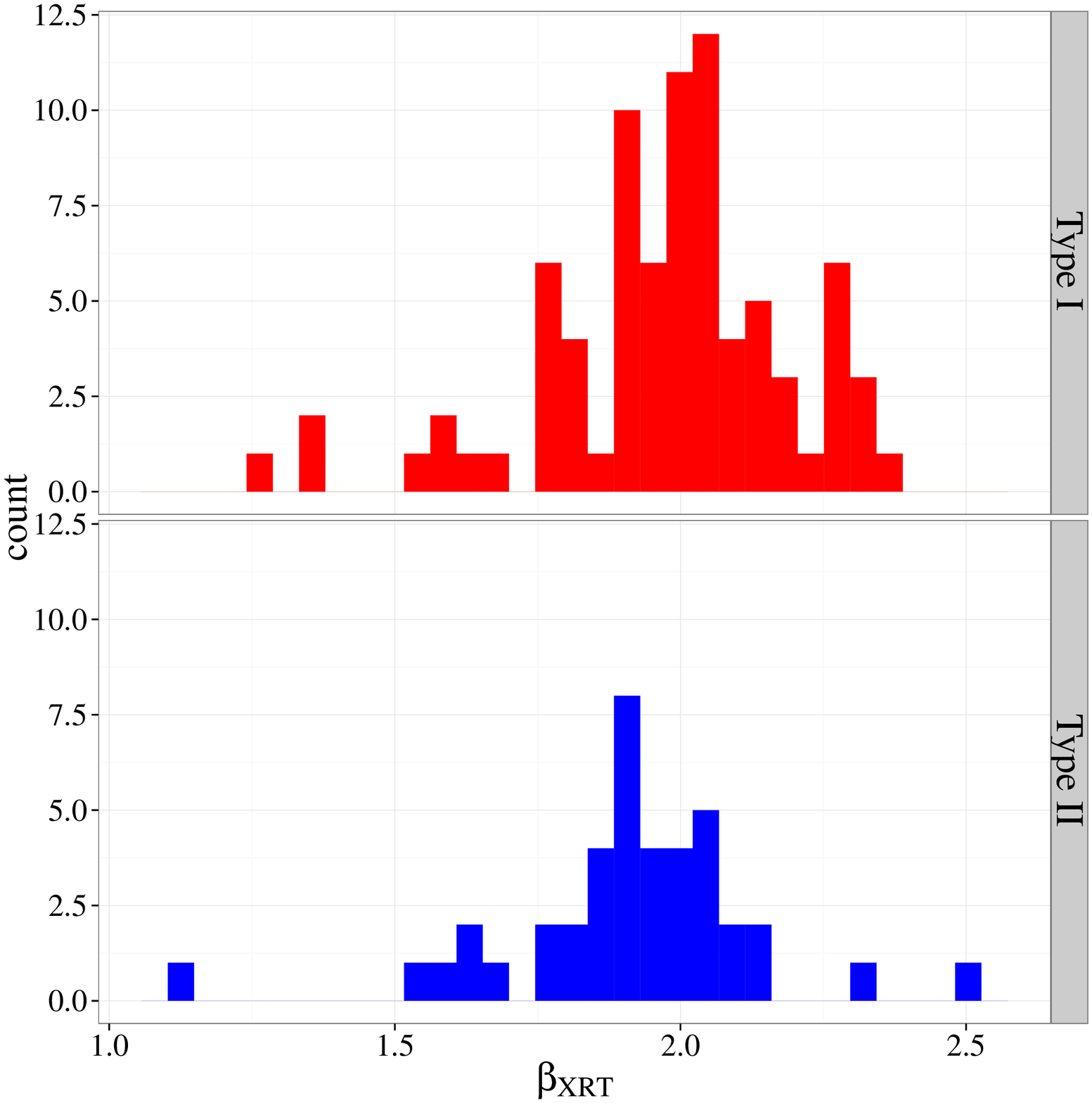}
   \end{center}
   \caption{The histograms of $\beta_{\rm XRT}$ for Type I (top) and Type II (bottom) LGRBs, respectively. }\label{fig:beta}
\end{figure}
\begin{figure}[h]
   \begin{center}
      \FigureFile(80mm,80mm){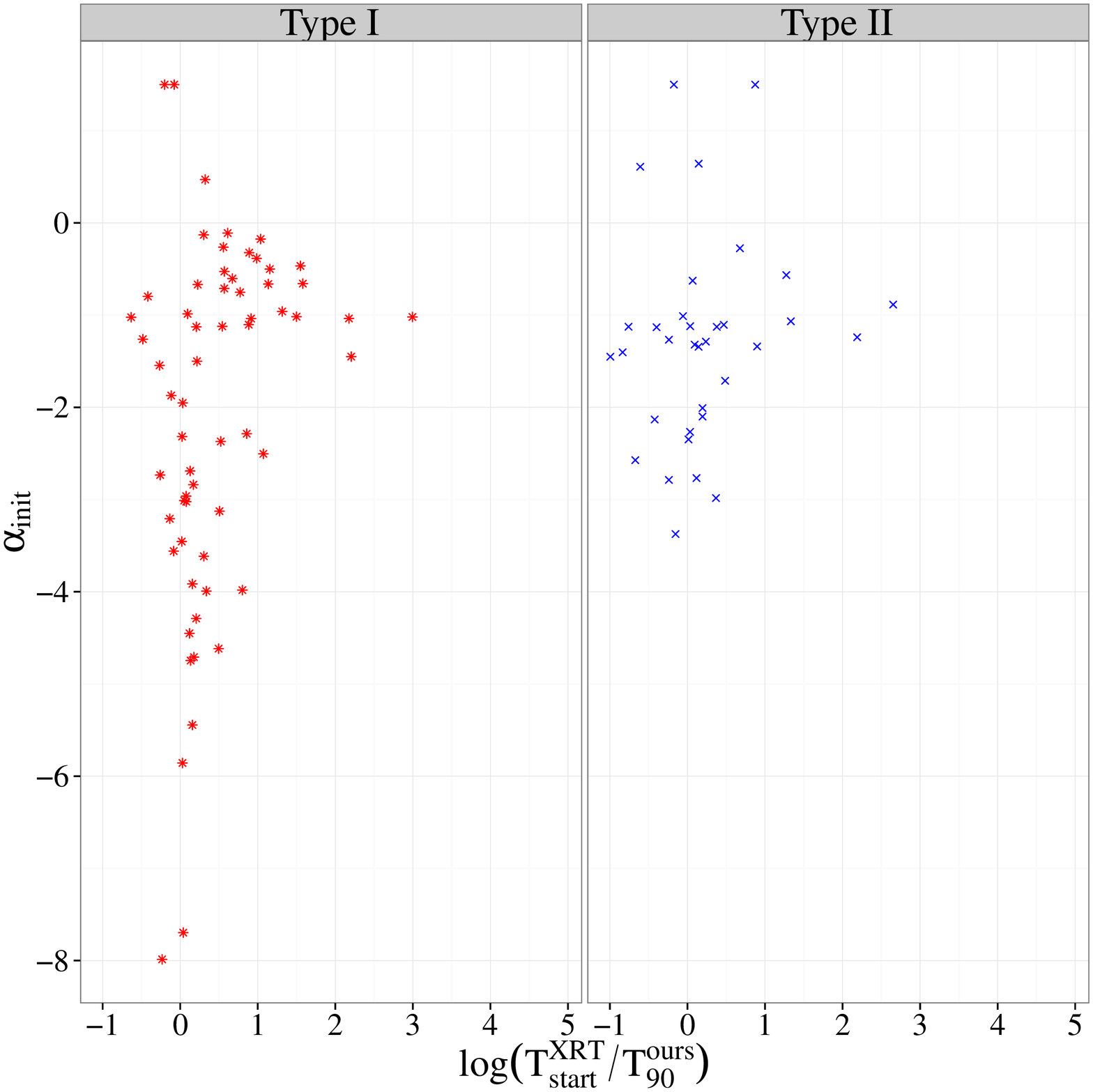}
   \end{center}
   \caption{The relationship between $\log(T_{\rm start}^{\rm XRT}/T_{90})$ and $\alpha_{\rm init}$}\label{fig:fig6}
\end{figure}

\section{Implications for central engines}
\label{sec:engines}

There are two prominent theoretical candidates for the central engine of LGRBs: a black hole with an accretion disk \citep{Woosley:1993}, hereafter the collapsar model, 
and a strongly magnetized proto neutron star \citep{Usov:1992,Duncan:1992}, the magnetar model.
Though these two models have been vigorously studied by many authors \citep{MacFadyen:1999,Proga:2003,Bucciantini:2008,Bucciantini:2009,Nagataki:2011,Metzger:2011}, 
 which engine is really responsible for LGRBs is still controversy.
In this section, we briefly discuss how we can connect our classification with these two different central engine models.

The features of Type I and Type II LGRBs are summarized as below: 
\begin{itemize}
\item The prompt emission of Type I LGRBs consists of several separated pulses with similar heights  and sharply decreases with decay index between $-3$ and $-6$.
\item The prompt emission of Type II LGRBs has a broad peak and slowly decreases with index between $-2$ and $-3$.
\end{itemize}
These light curves imply that  the central engine  of Type I LGRBs injects energies by several violent episodes with similar scales. 
An accreting black hole potentially makes such episodes because the accretion disk is formed  only when the specific angular momentum of the stat exceeds the critical value. 
\citet{2010ApJ...716.1308L} showed that the radial distribution of the specific angular momentum in the star can be reflected in the variability and quiescent time in LGRBs.

On the other hand, the central engine of a Type II LGRBs continuously inject energies by a single major event decaying inversely proportional to the square of time. 
From these light curve properties of Type II LGRBs, we suggest that magnetic dipole radiation from a proto-magnetar may be responsible for the decay part of their prompt emission.

\section{Summary and Discussion}\label{sec:summary}
Using the complete sample of the bright {\it Swift} LGRBs, we have confirmed the subclasses in LGRBs first discovered by \citet{Tsutsui:2013a}.  
We use parameters characterizing light curve shapes to avoid influence of nuisance parameters such as the distance and jet opening angle. We do not use the duration, hardness ratio
or fluence depending on nuisance parameters.

We should notify the readers that our classification method is not exactly the same with our previous works \citep{Tsutsui:2013a,Tsutsui:2013b}.
Although there are some fraction of LGRBs referred to as 'outlier' from the fundamental planes, which probably belong to unclassified types of LGRBs, we can not consider them in this paper because of the lack of  the redshifts and prompt spectral parameters.
To study statistical properties of Type I and Type II LGRBs more accurately, we need to somehow detect contamination of our sample by outliers and remove them from the analysis.
In \citet{Tsutsui:2013b}, we found that Type I and Type II LGRBs show a canonical X-ray afterglow, while outliers do not.
If we use the light-curve shape of X-ray afterglow instead of the fundamental planes, it seems to be possible to detect outliers contaminating Type I and II GRBs classified in this paper.
and it enables us to identify more accurate properties of X-ray afterglow of these subclasses.
The analysis of X-ray afterglows, however, is beyond the scope of this paper,  therefore we have roughly discussed only the difference in the X-ray afterglows.
A more detailed statistical study of the parameters of canonical light curves for Type I and Type II LGRBs is a subject in our future work (Tsutsui and Shigeyama in prep).

Before we summarize this paper, we need to stress the importance of the study of systematic errors in observations.
We use light curves in the fixed {\it Swift} $150$-$350$ keV energy band, which means that we use different energy bands in the rest frames of different GRBs. 
From this viewpoint, the bolometric flux light curve seems to be more appropriate to discuss the subclasses of LGRBs.  Though broad band observations are inevitable to determine the spectral shape of the prompt emission,  the band width of BAT detector onboard  {\it Swift} is too narrow to do so.
The future Chinese-French mission {\it SVOM} will provide a good opportunity to construct bolometric light curves by broader band observations with ECLAIR and GRM and to confirm or falsify our present classifications.

With the result of our previous paper \citep{Tsutsui:2013a}, LGRBs are divided into three subgroups as follows:
\begin{itemize}
\item Type I LGRBs exhibiting linearly increasing cumulative light curves of the prompt emission reside on a fundamental plane and show canonical light curves in their X-ray afterglow.  The central engine of this type of LGRBs seem to  originate from an accretion disk around a black hole.
\item Type II LGRBs with long tailed prompt emission reside on the other fundamental plane and show canonical light curve in their X-ray afterglow with a  steep decay phase slower than Type I 
and seem to originate from a magnetar.
\item outliers which do not reside on either of the fundamental planes and do not show a canonical light curve. 
\end{itemize}

Type I and Type II LGRBs on the separate fundamental planes can be used as distance indicators like other empirical correlations of GRBs 
\citep{Liang:2005,Ghirlanda:2006,Schaefer:2007,Kodama:2008,Liang:2008,Amati:2008,Cardone:2009,Tsutsui:2009b}.    
They are so tight that they provide much more accurate distance measurements than previous studies \citep{Tsutsui:2012b}.

\section*{Acknowledgments}
 This work made use of data supplied by the UK Swift Science Data Centre at the University of Leicester.
 This work is supported in part by the Grant-in-Aid for Young Scientists (B) 
from the Japan Society for Promotion of Science (JSPS), No.24740116(RT).

\appendix
\section{Statistical method}
We briefly explain the statistical method used in the section~\ref{sec:prompt}.
We explain the Gaussian mixture model and the EM algorithm to select optimal parameters with a fixed number of clusters first and 
then explain the Bayesian Information Criterion (BIC) to select the optimal number of clusters.
For  details of the statistical method, we refer the reader to an excellent text book \cite{Bishop:2006}
(Chapter 9 for Gaussian mixture model and EM algorithm and Chapter 4 for BIC).  
  
\subsection{Gaussian mixture model and EM algorithm}
Let  us start with an assumption that a data vector comes from a Gaussian mixture model with the density 
\begin{equation}
p({\vector x}) =\sum_{k=1}^{G}\pi_k {\mathcal N}({\vector x}|{\vector \mu_k}, {\vector \Sigma_k}),\\
\end{equation}
where ${\vector x}$ is a $D$-dimensional data vector,  $G$ is the number of clusters, $\pi_k$ is a mixing probability of a data  element $\vector x$ 
belonging to the $k$th component, and ${\mathcal N}({\vector x}|{\vector \mu_k}, {\vector \Sigma_k})$ is a $D$-dimensional Gaussian density of the $k$ component given by 
\begin{equation}
{\mathcal N}({\vector x}|{\vector \mu_k}, {\vector\Sigma_k}) = \frac{\exp \left\{-\frac{1}{2} ({\vector x}-{\vector \mu_k})^{T} \Sigma_k^{-1} ({\vector x}-{\vector \mu_k})\right\}}{(2\pi)^{D/2}|\Sigma_k|^{1/2}} 
\end{equation}
where ${\vector \mu_k}$ and ${\Sigma_k}$ are the mean vector and the variance matrix for the $k$th component, respectively.
The mixing probability must satisfy the constraint that $\sum_{k=1}^{G} \pi_k =1$, and $\pi_k\ge0$.  

Given a sequence of  independent data $X=\{ {\vector x_1},{\vector x_2}, ...,{\vector x_N}\}$, we want to maximize the following log-likelihood function 
\begin{equation}
\label{eq:likelihood}
\ln p(\vector X|\vector \pi, \vector \mu,\vector \Sigma) = \sum_{i=1}^{N} \ln \left\{ \sum_{k=1}^G \pi_k{\mathcal N}({\vector x}_i |{\vector \mu_k}, {\vector \Sigma_k}) \right\}.
\end{equation}
Taking a derivative of equation (\ref{eq:likelihood}) with respect to $\vector \mu_{k}$ and equating it to 0, we obtain the following equation 
\begin{equation}
\label{eq:mean}
\vector \mu_k=\frac{1}{N_k}\sum_{i=1}^{N}\gamma_{ik}x_i, 
\end{equation} 
where $\gamma_{ik}$ is called the responsibility and given by the following equation, 
\begin{equation}
\label{eq:res}
\vector \gamma_{ik}=\frac{\pi_k {\mathcal N}(\vector x_i|\vector \mu_k, \vector \Sigma_k)}{\sum_{j=1}^G \pi_j {\mathcal N(\vector x_i|\vector \mu_j, \vector \Sigma_j)}},
\end{equation}
and 
\begin{equation}
N_k=\sum_{i=1}^N\gamma_{ik}.
\end{equation}
Similarly, we can obtain the form of the variance matrix by taking the derivative of equation (\ref{eq:likelihood}) with respect to $\vector \Sigma_k$ and equating it to 0,
\begin{equation}
\label{eq:var}
\vector \Sigma_k=\frac{1}{N_k}\sum_{i=1}^{N}\gamma_{ik}(\vector x_i-\vector \mu_k) (\vector x_i-\vector \mu_k)^T.
\end{equation}
To obtain the form of the mixing probability $\pi_k$, we must take into account the constraint that $\sum \pi_k=1$. Using the Lagrange multiplier method, we obtain the following equation
\begin{equation}
\label{eq:pi}
\pi_k=\frac{N_k}{N}.
\end{equation}

Because $\gamma_{ik}$ itself is a function of the model parameters, equations (\ref{eq:mean}), (\ref{eq:var}), (\ref{eq:pi}) do not give solutions. 
To obtain optimal solutions for these parameters, we need to follow an iterative procedure called the EM algorithm as below:
\begin{enumerate}
\item Setting initial guesses for  $\vector \mu_k$, $\vector \Sigma_k$, and $\pi_k$ and calculate the initial value of the log-likelihood function.
\item Using the current value of parameters, calculate the responsibility $\gamma_{ik}$ from equation (\ref{eq:res}). (E step)
\item Using the current value of the responsibility $\gamma_{ik}$, calculate parameters from equations (\ref{eq:mean}), (\ref{eq:var}), (\ref{eq:pi}). (M step) 
\item Calculate the log-likelihood function, and check the convergence of the parameters or log-likelihood. If convergence criteria are not satisfied, return to step 2.  
\end{enumerate}

Given the optimal Gaussian mixture model density and a data vector $\vector x$, the class K and the uncertainty of classification $p_{\rm mis}$ of the data are determined by, 
\begin{equation}
K = {\rm which.max}_k[ \pi_k {\mathcal N}({\vector x}|{\vector \mu_k}, {\vector \Sigma_k})],
\end{equation}
 and 
\begin{equation}
p_{\rm mis} = 1 -  \pi_K {\mathcal N}({\vector x}|{\vector \mu_K}, {\vector \Sigma_K})],
\end{equation}
respectively. Here which.max$_k[X_k]$ is a function that returns the value of the index $k$ yielding the maximum value of $X_k$.

\subsection{Bayesian Information Criterion BIC}
To select the optimal number of clusters, {\tt mclust} package uses the Bayesian Informative Criterion ($BIC$)
\begin{equation}
BIC = 2 \ln p_{\mathcal M}(\vector x|\vector \pi^*, \vector \mu^*,\vector \Sigma^*)- N_{\rm p} \ln(N), 
\end{equation}
where $\ln p_{\mathcal M}(\vector x|\vector \pi^*, \vector \mu^*,\vector \Sigma^*)$ is the maximized log-likelihood function for the model and data, 
$N_{\rm p}$ is the number of independent parameters for the model ${\mathcal M}$.
The log-likelihood increases with increasing number of the model parameters  and is likely to result in overfitting. 
The BIC selects the optimal number of clusters by giving a penalty for adding parameters.

\section{Four component model}
\label{sec:appendix2}
\begin{figure*}[htb]
\begin{tabular}{cc}
\begin{minipage}{0.50\hsize}
\begin{center}
    \FigureFile(80mm,80mm){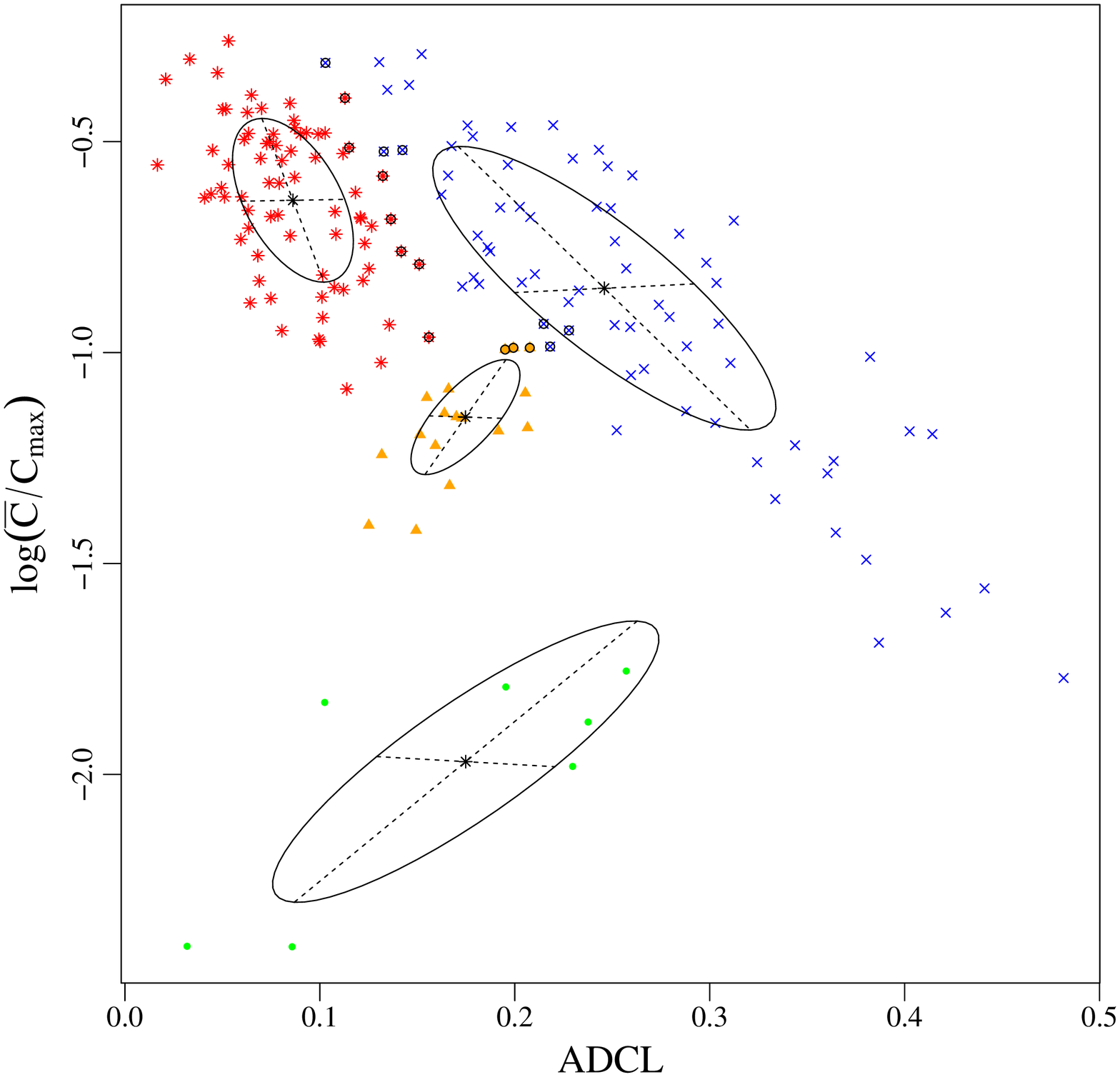}
\end{center}
\end{minipage}
\begin{minipage}{0.50\hsize}
\begin{center}
    \FigureFile(80mm,80mm){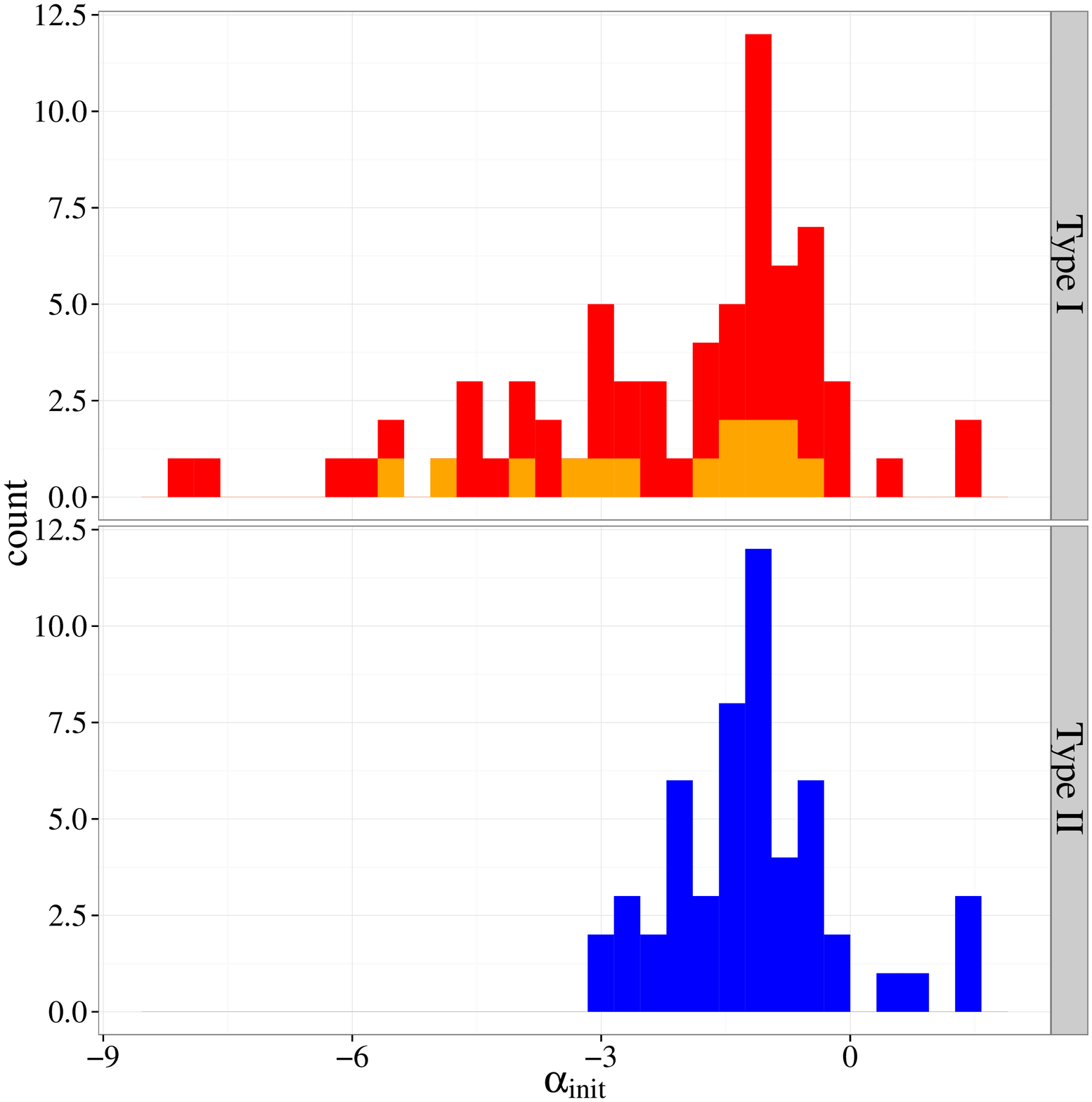}
 \end{center} 
\end{minipage} \\
\begin{minipage}{0.50\hsize}
\begin{center}
    \FigureFile(80mm,80mm){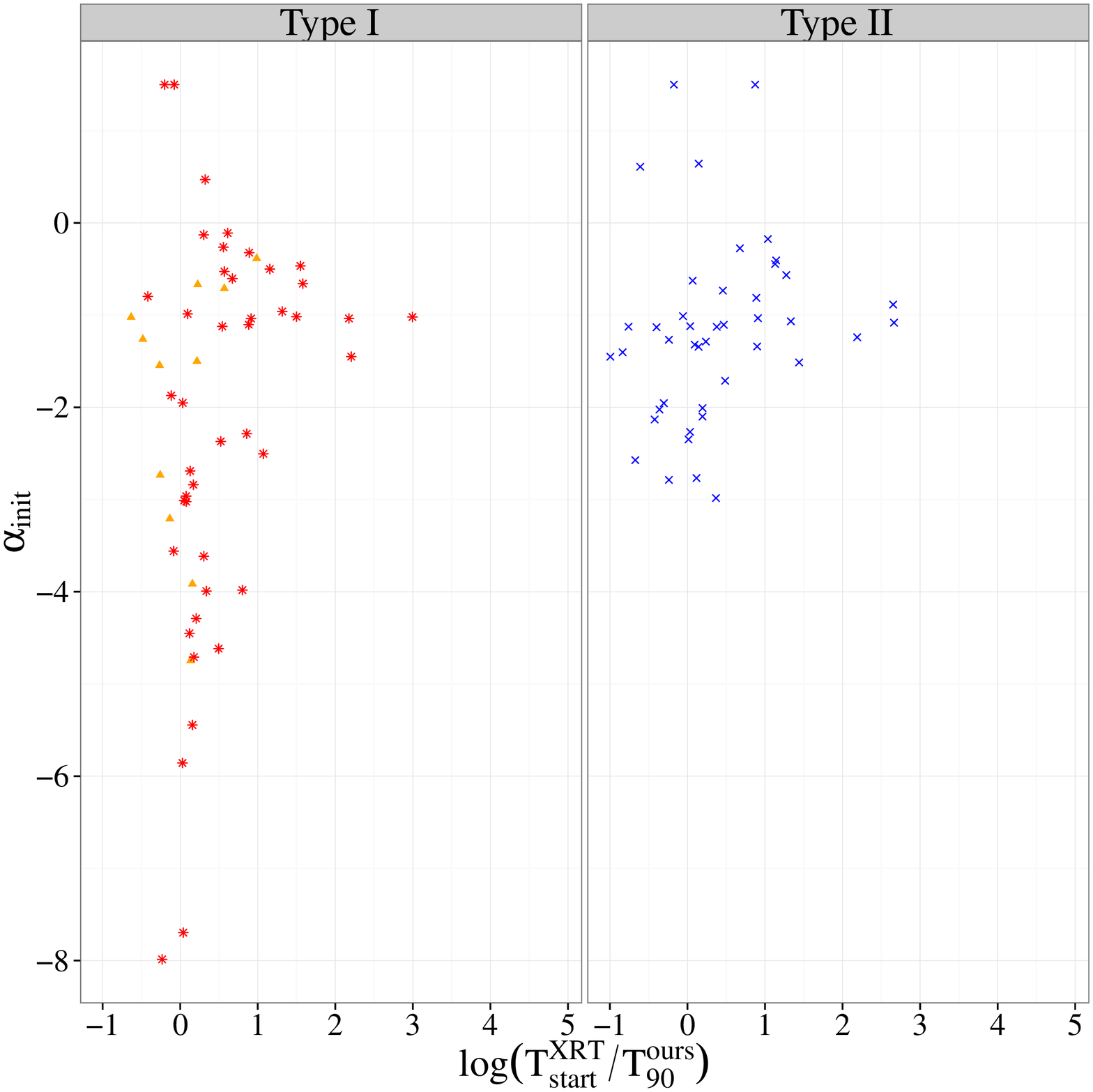}
\end{center}
\end{minipage}
\begin{minipage}{0.50\hsize}
\begin{center}
    \FigureFile(80mm,80mm){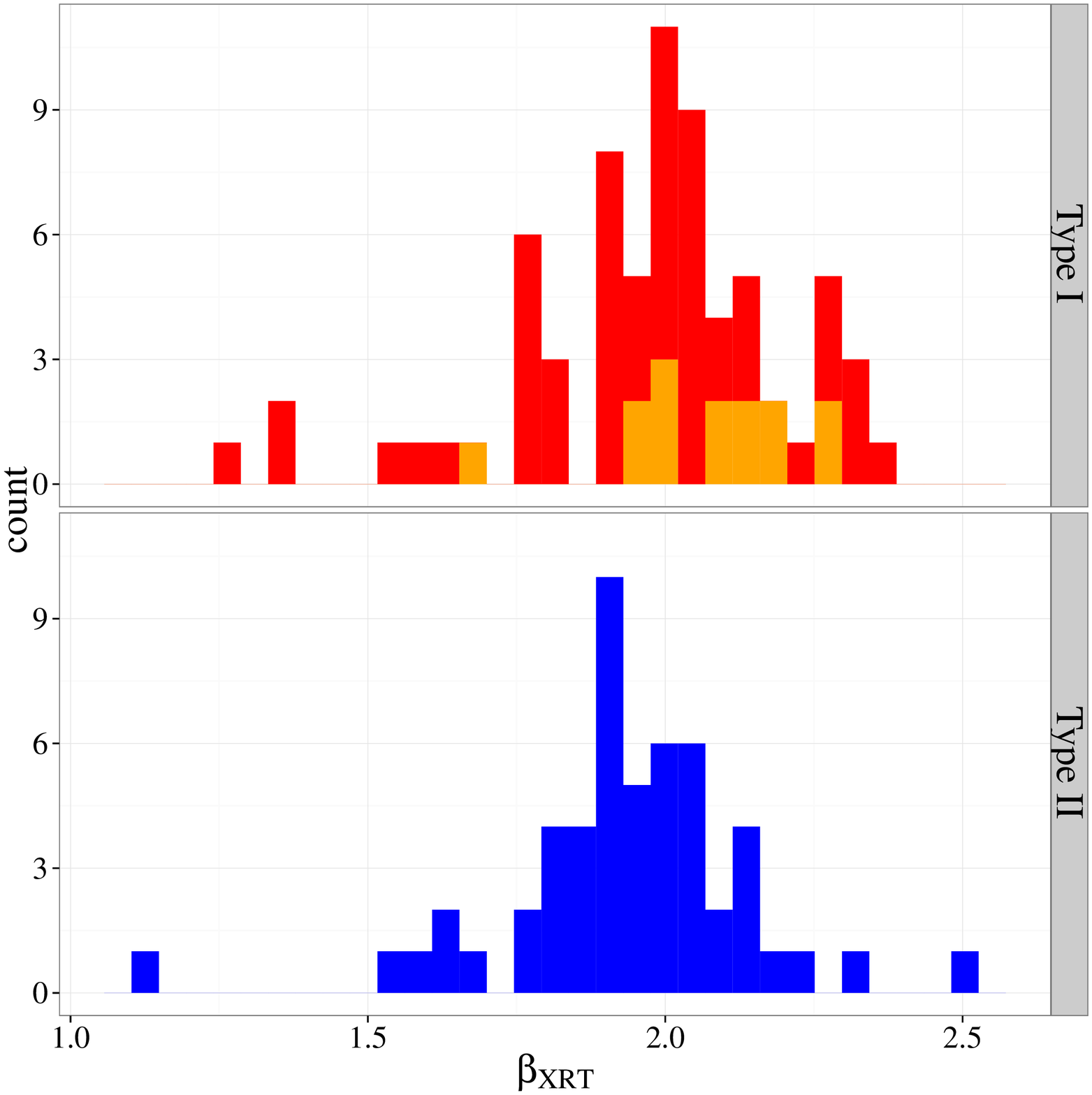}
 \end{center}
\end{minipage} 
\end{tabular}
   \caption{The same figures as Figure \ref{fig:2DCluster} -- \ref{fig:fig6}, but we assumed four-component model in the $ADCL$ -- $\log(\bar C/C_{\rm max})$ diagram.}
   \label{fig:Appendix}
 \end{figure*}

In this section, we show the results when we assume four-component model in the $ADCL$-$\log(\bar C/C_{\rm max})$ diagram.
The upper left panel of Figure~\ref{fig:Appendix} shows the four-component model  divides Type I LGRBs into two subclasses (red asterisks and orange triangles).
However, the other panels indicate that X-ray afterglow properties of these two subclasses are hardly distinguishable.
This is why we adopted three-component model in the main discussion.

%%%%%%%%%%%%%%%%%%%%%
%
% Light curve figures
%

\begin{figure*}[htb]
\begin{tabular}{cccc}
\begin{minipage}{0.25\hsize}
\begin{center}
    \FigureFile(40mm,40mm){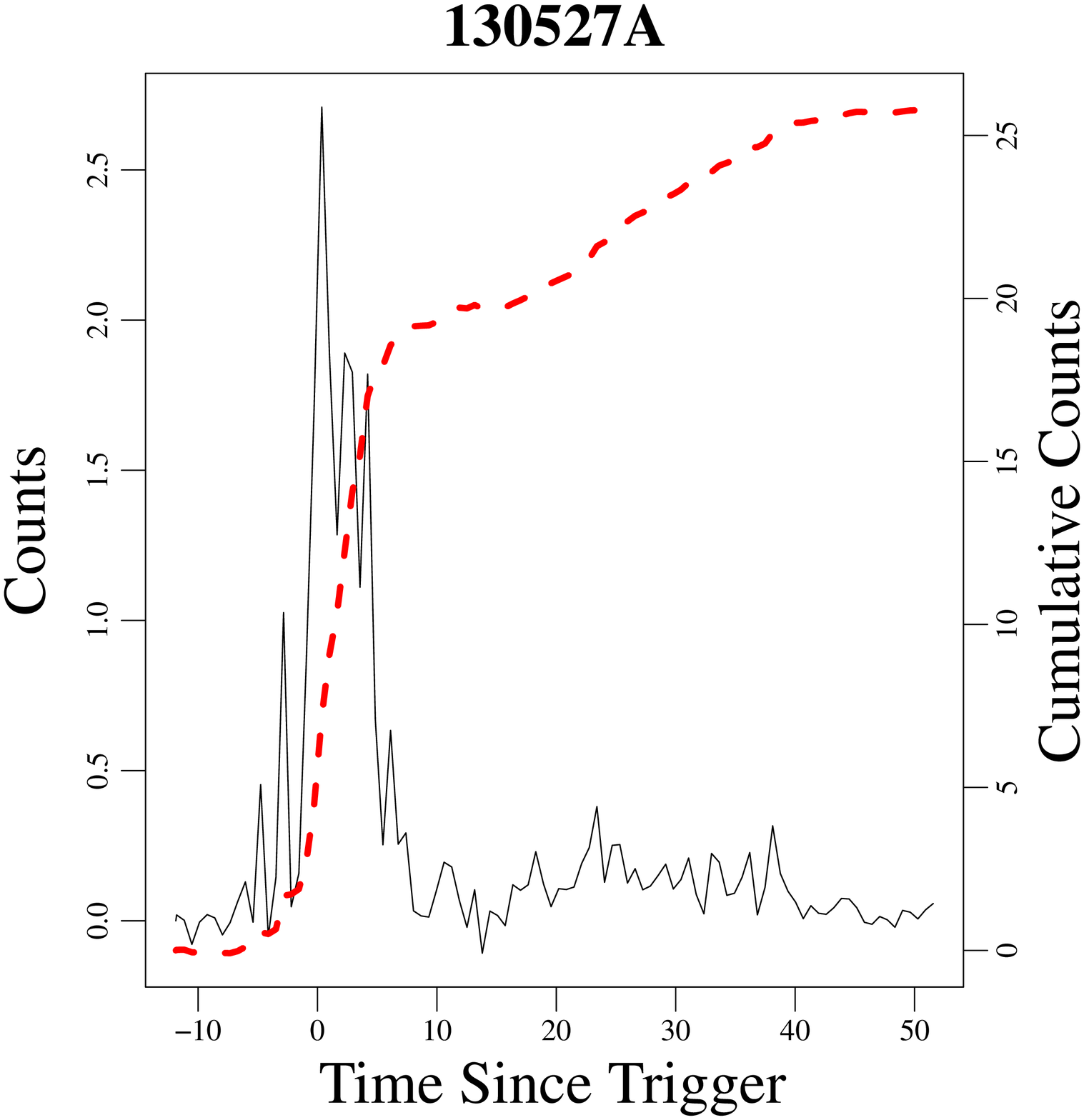}
\end{center}
\end{minipage}
\begin{minipage}{0.25\hsize}
\begin{center}
    \FigureFile(40mm,40mm){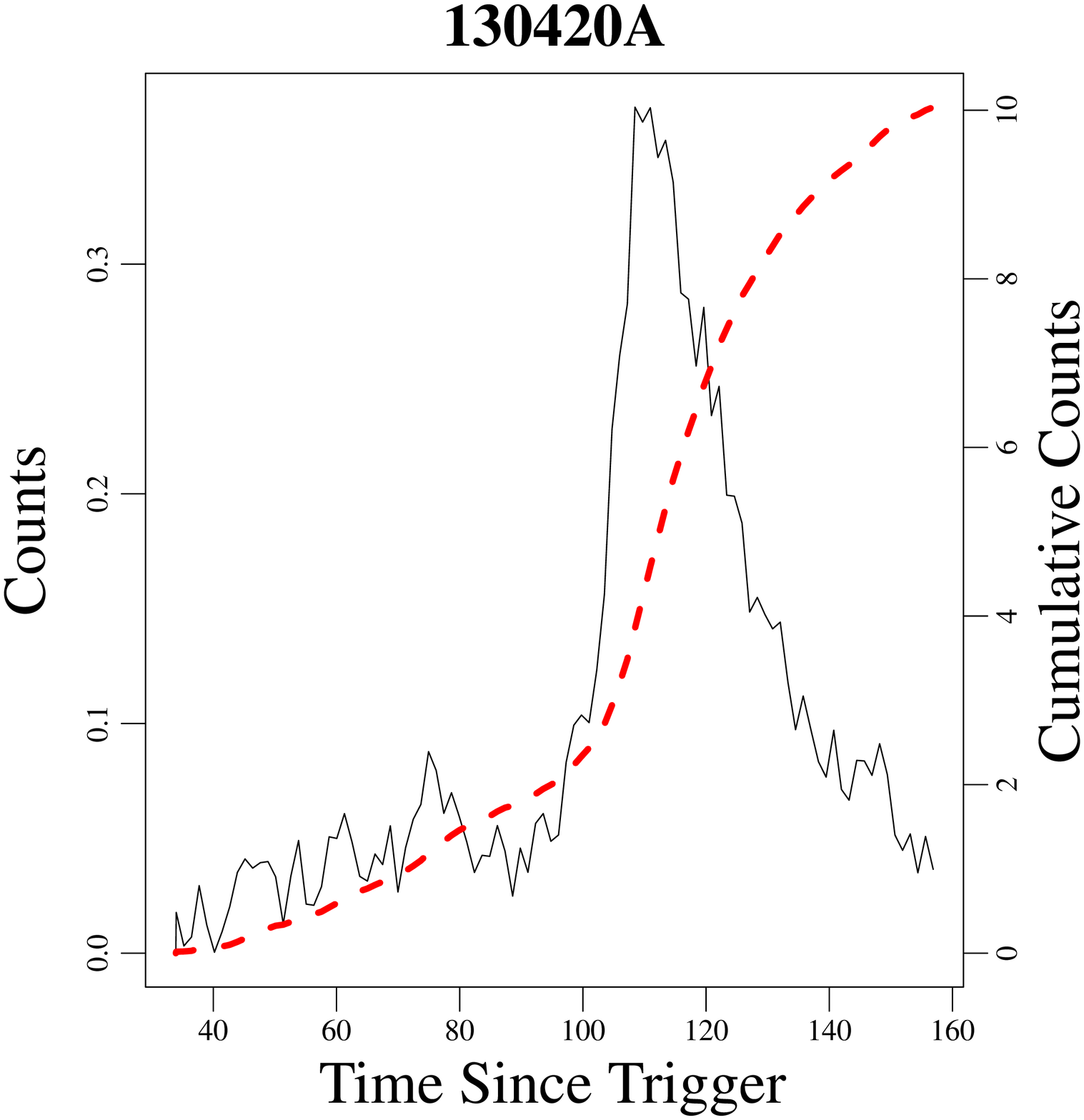}
 \end{center}
\end{minipage}
\begin{minipage}{0.25\hsize}
\begin{center}
    \FigureFile(40mm,40mm){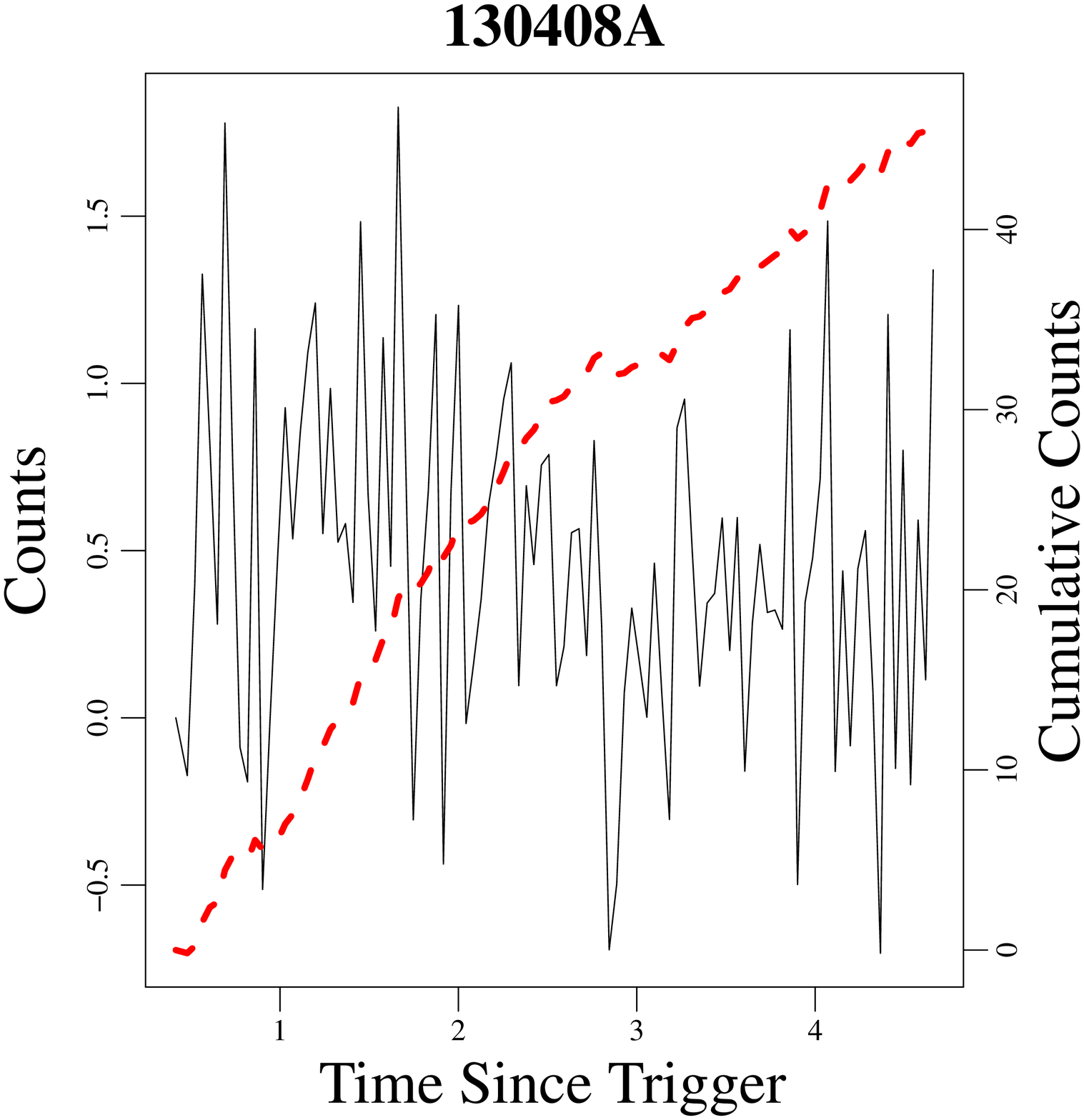}
\end{center}
\end{minipage}
\begin{minipage}{0.25\hsize}
\begin{center}
    \FigureFile(40mm,40mm){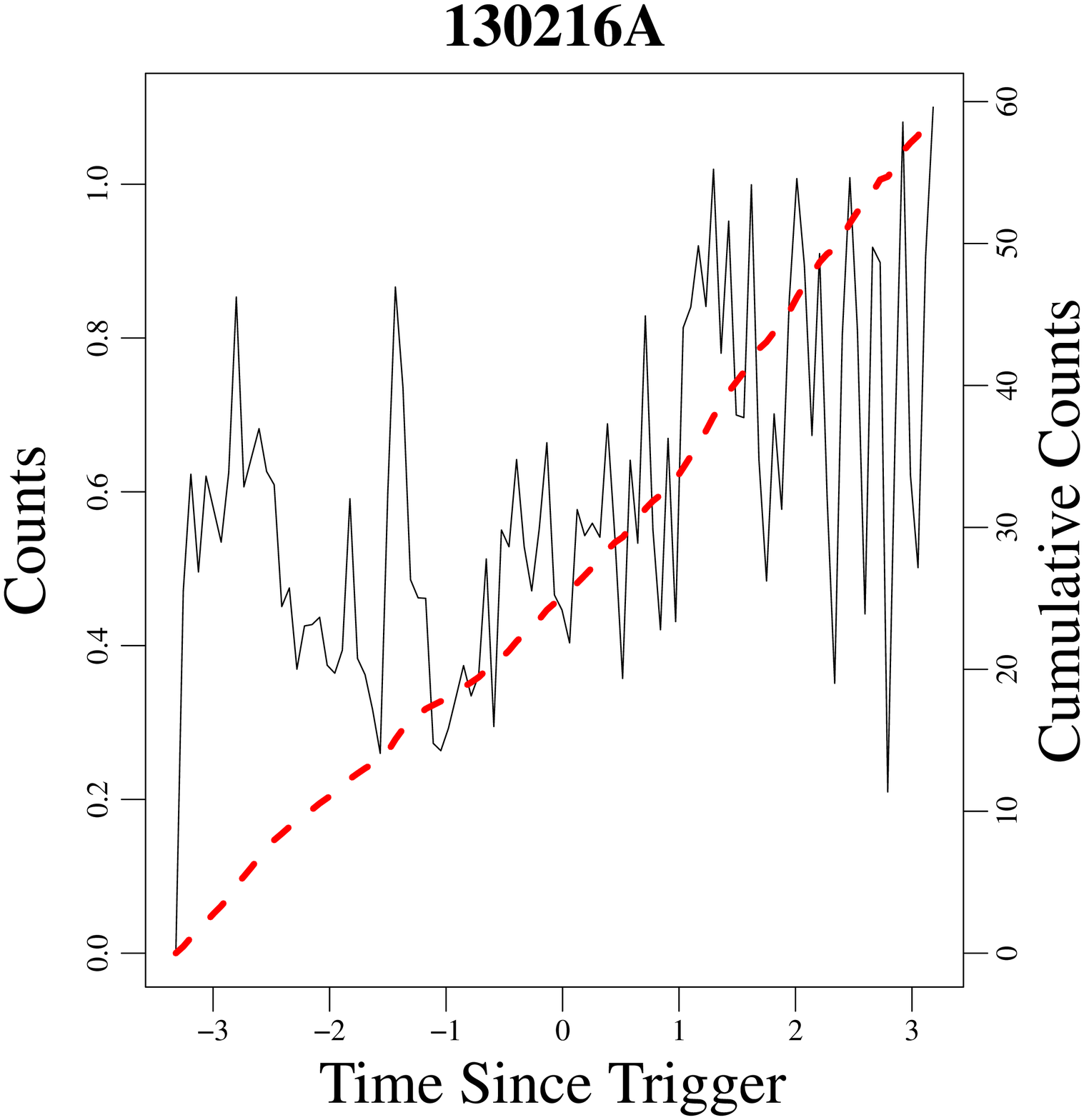}
 \end{center}
\end{minipage}\\
\begin{minipage}{0.25\hsize}
\begin{center}
    \FigureFile(40mm,40mm){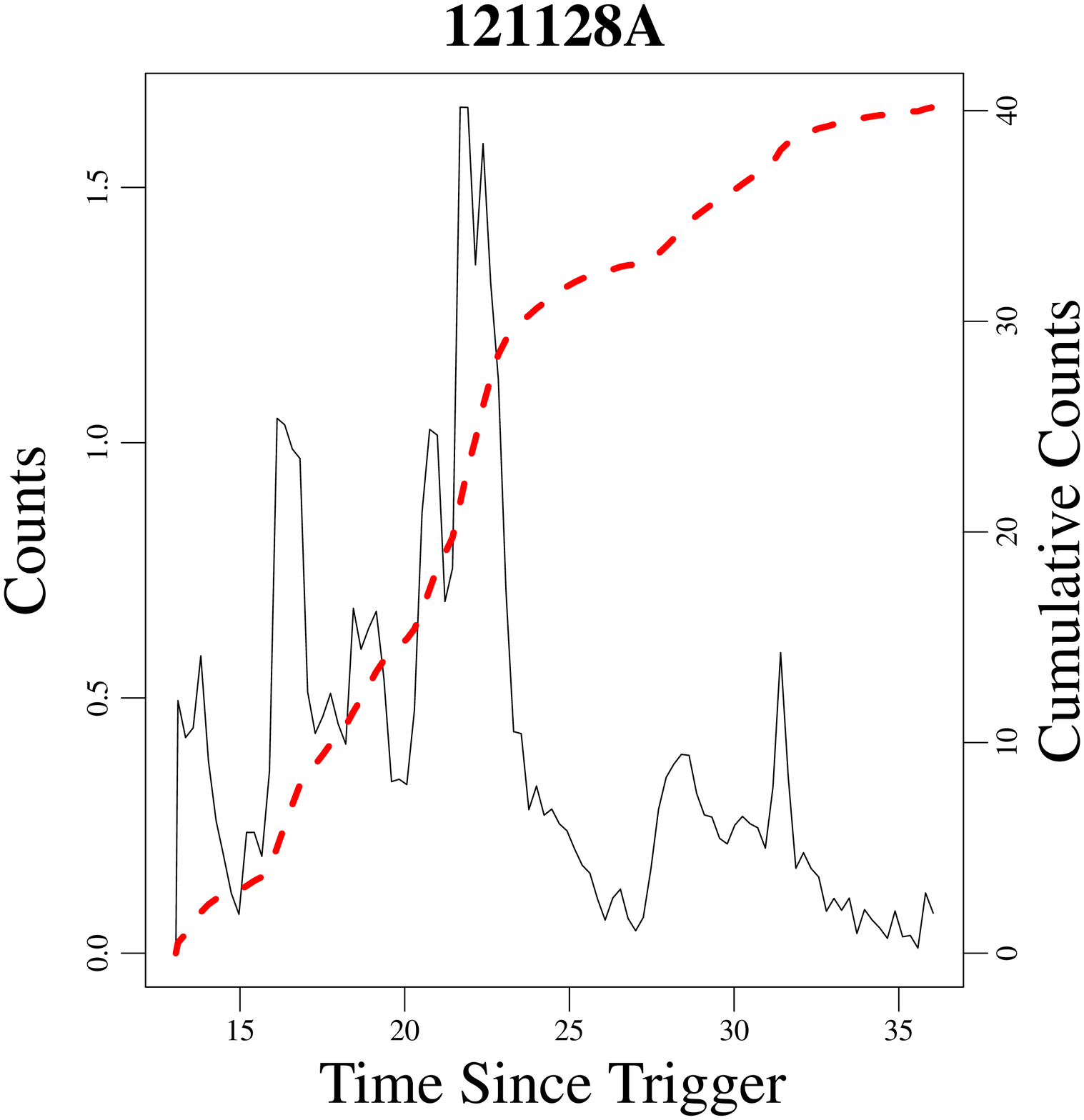}
\end{center}
\end{minipage}
\begin{minipage}{0.25\hsize}
\begin{center}
    \FigureFile(40mm,40mm){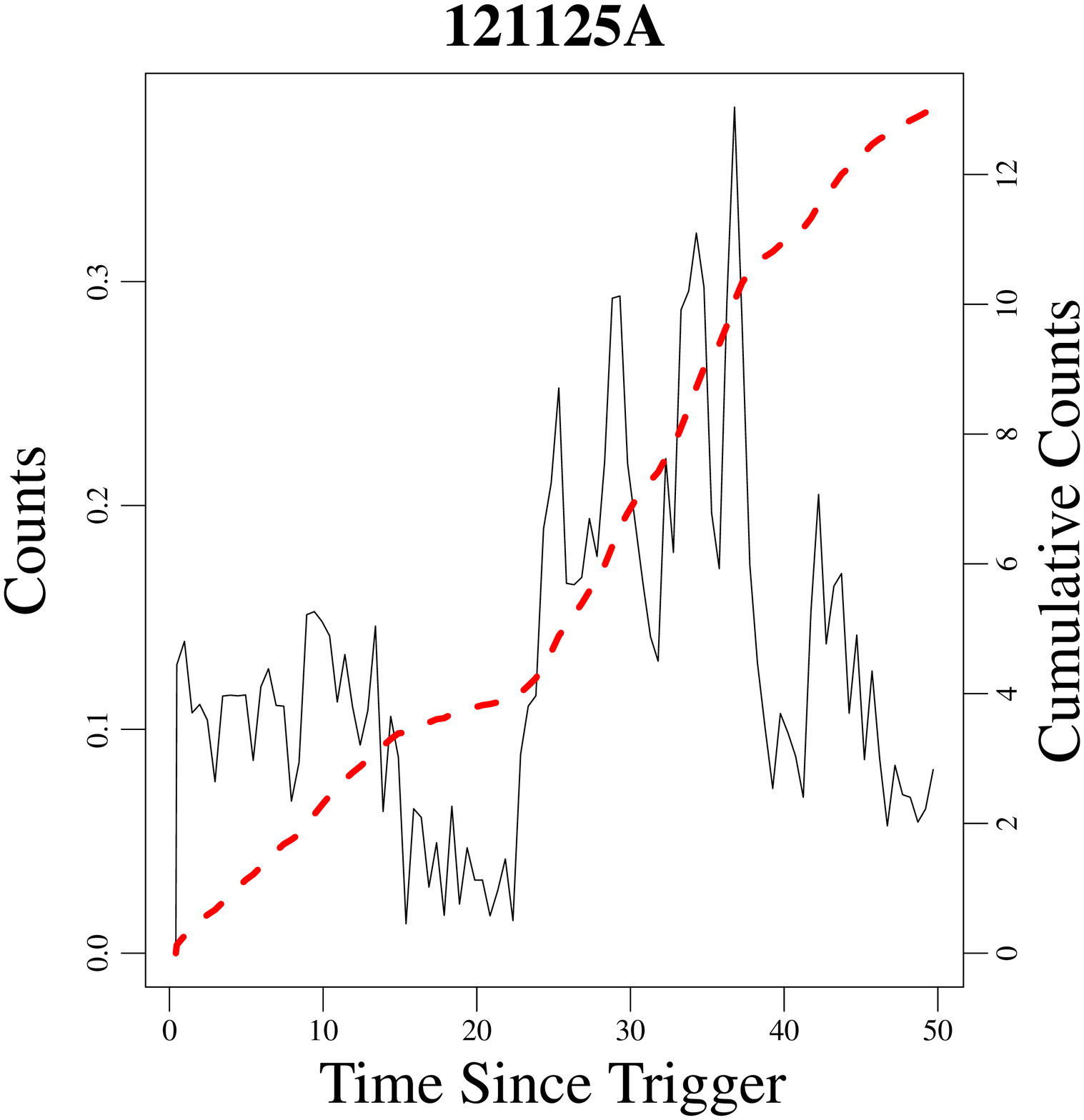}
 \end{center}
\end{minipage}
\begin{minipage}{0.25\hsize}
\begin{center}
    \FigureFile(40mm,40mm){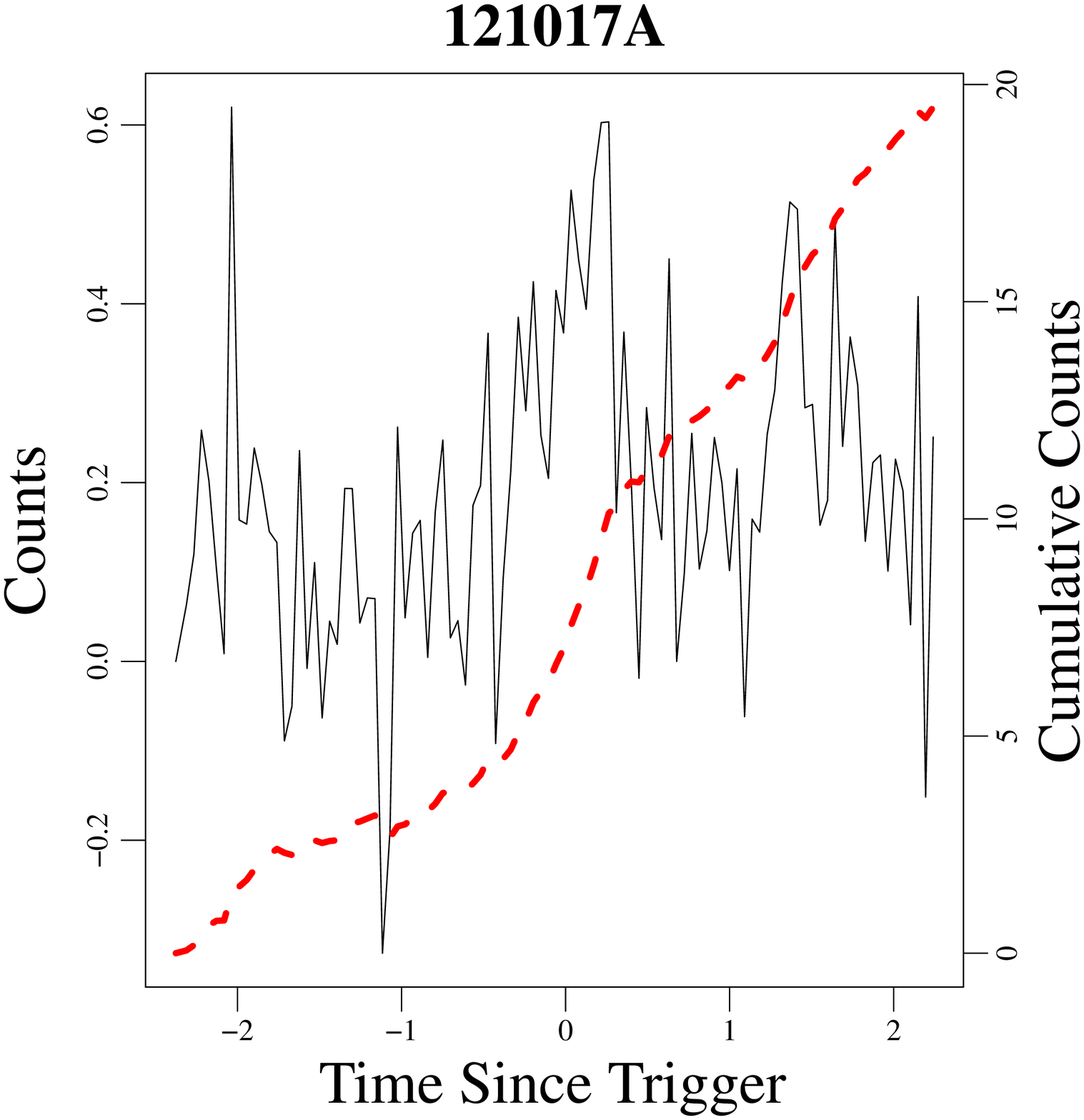}
\end{center}
\end{minipage}
\begin{minipage}{0.25\hsize}
\begin{center}
    \FigureFile(40mm,40mm){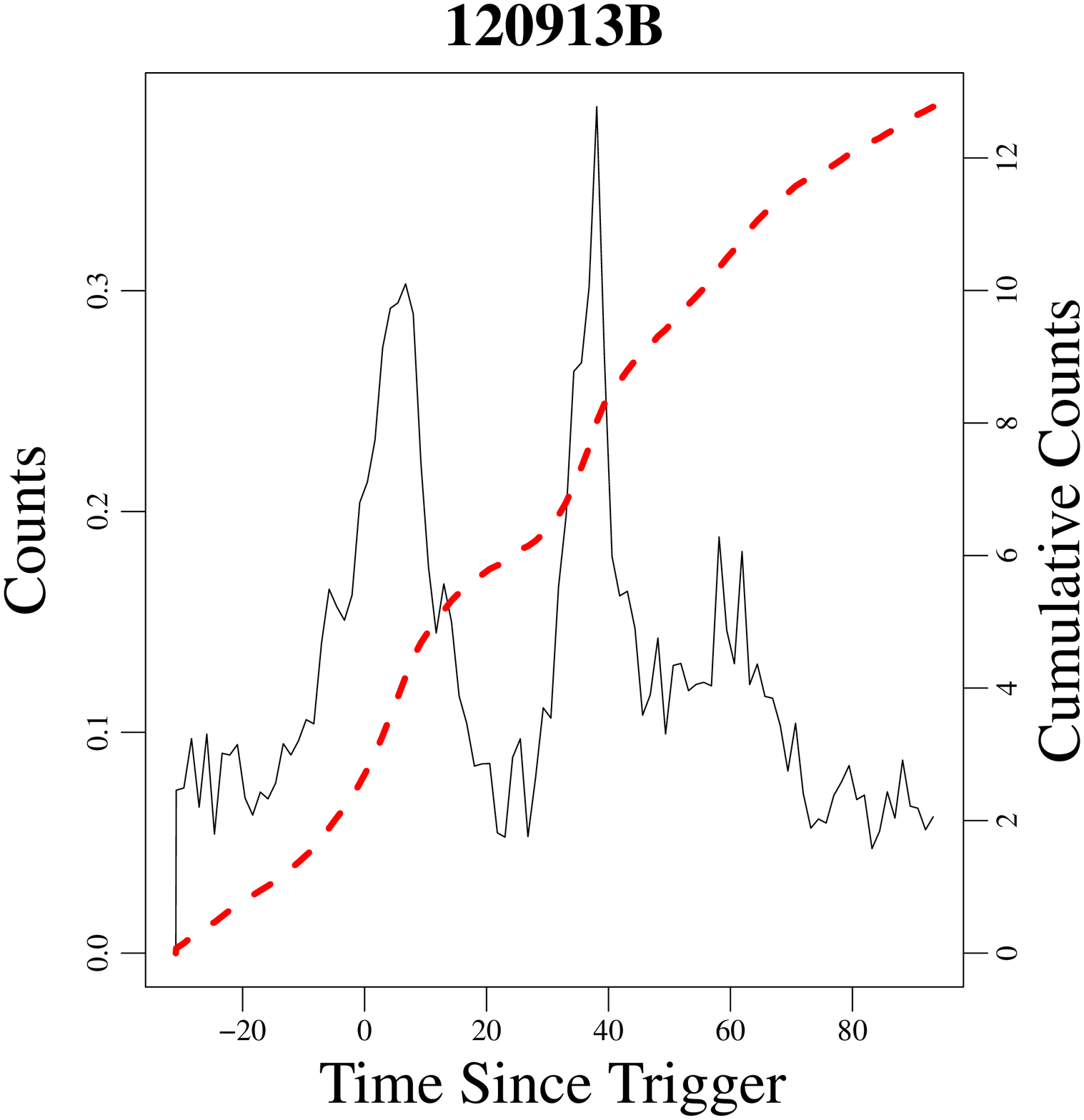}
 \end{center}
\end{minipage}\\
\begin{minipage}{0.25\hsize}
\begin{center}
    \FigureFile(40mm,40mm){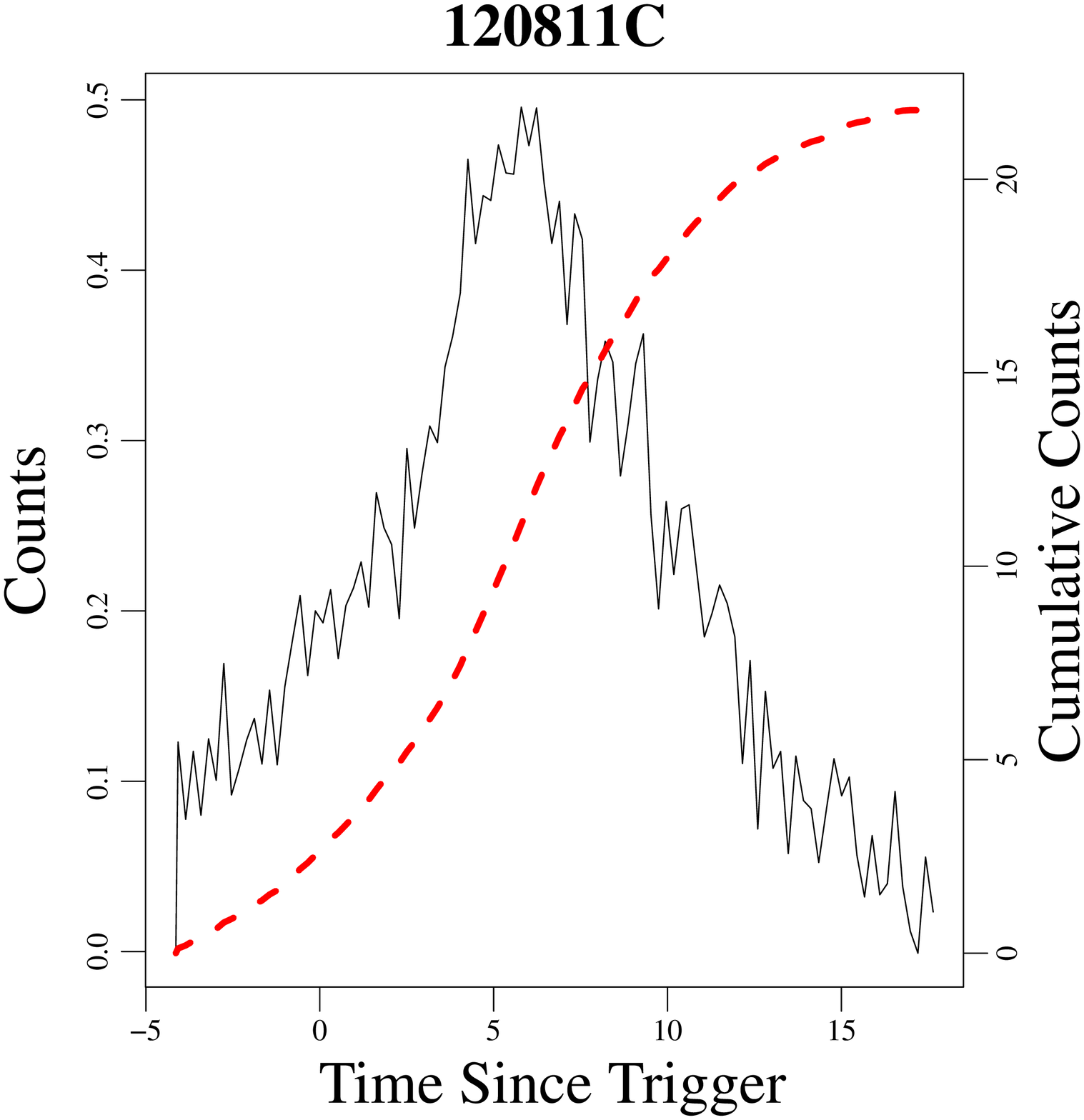}
\end{center}
\end{minipage}
\begin{minipage}{0.25\hsize}
\begin{center}
    \FigureFile(40mm,40mm){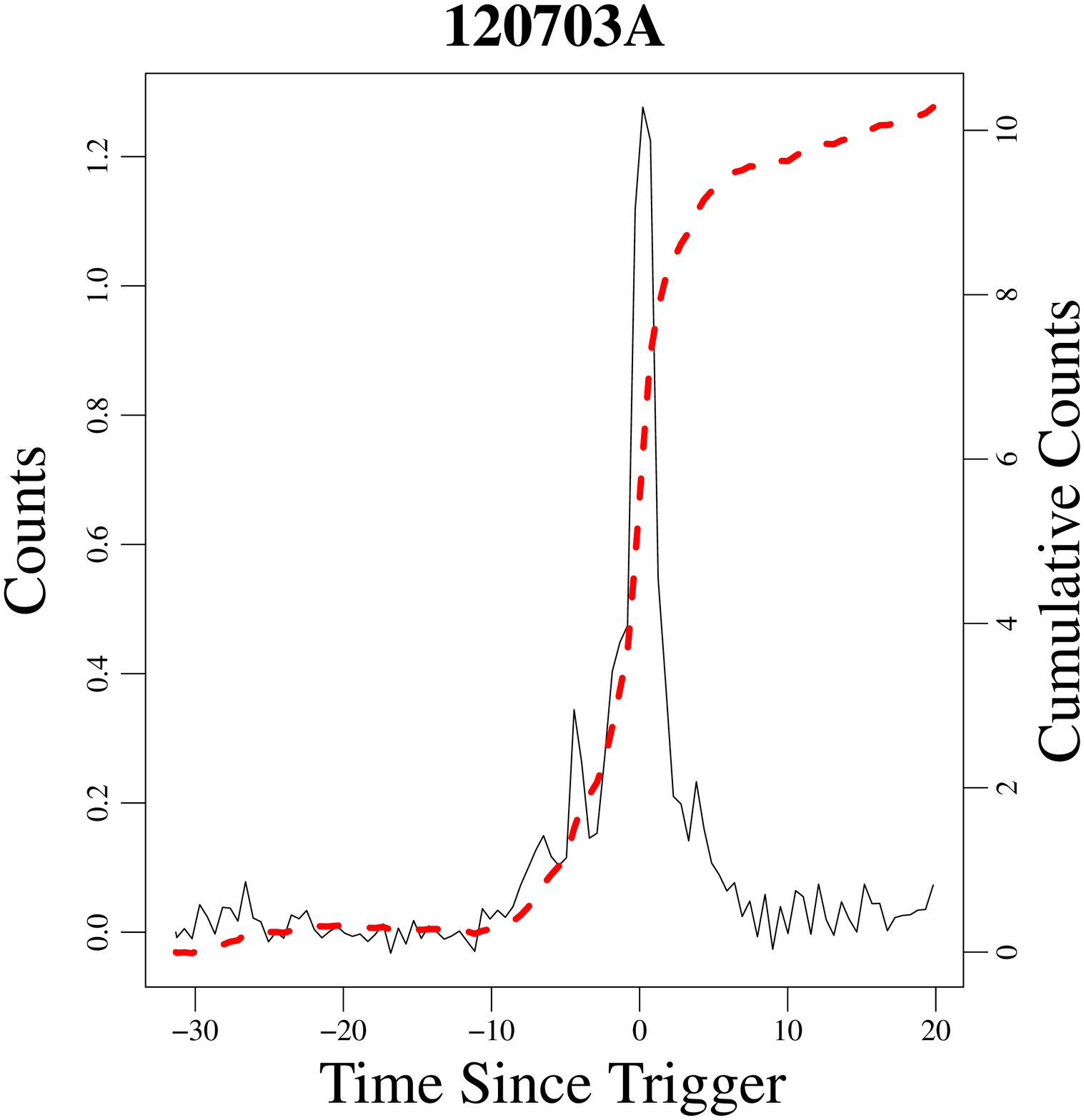}
 \end{center}
\end{minipage}
\begin{minipage}{0.25\hsize}
\begin{center}
    \FigureFile(40mm,40mm){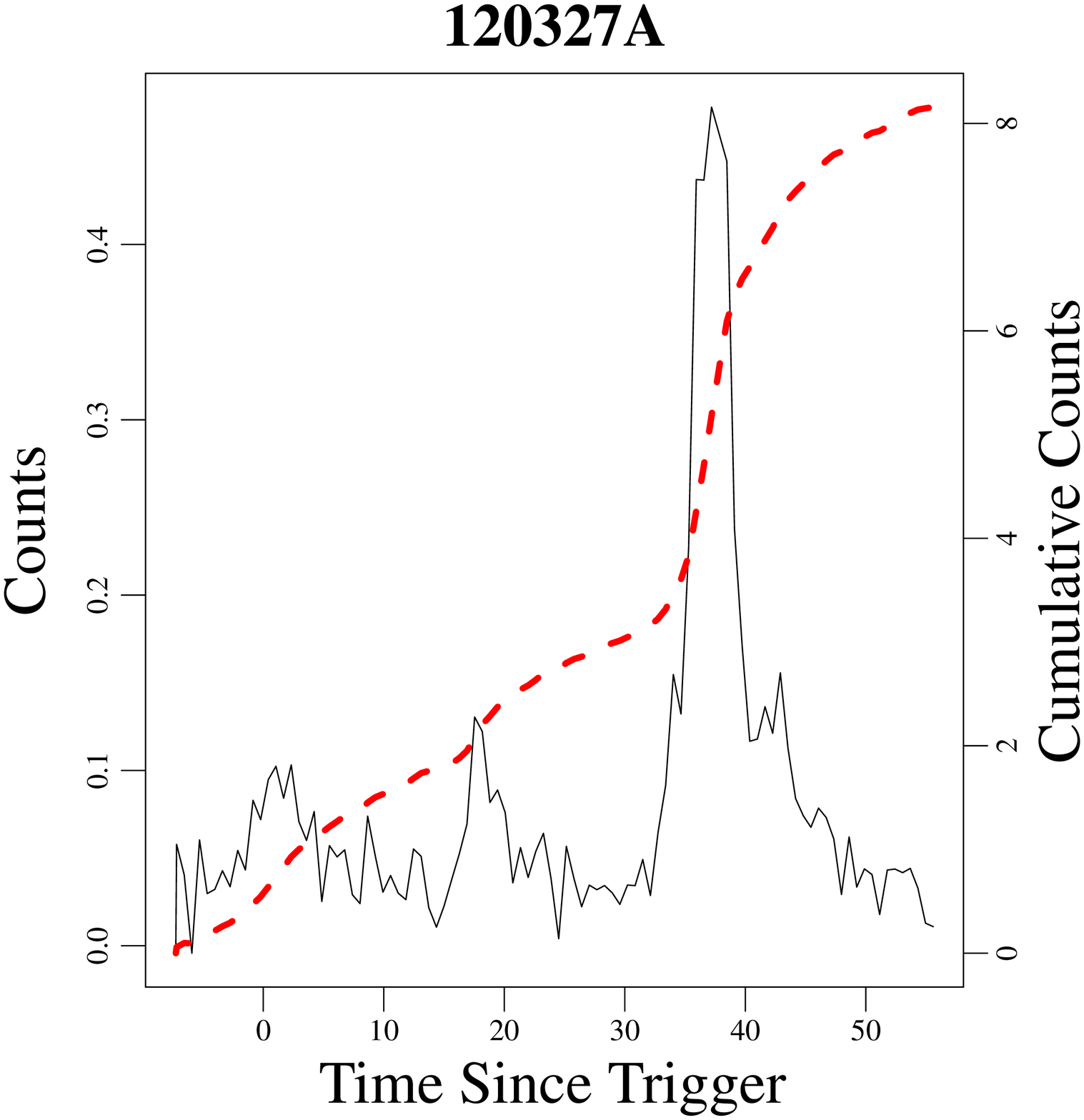}
\end{center}
\end{minipage}
\begin{minipage}{0.25\hsize}
\begin{center}
    \FigureFile(40mm,40mm){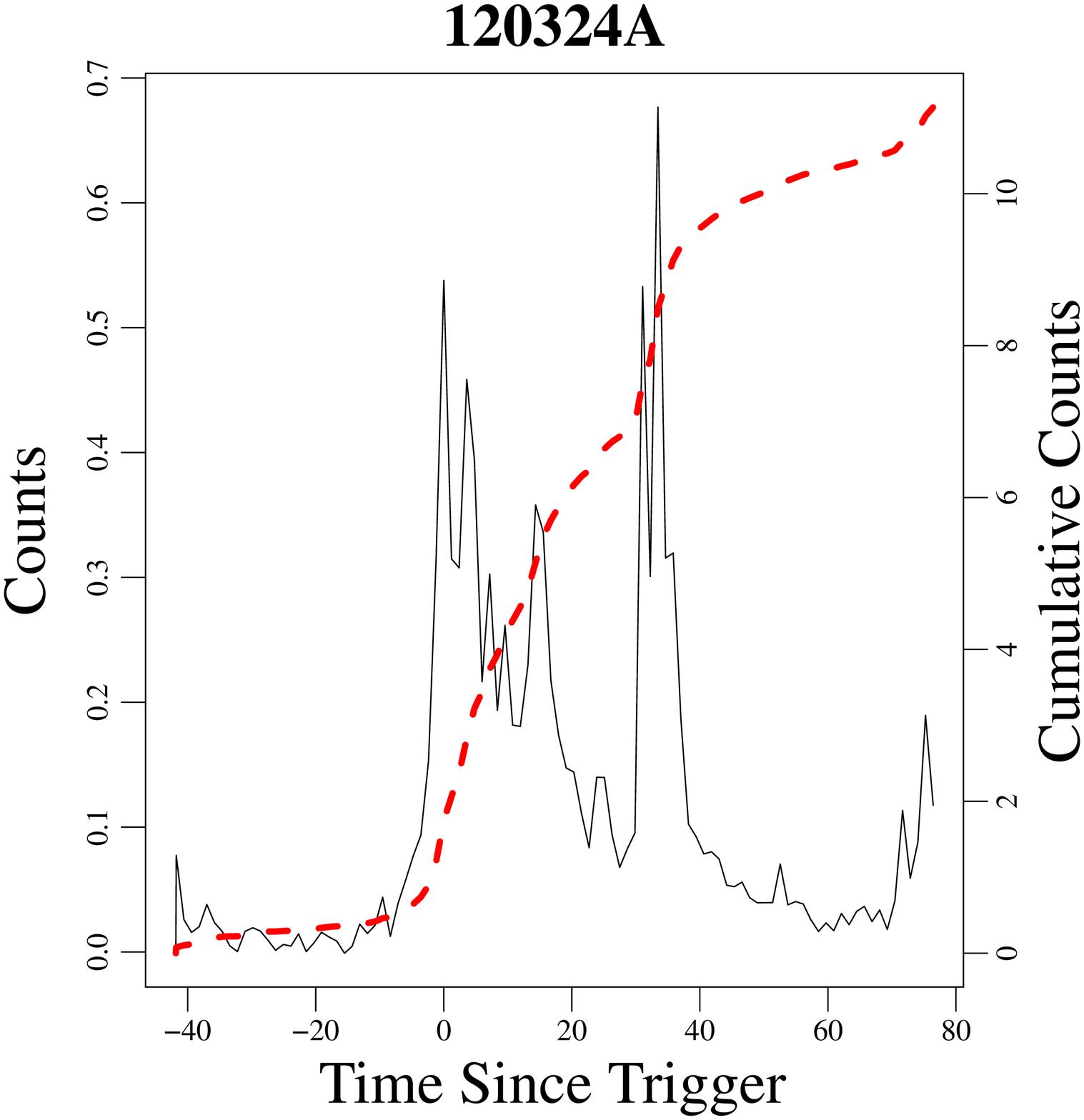}
 \end{center}
\end{minipage}\\
\begin{minipage}{0.25\hsize}
\begin{center}
    \FigureFile(40mm,40mm){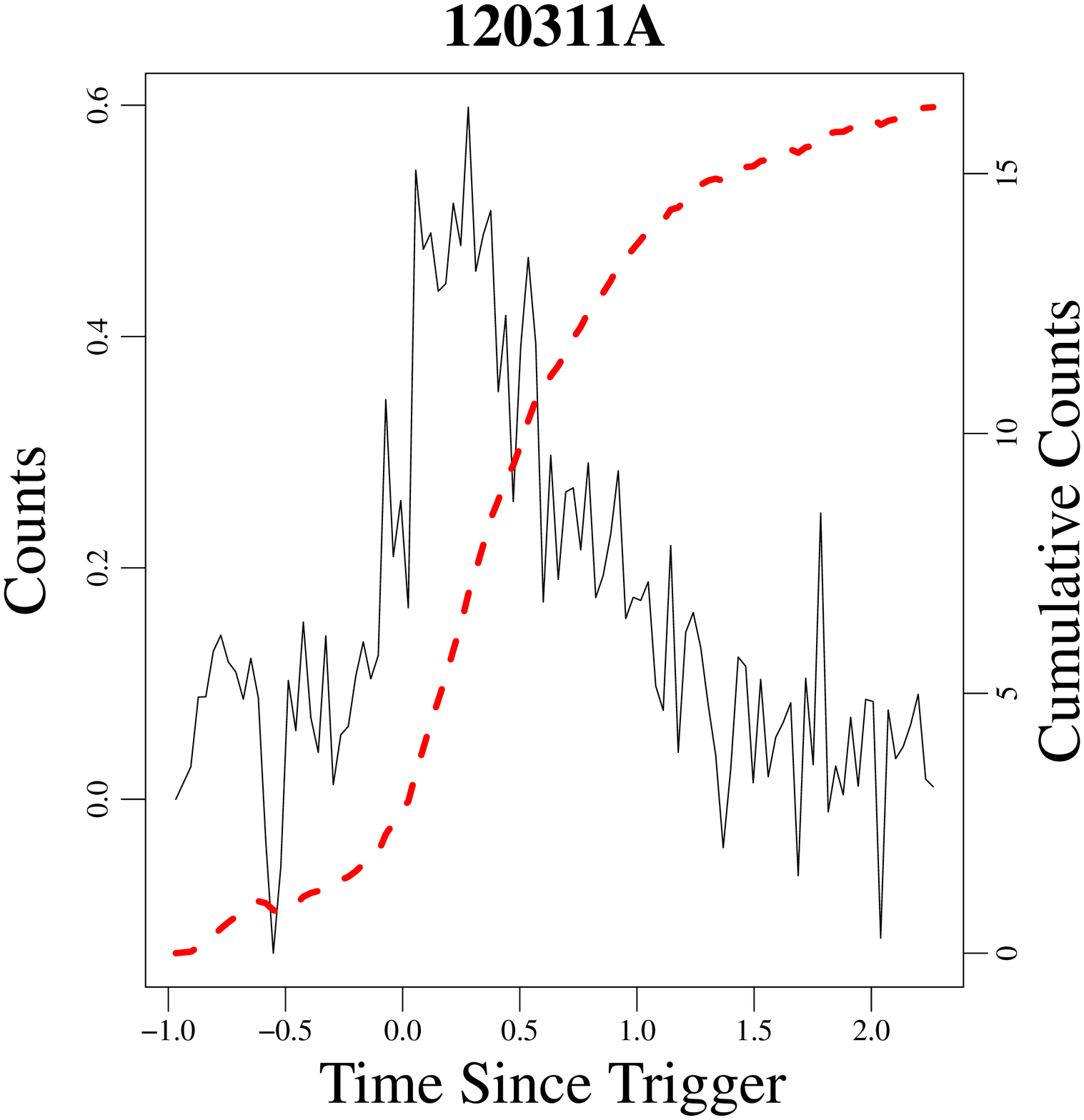}
\end{center}
\end{minipage}
\begin{minipage}{0.25\hsize}
\begin{center}
    \FigureFile(40mm,40mm){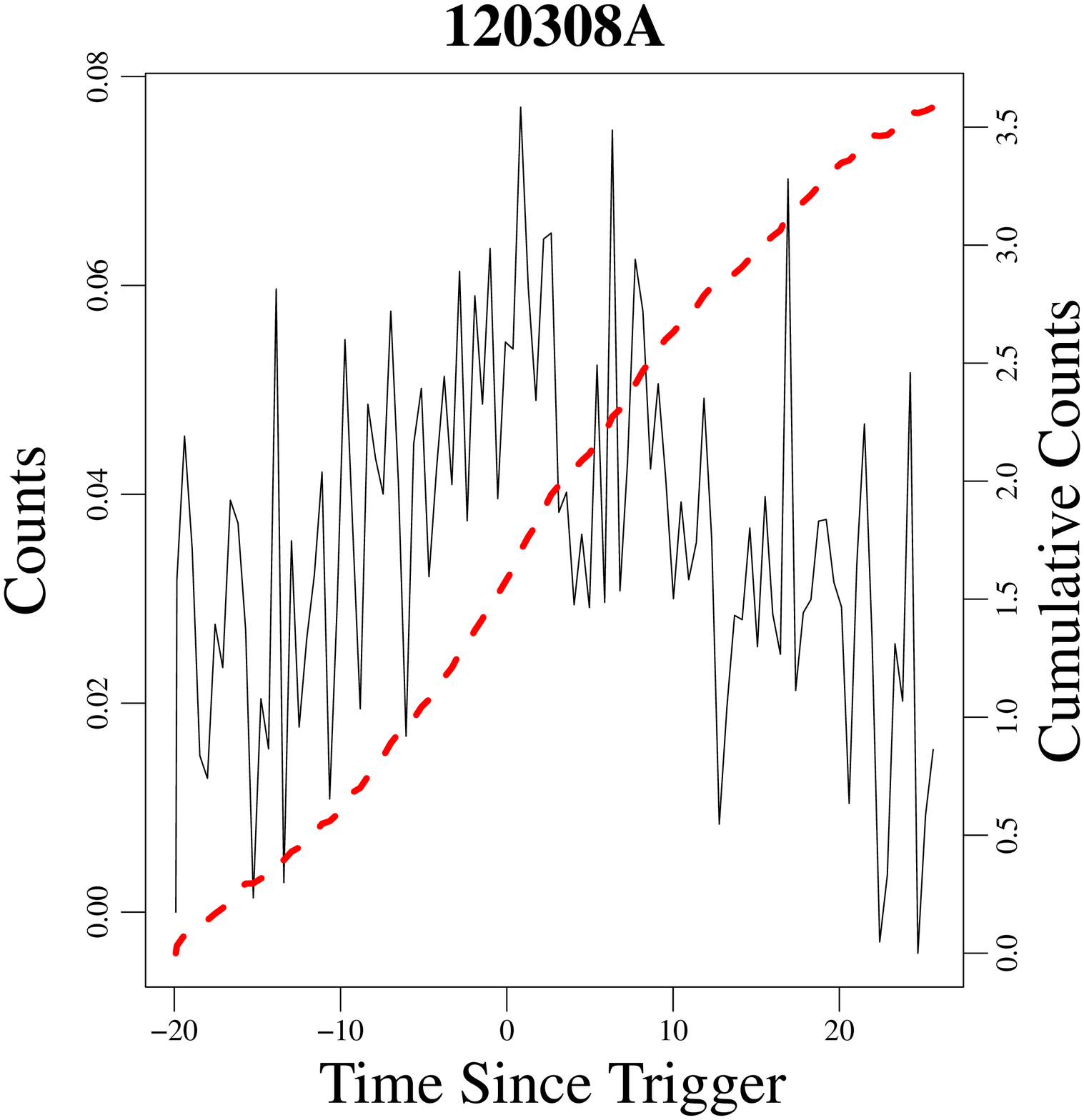}
 \end{center}
\end{minipage}
\begin{minipage}{0.25\hsize}
\begin{center}
    \FigureFile(40mm,40mm){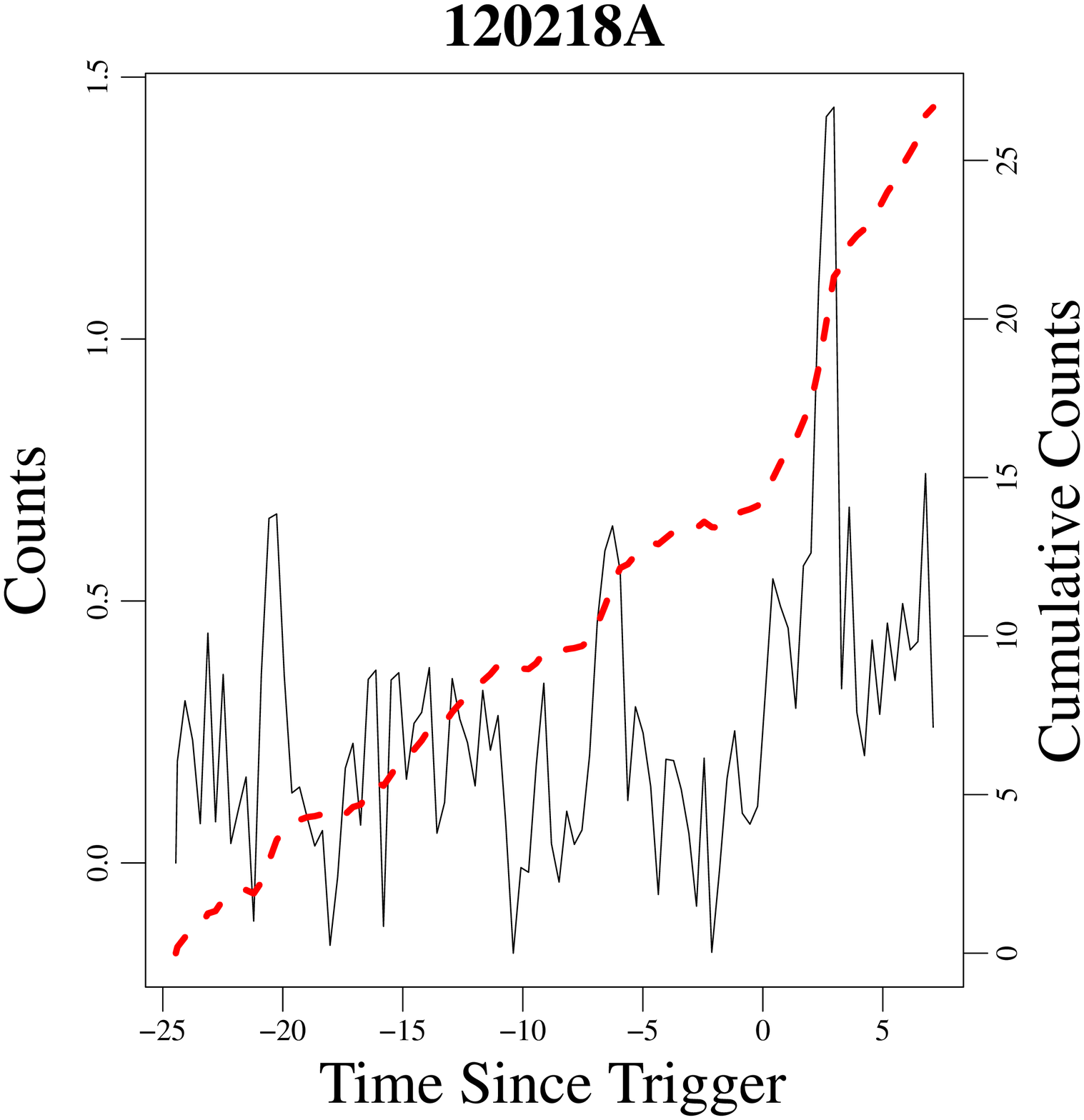}
\end{center}
\end{minipage}
\begin{minipage}{0.25\hsize}
\begin{center}
    \FigureFile(40mm,40mm){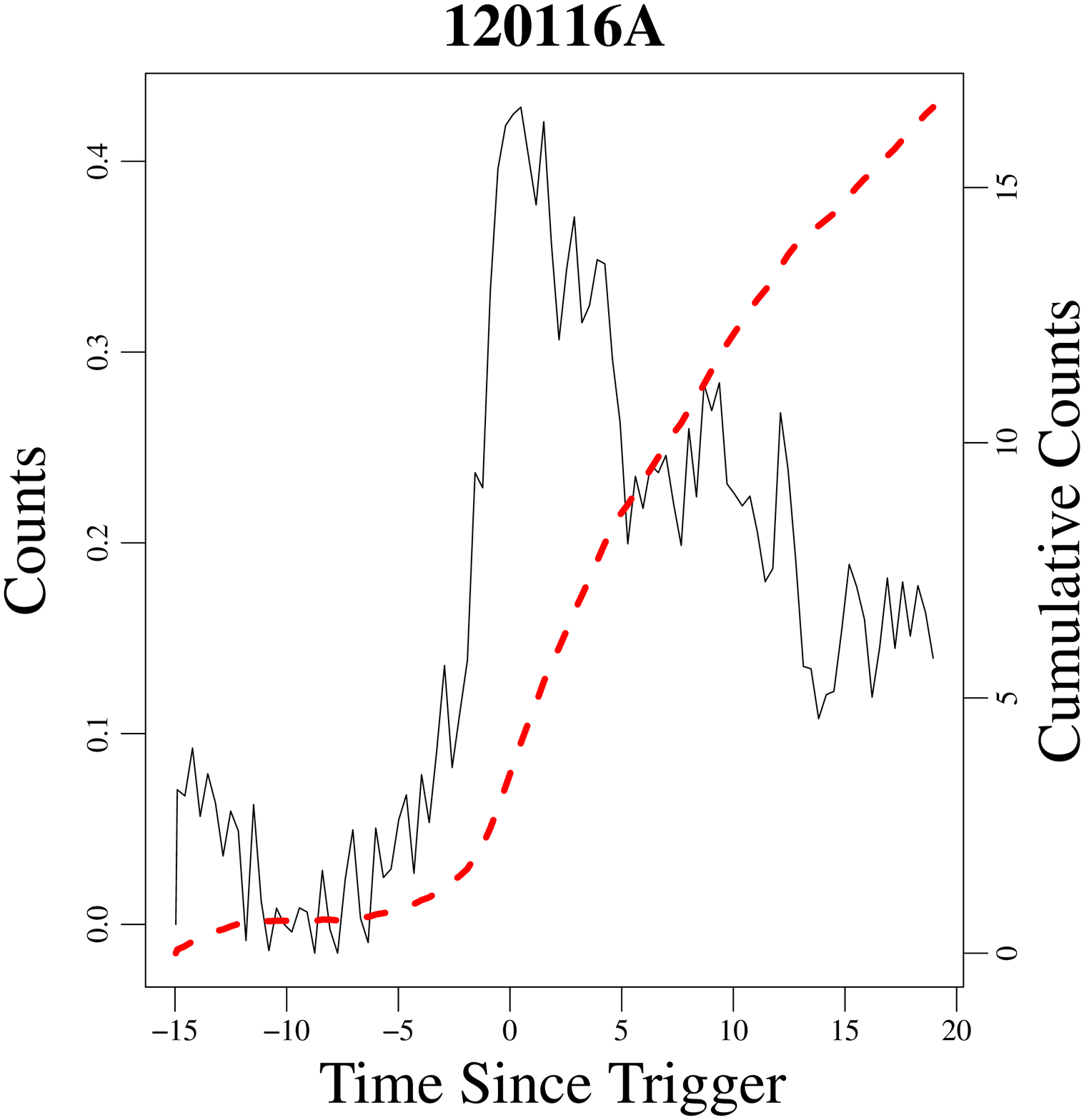}
 \end{center}
\end{minipage}\\
\begin{minipage}{0.25\hsize}
\begin{center}
    \FigureFile(40mm,40mm){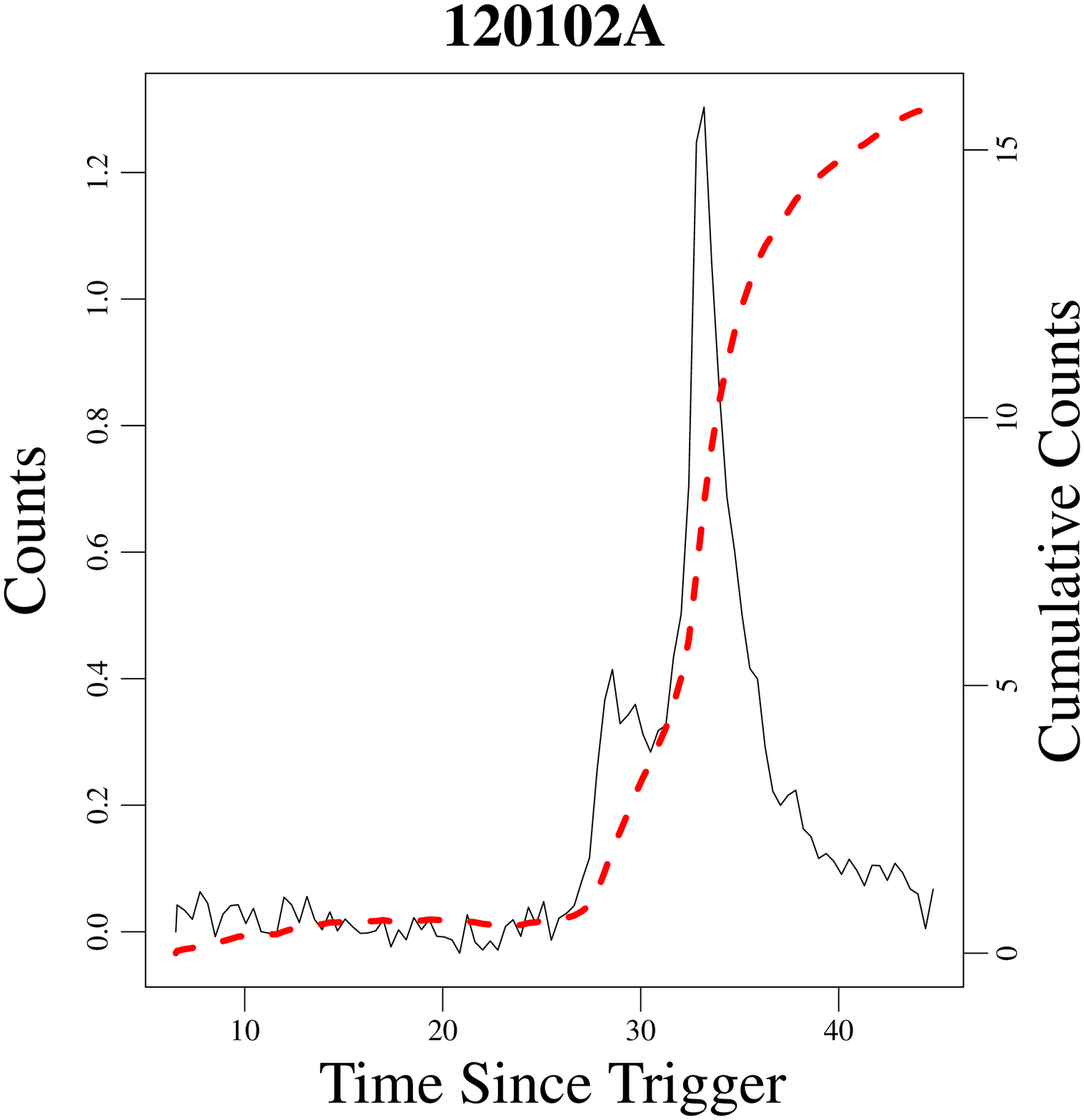}
\end{center}
\end{minipage}
\begin{minipage}{0.25\hsize}
\begin{center}
    \FigureFile(40mm,40mm){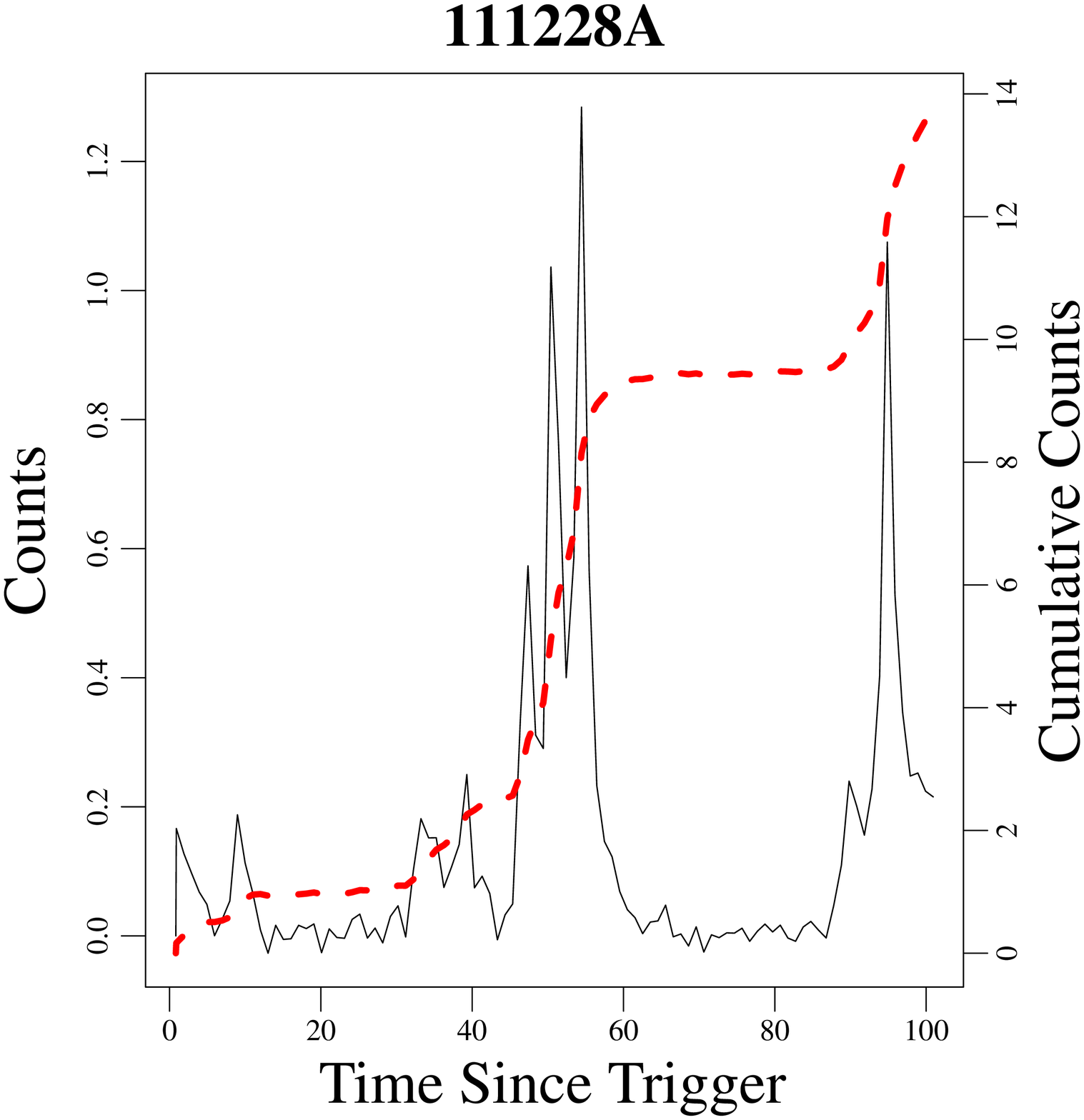}
 \end{center}
\end{minipage}
\begin{minipage}{0.25\hsize}
\begin{center}
    \FigureFile(40mm,40mm){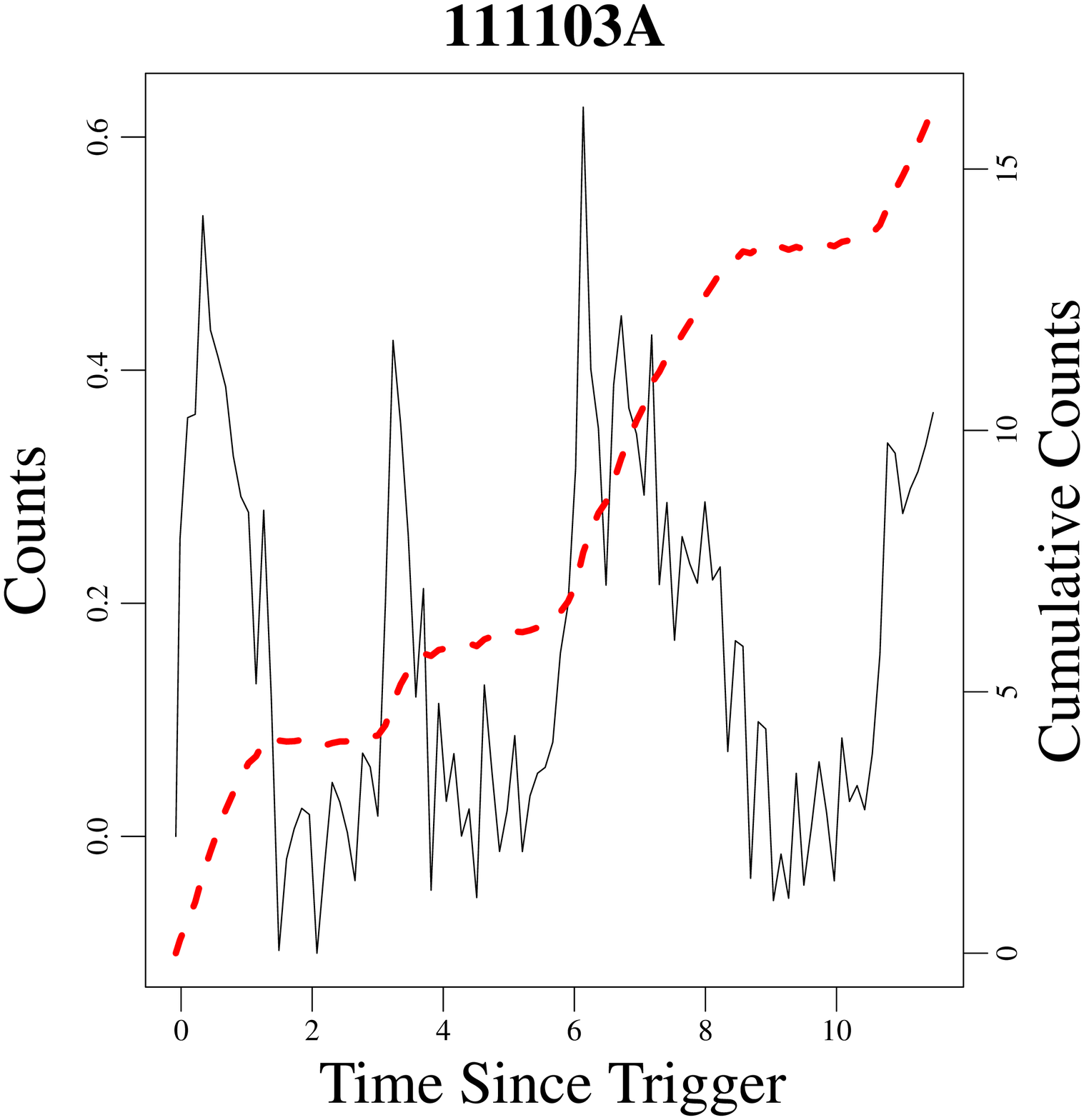}
\end{center}
\end{minipage}
\begin{minipage}{0.25\hsize}
\begin{center}
    \FigureFile(40mm,40mm){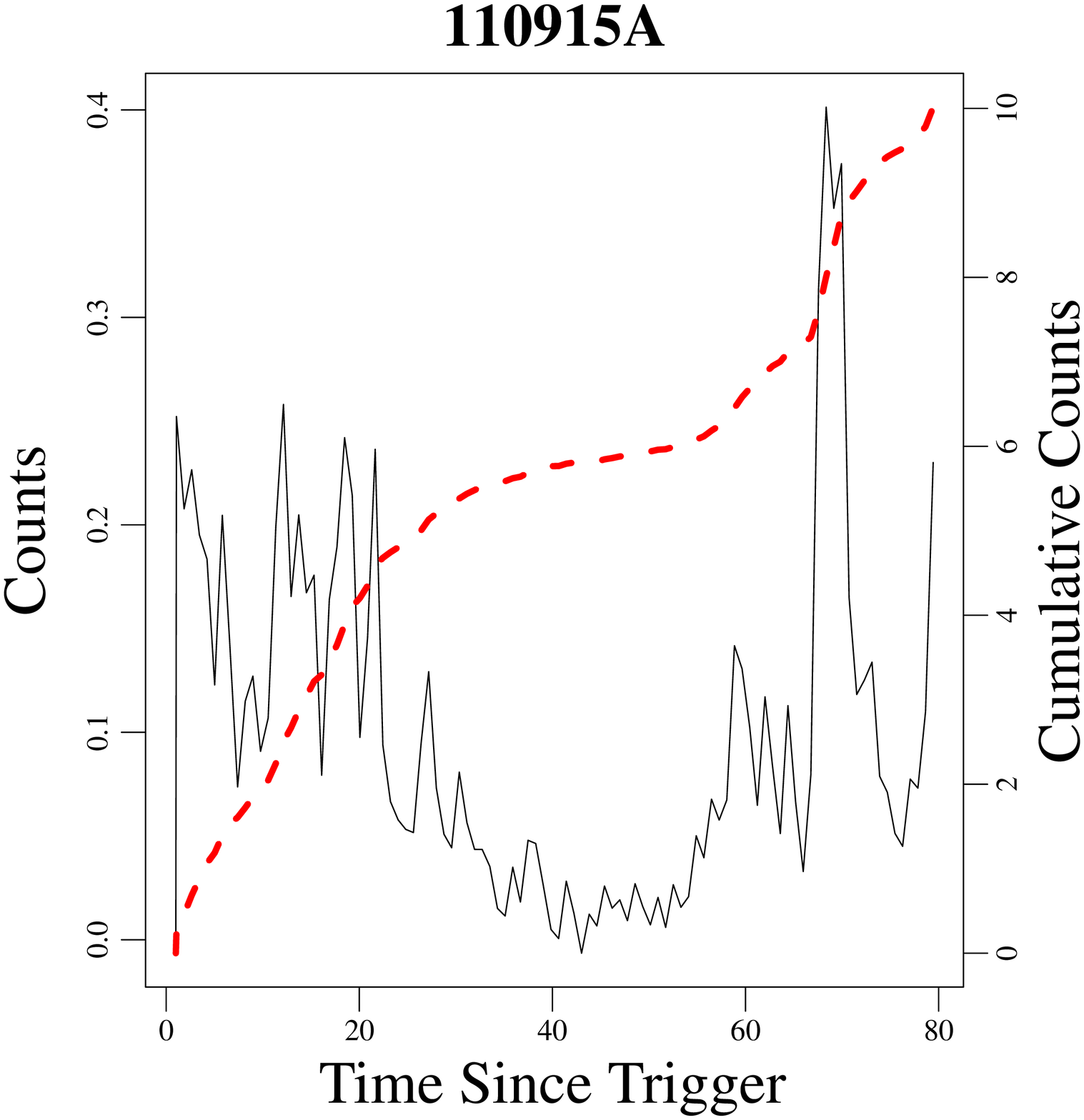}
 \end{center}
\end{minipage}\\
\begin{minipage}{0.25\hsize}
\begin{center}
    \FigureFile(40mm,40mm){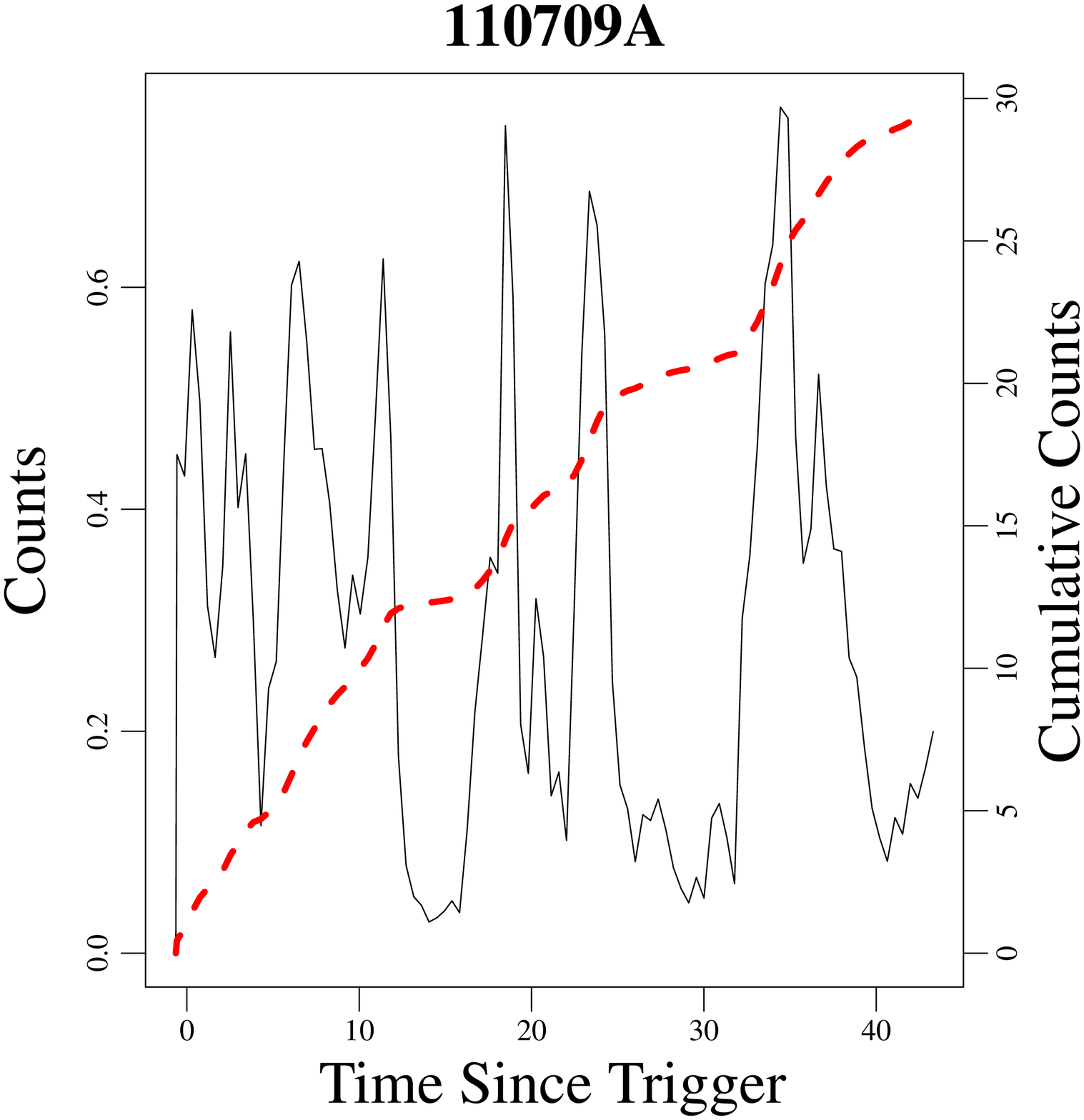}
\end{center}
\end{minipage}
\begin{minipage}{0.25\hsize}
\begin{center}
    \FigureFile(40mm,40mm){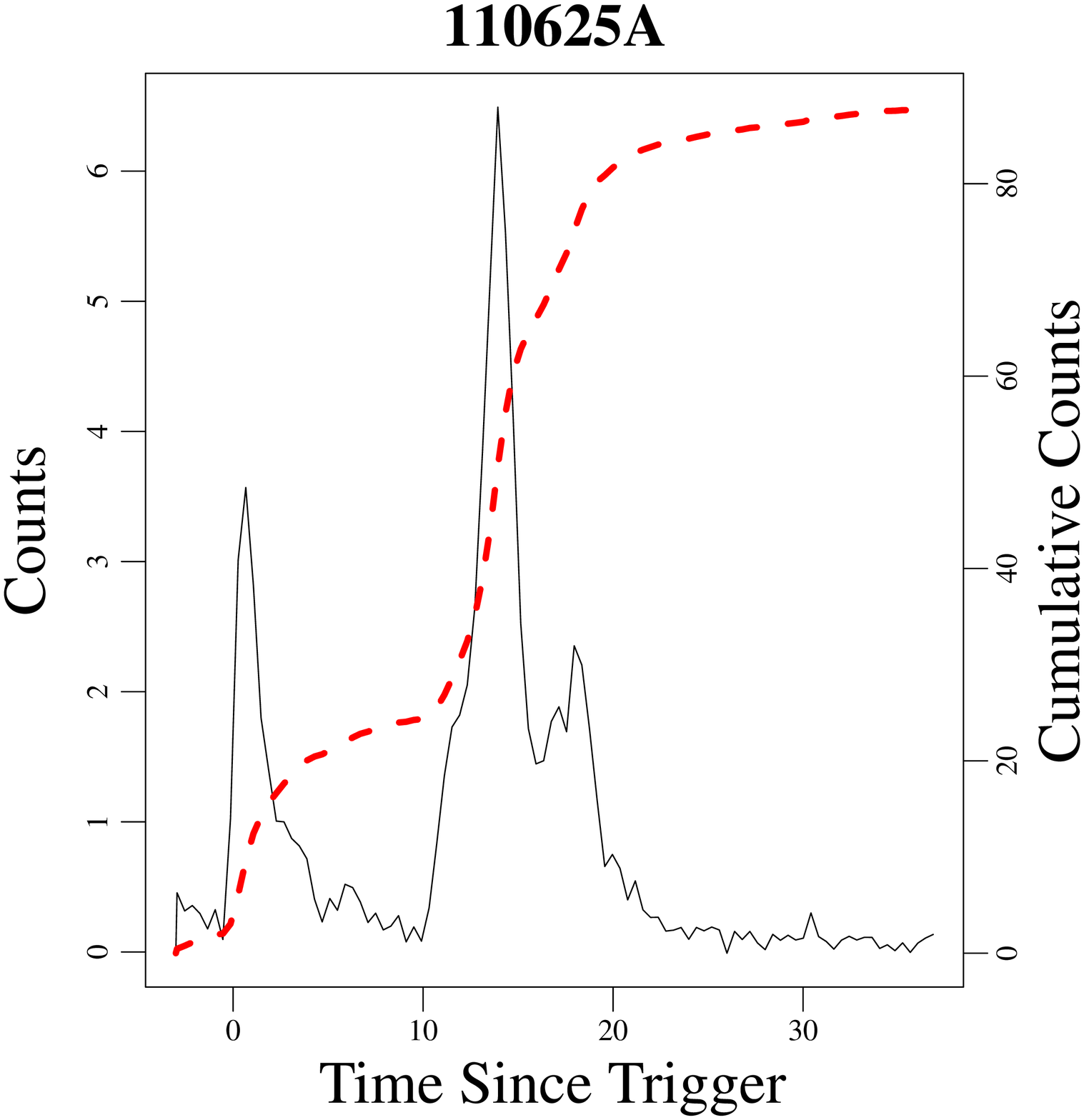}
 \end{center}
\end{minipage}
\begin{minipage}{0.25\hsize}
\begin{center}
    \FigureFile(40mm,40mm){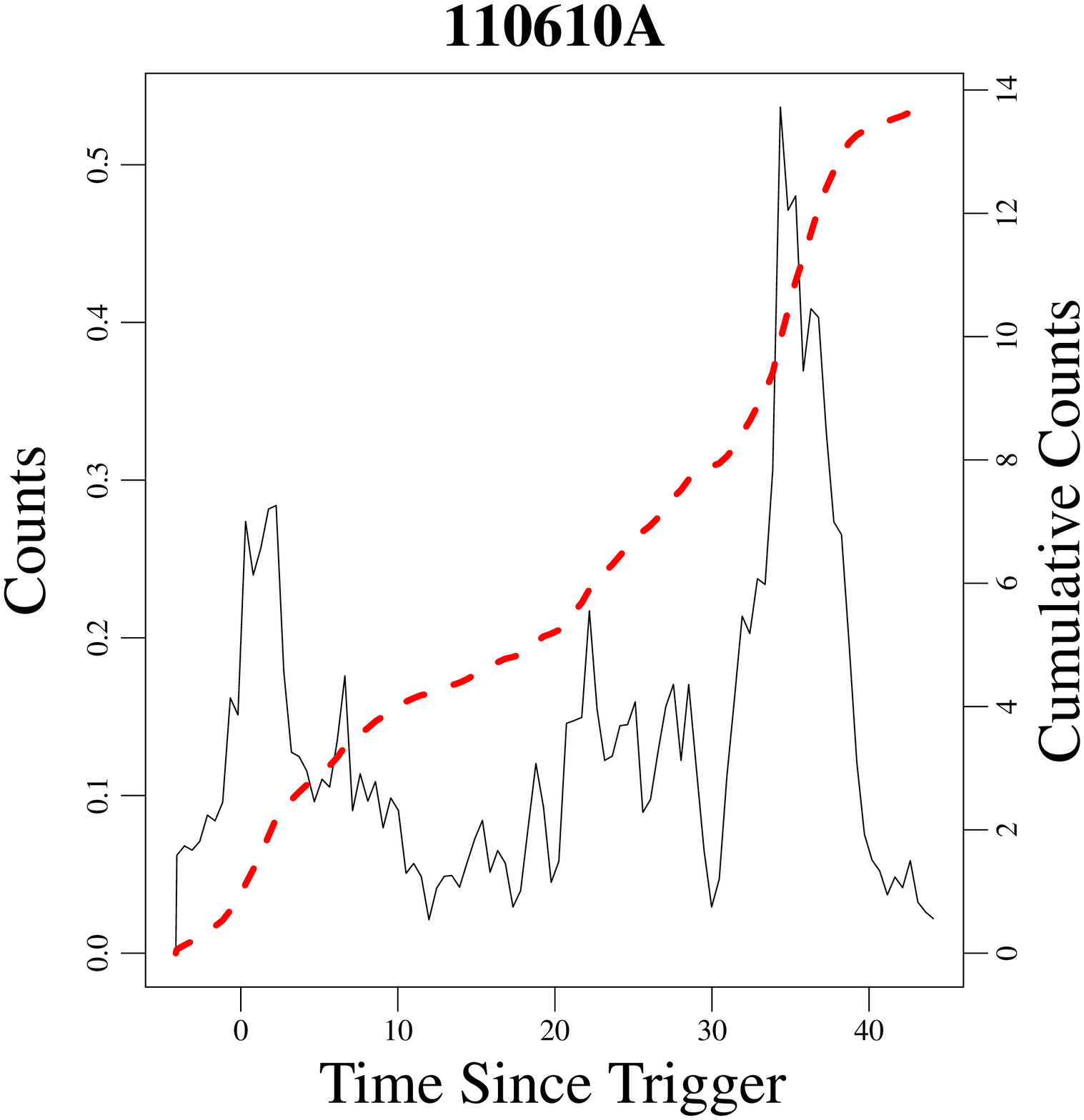}
\end{center}
\end{minipage}
\begin{minipage}{0.25\hsize}
\begin{center}
    \FigureFile(40mm,40mm){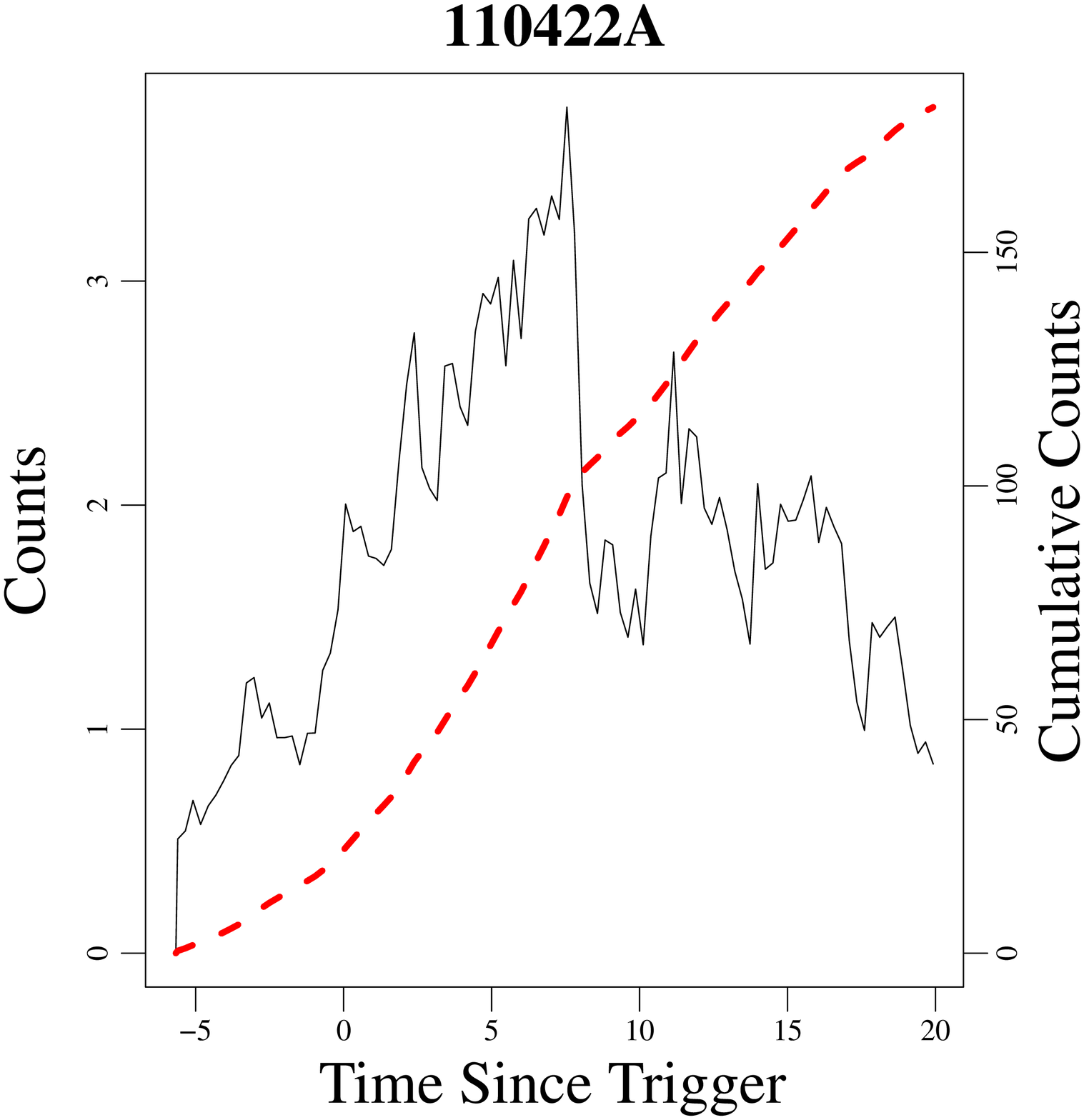}
 \end{center}
\end{minipage}\\

\end{tabular}
\caption{Light curves (black solid) and cumulative light curves (red doted) of Type I LGRBs.}\label{fig:A1-1}
\end{figure*}
%%%%%%%%%%%%%%%%%%%%
\begin{figure*}[htb]
\begin{tabular}{cccc}
\begin{minipage}{0.25\hsize}
\begin{center}
    \FigureFile(40mm,40mm){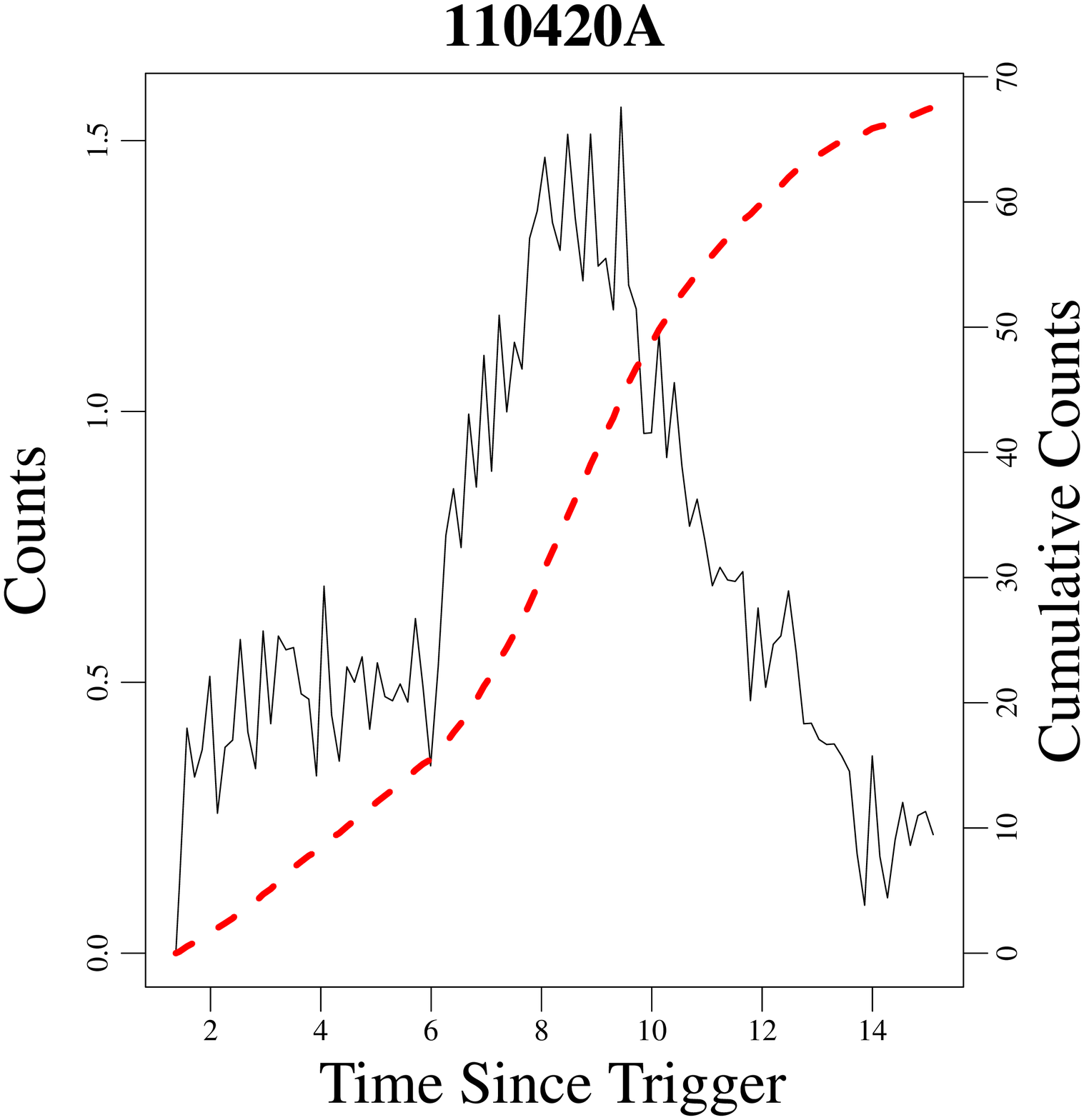}
\end{center}
\end{minipage}
\begin{minipage}{0.25\hsize}
\begin{center}
    \FigureFile(40mm,40mm){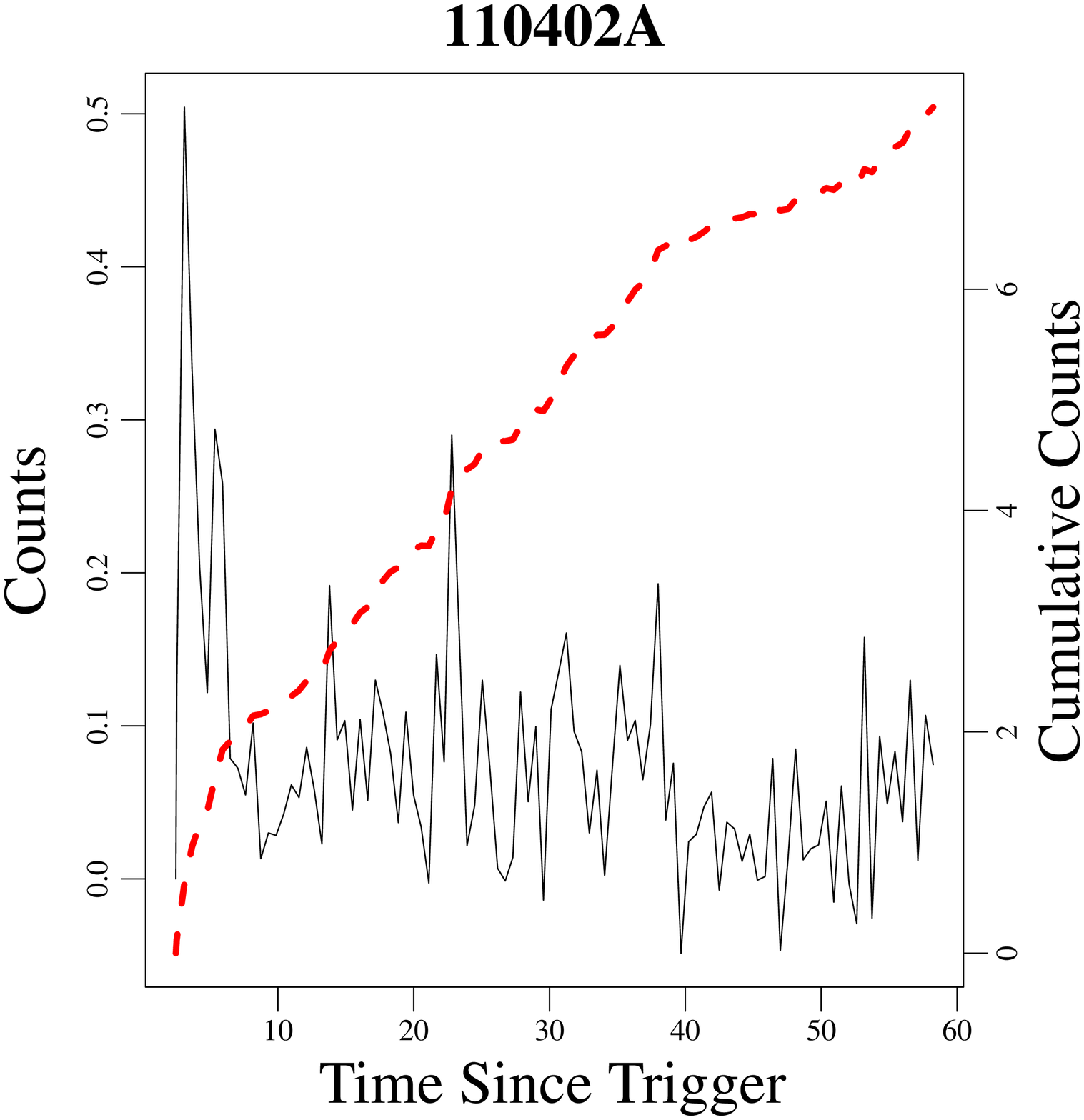}
 \end{center}
\end{minipage}
\begin{minipage}{0.25\hsize}
\begin{center}
    \FigureFile(40mm,40mm){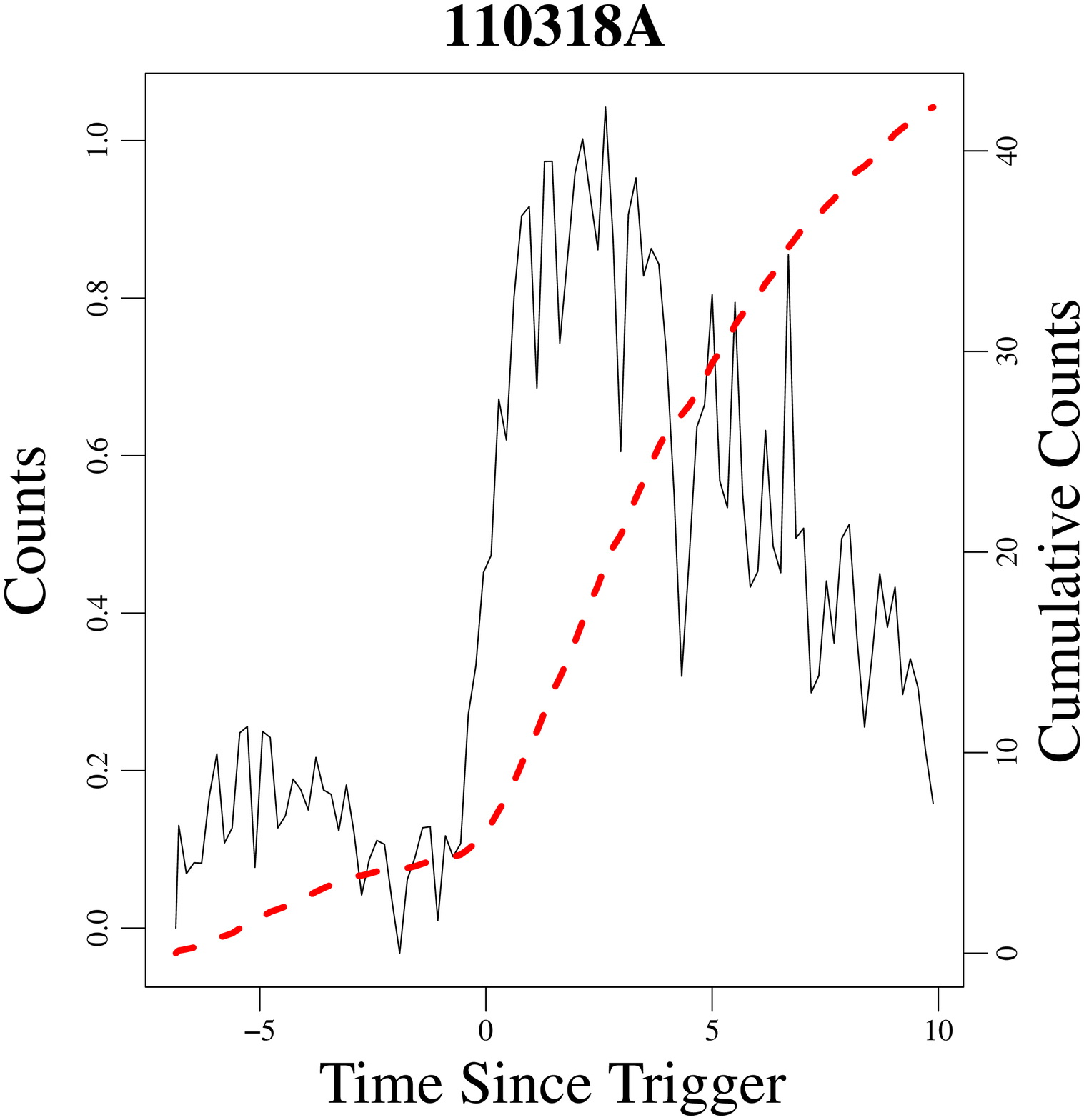}
\end{center}
\end{minipage}
\begin{minipage}{0.25\hsize}
\begin{center}
    \FigureFile(40mm,40mm){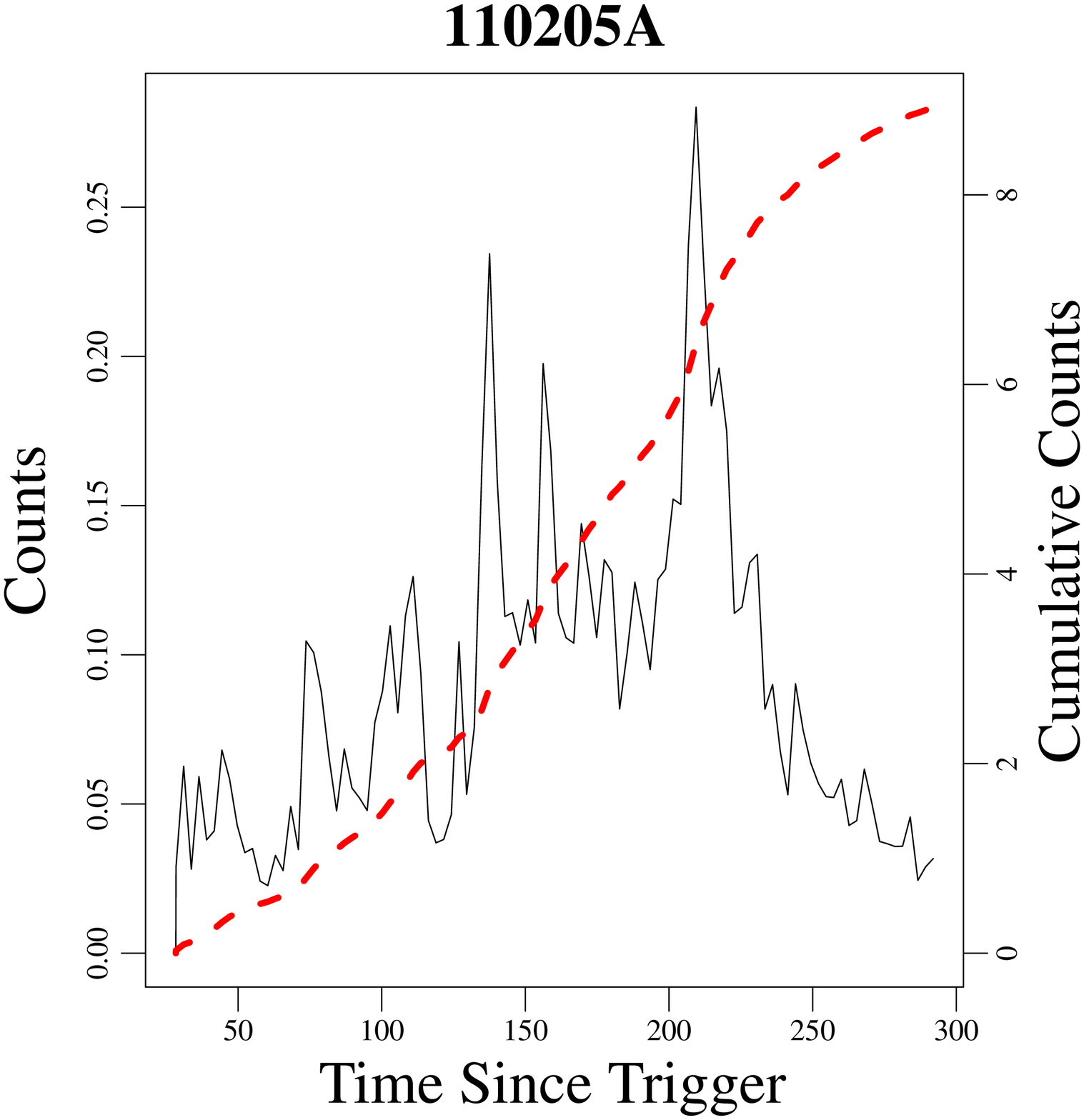}
 \end{center}
\end{minipage}\\
\begin{minipage}{0.25\hsize}
\begin{center}
    \FigureFile(40mm,40mm){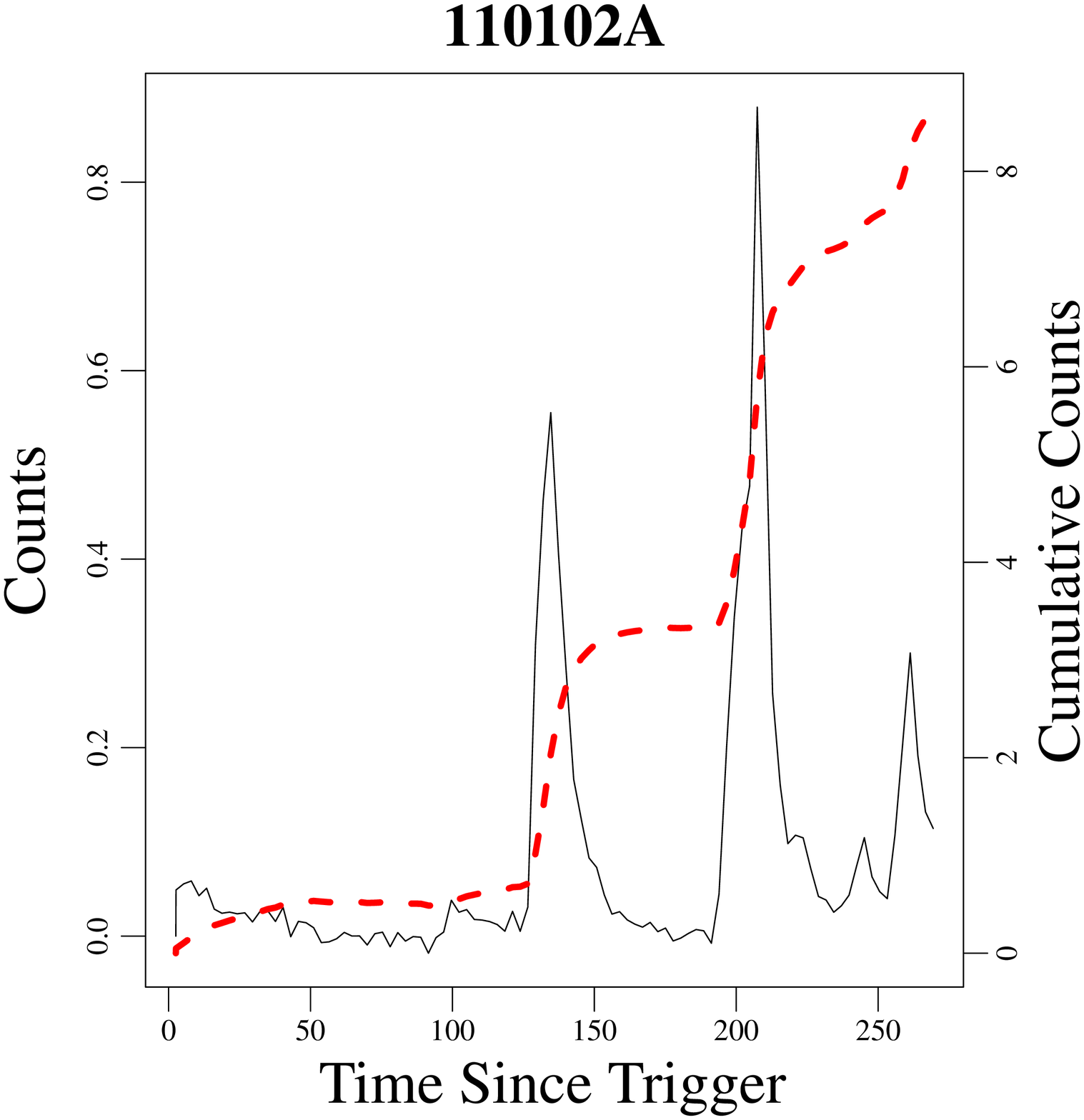}
\end{center}
\end{minipage}
\begin{minipage}{0.25\hsize}
\begin{center}
    \FigureFile(40mm,40mm){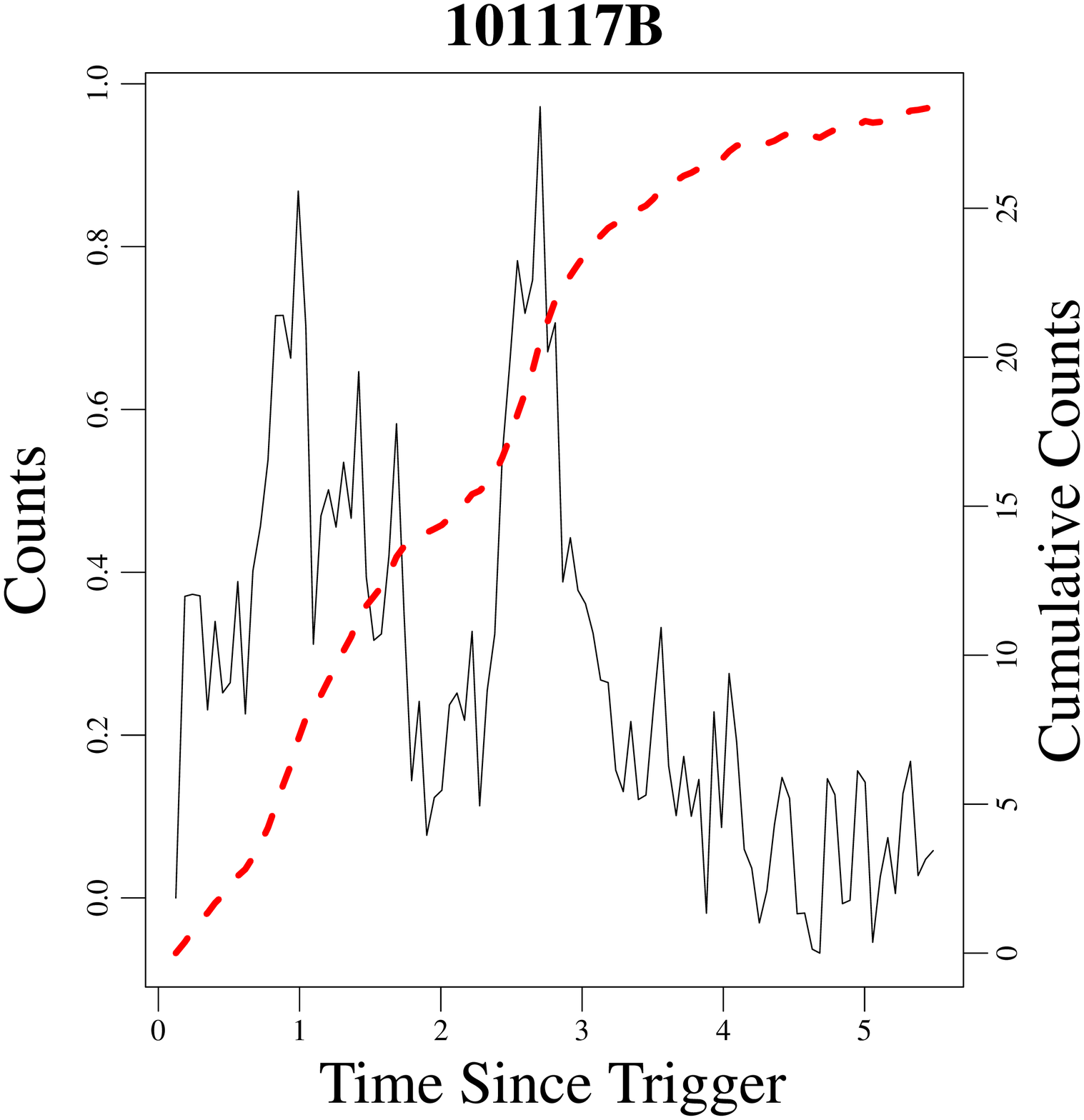}
 \end{center}
\end{minipage}
\begin{minipage}{0.25\hsize}
\begin{center}
    \FigureFile(40mm,40mm){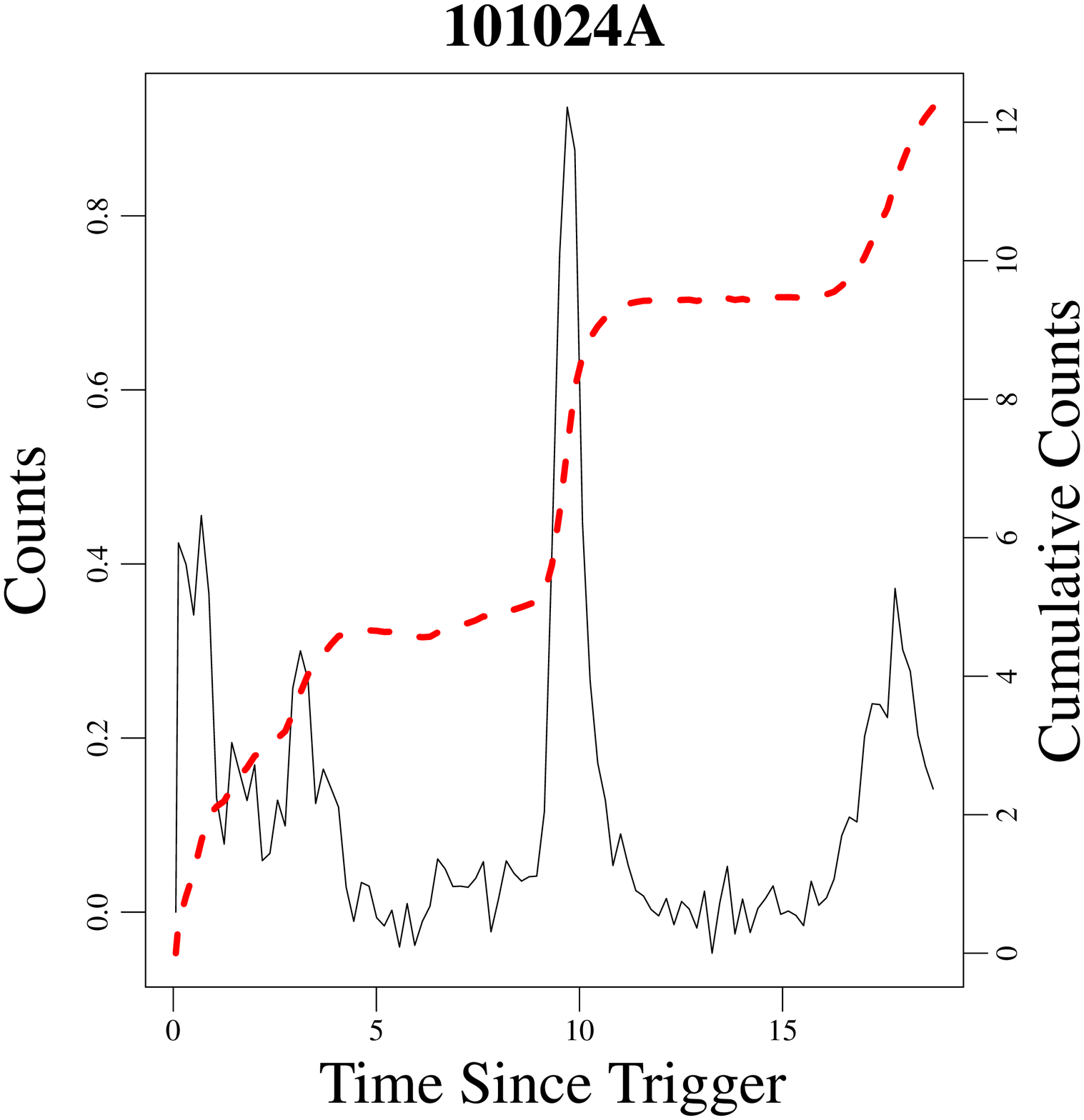}
\end{center}
\end{minipage}
\begin{minipage}{0.25\hsize}
\begin{center}
    \FigureFile(40mm,40mm){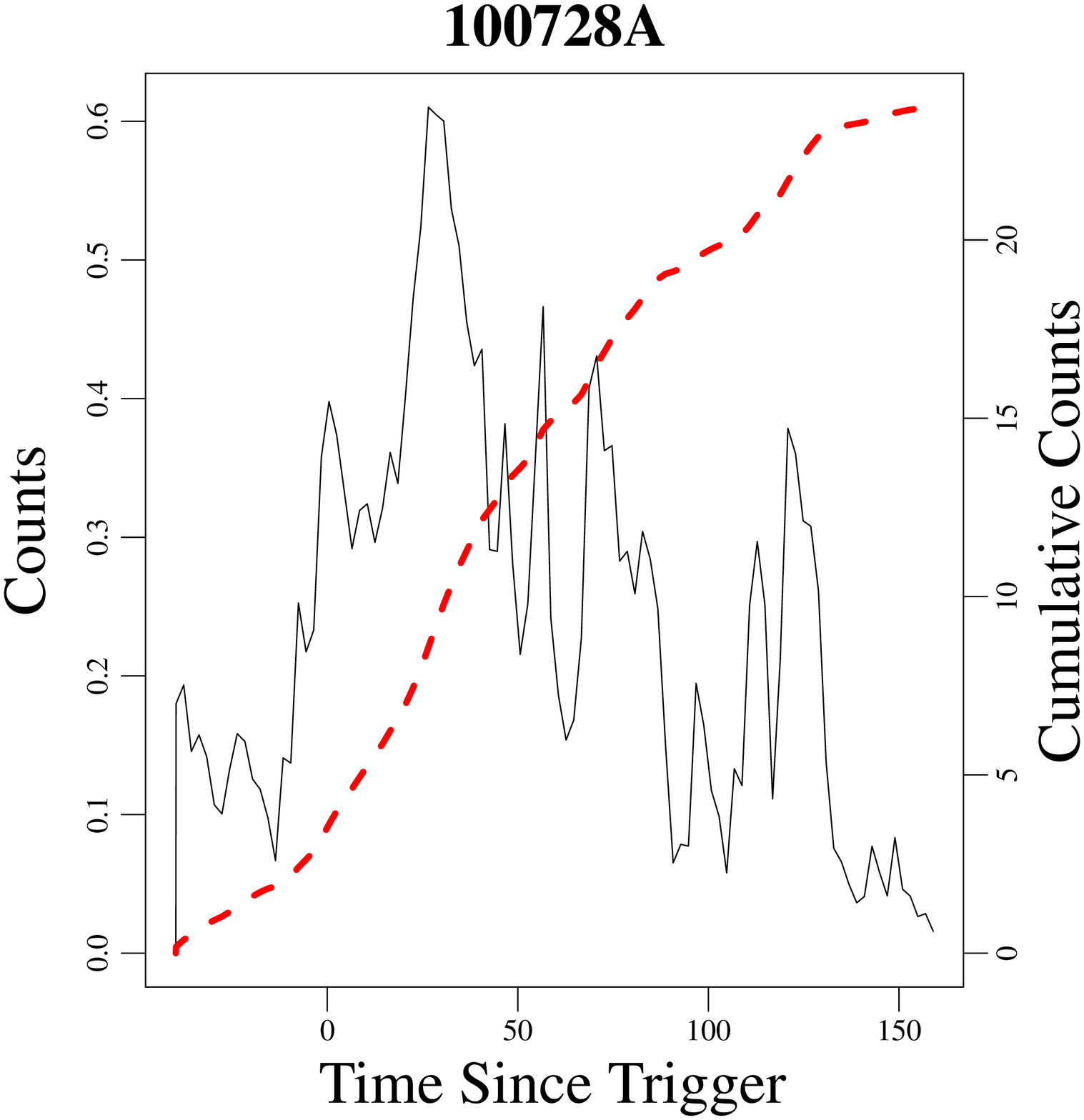}
 \end{center}
\end{minipage}\\
\begin{minipage}{0.25\hsize}
\begin{center}
    \FigureFile(40mm,40mm){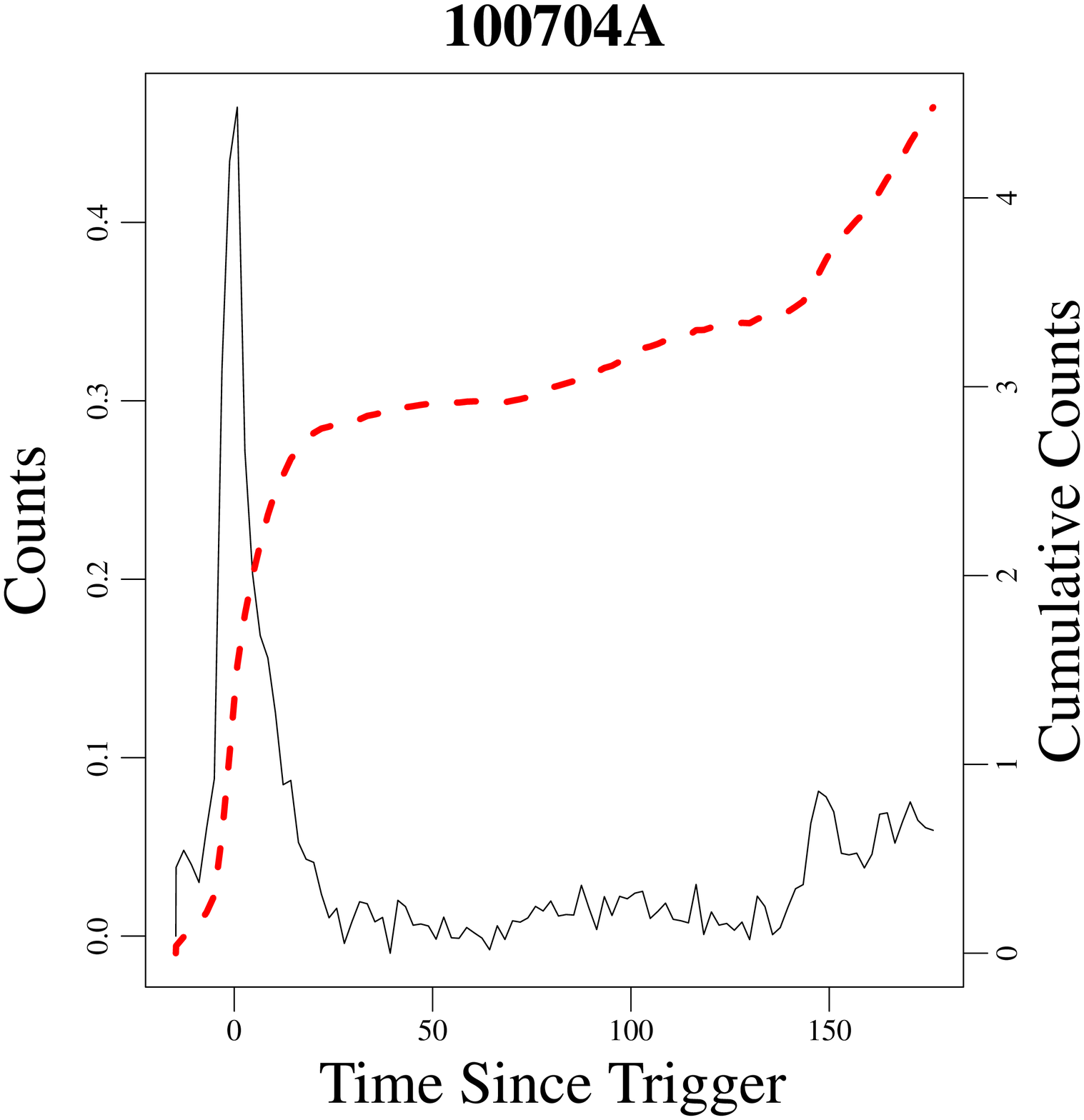}
\end{center}
\end{minipage}
\begin{minipage}{0.25\hsize}
\begin{center}
    \FigureFile(40mm,40mm){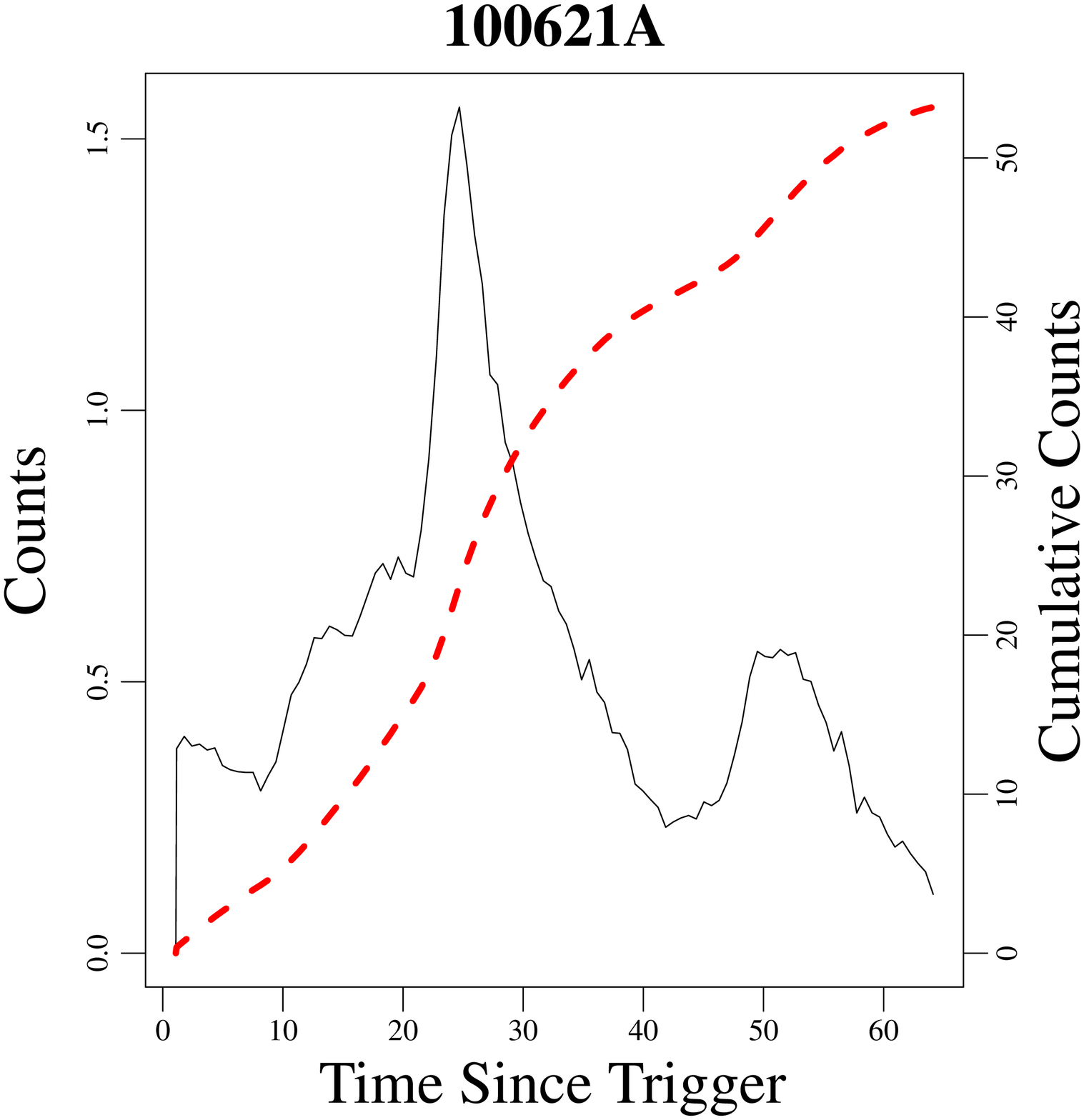}
 \end{center}
\end{minipage}
\begin{minipage}{0.25\hsize}
\begin{center}
    \FigureFile(40mm,40mm){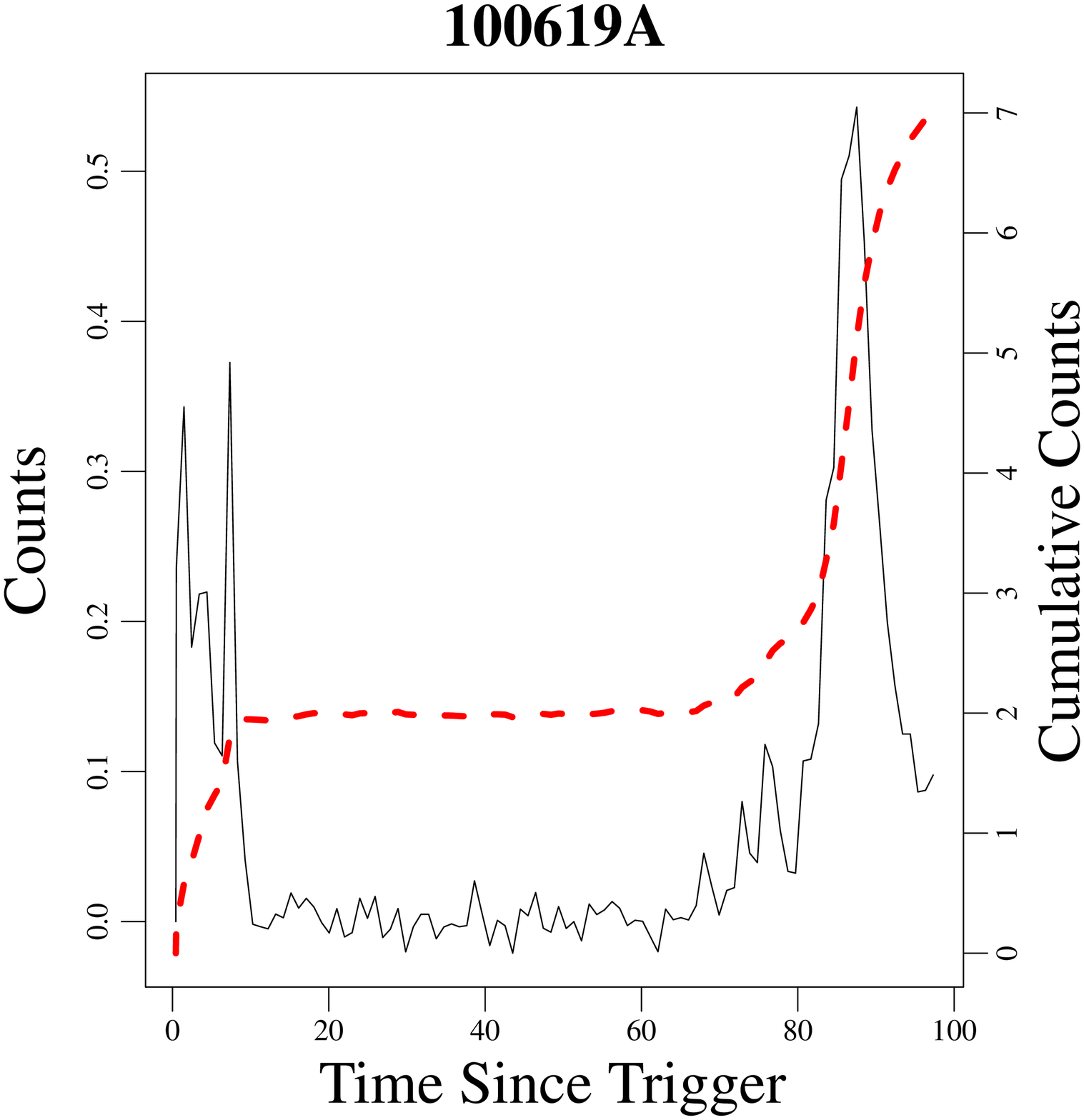}
\end{center}
\end{minipage}
\begin{minipage}{0.25\hsize}
\begin{center}
    \FigureFile(40mm,40mm){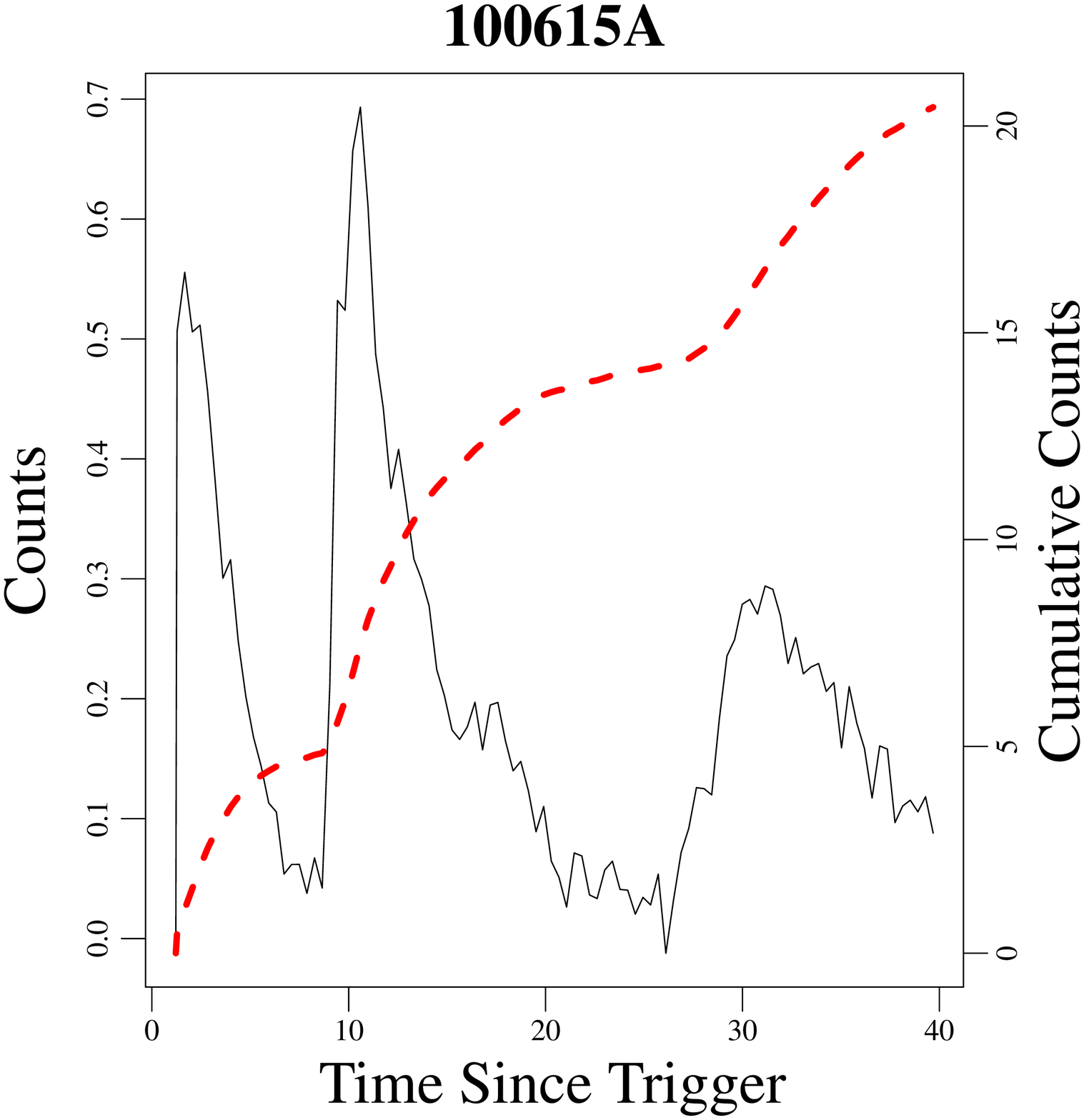}
 \end{center}
\end{minipage}\\
\begin{minipage}{0.25\hsize}
\begin{center}
    \FigureFile(40mm,40mm){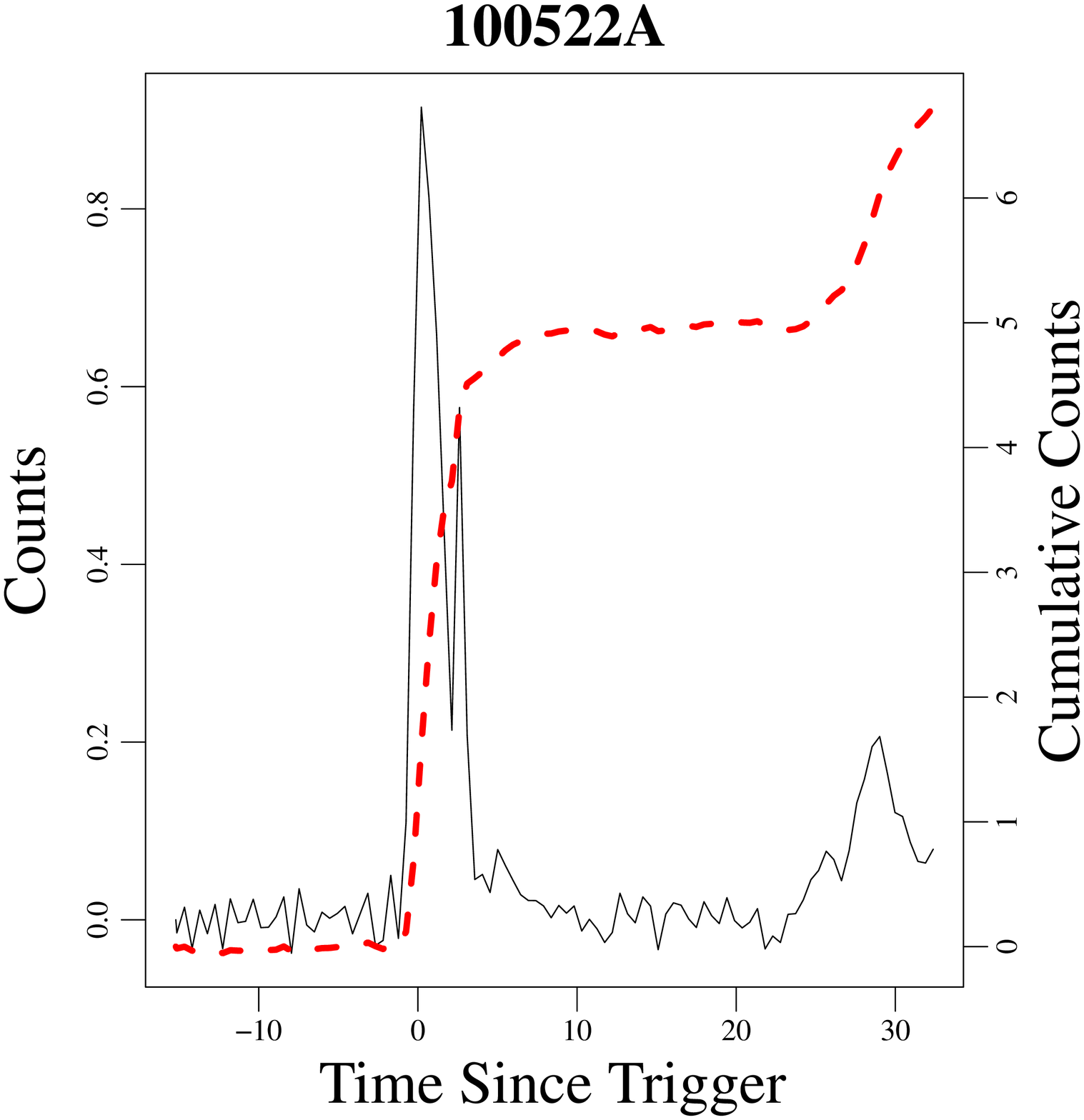}
\end{center}
\end{minipage}
\begin{minipage}{0.25\hsize}
\begin{center}
    \FigureFile(40mm,40mm){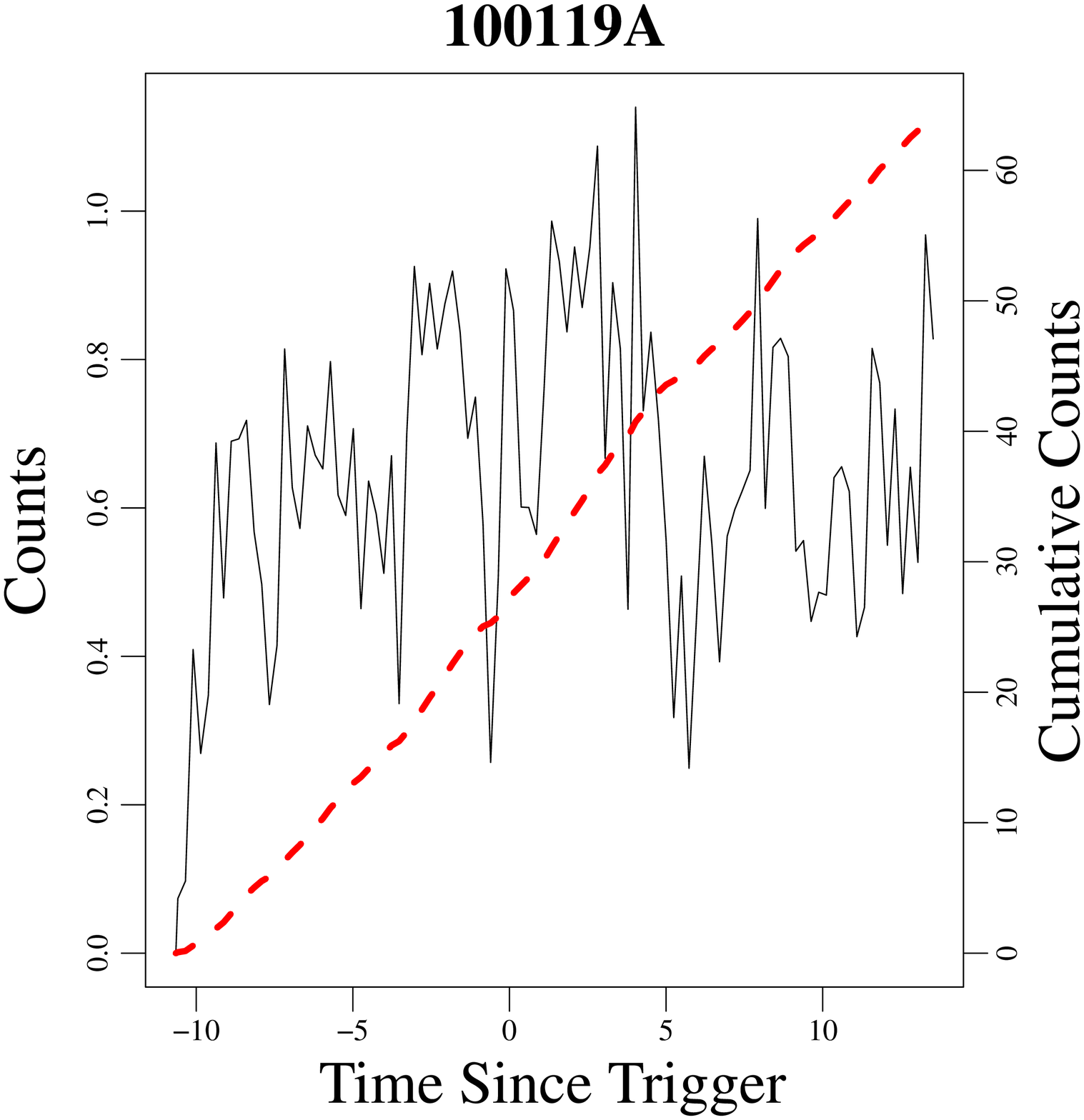}
 \end{center}
\end{minipage}
\begin{minipage}{0.25\hsize}
\begin{center}
    \FigureFile(40mm,40mm){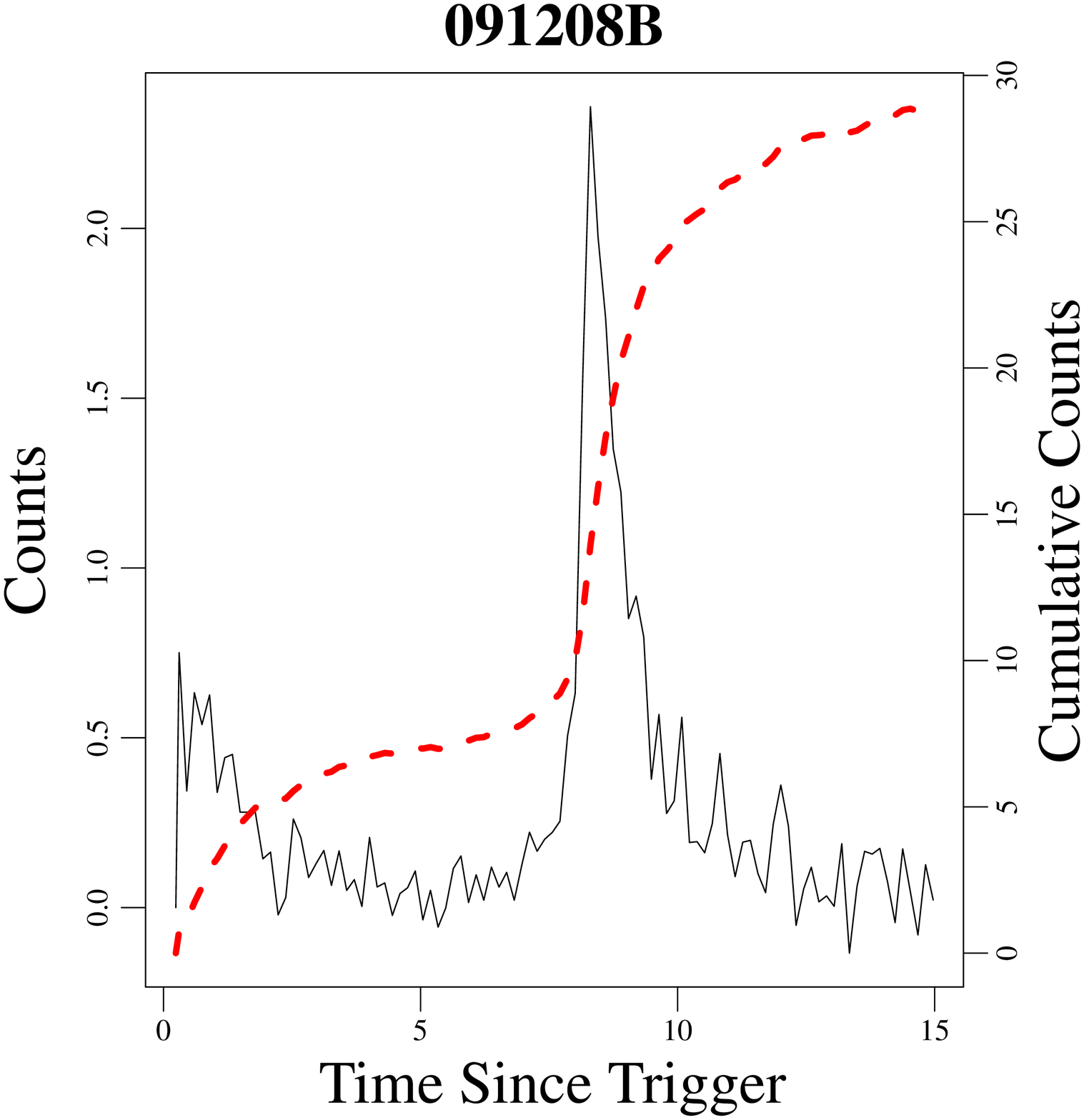}
\end{center}
\end{minipage}
\begin{minipage}{0.25\hsize}
\begin{center}
    \FigureFile(40mm,40mm){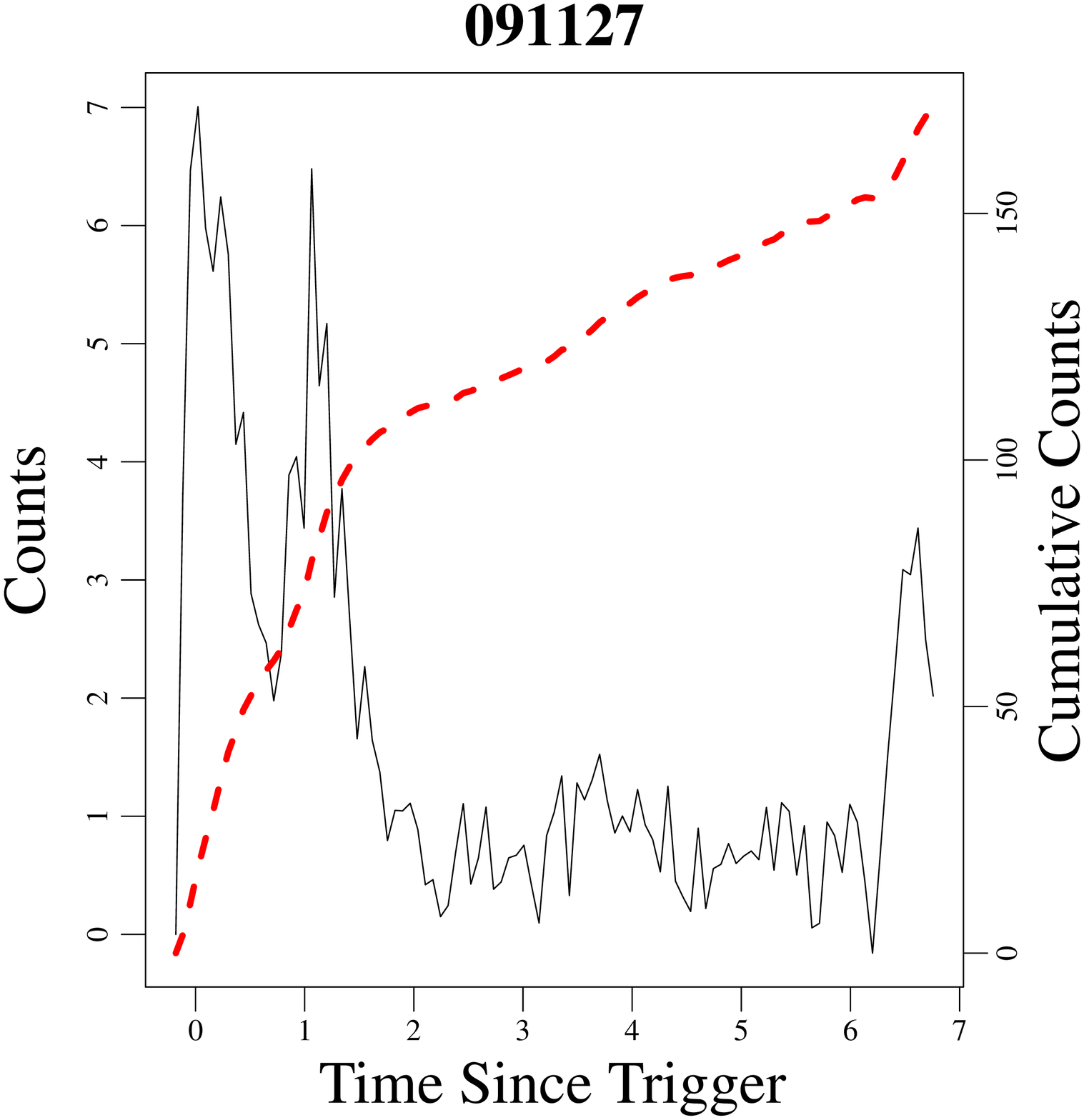}
 \end{center}
\end{minipage}\\
\begin{minipage}{0.25\hsize}
\begin{center}
    \FigureFile(40mm,40mm){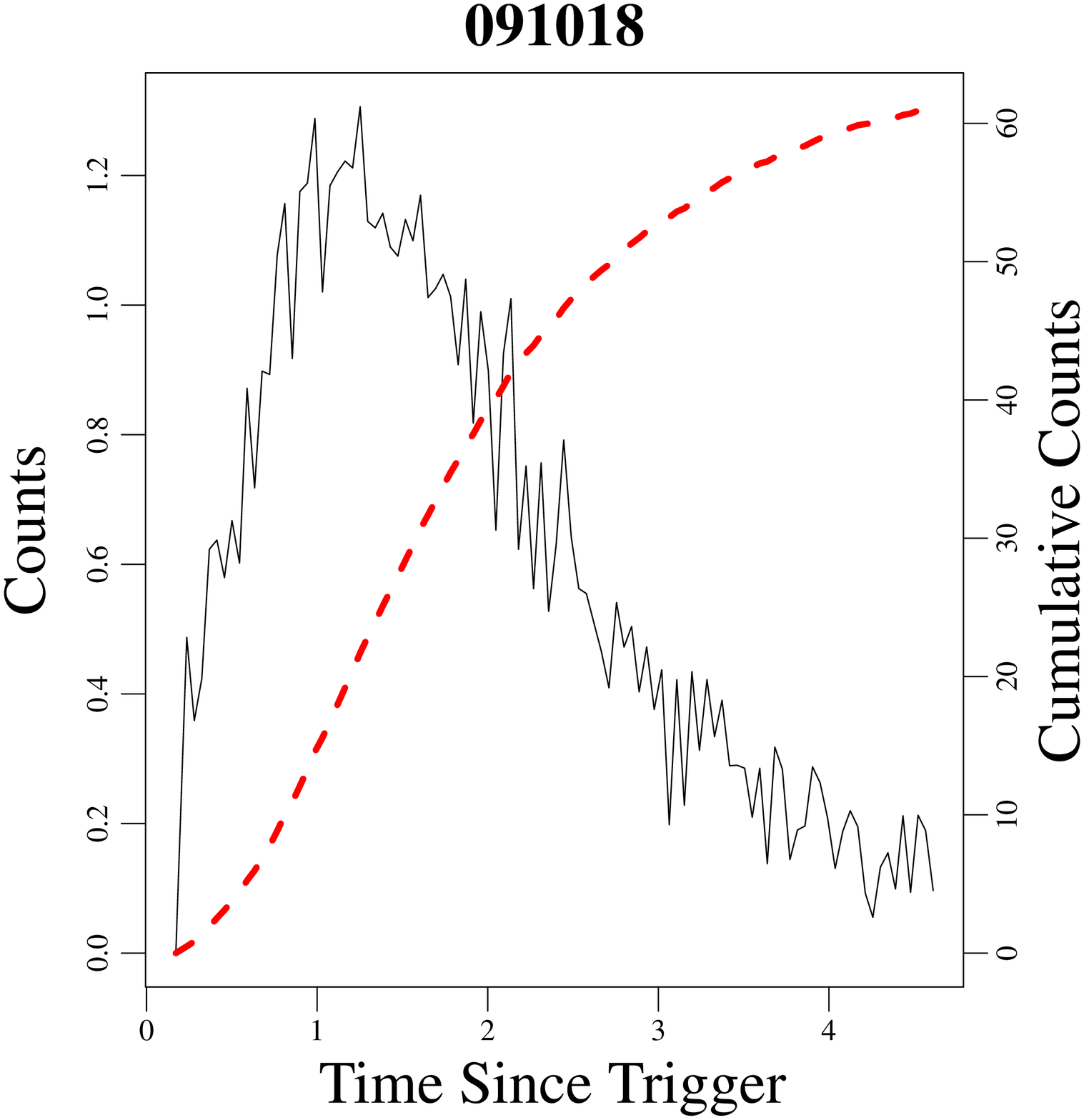}
\end{center}
\end{minipage}
\begin{minipage}{0.25\hsize}
\begin{center}
    \FigureFile(40mm,40mm){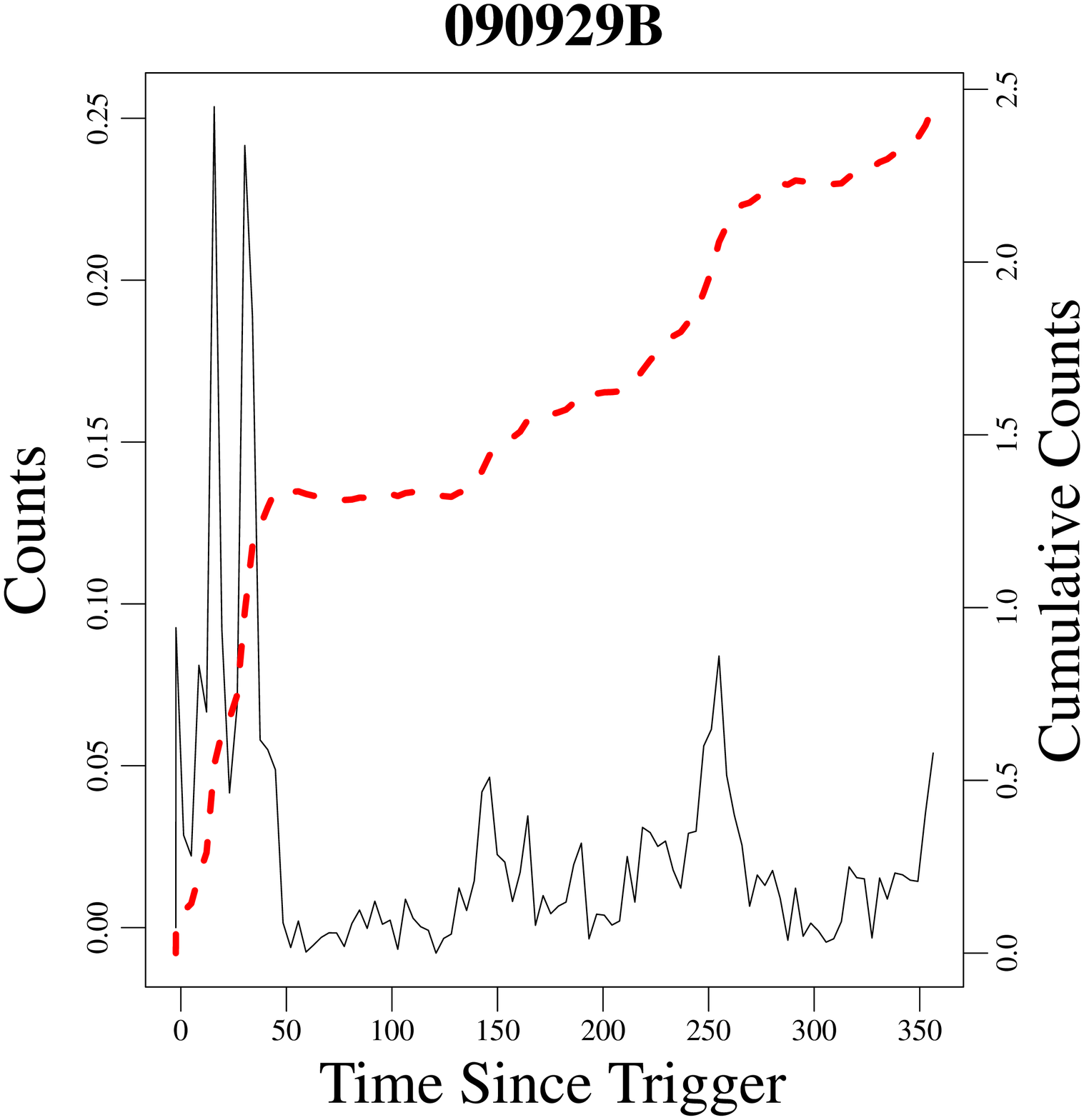}
 \end{center}
\end{minipage}
\begin{minipage}{0.25\hsize}
\begin{center}
    \FigureFile(40mm,40mm){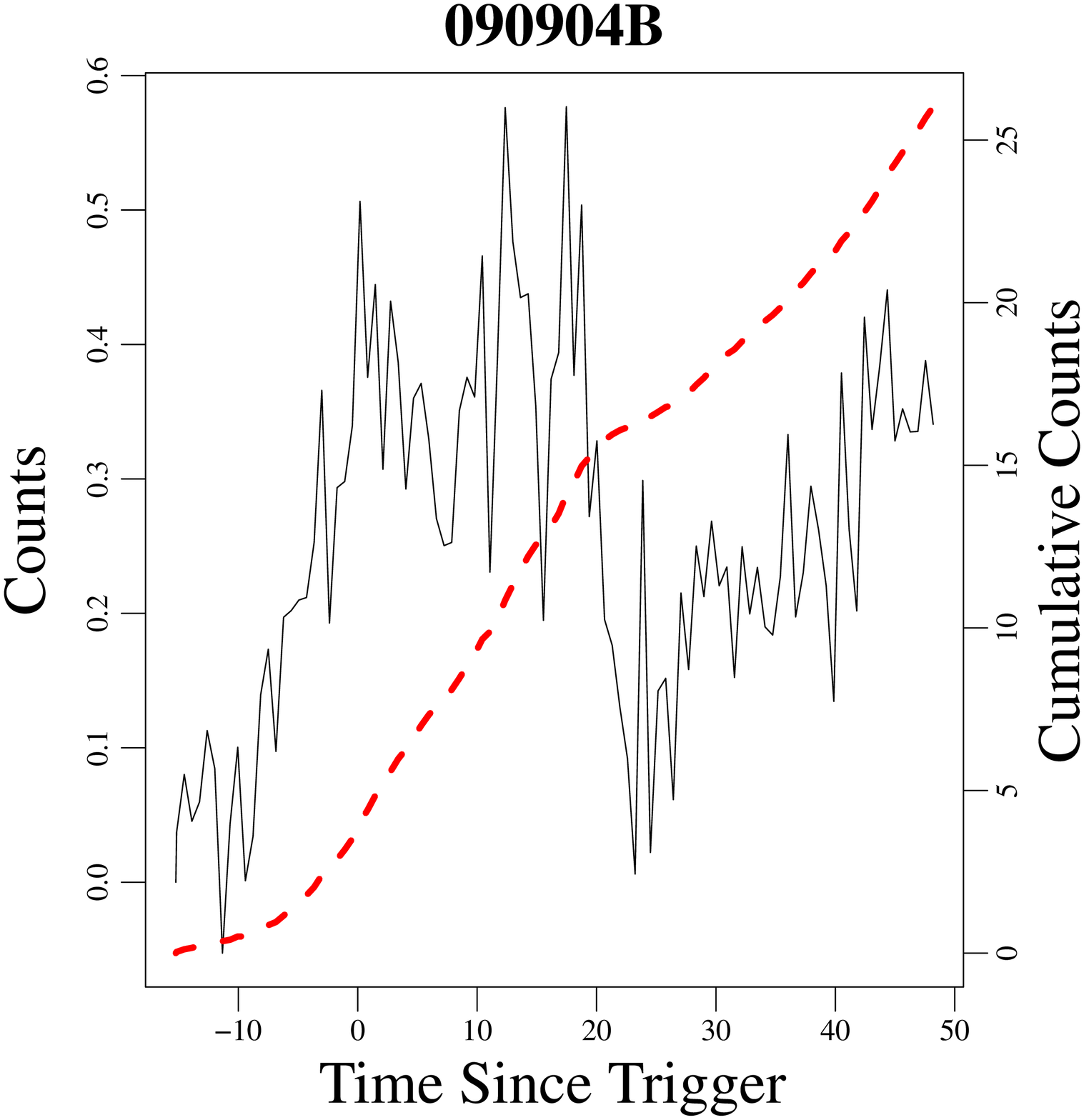}
\end{center}
\end{minipage}
\begin{minipage}{0.25\hsize}
\begin{center}
    \FigureFile(40mm,40mm){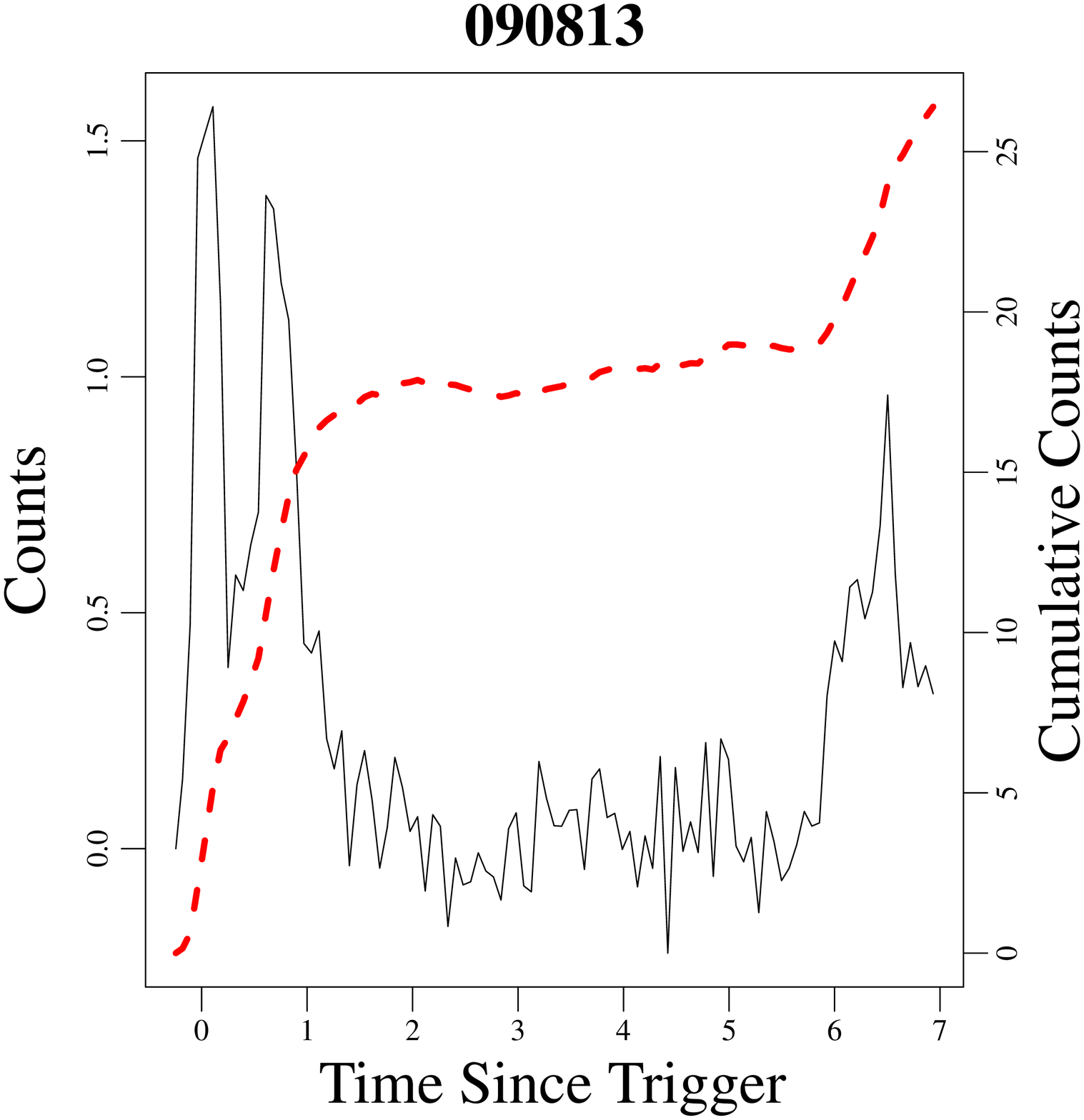}
 \end{center}
\end{minipage}\\
\begin{minipage}{0.25\hsize}
\begin{center}
    \FigureFile(40mm,40mm){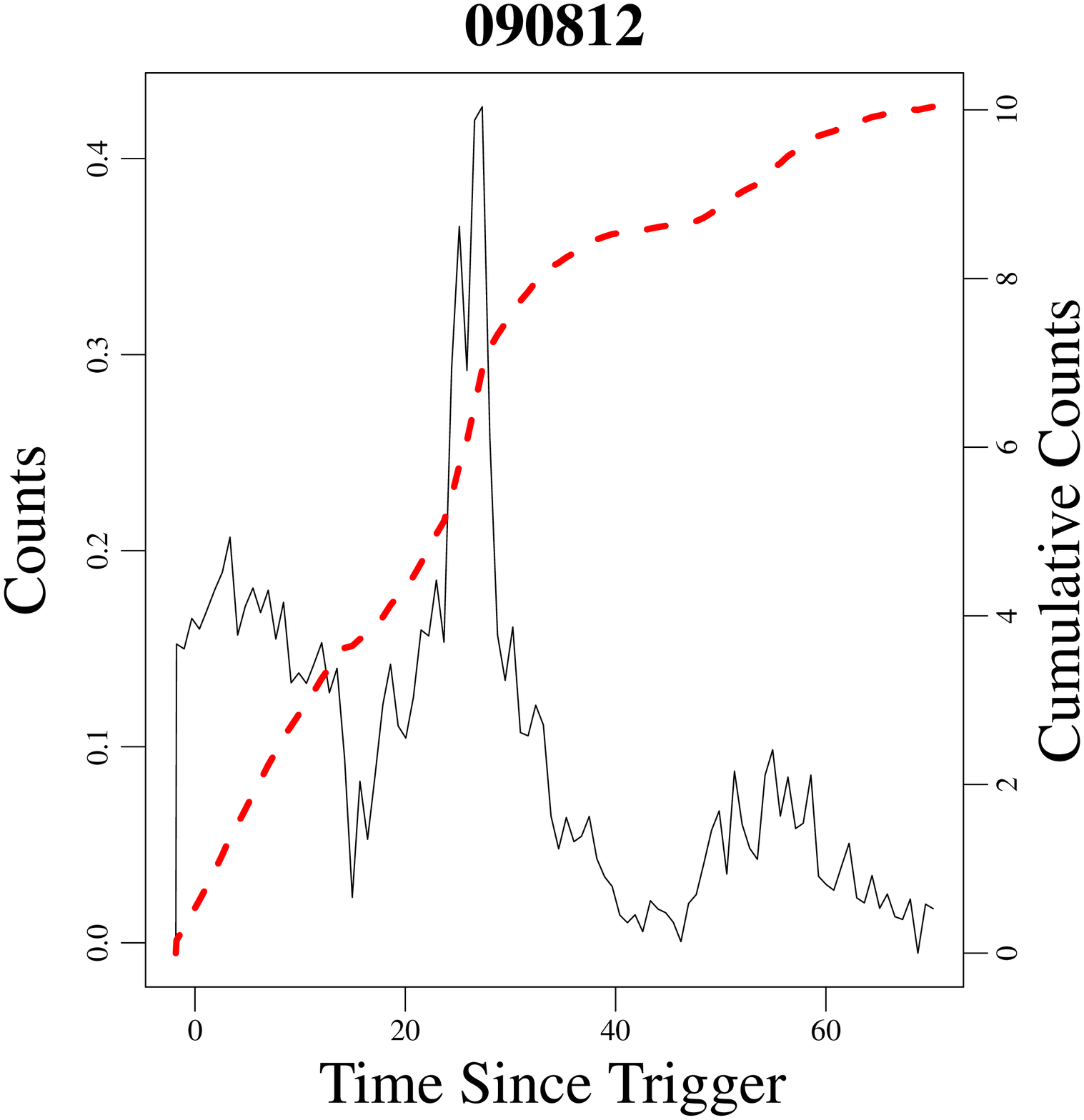}
\end{center}
\end{minipage}
\begin{minipage}{0.25\hsize}
\begin{center}
    \FigureFile(40mm,40mm){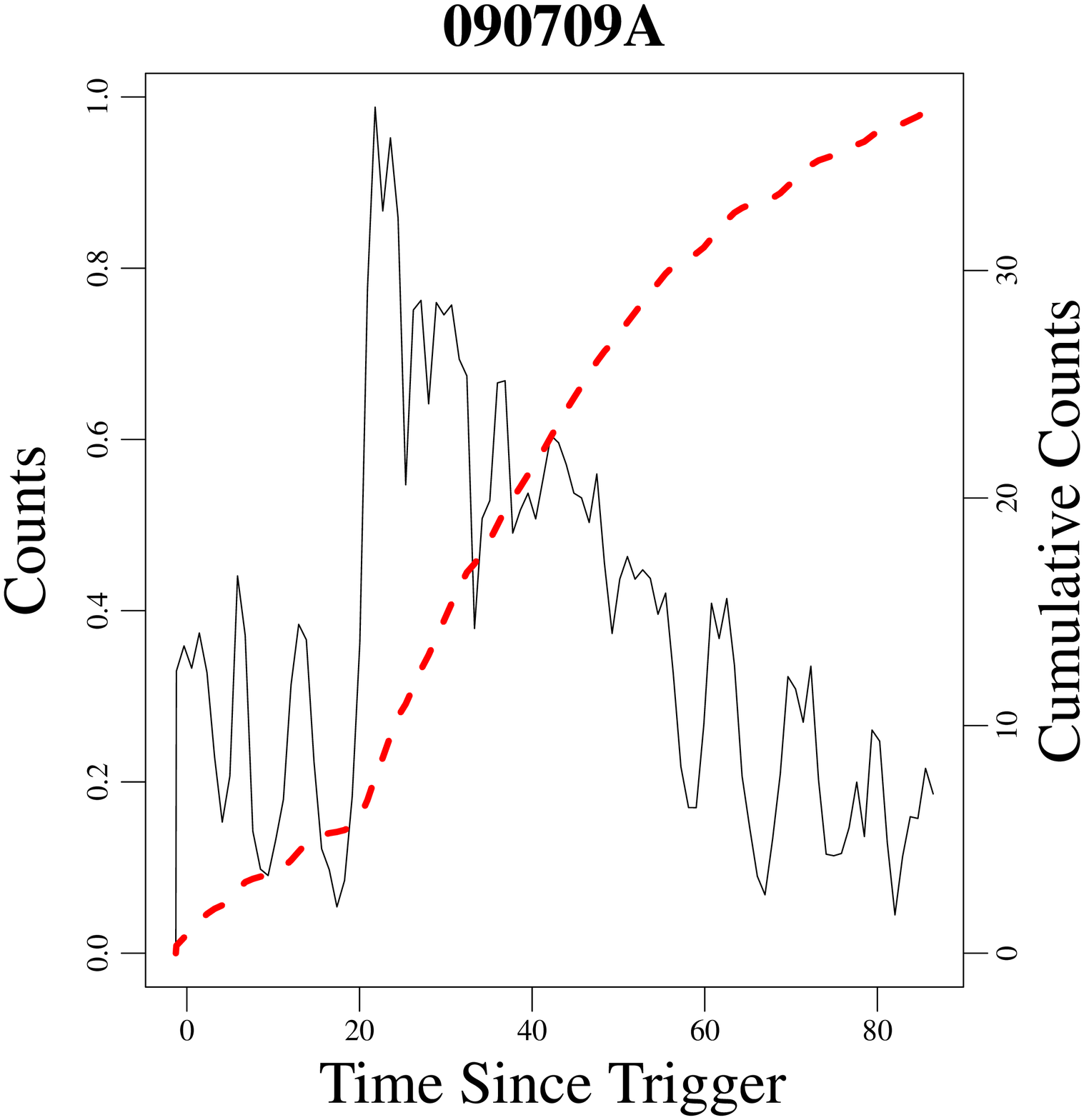}
 \end{center}
\end{minipage}
\begin{minipage}{0.25\hsize}
\begin{center}
    \FigureFile(40mm,40mm){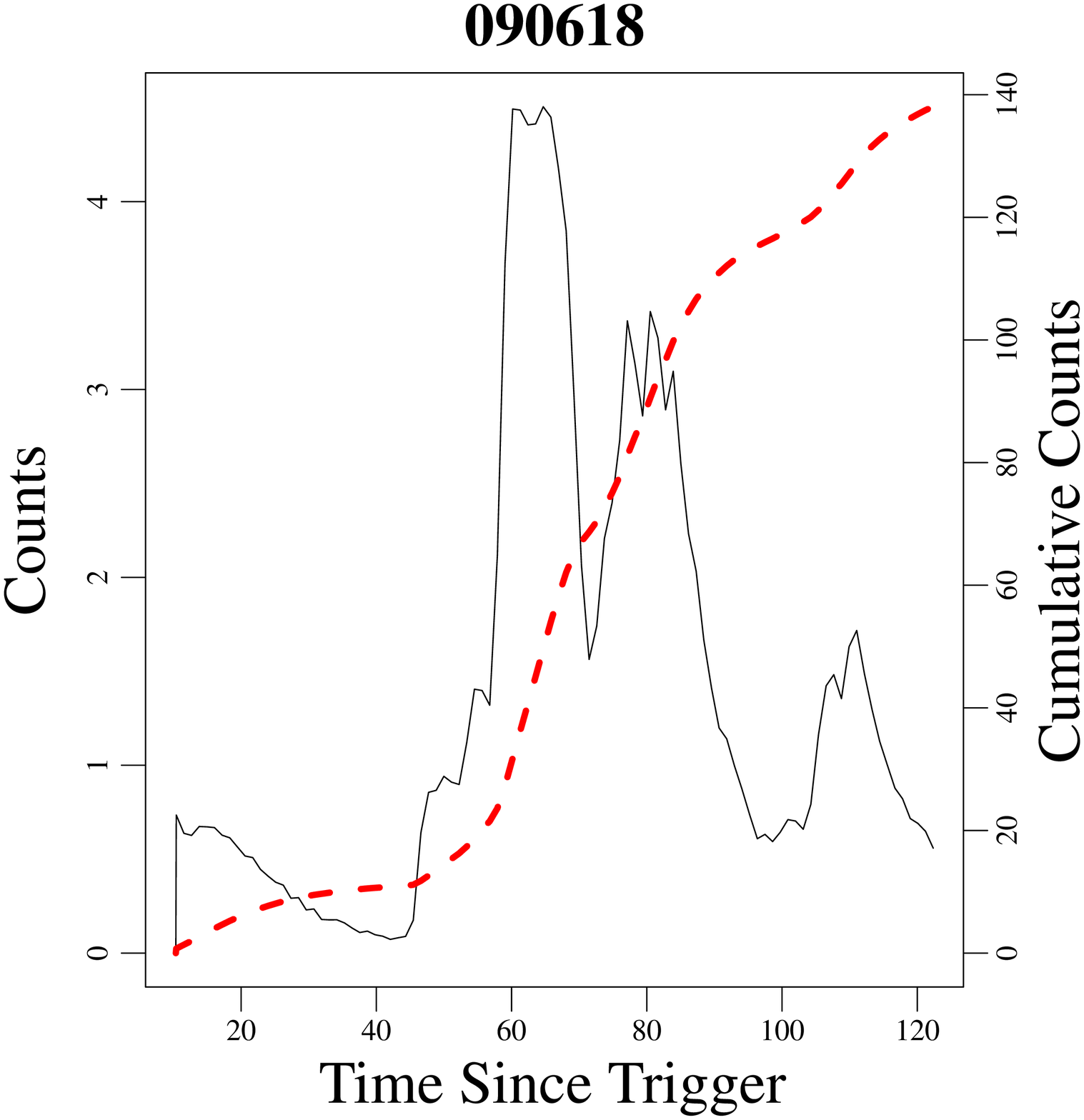}
\end{center}
\end{minipage}
\begin{minipage}{0.25\hsize}
\begin{center}
    \FigureFile(40mm,40mm){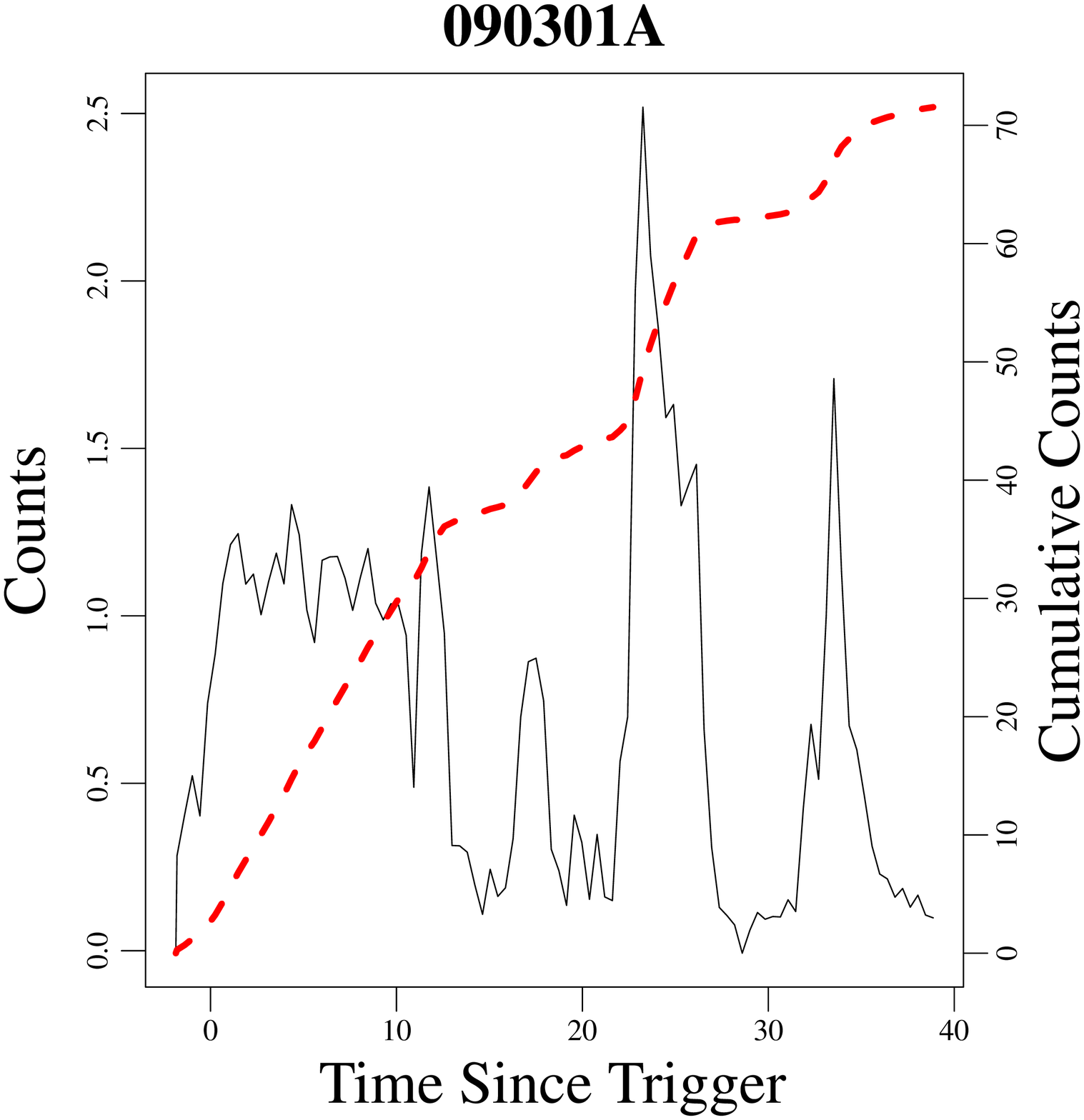}
 \end{center}
\end{minipage}\\
\end{tabular}
 \caption{Light curves (black solid) and cumulative light curves (red doted) of Type I LGRBs.}\label{fig:A1-2}
\end{figure*}
%%%%%%%%%%%%%%%%%%%%%%
\begin{figure*}[htb]
\begin{tabular}{cccc}
\begin{minipage}{0.25\hsize}
\begin{center}
    \FigureFile(40mm,40mm){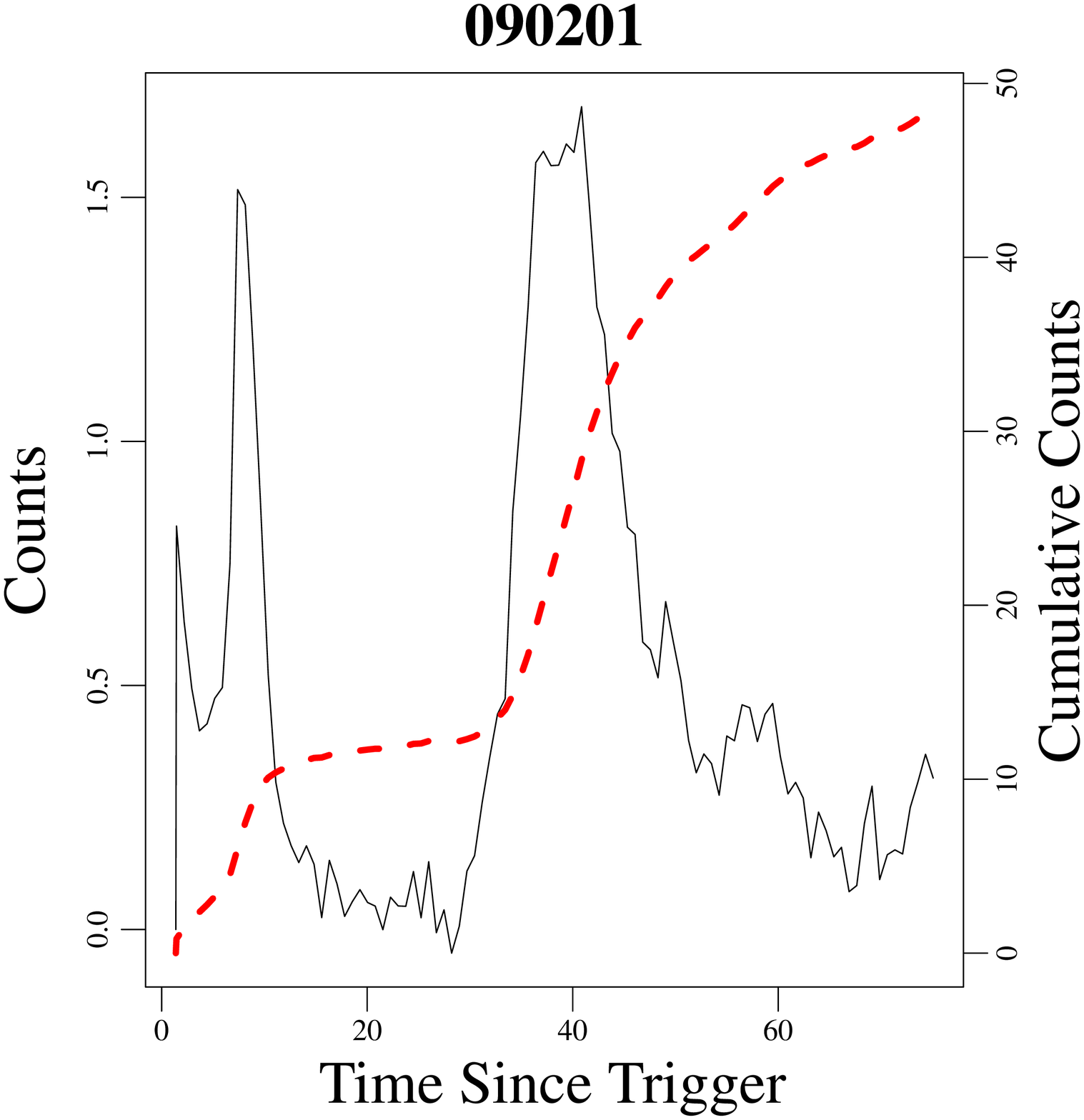}
\end{center}
\end{minipage}
\begin{minipage}{0.25\hsize}
\begin{center}
    \FigureFile(40mm,40mm){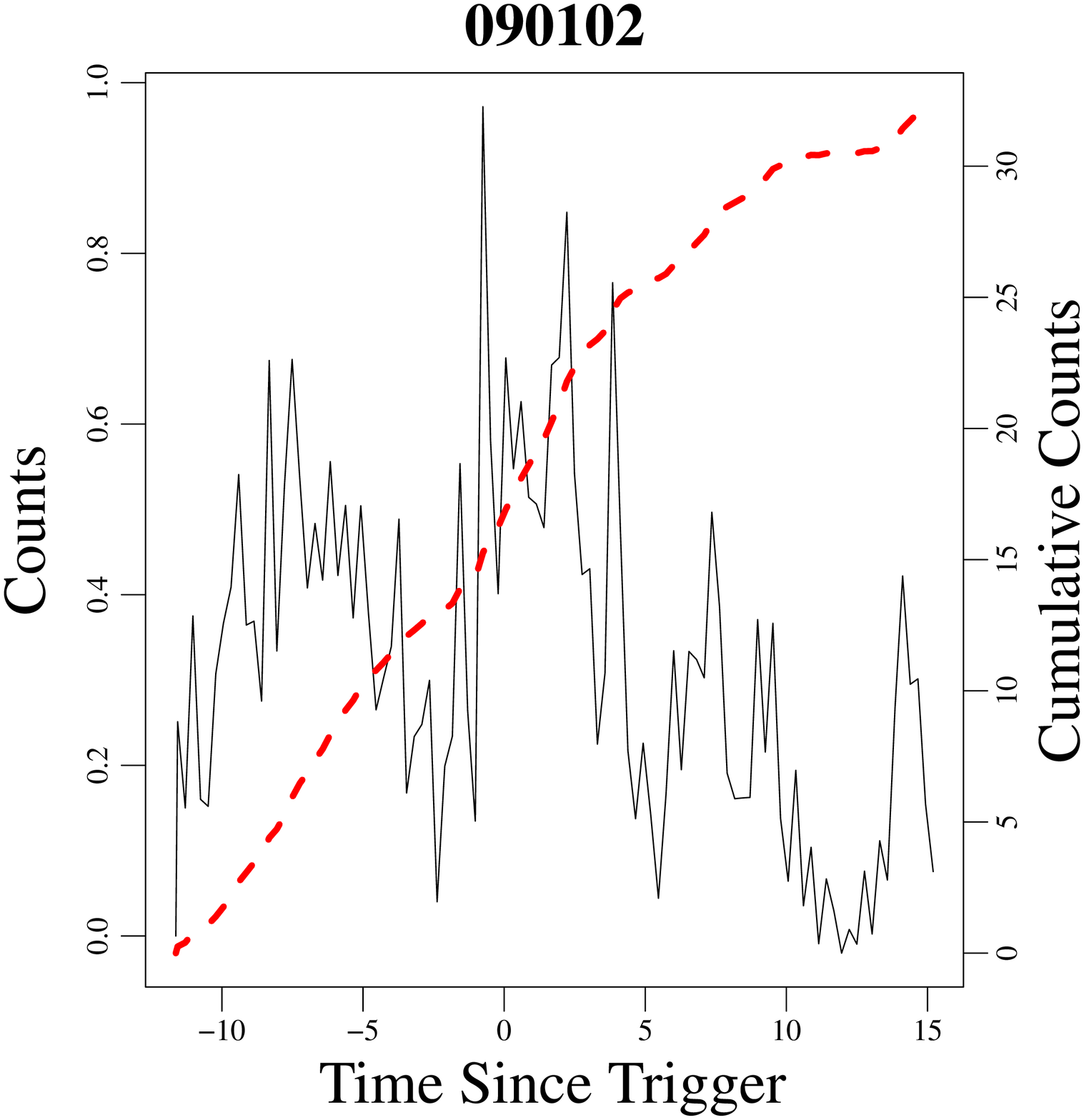}
 \end{center}
\end{minipage}
\begin{minipage}{0.25\hsize}
\begin{center}
    \FigureFile(40mm,40mm){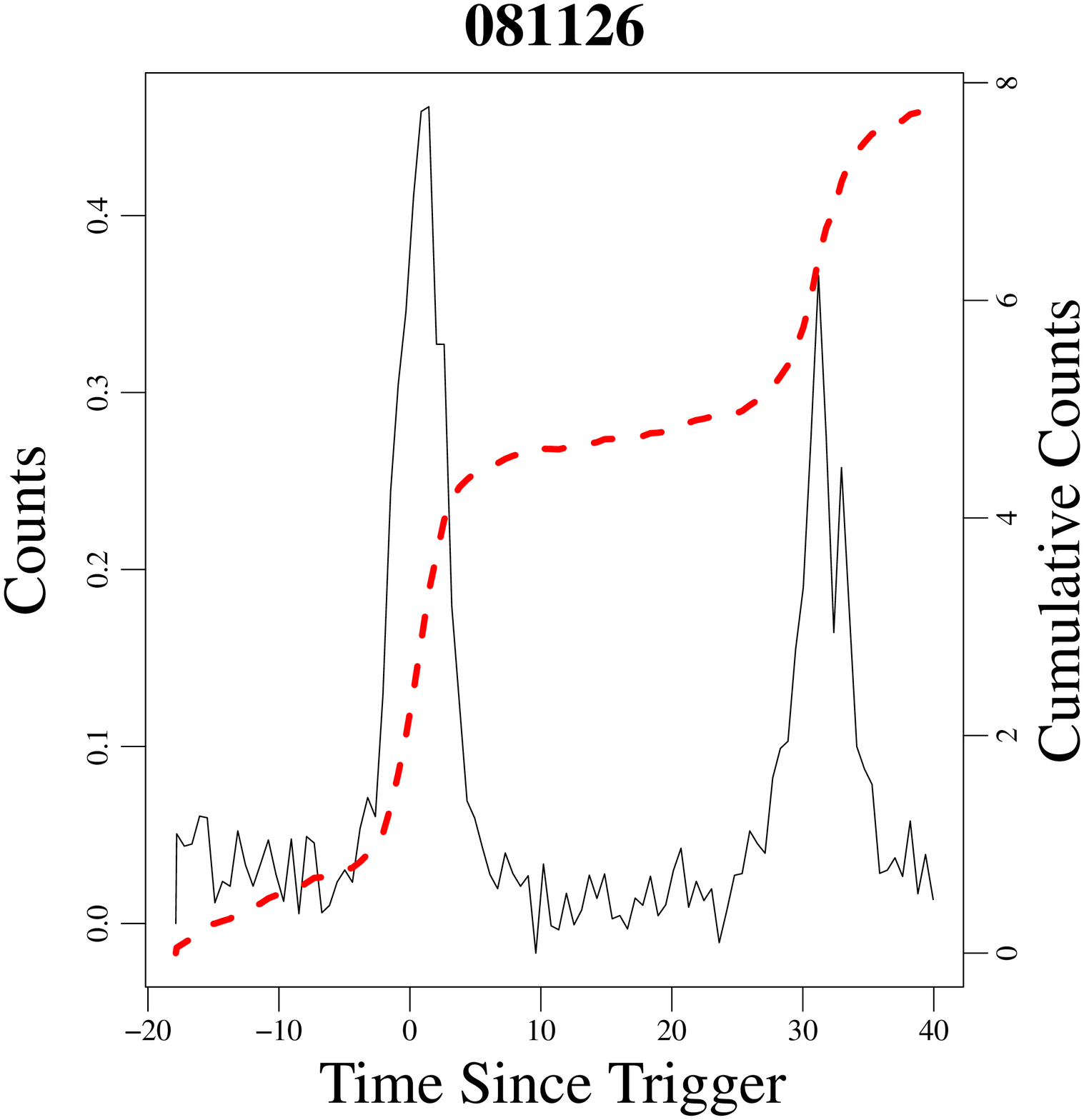}
\end{center}
\end{minipage}
\begin{minipage}{0.25\hsize}
\begin{center}
    \FigureFile(40mm,40mm){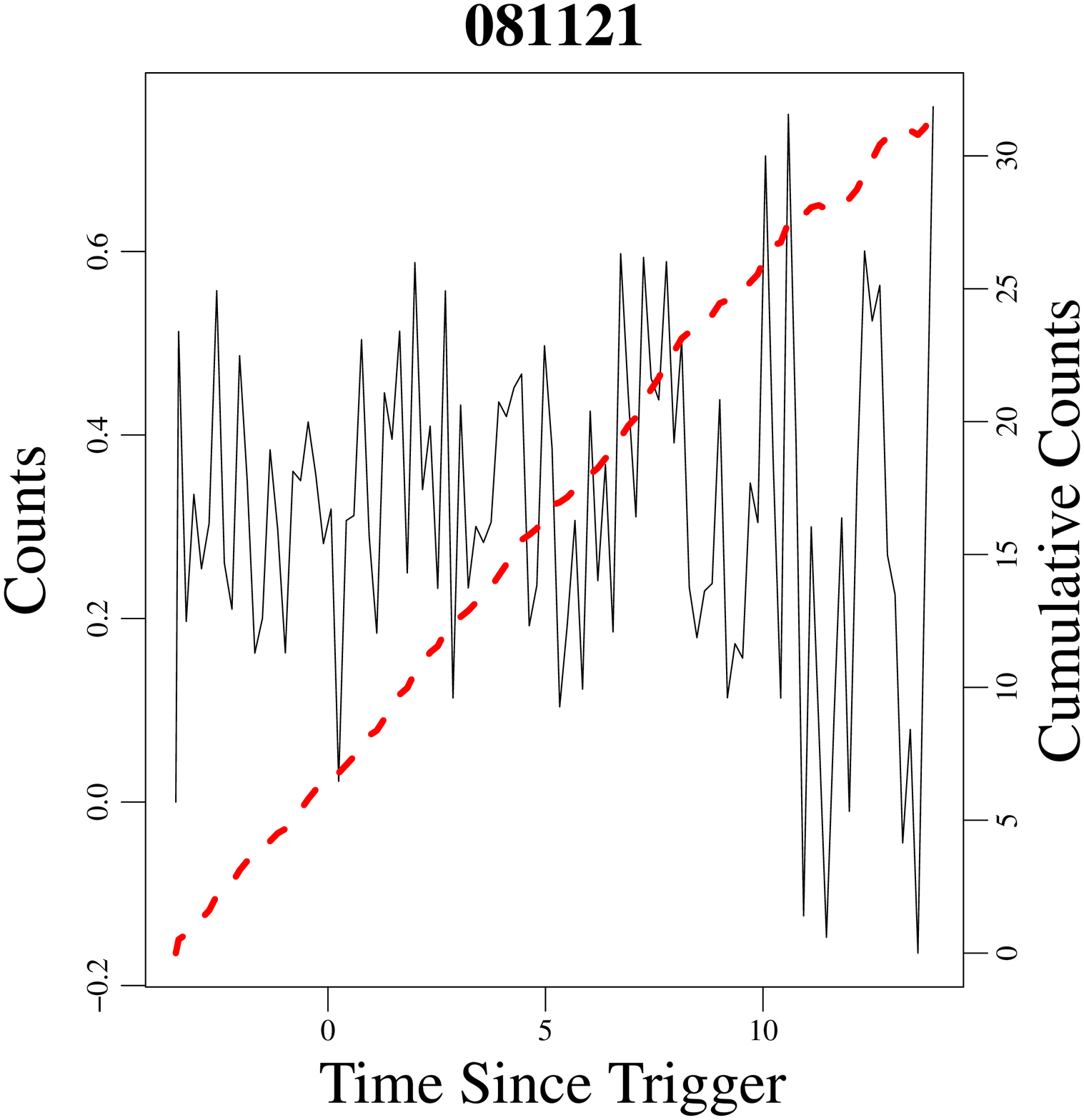}
 \end{center}
\end{minipage}\\
\begin{minipage}{0.25\hsize}
\begin{center}
    \FigureFile(40mm,40mm){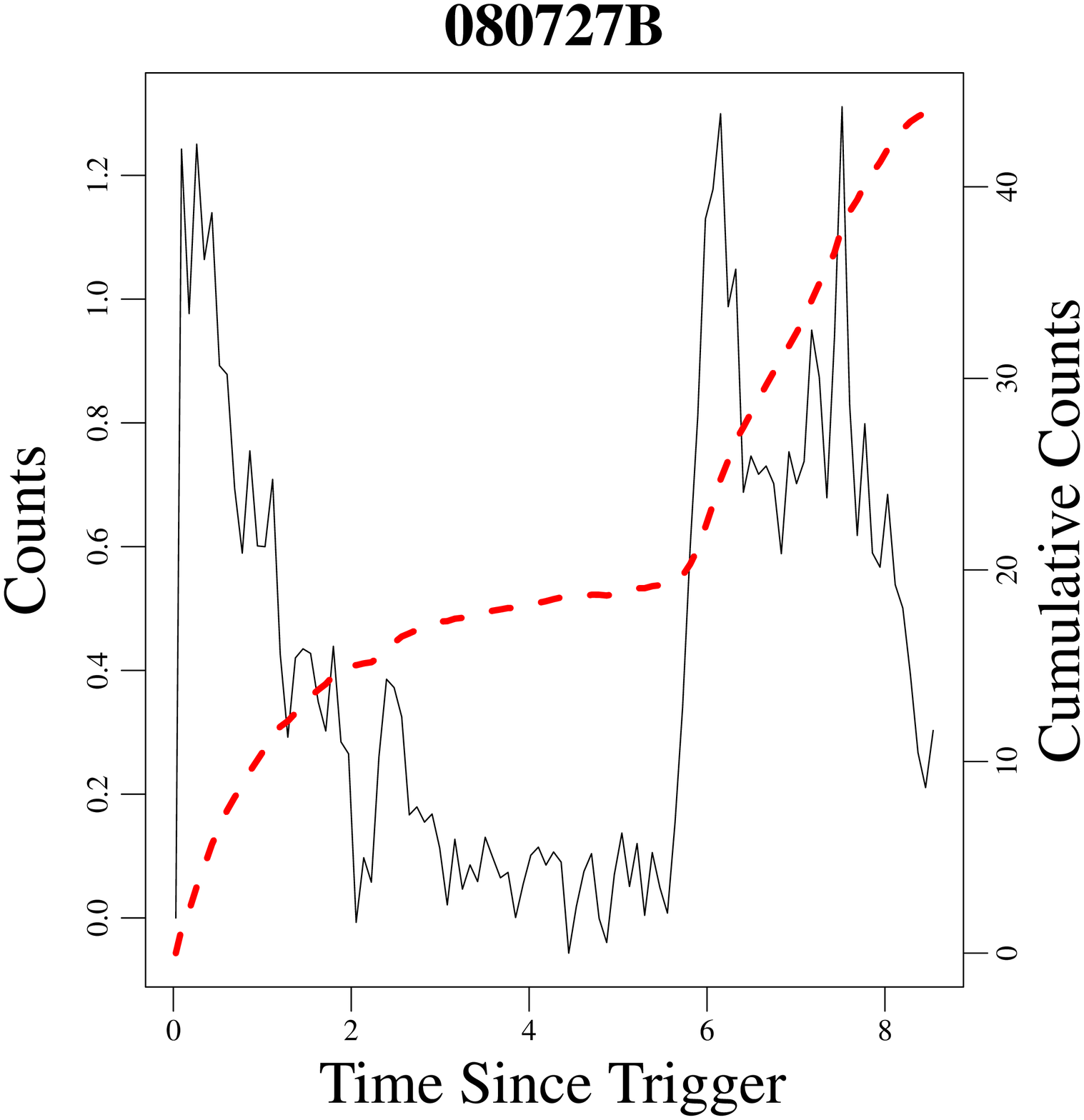}
\end{center}
\end{minipage}
\begin{minipage}{0.25\hsize}
\begin{center}
    \FigureFile(40mm,40mm){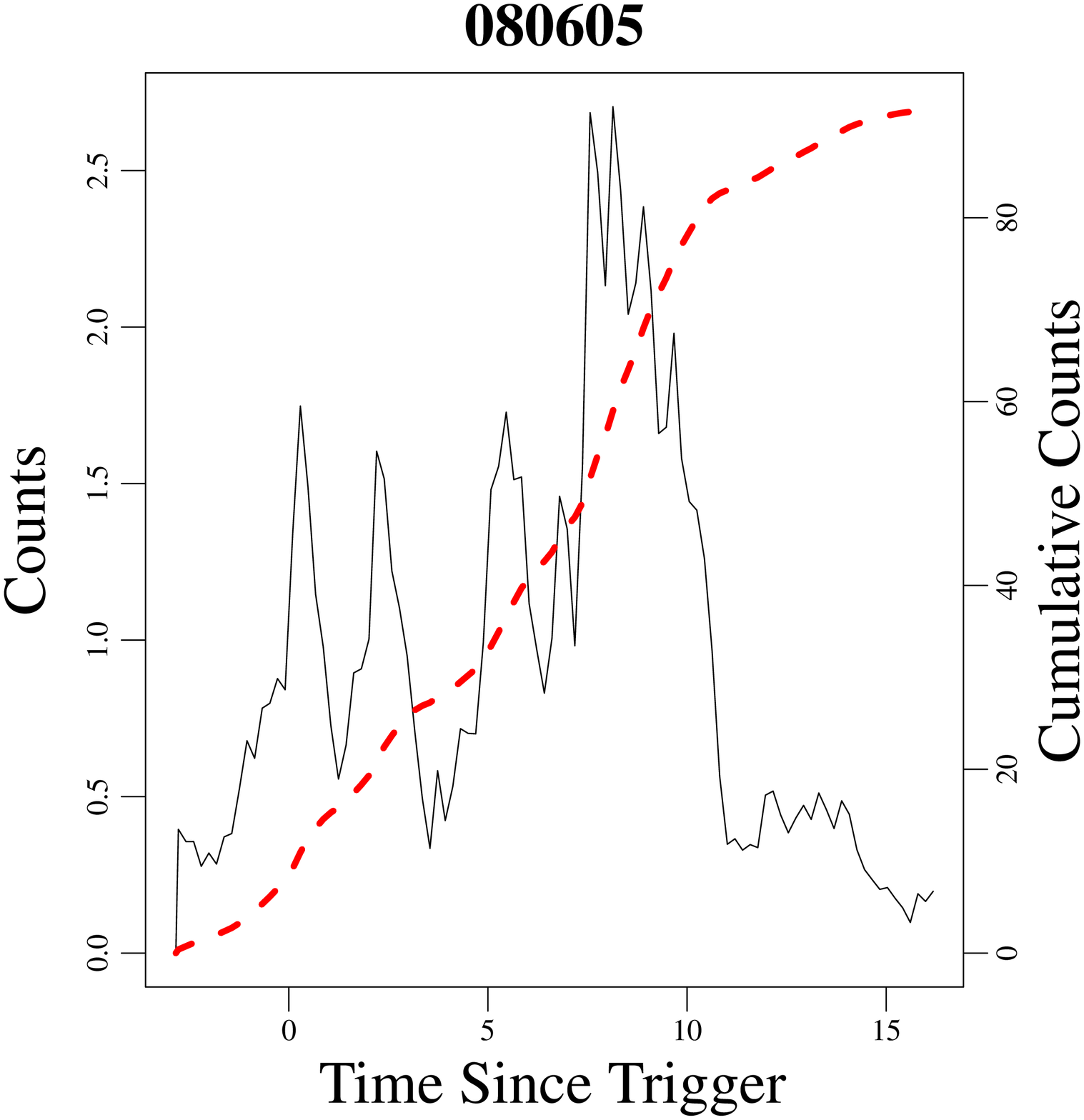}
 \end{center}
\end{minipage}
\begin{minipage}{0.25\hsize}
\begin{center}
    \FigureFile(40mm,40mm){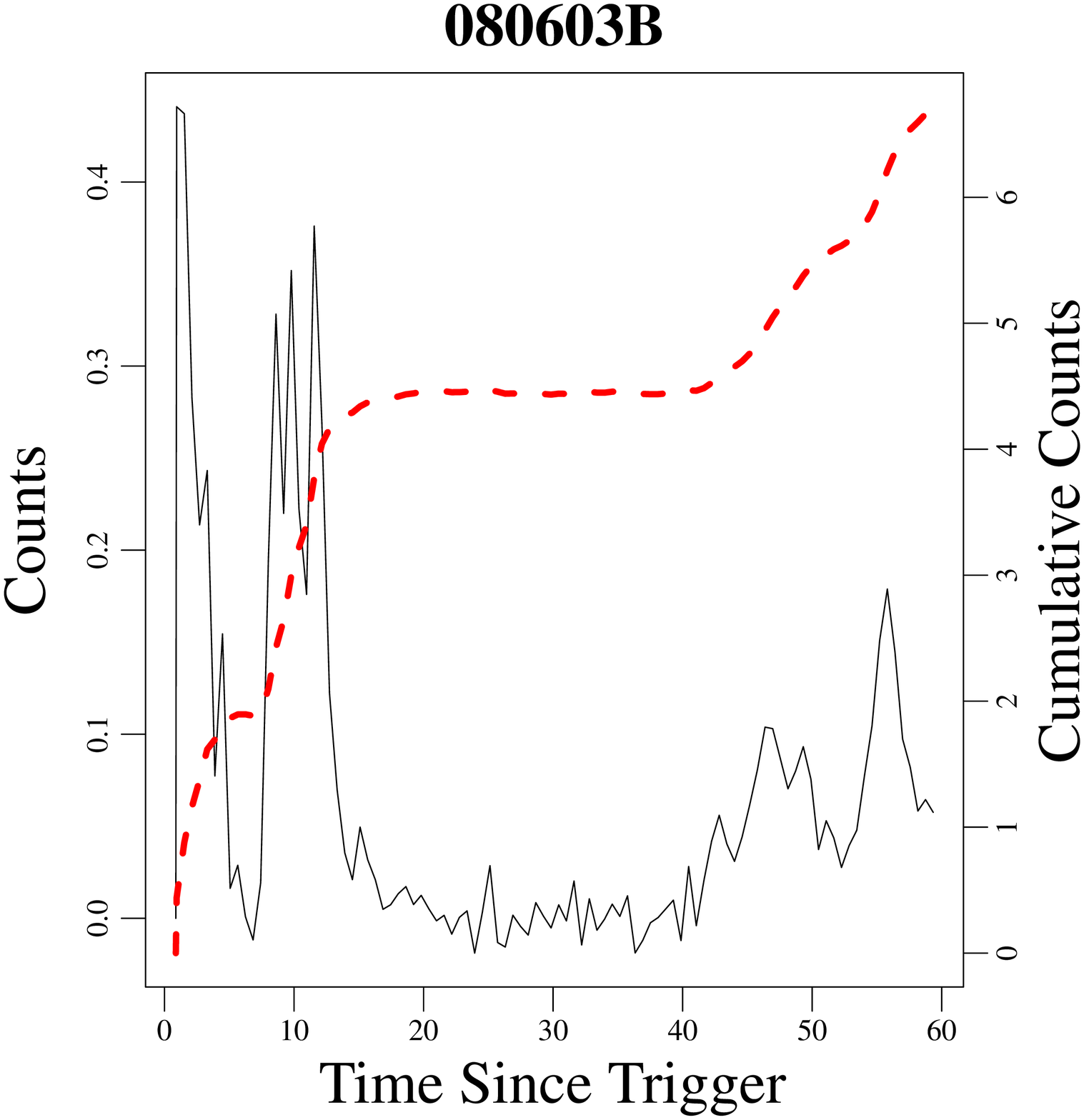}
\end{center}
\end{minipage}
\begin{minipage}{0.25\hsize}
\begin{center}
    \FigureFile(40mm,40mm){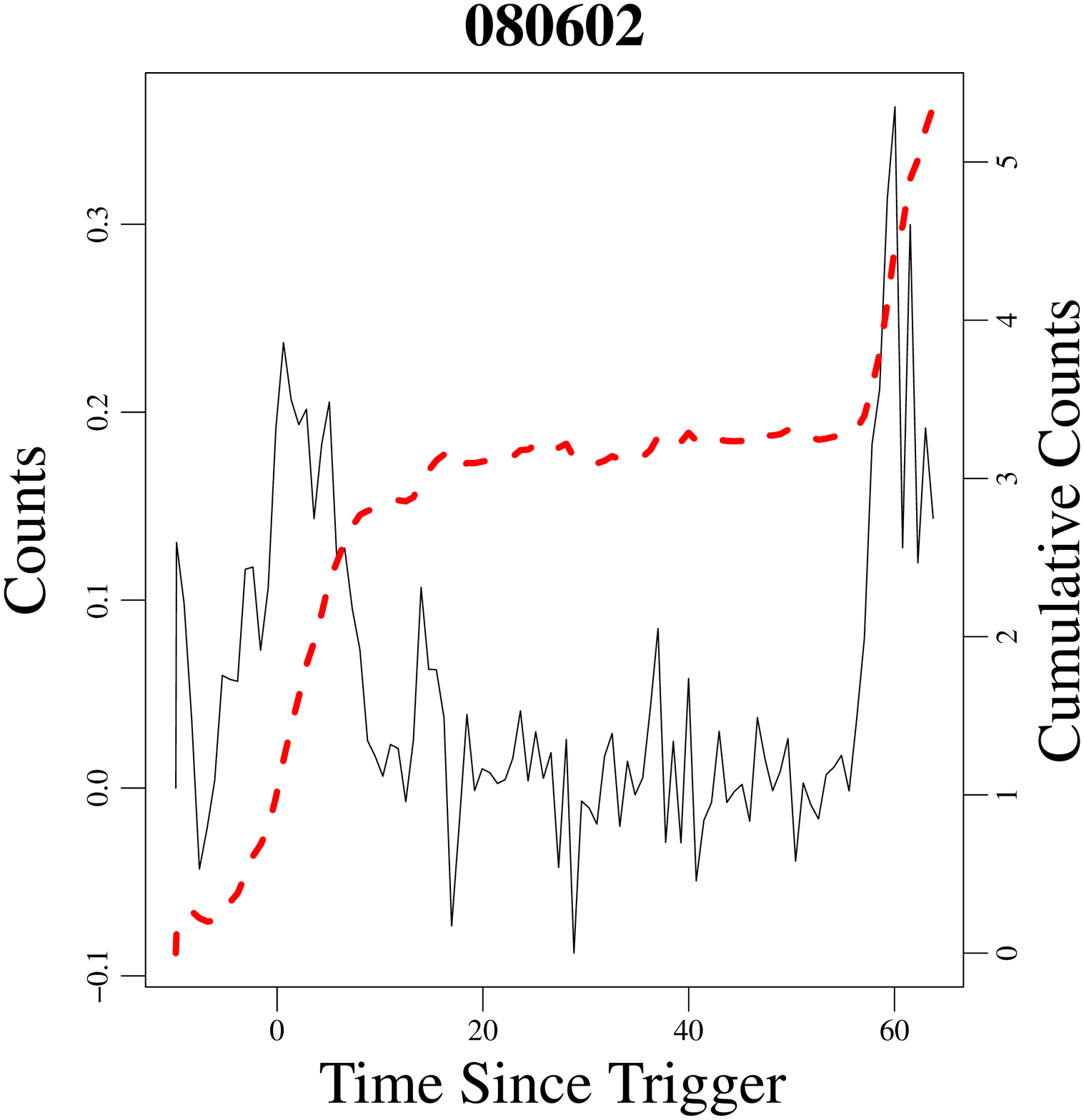}
 \end{center}
\end{minipage}\\
\begin{minipage}{0.25\hsize}
\begin{center}
    \FigureFile(40mm,40mm){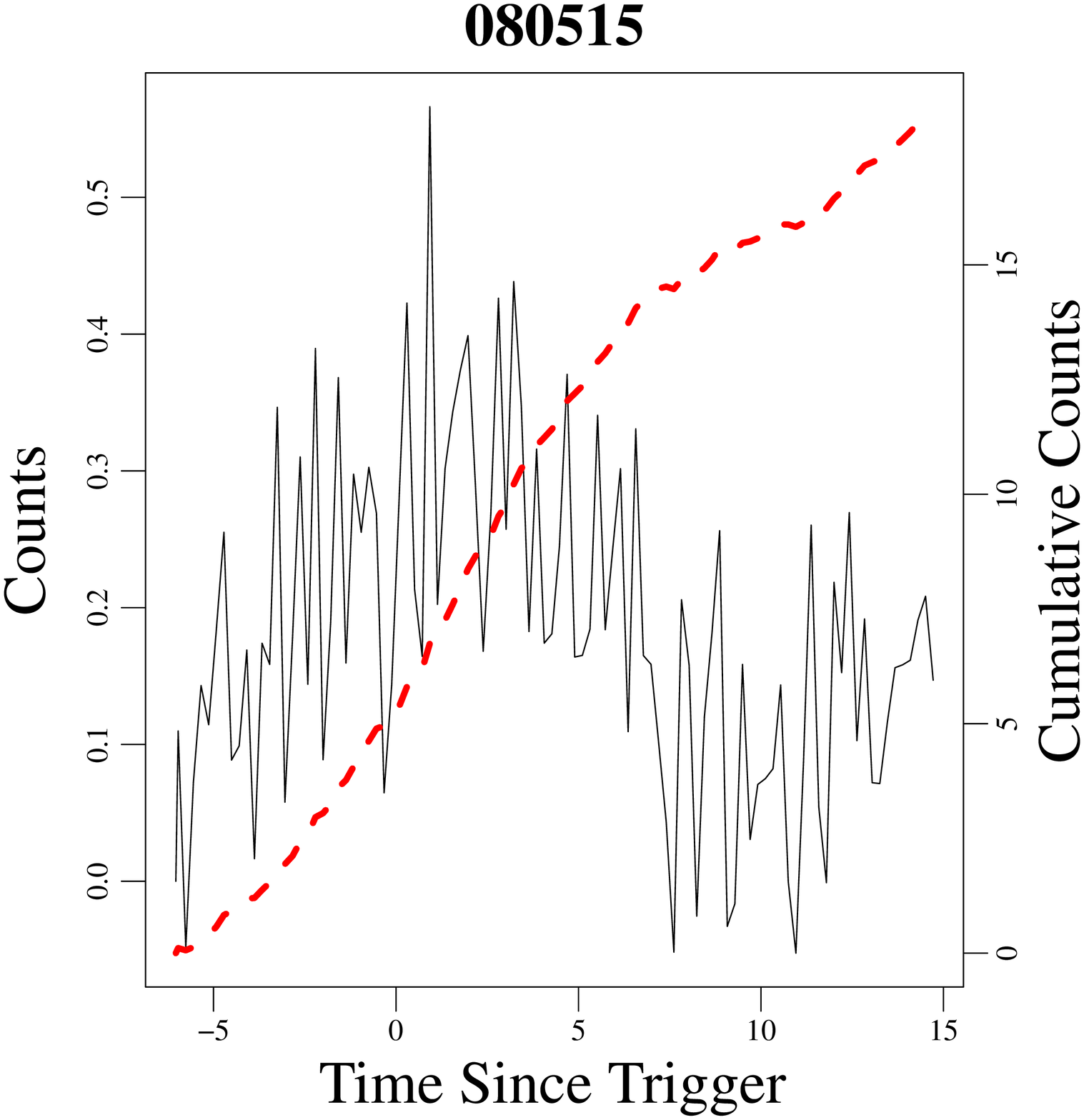}
\end{center}
\end{minipage}
\begin{minipage}{0.25\hsize}
\begin{center}
    \FigureFile(40mm,40mm){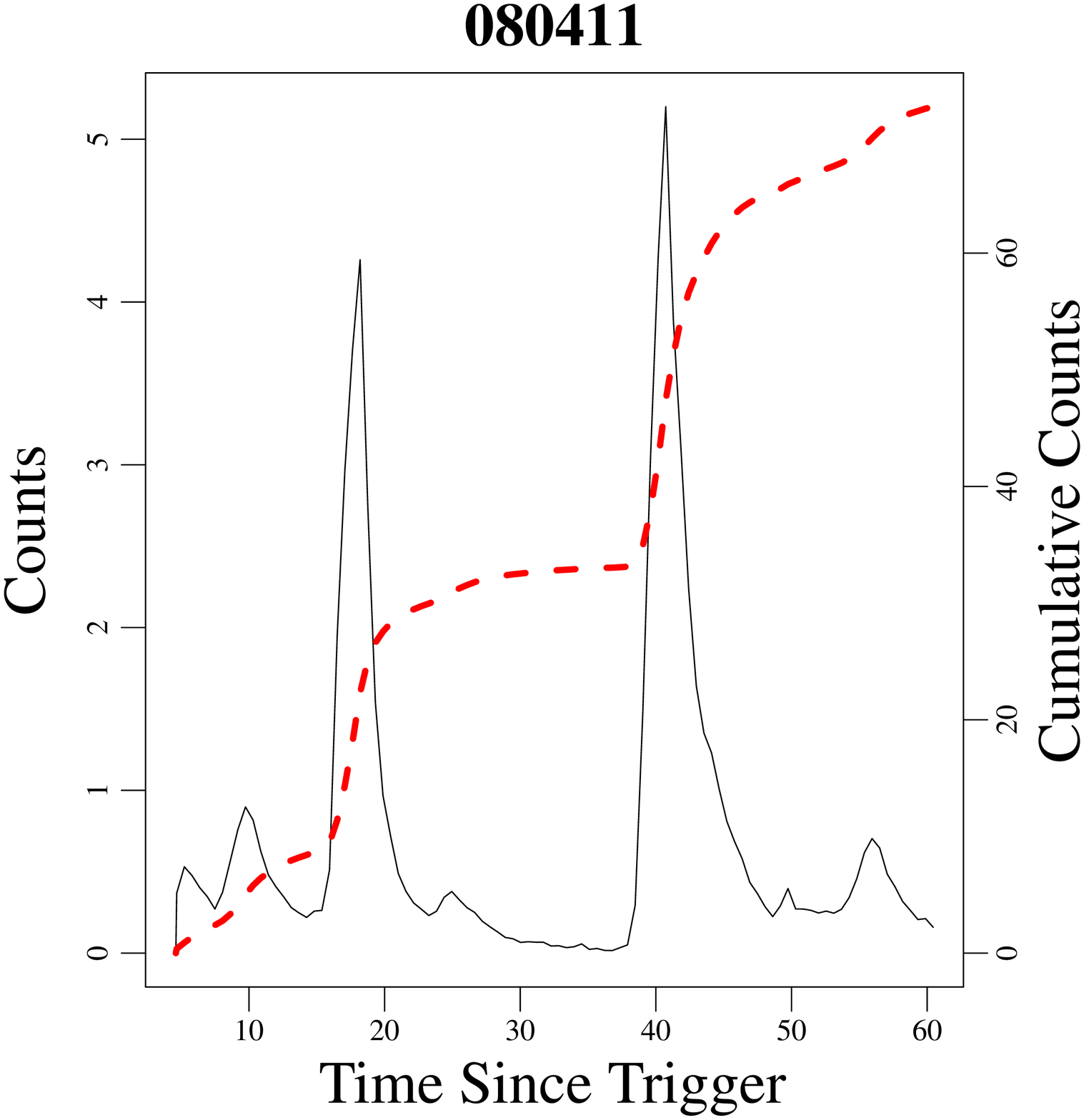}
 \end{center}
\end{minipage}
\begin{minipage}{0.25\hsize}
\begin{center}
    \FigureFile(40mm,40mm){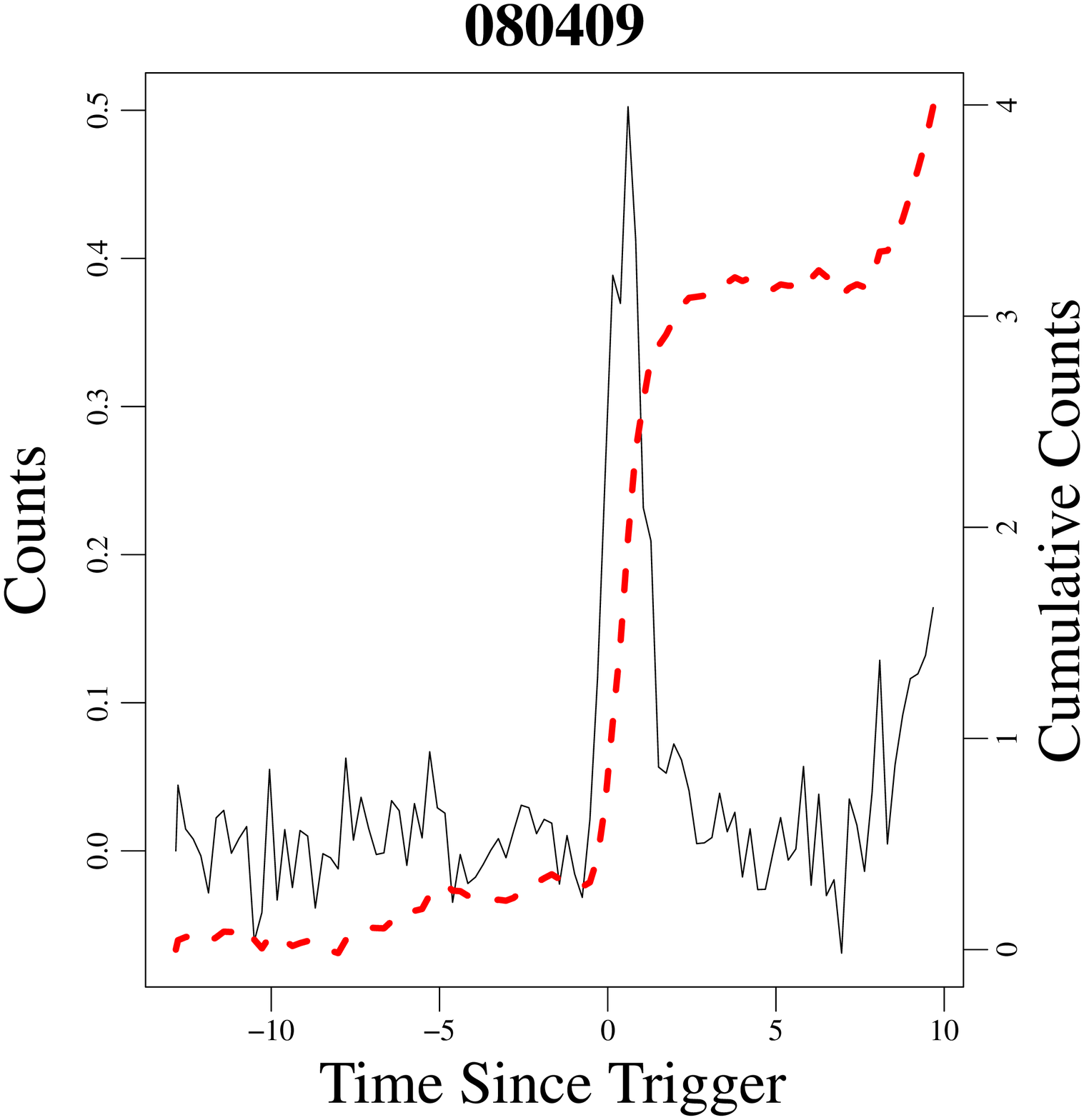}
\end{center}
\end{minipage}
\begin{minipage}{0.25\hsize}
\begin{center}
    \FigureFile(40mm,40mm){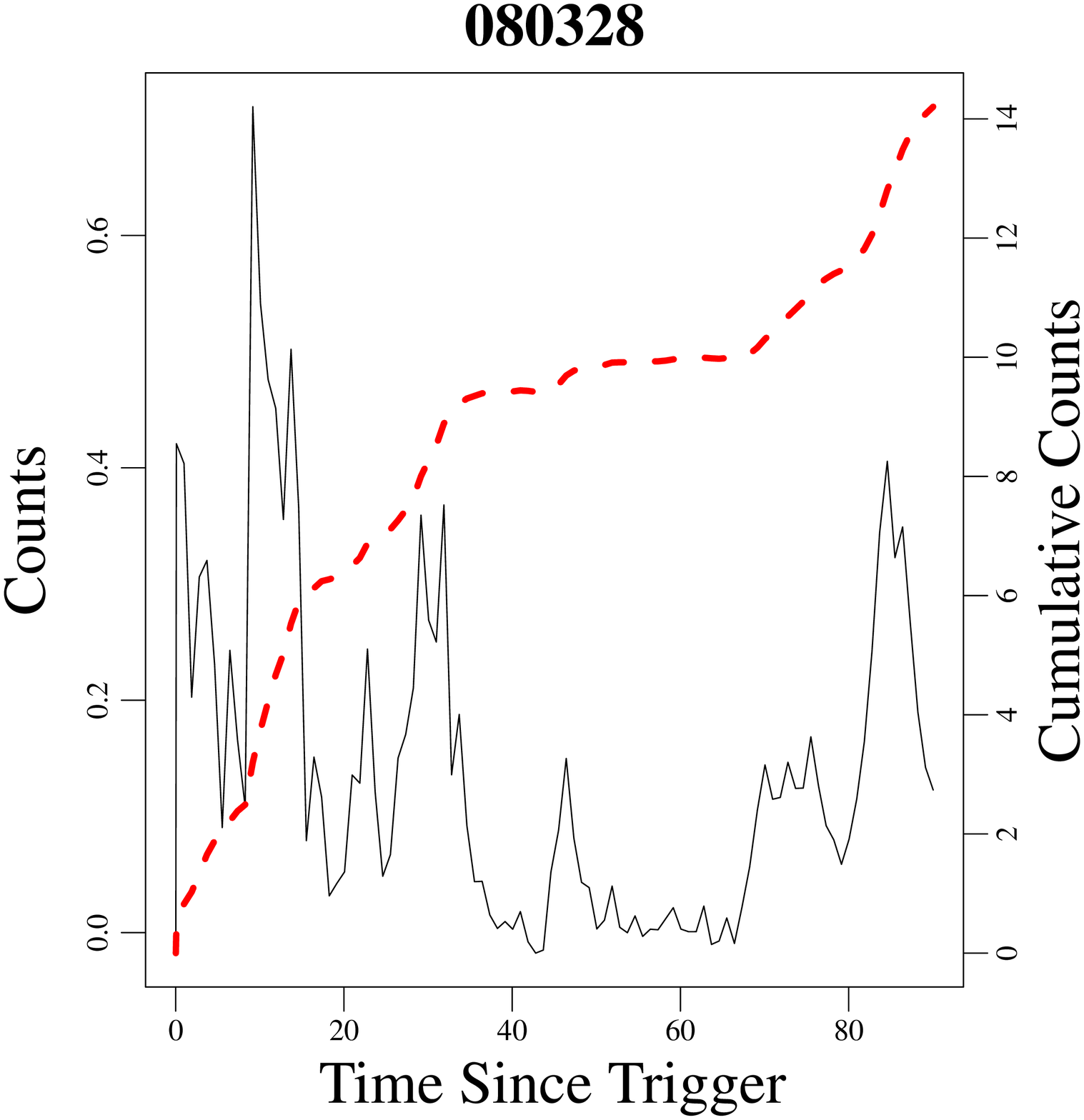}
 \end{center}
\end{minipage}\\
\begin{minipage}{0.25\hsize}
\begin{center}
    \FigureFile(40mm,40mm){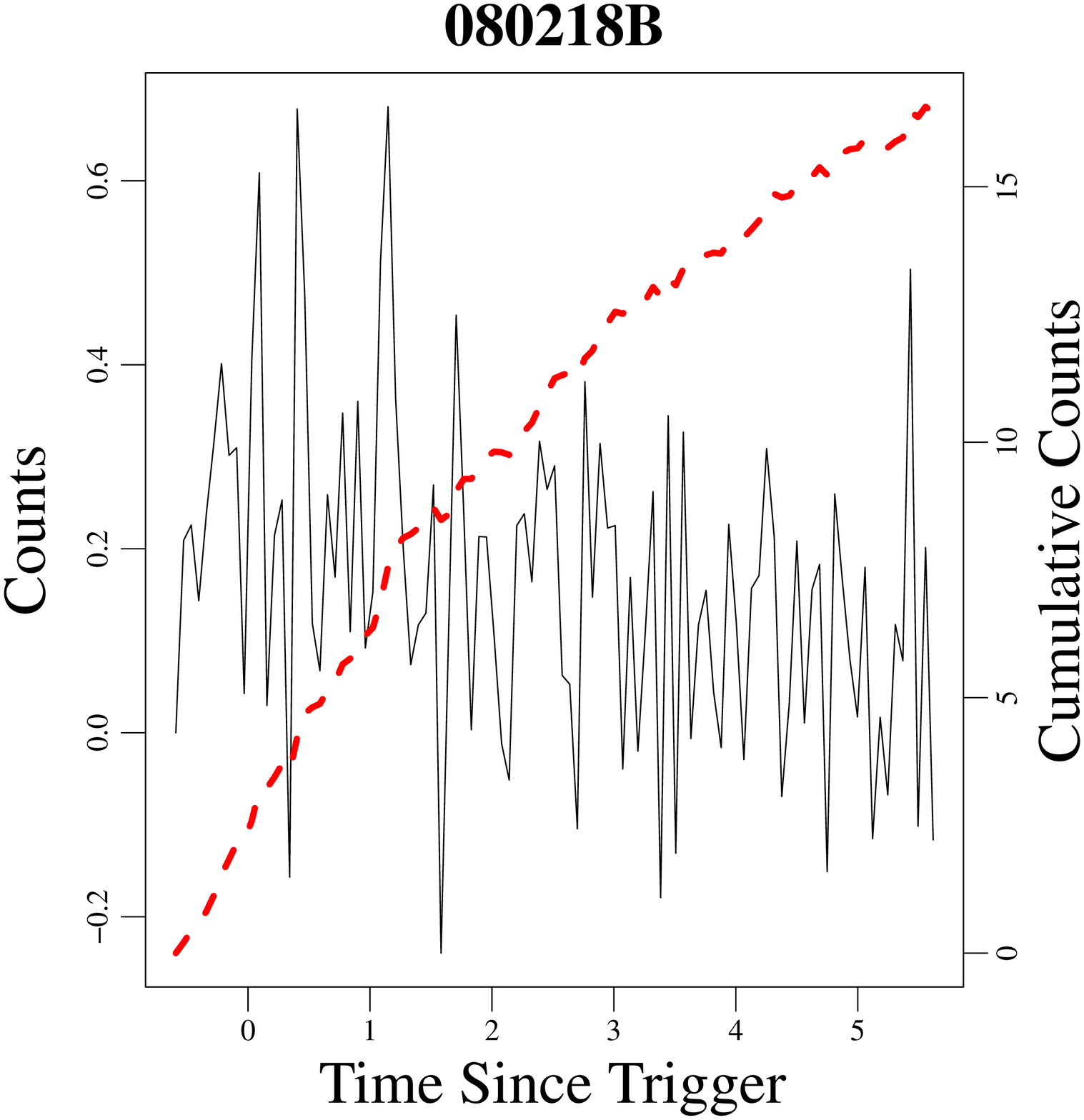}
\end{center}
\end{minipage}
\begin{minipage}{0.25\hsize}
\begin{center}
    \FigureFile(40mm,40mm){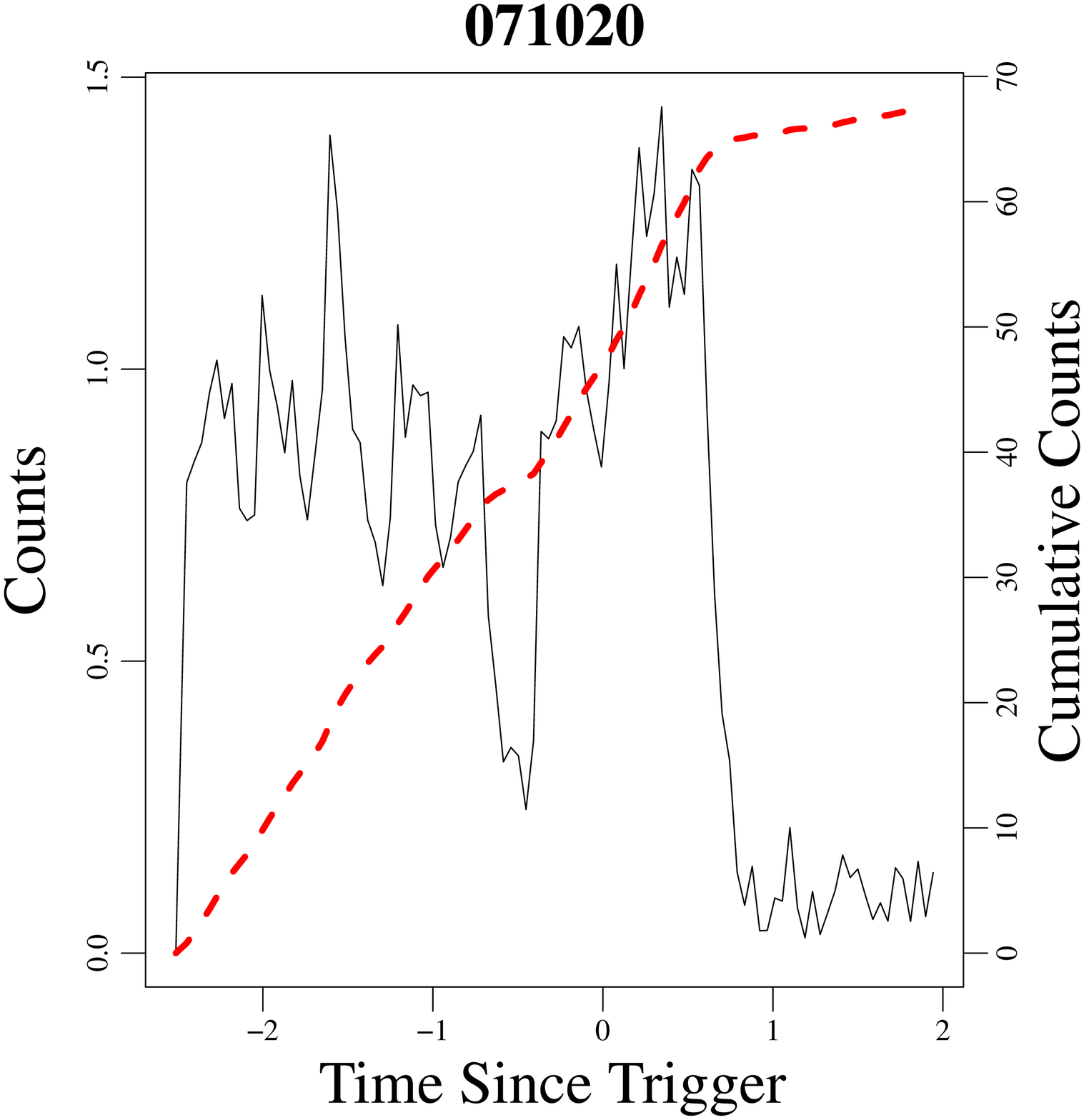}
 \end{center}
\end{minipage}
\begin{minipage}{0.25\hsize}
\begin{center}
    \FigureFile(40mm,40mm){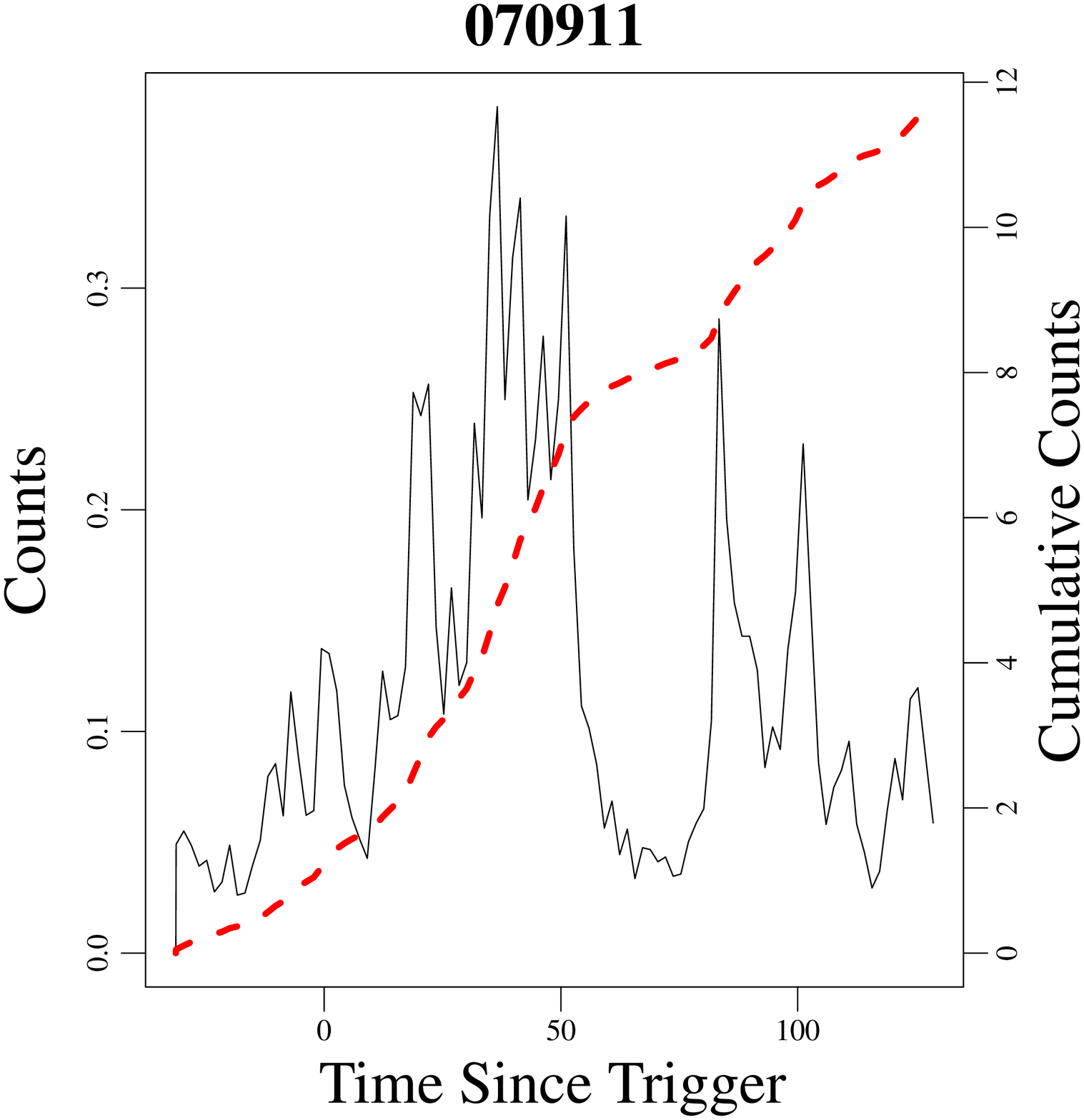}
\end{center}
\end{minipage}
\begin{minipage}{0.25\hsize}
\begin{center}
    \FigureFile(40mm,40mm){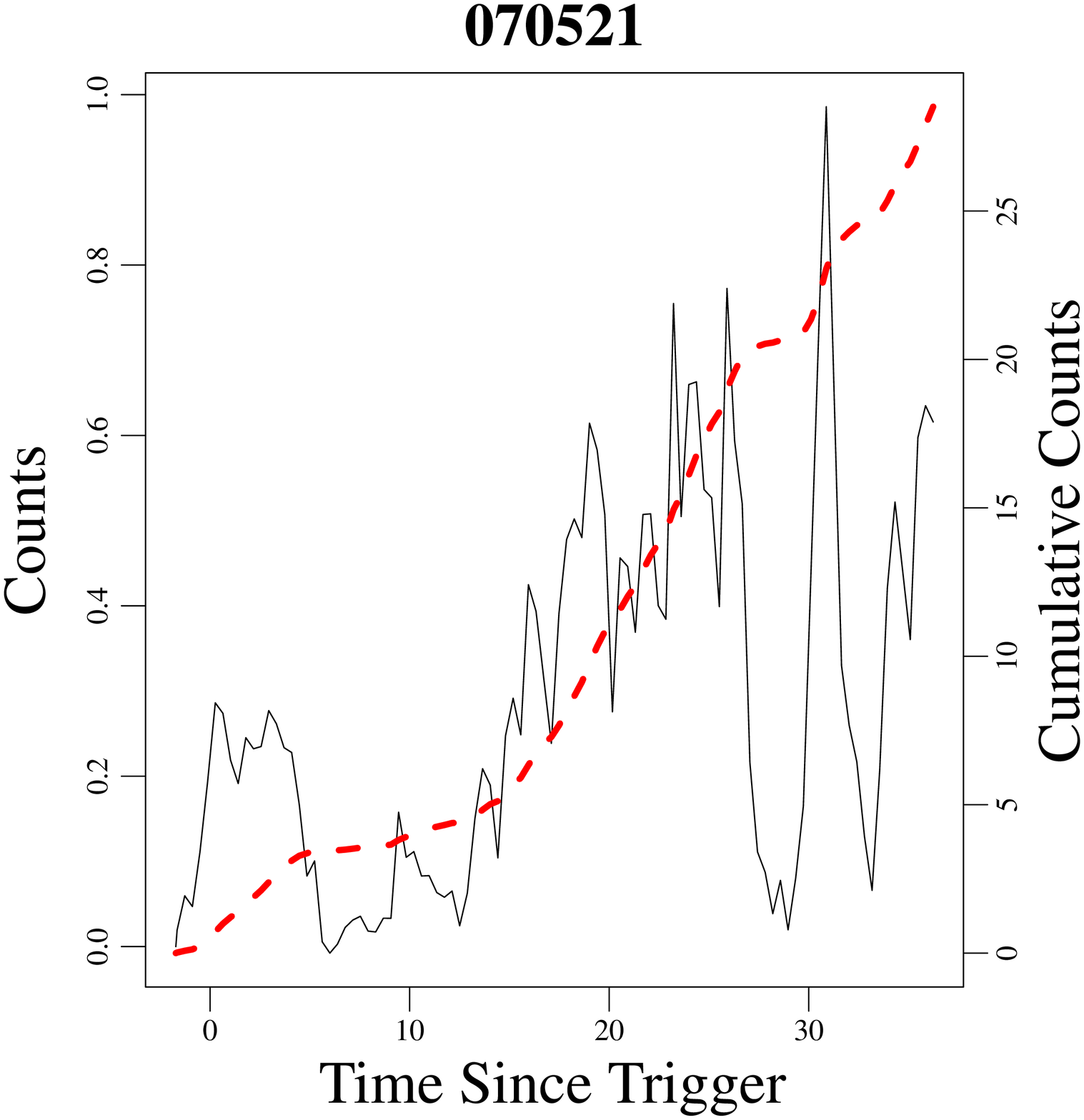}
 \end{center}
\end{minipage}\\
\begin{minipage}{0.25\hsize}
\begin{center}
    \FigureFile(40mm,40mm){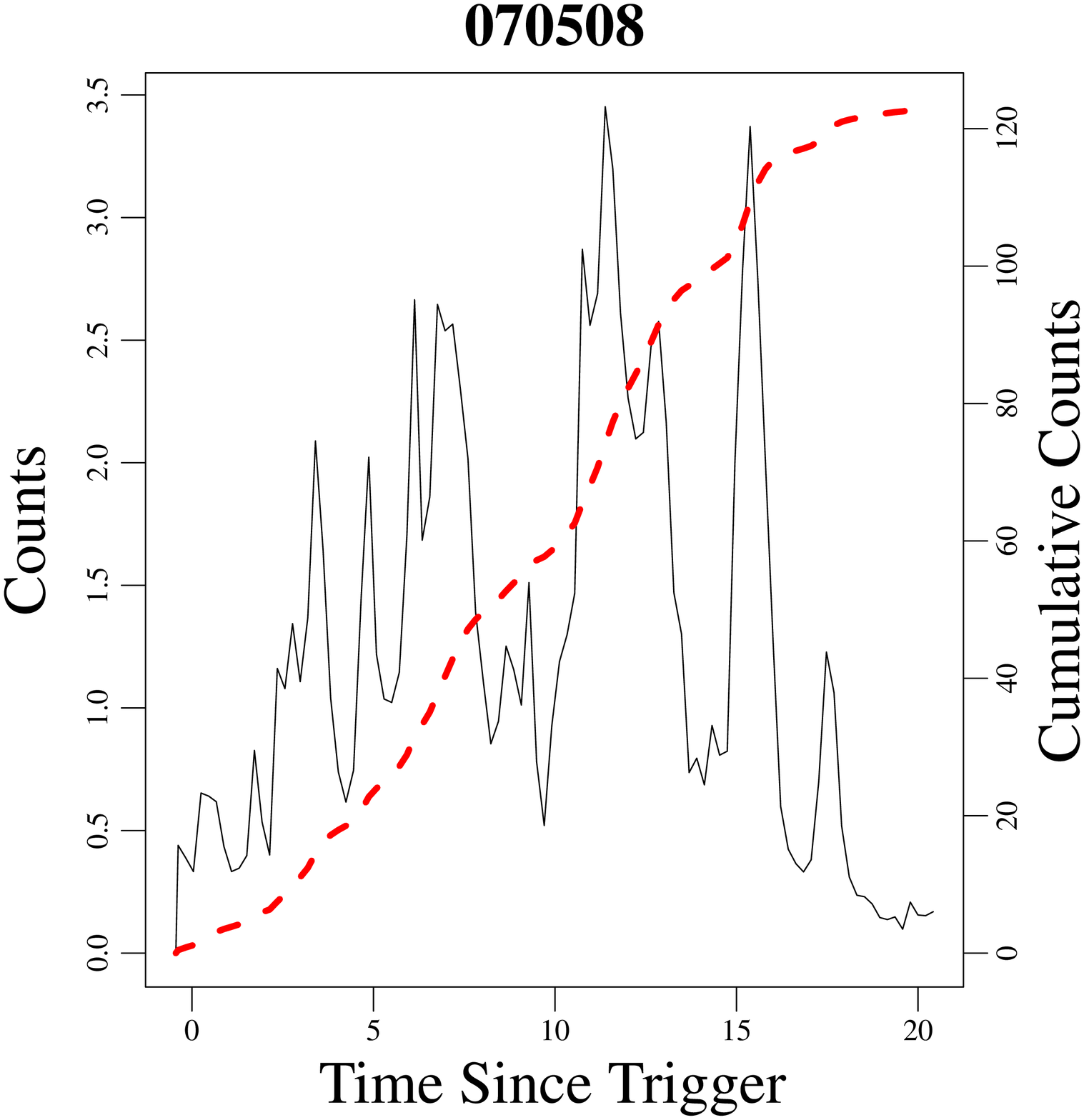}
\end{center}
\end{minipage}
\begin{minipage}{0.25\hsize}
\begin{center}
    \FigureFile(40mm,40mm){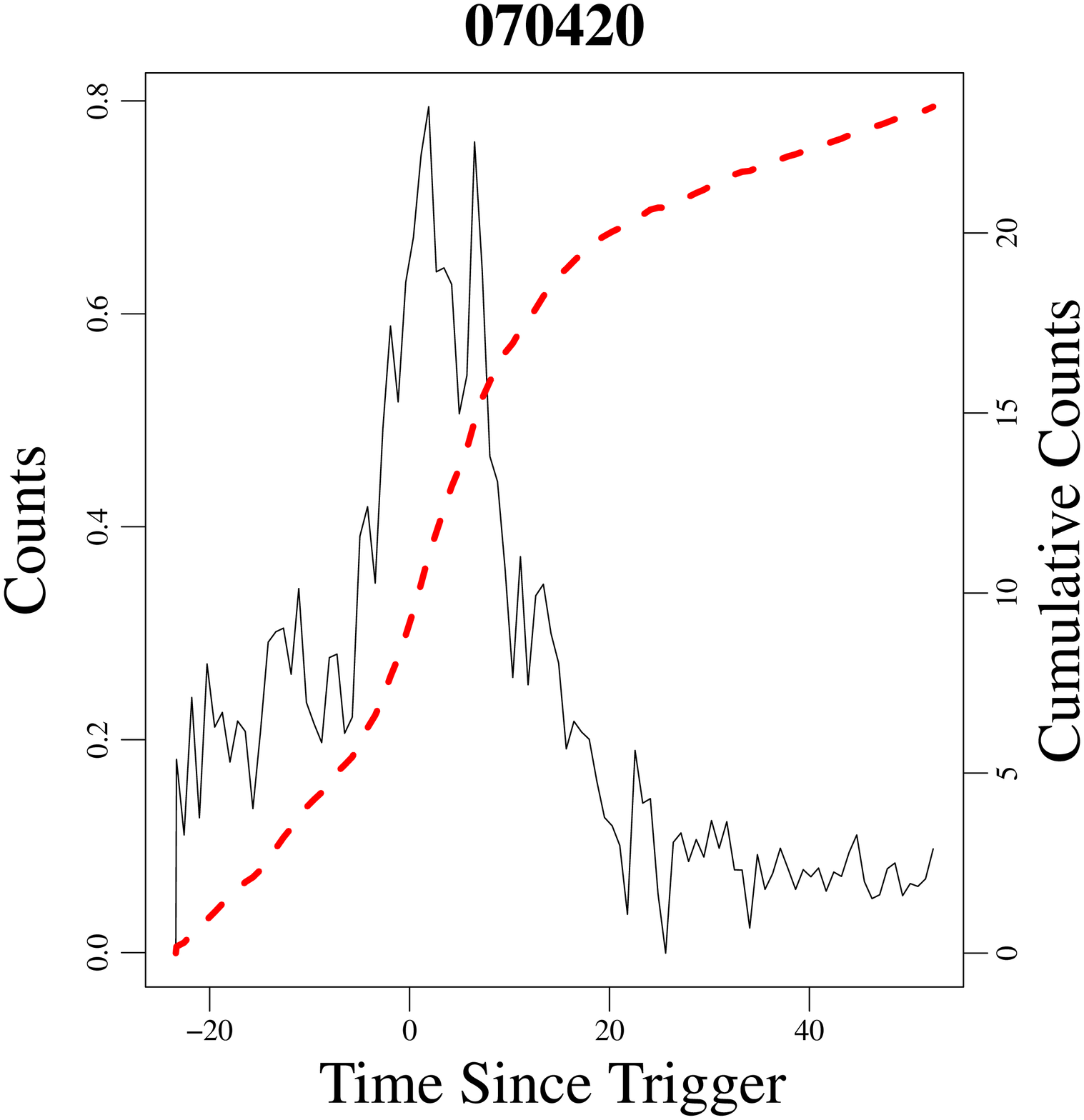}
 \end{center}
\end{minipage}
\begin{minipage}{0.25\hsize}
\begin{center}
    \FigureFile(40mm,40mm){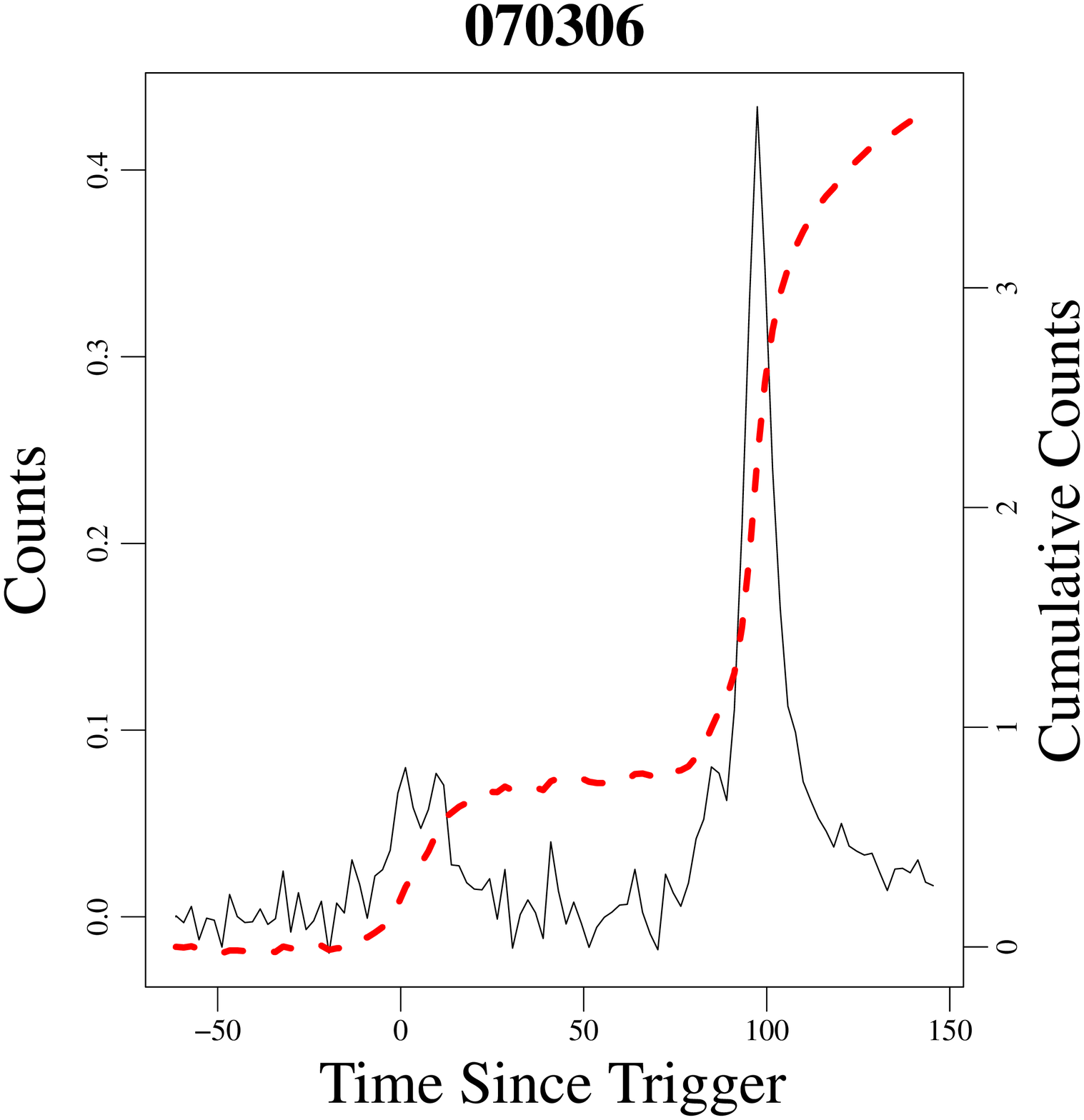}
\end{center}
\end{minipage}
\begin{minipage}{0.25\hsize}
\begin{center}
    \FigureFile(40mm,40mm){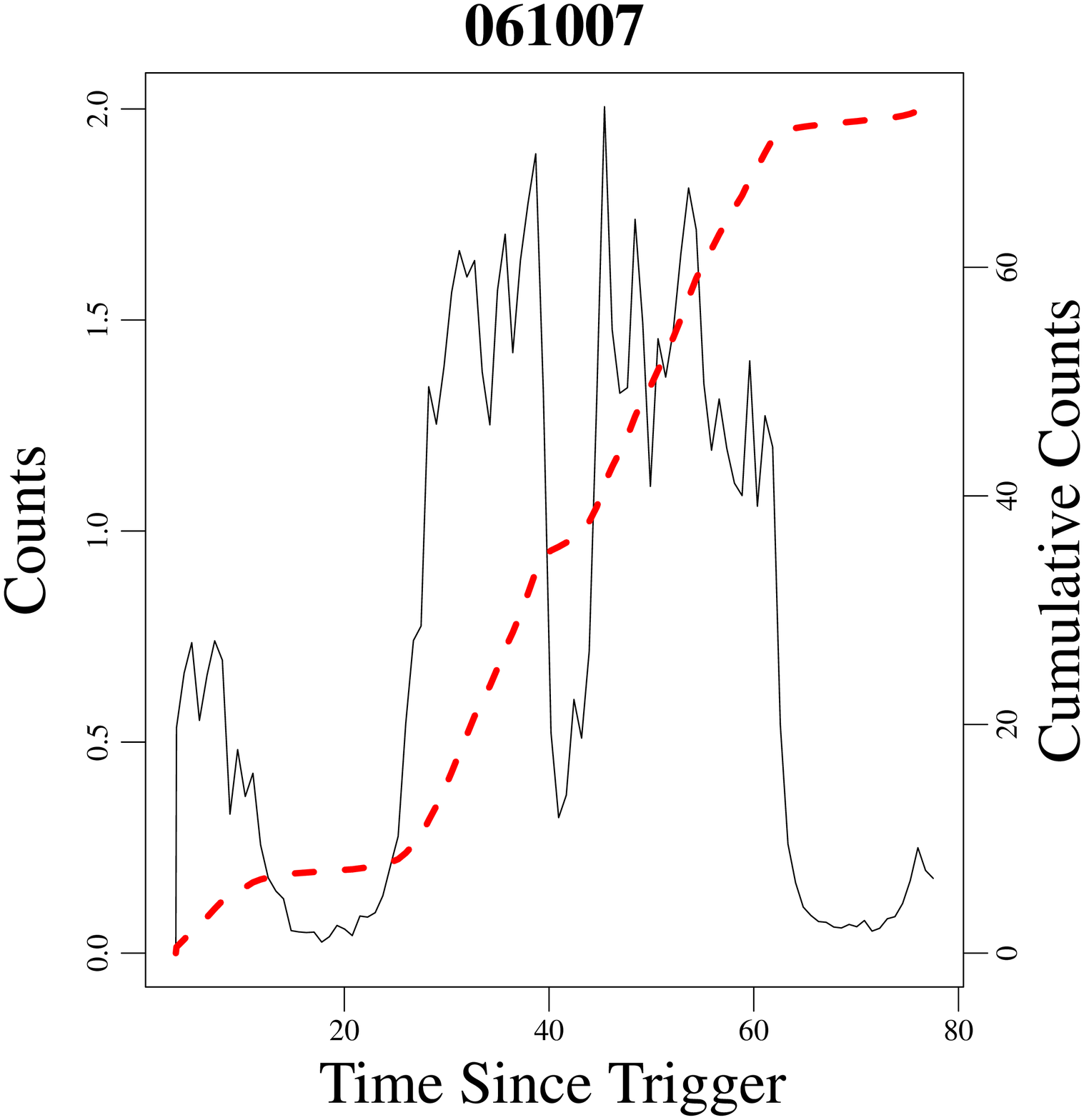}
 \end{center}
\end{minipage}\\
\begin{minipage}{0.25\hsize}
\begin{center}
    \FigureFile(40mm,40mm){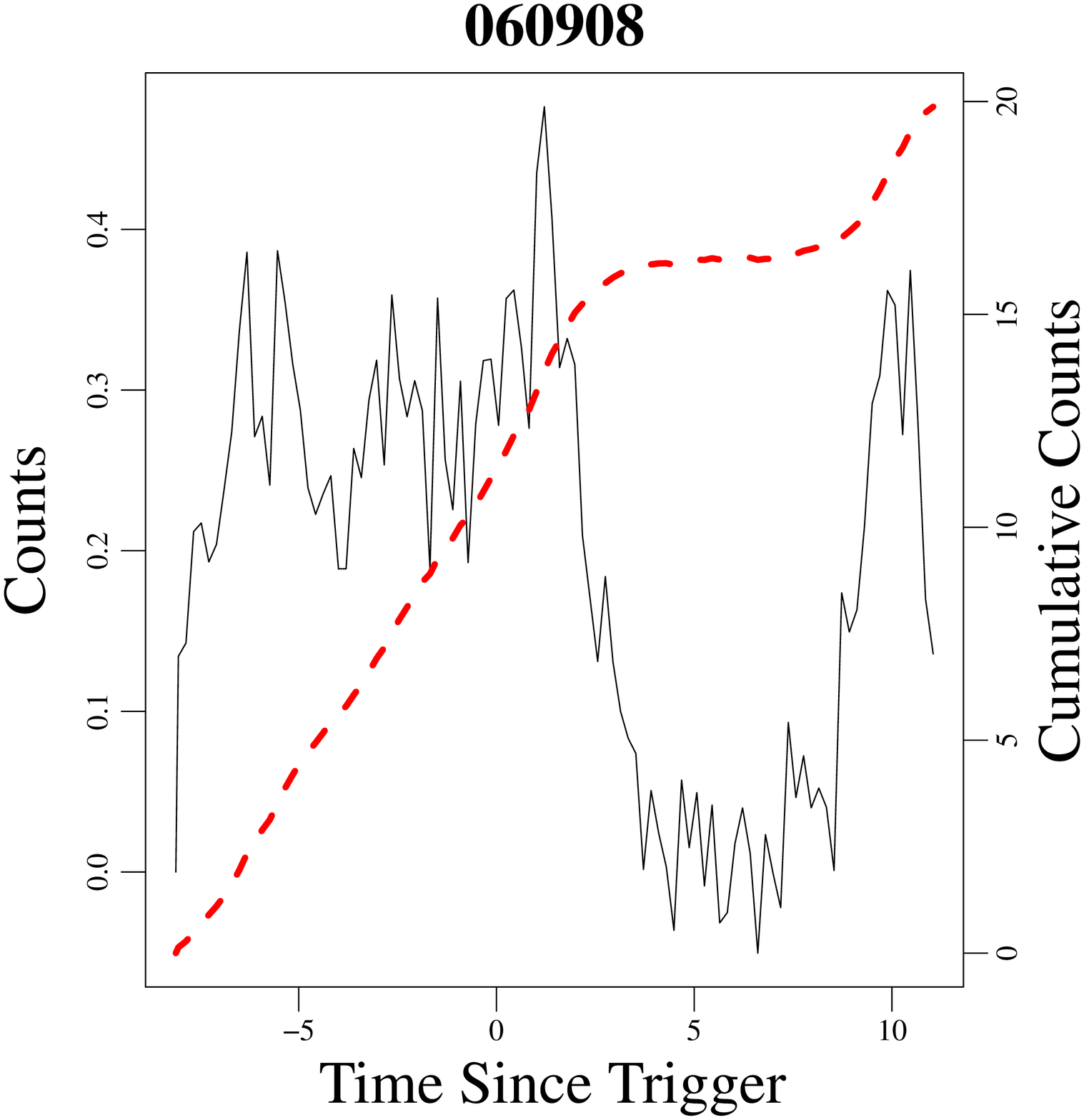}
\end{center}
\end{minipage}
\begin{minipage}{0.25\hsize}
\begin{center}
    \FigureFile(40mm,40mm){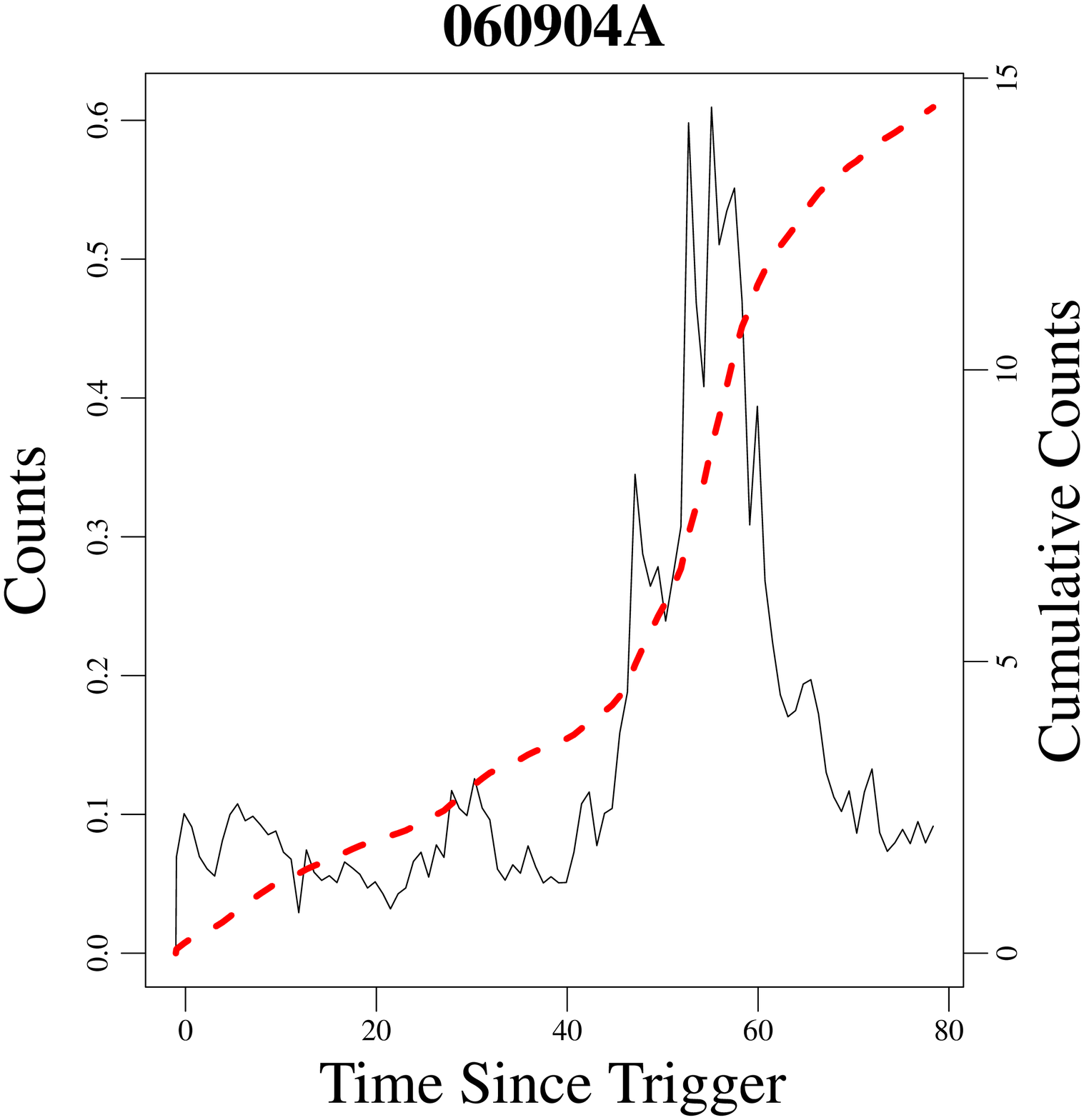}
 \end{center}
\end{minipage}
\begin{minipage}{0.25\hsize}
\begin{center}
    \FigureFile(40mm,40mm){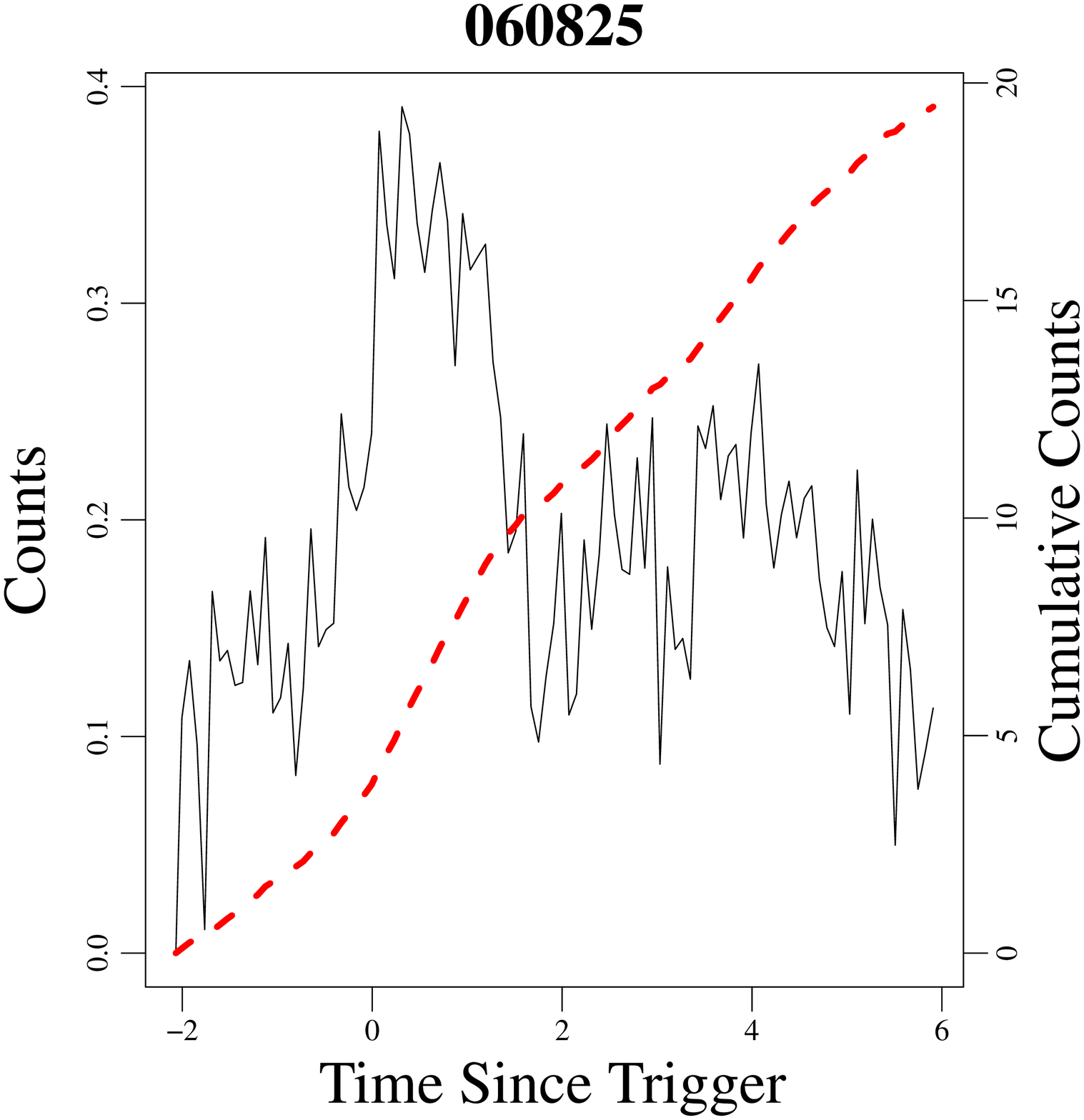}
\end{center}
\end{minipage}
\begin{minipage}{0.25\hsize}
\begin{center}
    \FigureFile(40mm,40mm){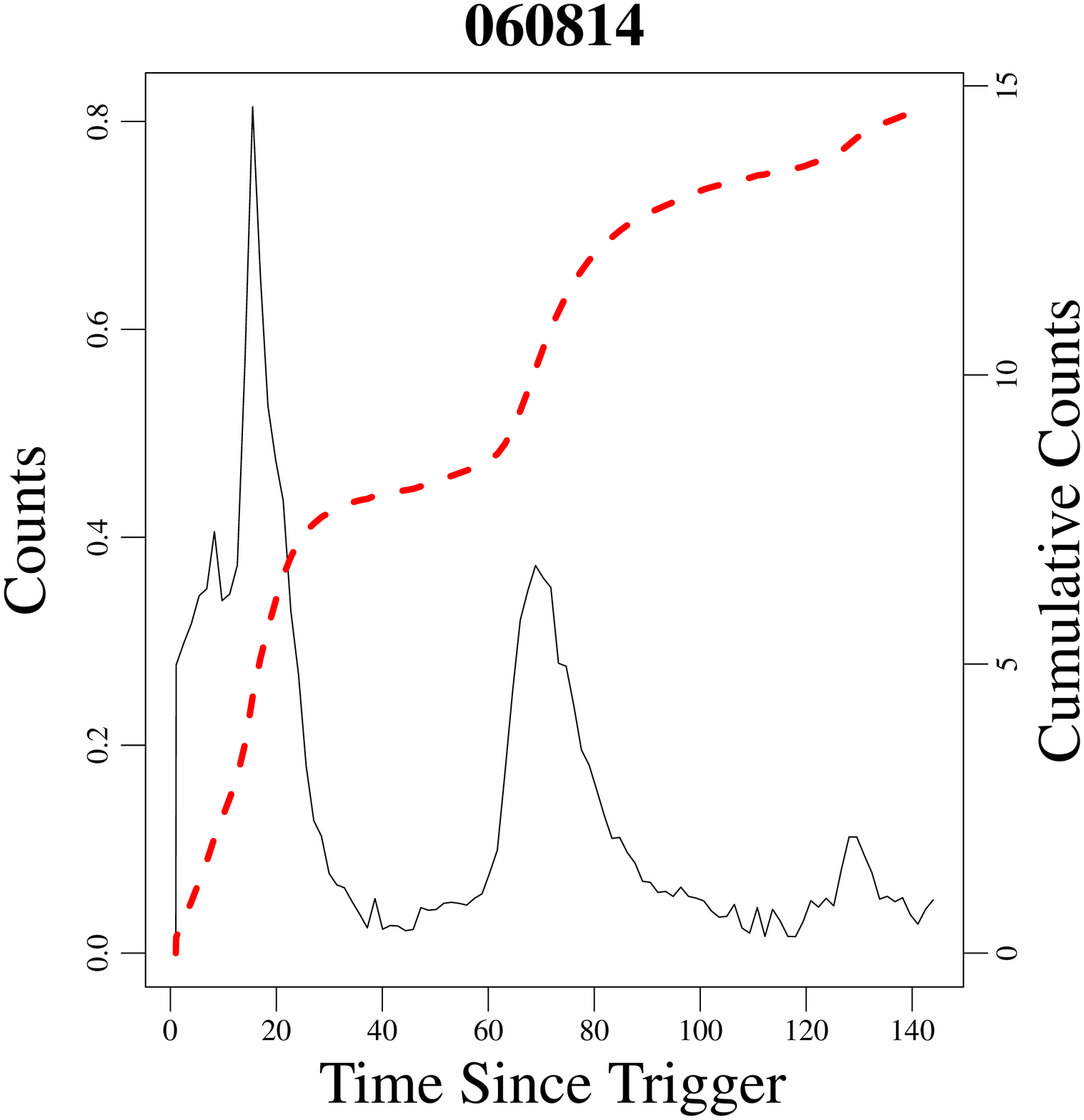}
 \end{center}
\end{minipage}\\
\end{tabular}
   \caption{Light curves (black solid) and cumulative light curves (red doted) of Type I LGRBs.}\label{fig:A1-3}
\end{figure*}

\begin{figure*}[htb]
\begin{tabular}{cccc}
\begin{minipage}{0.25\hsize}
\begin{center}
    \FigureFile(40mm,40mm){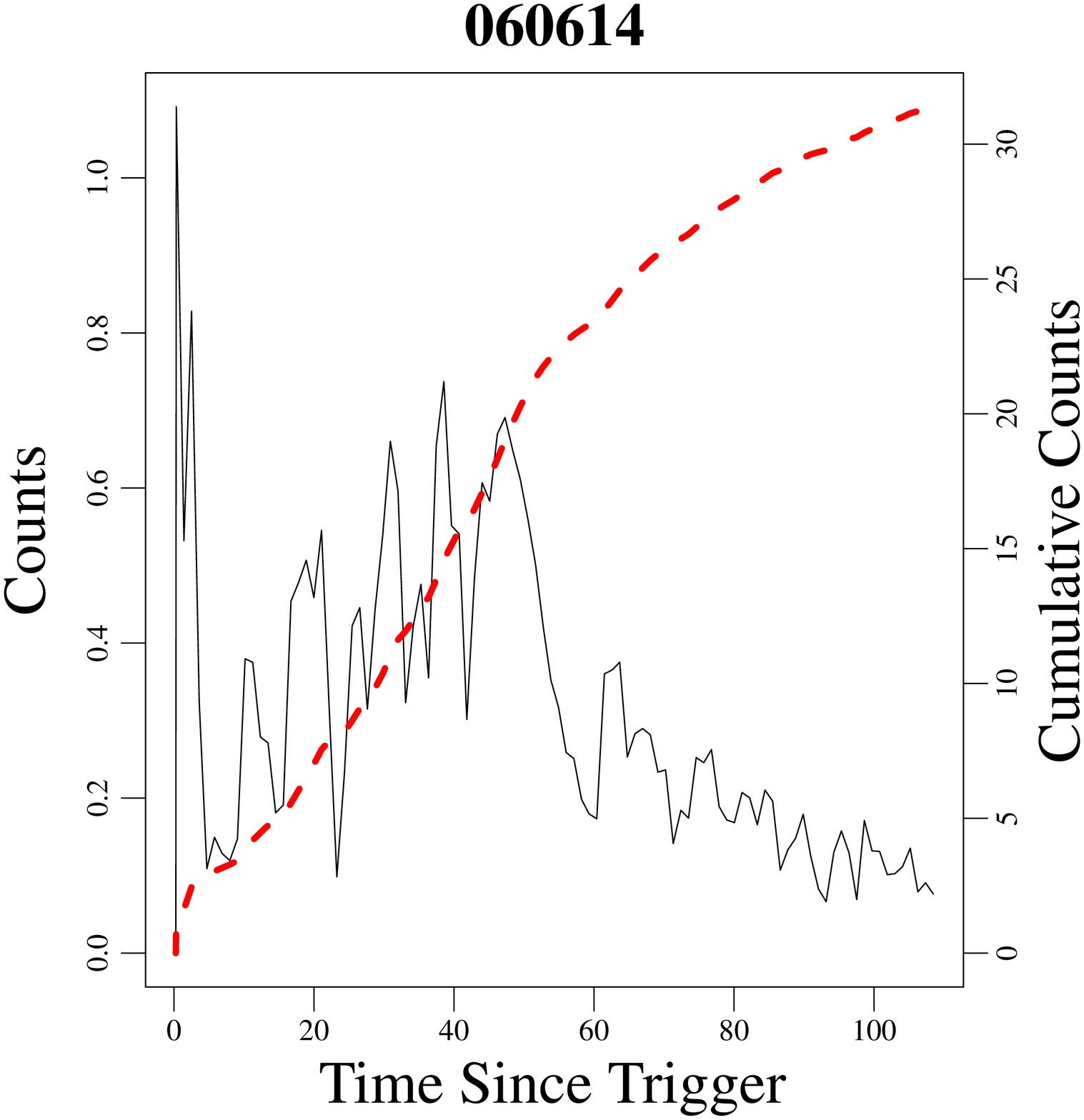}
\end{center}
\end{minipage}
\begin{minipage}{0.25\hsize}
\begin{center}
    \FigureFile(40mm,40mm){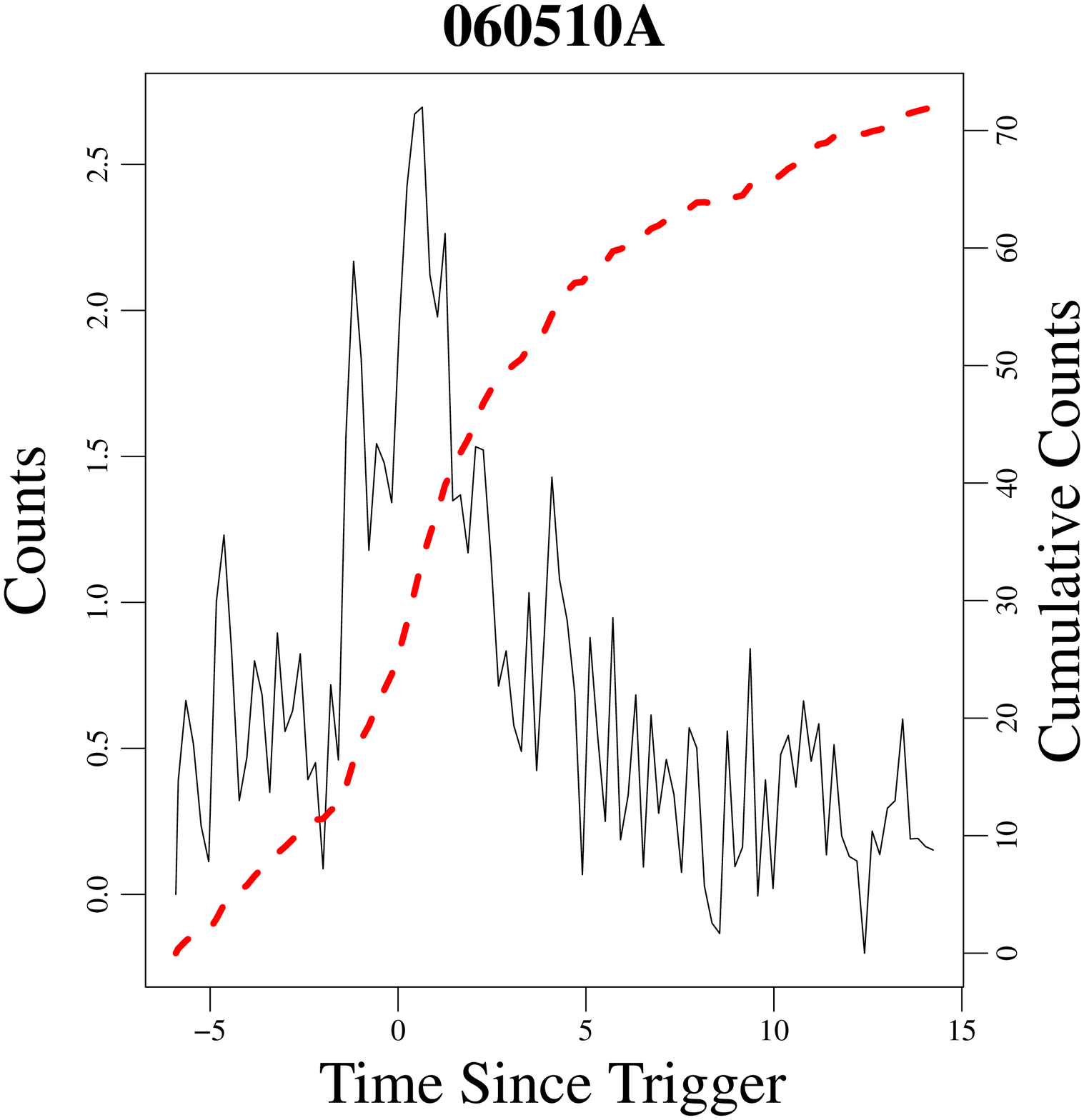}
 \end{center}
\end{minipage}
\begin{minipage}{0.25\hsize}
\begin{center}
    \FigureFile(40mm,40mm){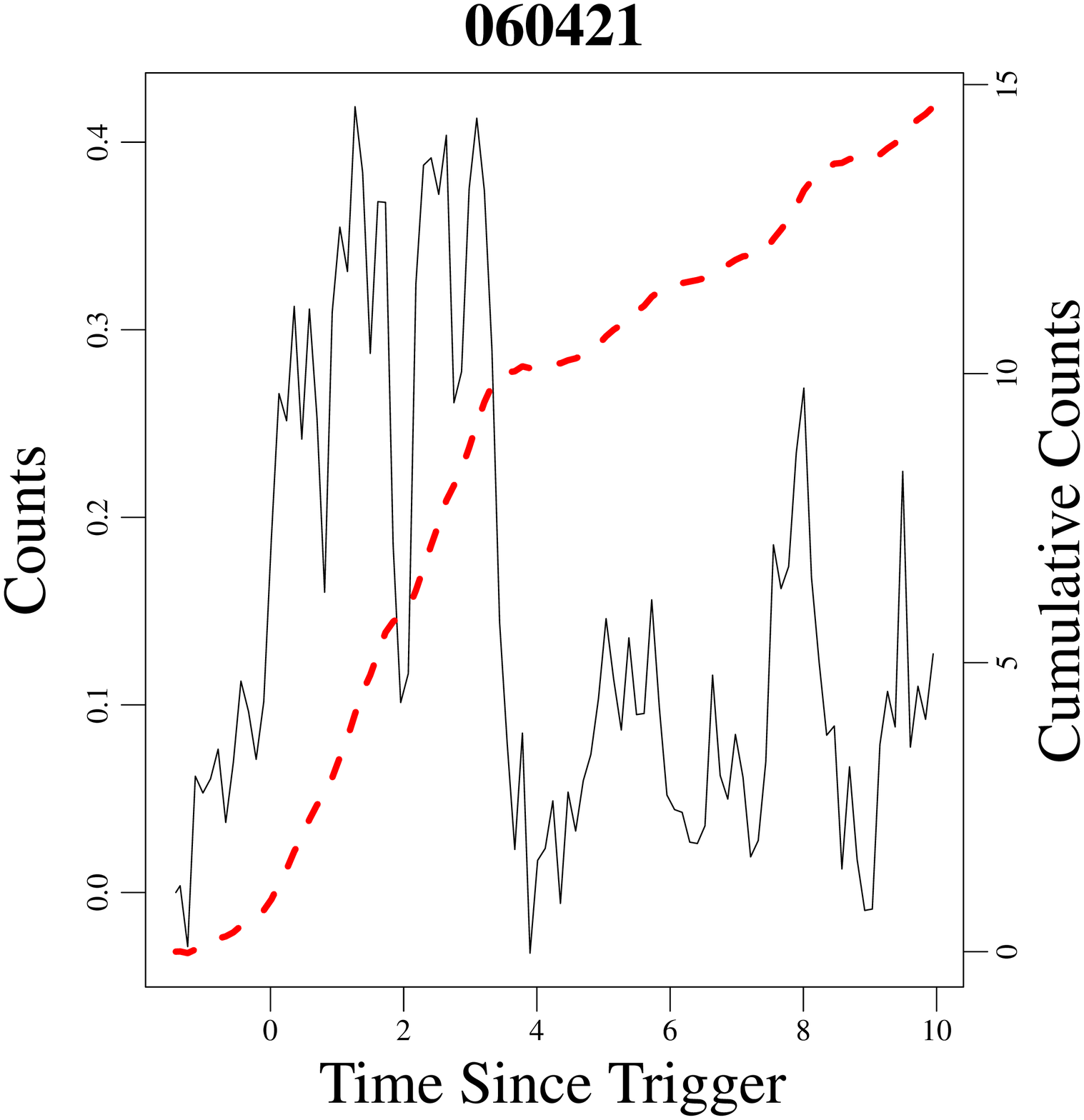}
\end{center}
\end{minipage}
\begin{minipage}{0.25\hsize}
\begin{center}
    \FigureFile(40mm,40mm){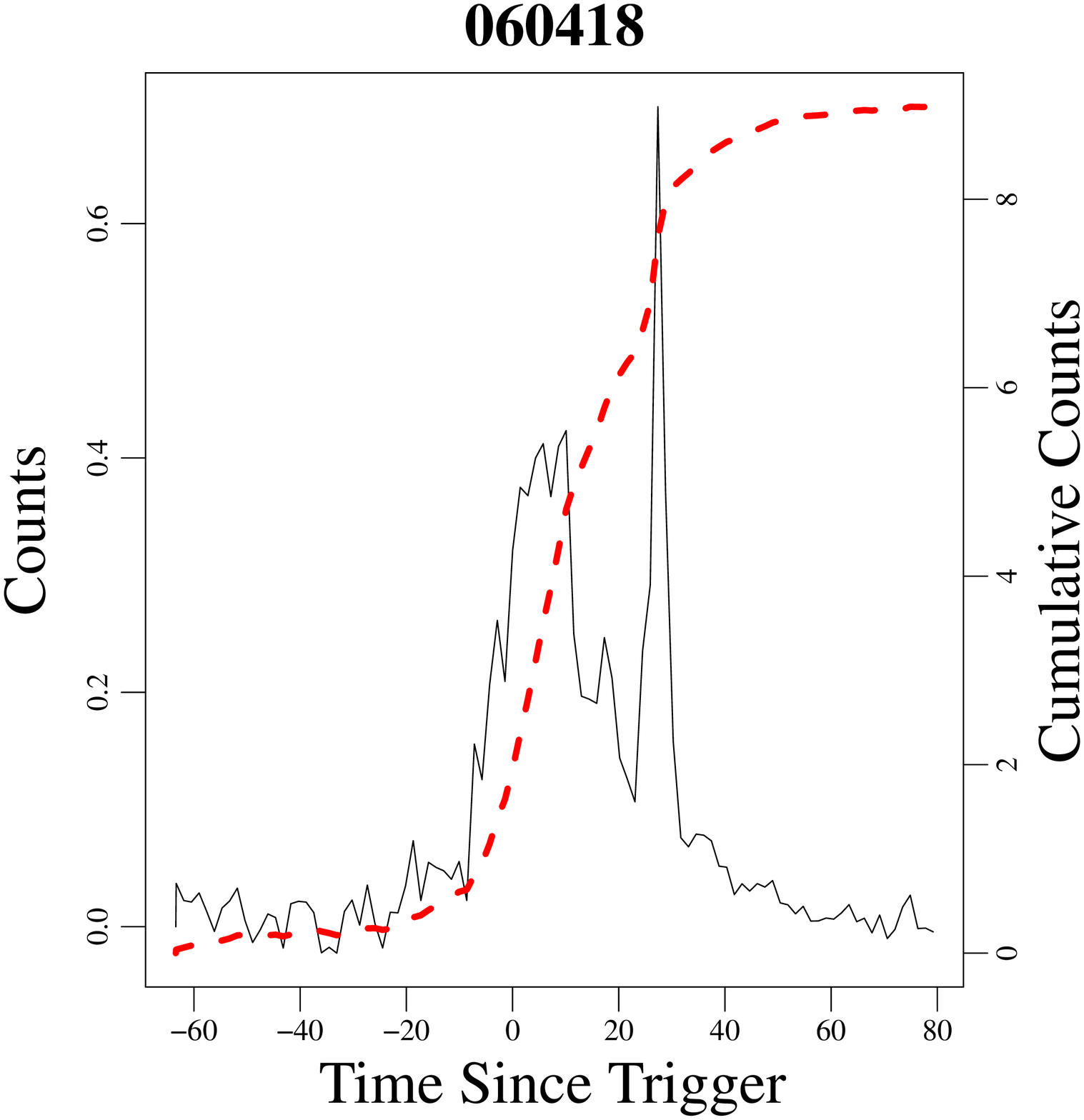}
 \end{center}
\end{minipage}\\
\begin{minipage}{0.25\hsize}
\begin{center}
    \FigureFile(40mm,40mm){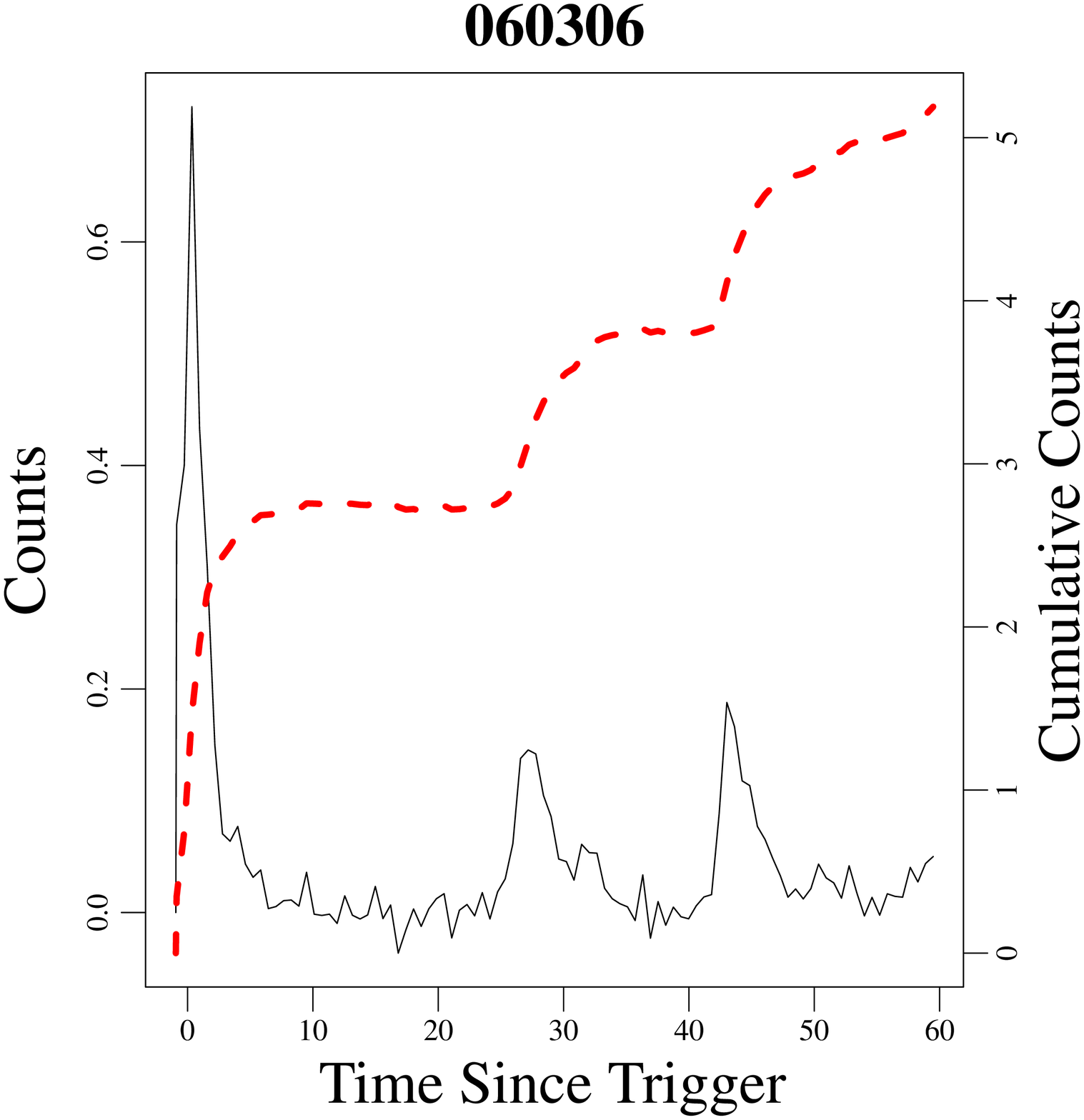}
\end{center}
\end{minipage}
\begin{minipage}{0.25\hsize}
\begin{center}
    \FigureFile(40mm,40mm){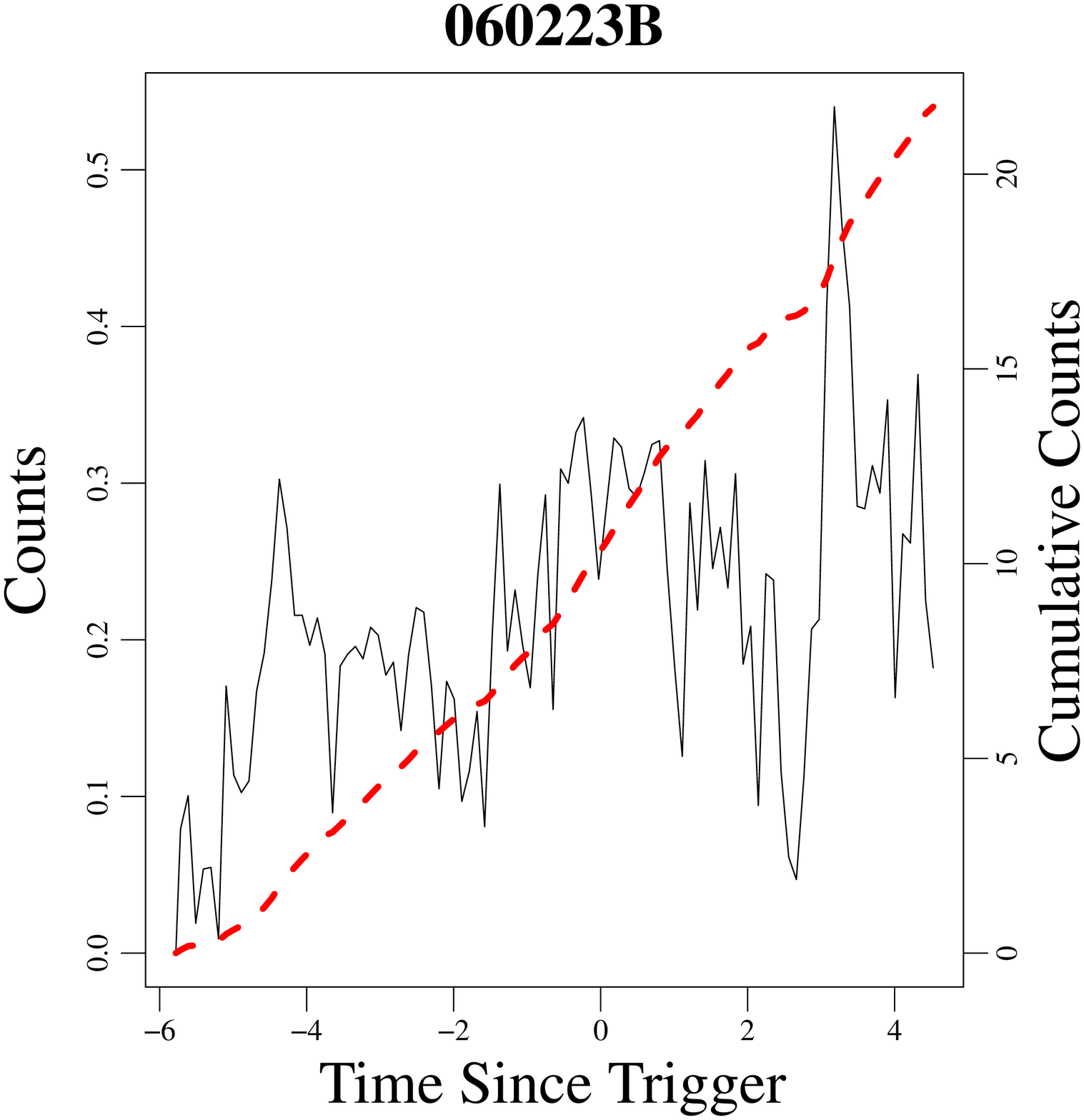}
 \end{center}
\end{minipage}
\begin{minipage}{0.25\hsize}
\begin{center}
    \FigureFile(40mm,40mm){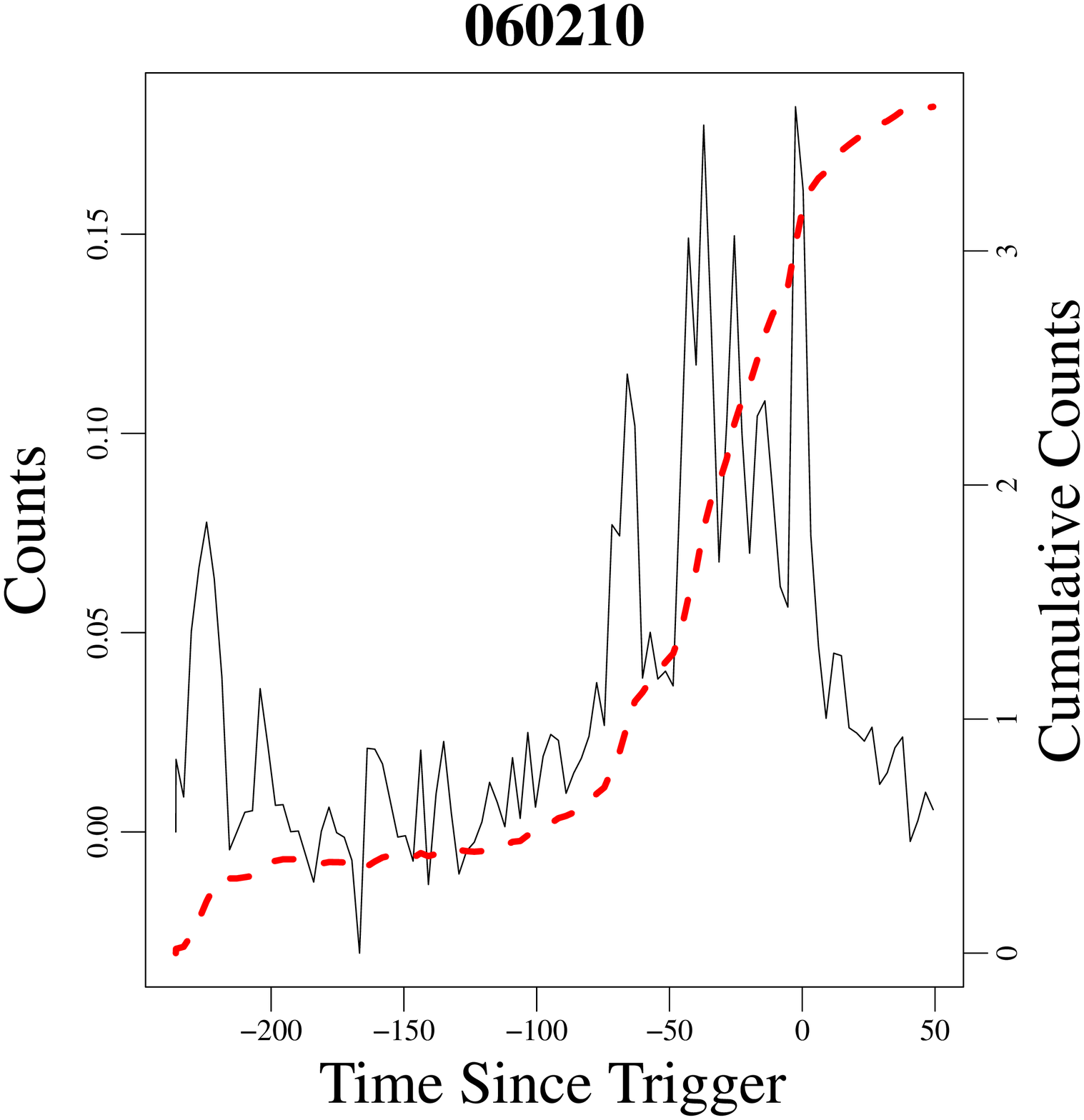}
\end{center}
\end{minipage}
\begin{minipage}{0.25\hsize}
\begin{center}
    \FigureFile(40mm,40mm){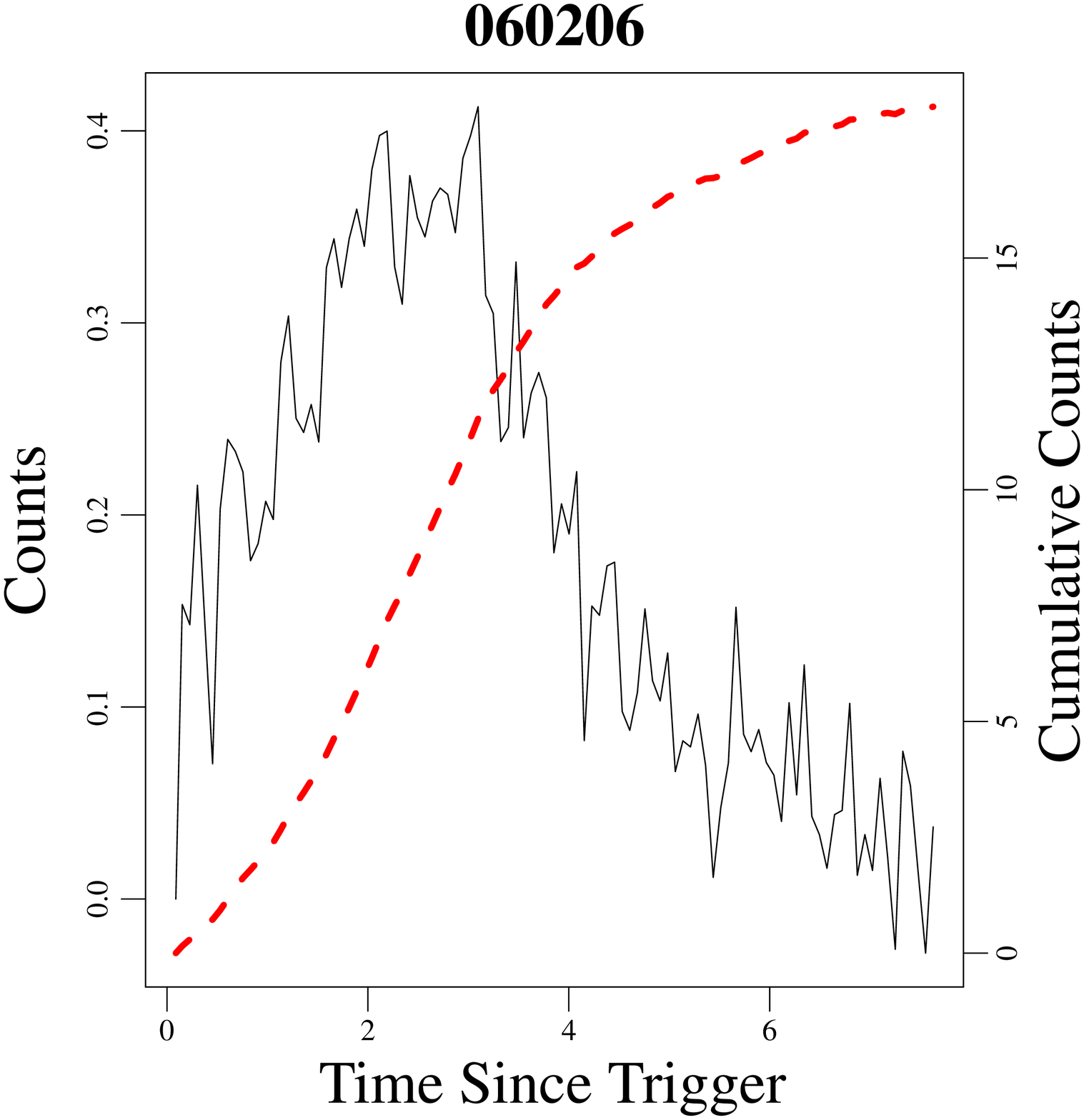}
 \end{center}
\end{minipage}\\
\begin{minipage}{0.25\hsize}
\begin{center}
    \FigureFile(40mm,40mm){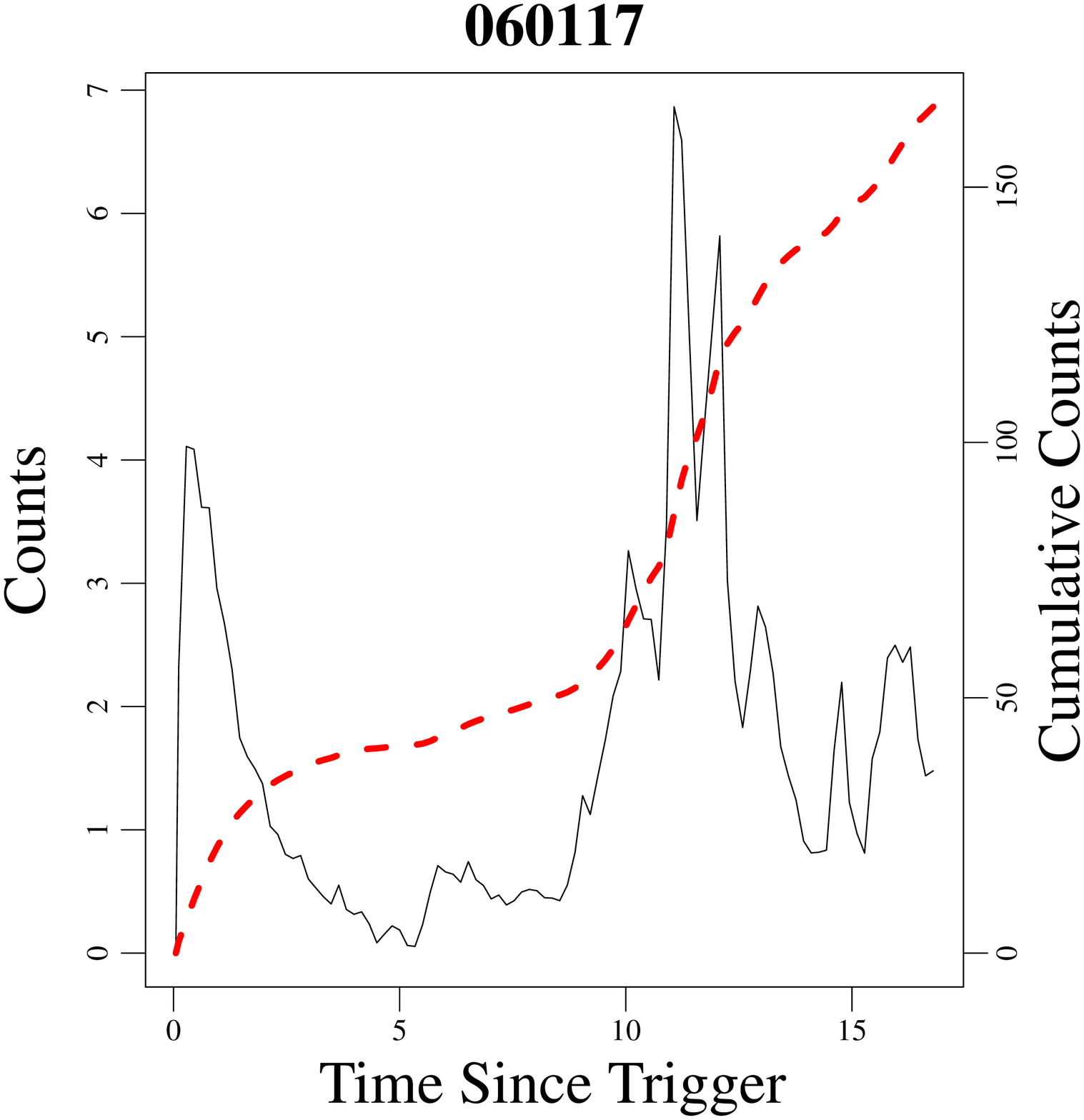}
\end{center}
\end{minipage}
\begin{minipage}{0.25\hsize}
\begin{center}
    \FigureFile(40mm,40mm){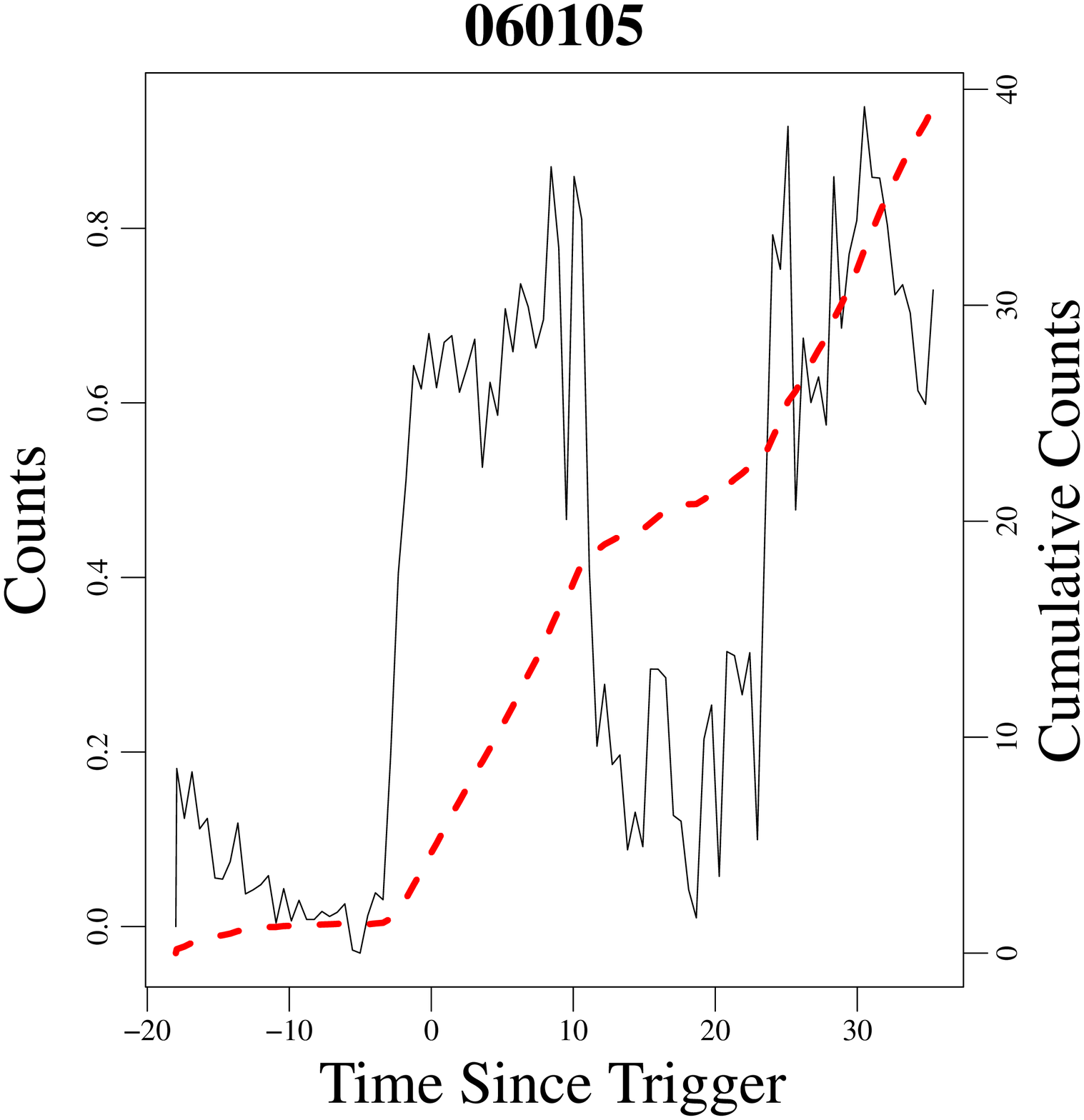}
 \end{center}
\end{minipage}
\begin{minipage}{0.25\hsize}
\begin{center}
    \FigureFile(40mm,40mm){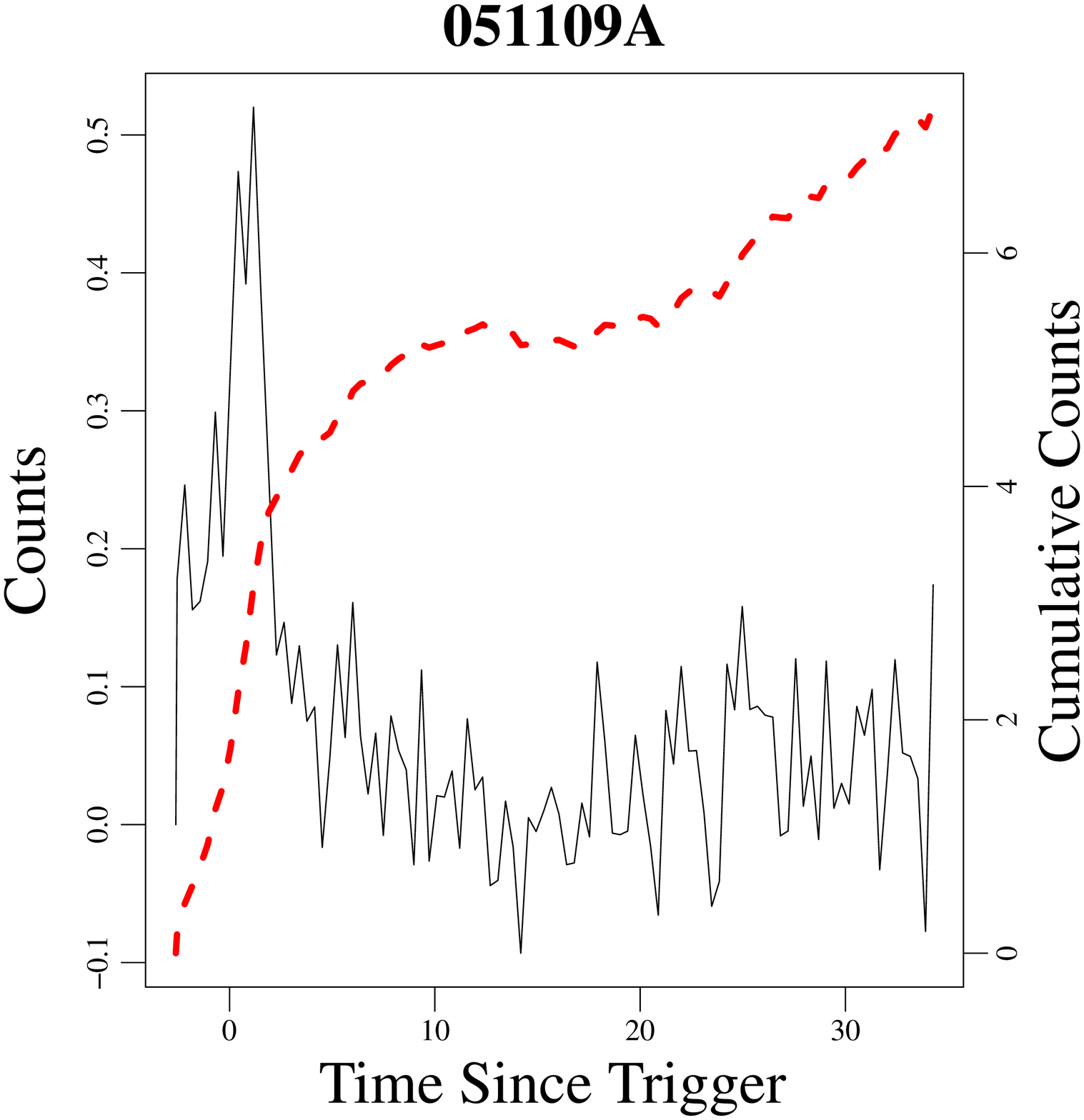}
\end{center}
\end{minipage}
\begin{minipage}{0.25\hsize}
\begin{center}
    \FigureFile(40mm,40mm){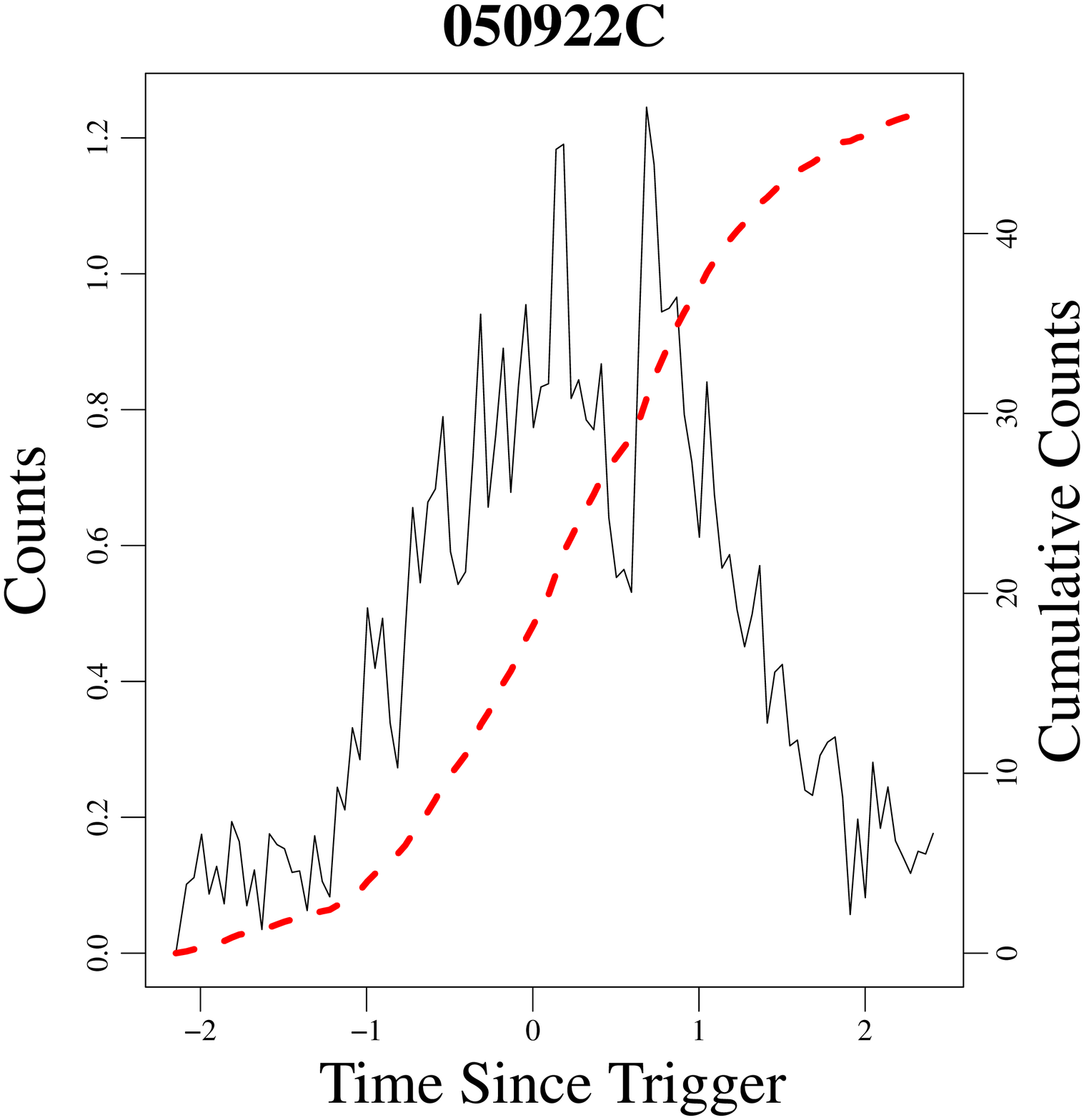}
 \end{center}
\end{minipage}\\
\begin{minipage}{0.25\hsize}
\begin{center}
    \FigureFile(40mm,40mm){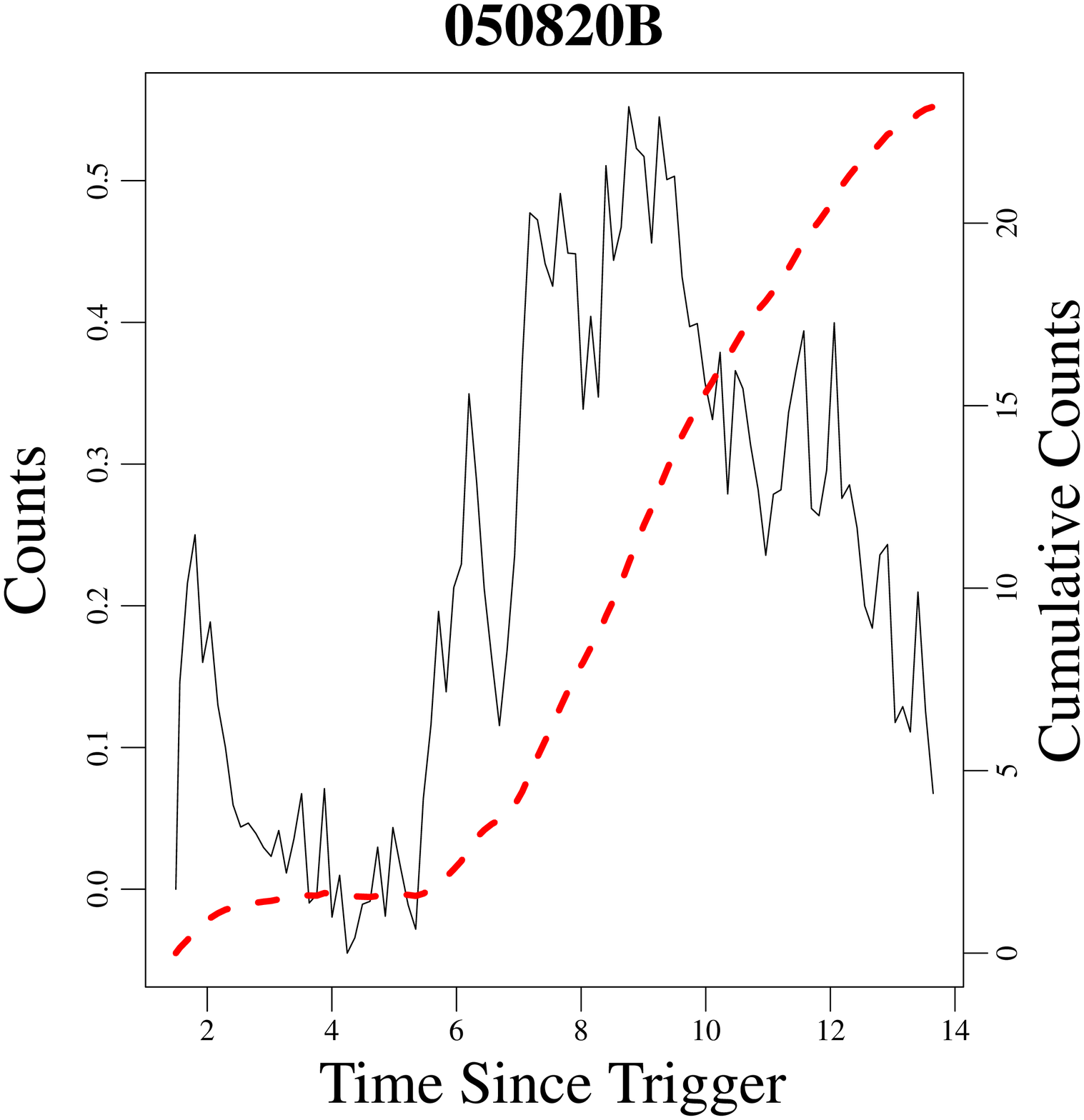}
\end{center}
\end{minipage}
\begin{minipage}{0.25\hsize}
\begin{center}
    \FigureFile(40mm,40mm){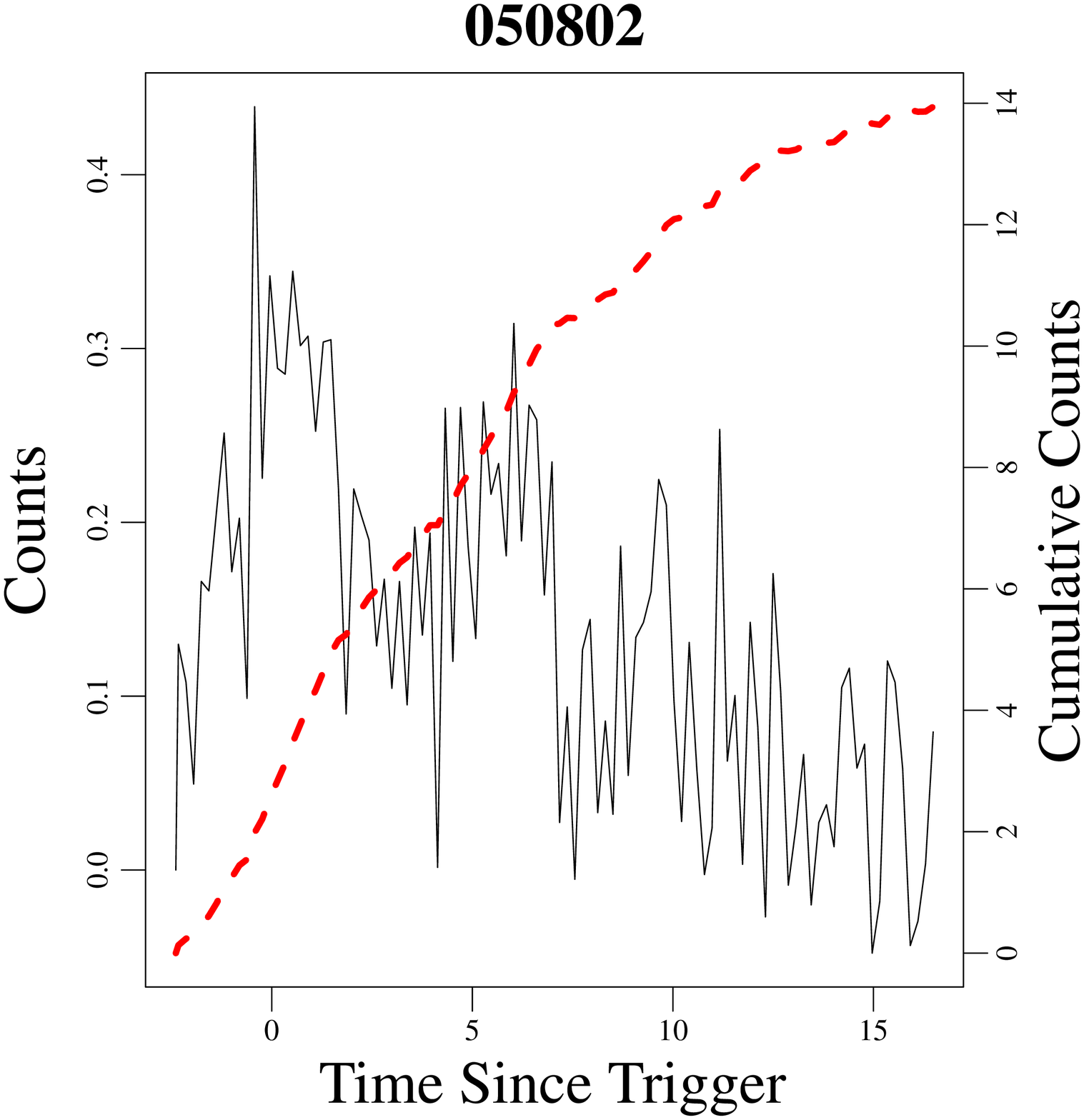}
 \end{center}
\end{minipage}
\begin{minipage}{0.25\hsize}
\begin{center}
    \FigureFile(40mm,40mm){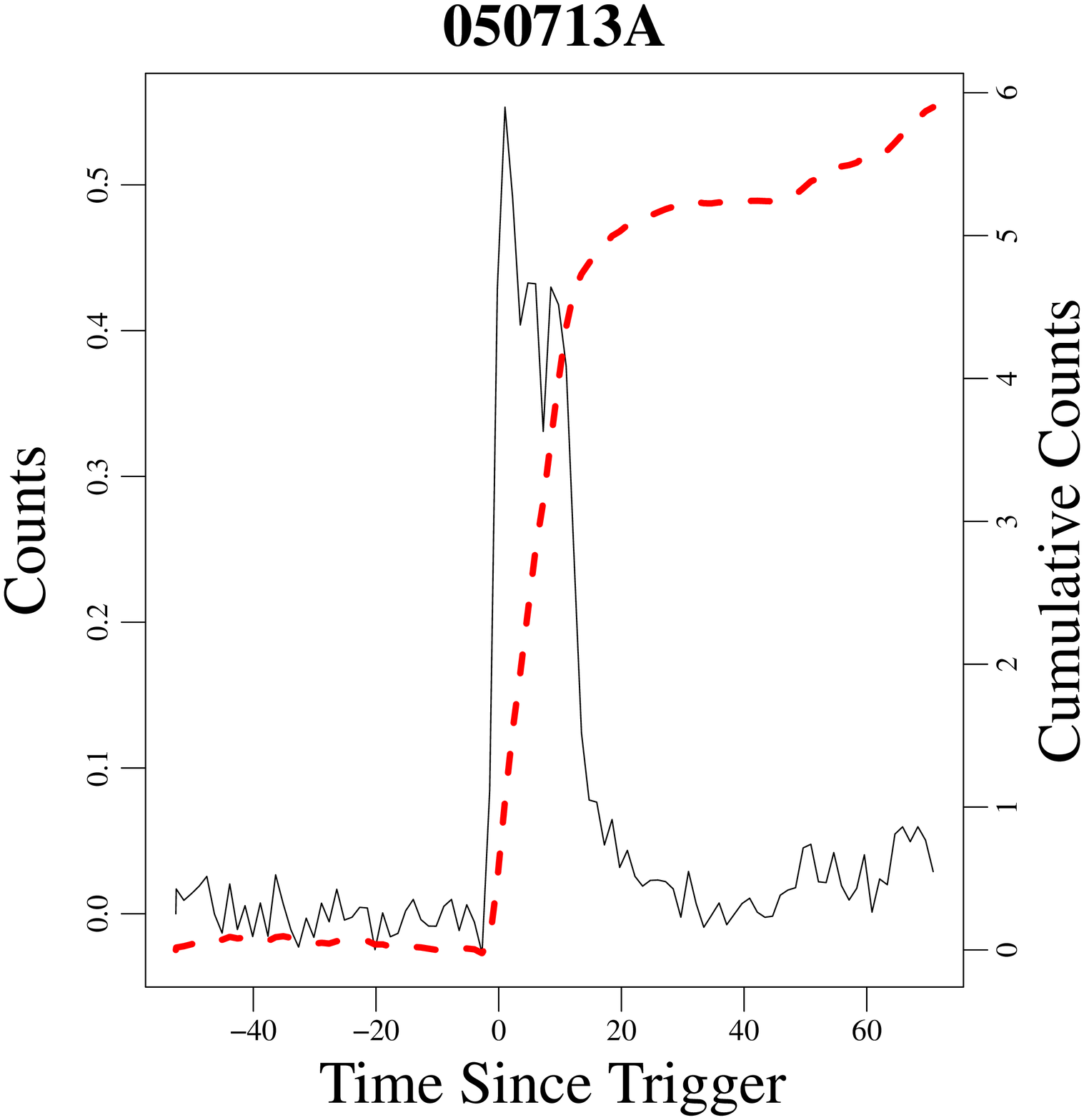}
\end{center}
\end{minipage}
\begin{minipage}{0.25\hsize}
\begin{center}
    \FigureFile(40mm,40mm){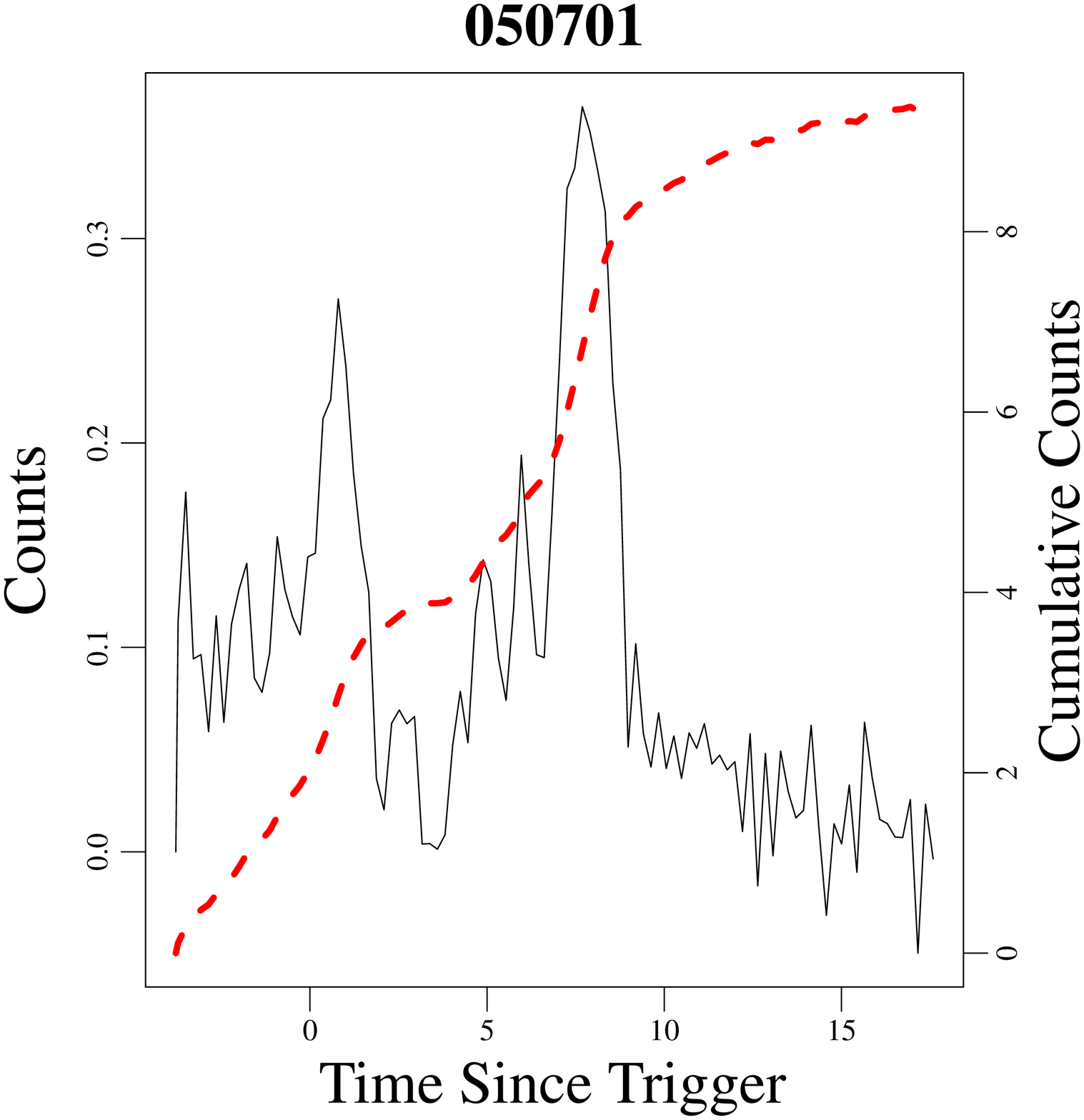}
 \end{center}
\end{minipage}\\
\begin{minipage}{0.25\hsize}
\begin{center}
    \FigureFile(40mm,40mm){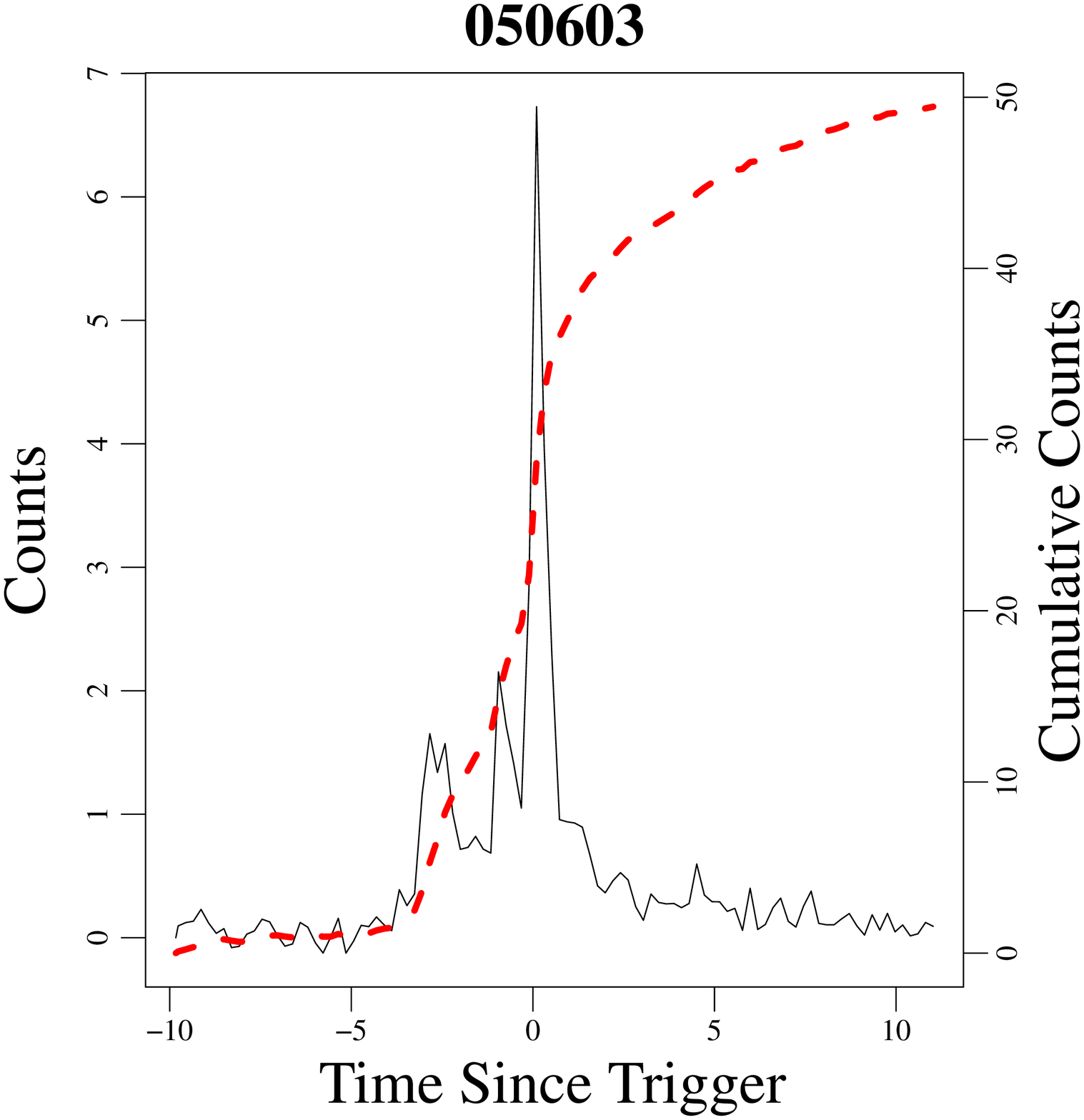}
\end{center}
\end{minipage}
\begin{minipage}{0.25\hsize}
\begin{center}
    \FigureFile(40mm,40mm){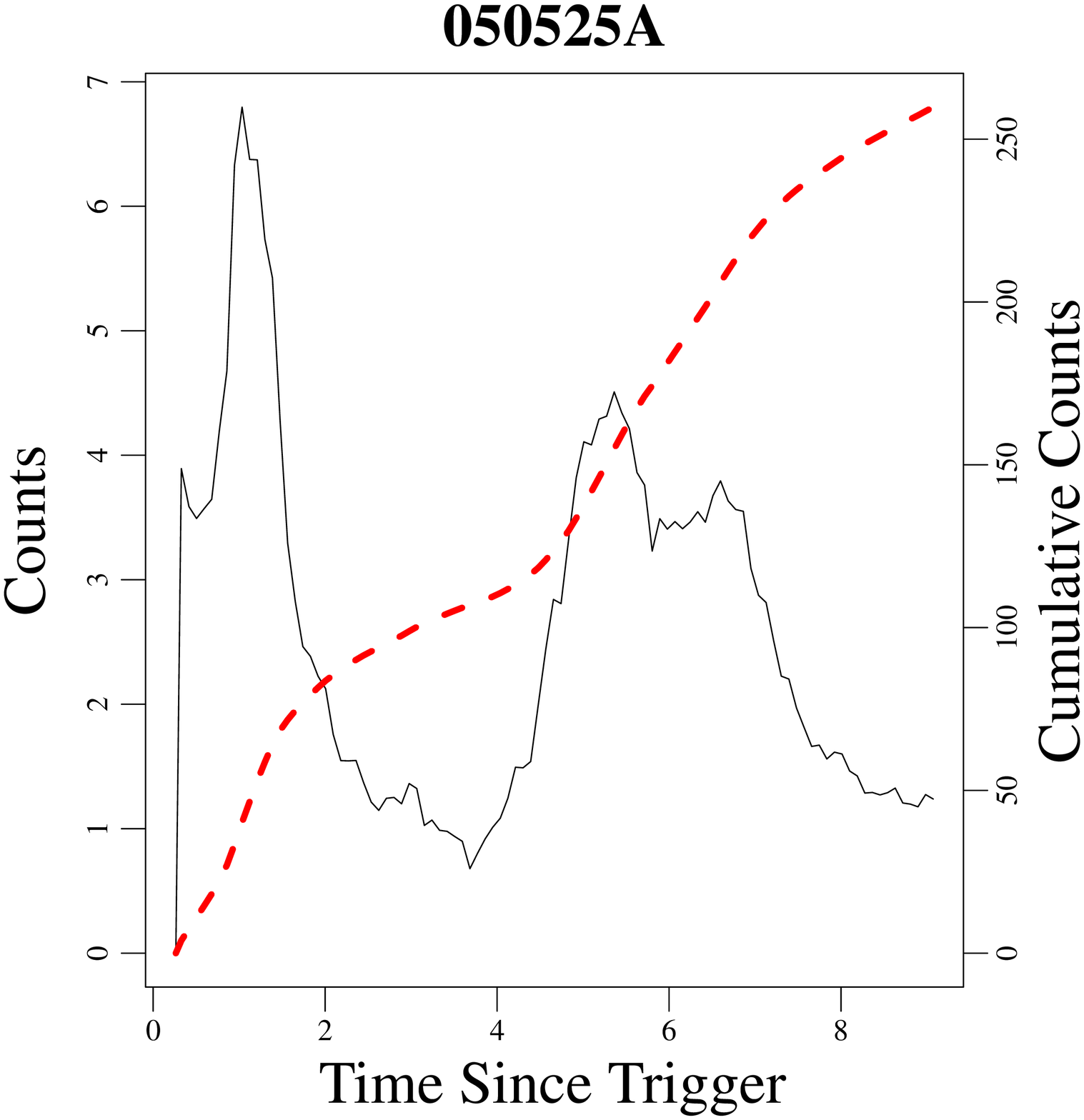}
 \end{center}
\end{minipage}
\begin{minipage}{0.25\hsize}
\begin{center}
    \FigureFile(40mm,40mm){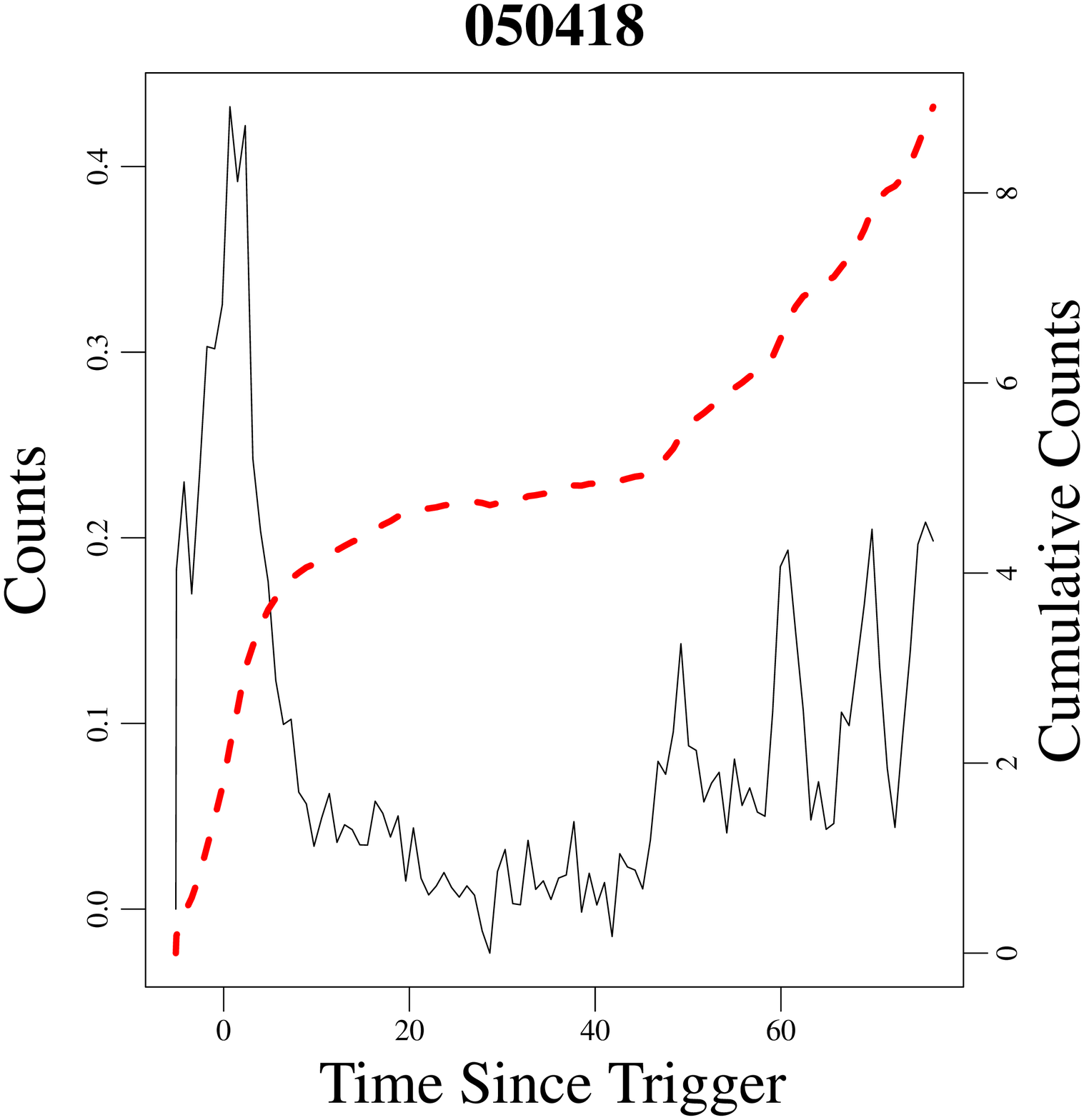}
\end{center}
\end{minipage}
\begin{minipage}{0.25\hsize}
\begin{center}
    \FigureFile(40mm,40mm){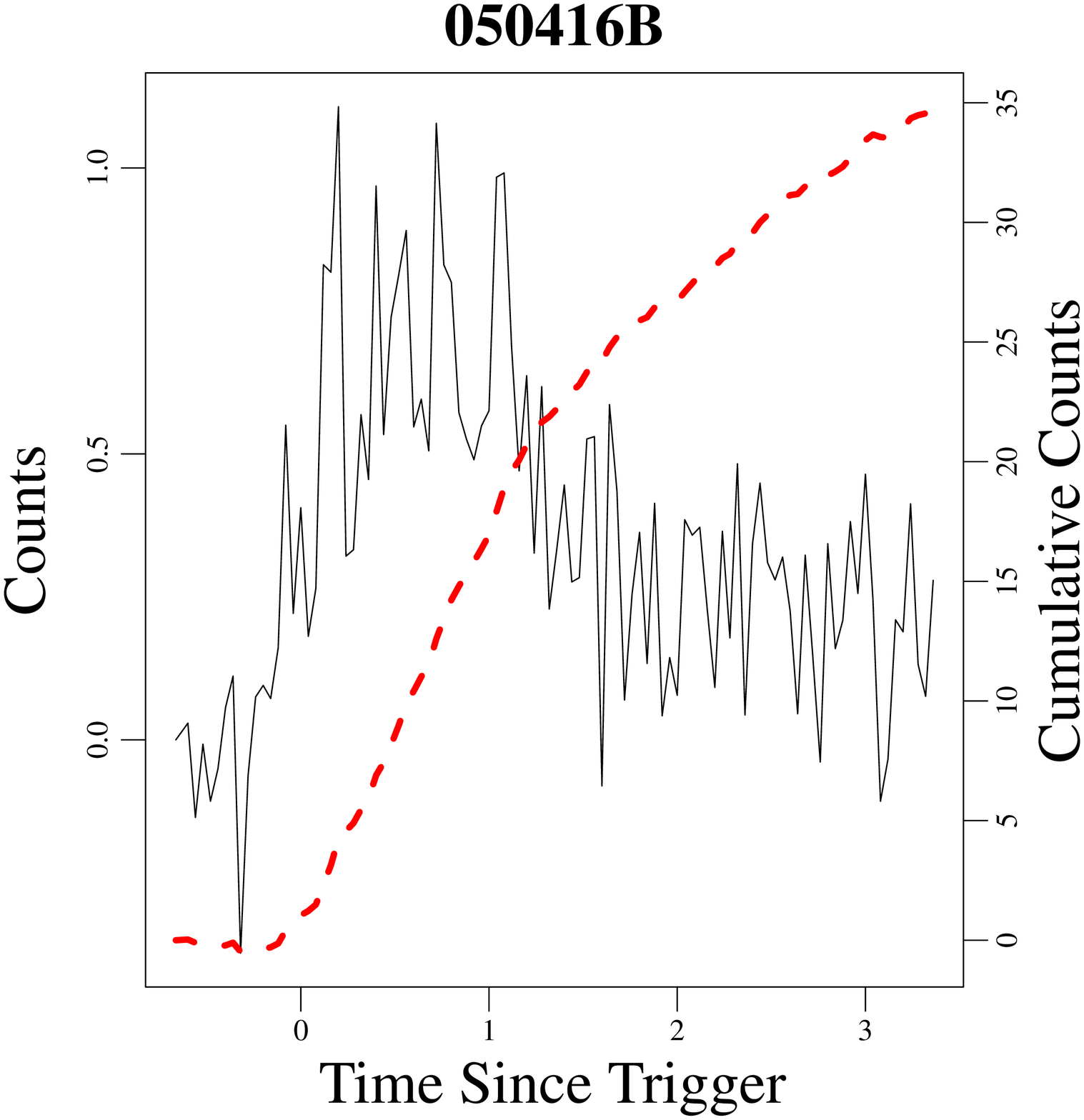}
 \end{center}
\end{minipage}\\
\begin{minipage}{0.25\hsize}
\begin{center}
    \FigureFile(40mm,40mm){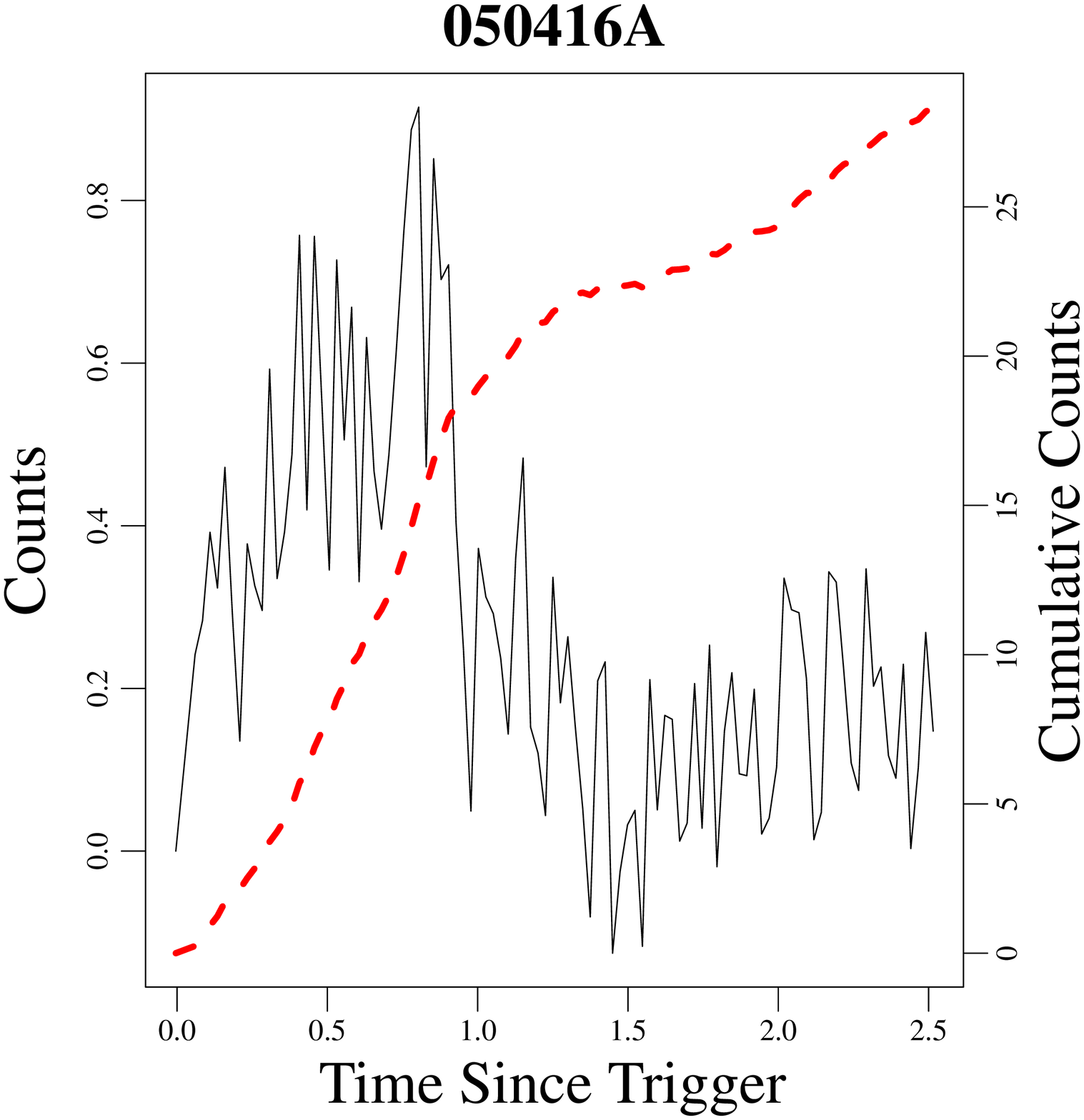}
\end{center}
\end{minipage}
\begin{minipage}{0.25\hsize}
\begin{center}
    \FigureFile(40mm,40mm){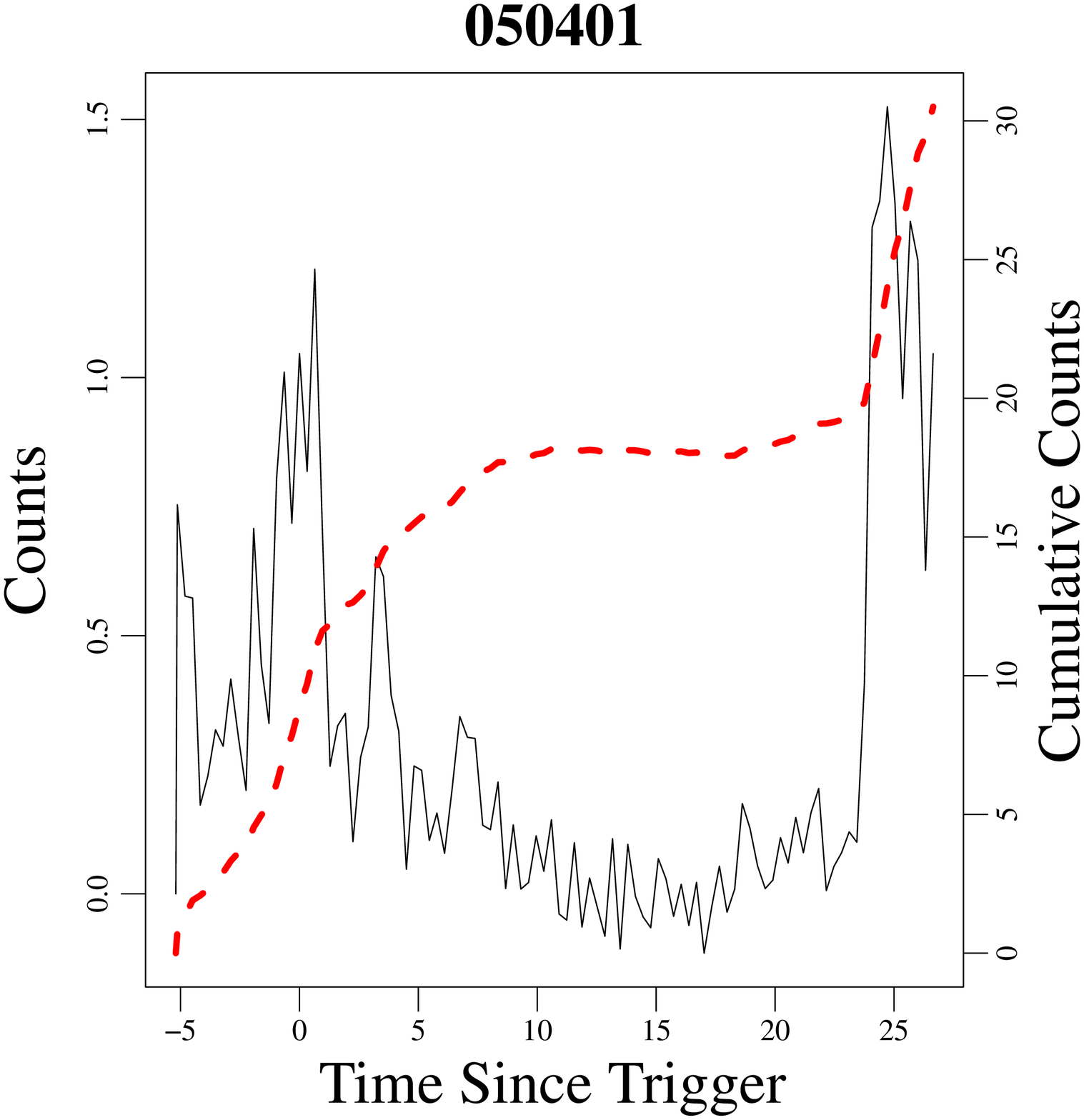}
 \end{center}
\end{minipage}
\begin{minipage}{0.25\hsize}
\begin{center}
    \FigureFile(40mm,40mm){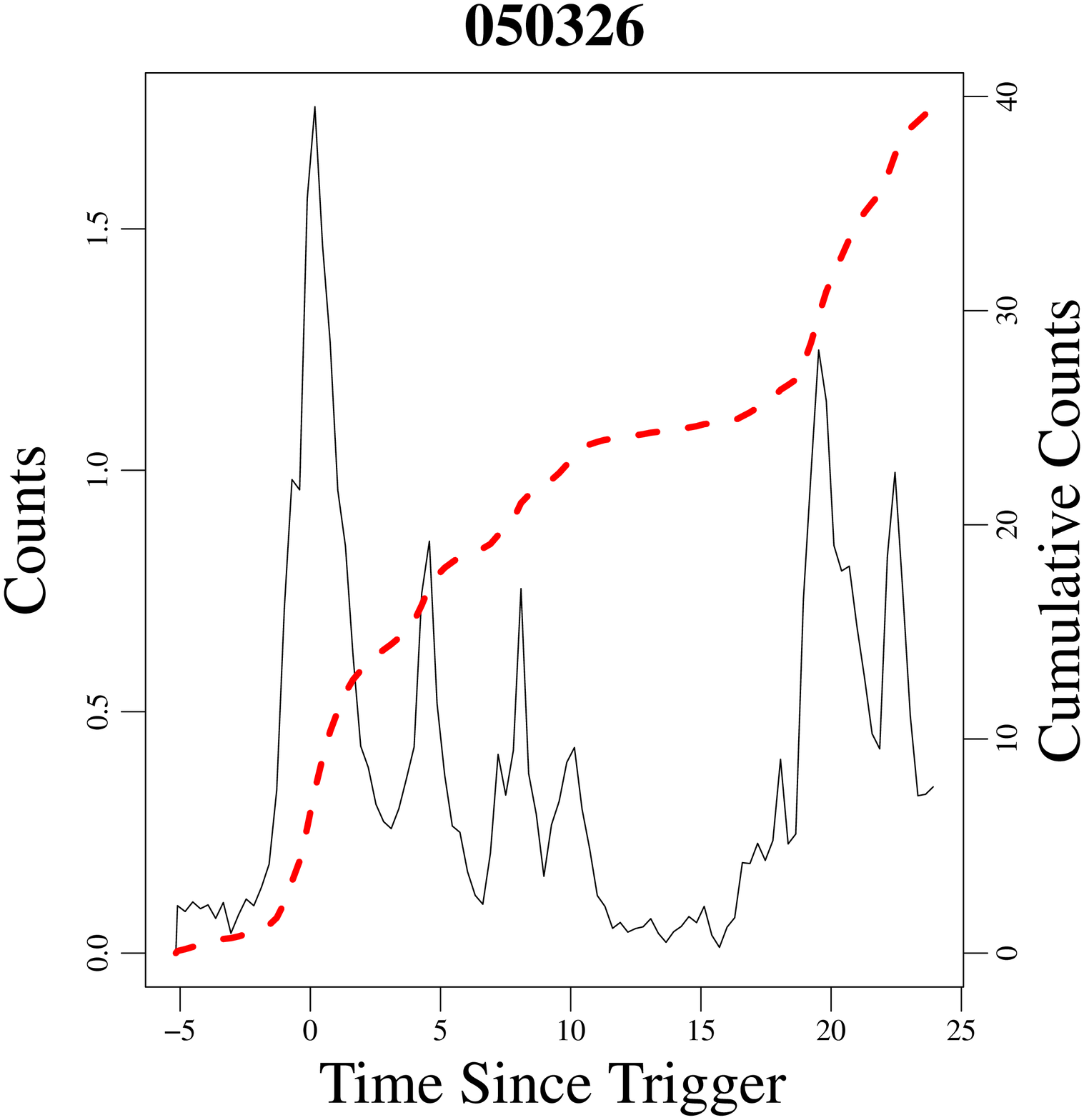}
\end{center}
\end{minipage}
\begin{minipage}{0.25\hsize}
\begin{center}
    \FigureFile(40mm,40mm){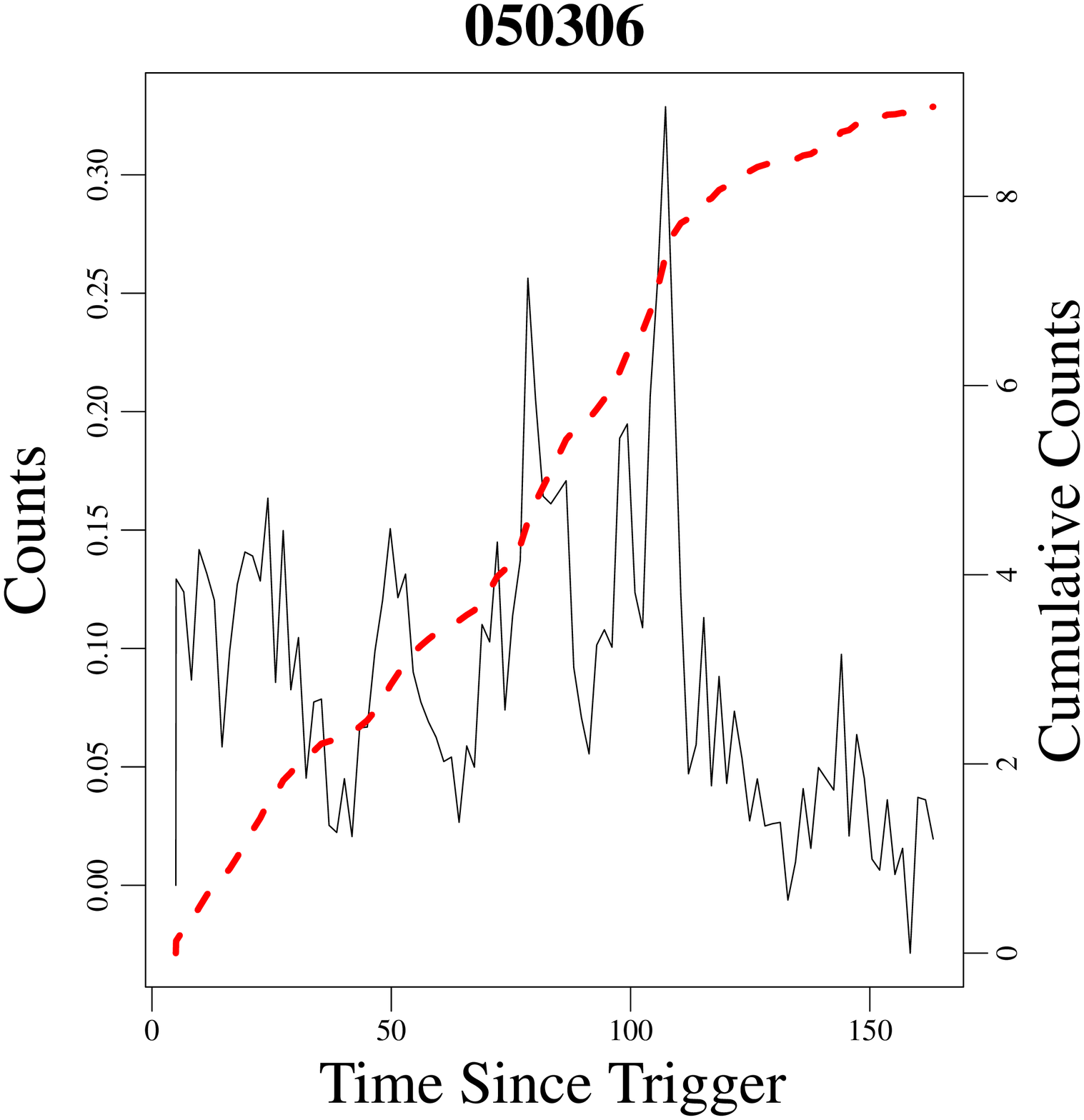}
 \end{center}
\end{minipage}\\
\end{tabular}
   \caption{Light curves (black solid) and cumulative light curves (red doted) of Type I LGRBs.}\label{fig:A1-4}
\end{figure*}

\begin{figure*}[htb]
\begin{tabular}{cccc}
\begin{minipage}{0.25\hsize}
\begin{center}
    \FigureFile(40mm,40mm){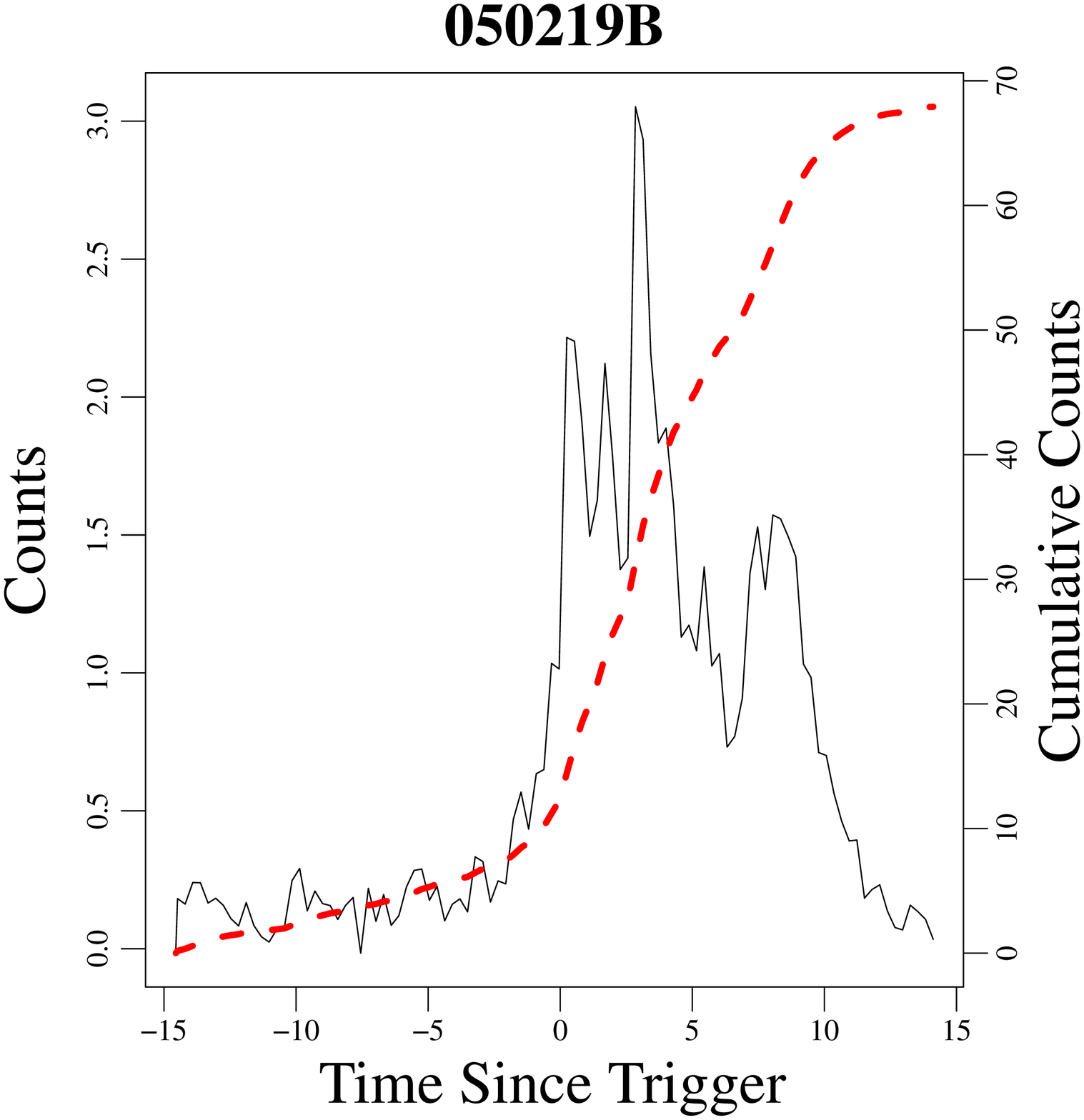}
\end{center}
\end{minipage}
\begin{minipage}{0.25\hsize}
\begin{center}
    \FigureFile(40mm,40mm){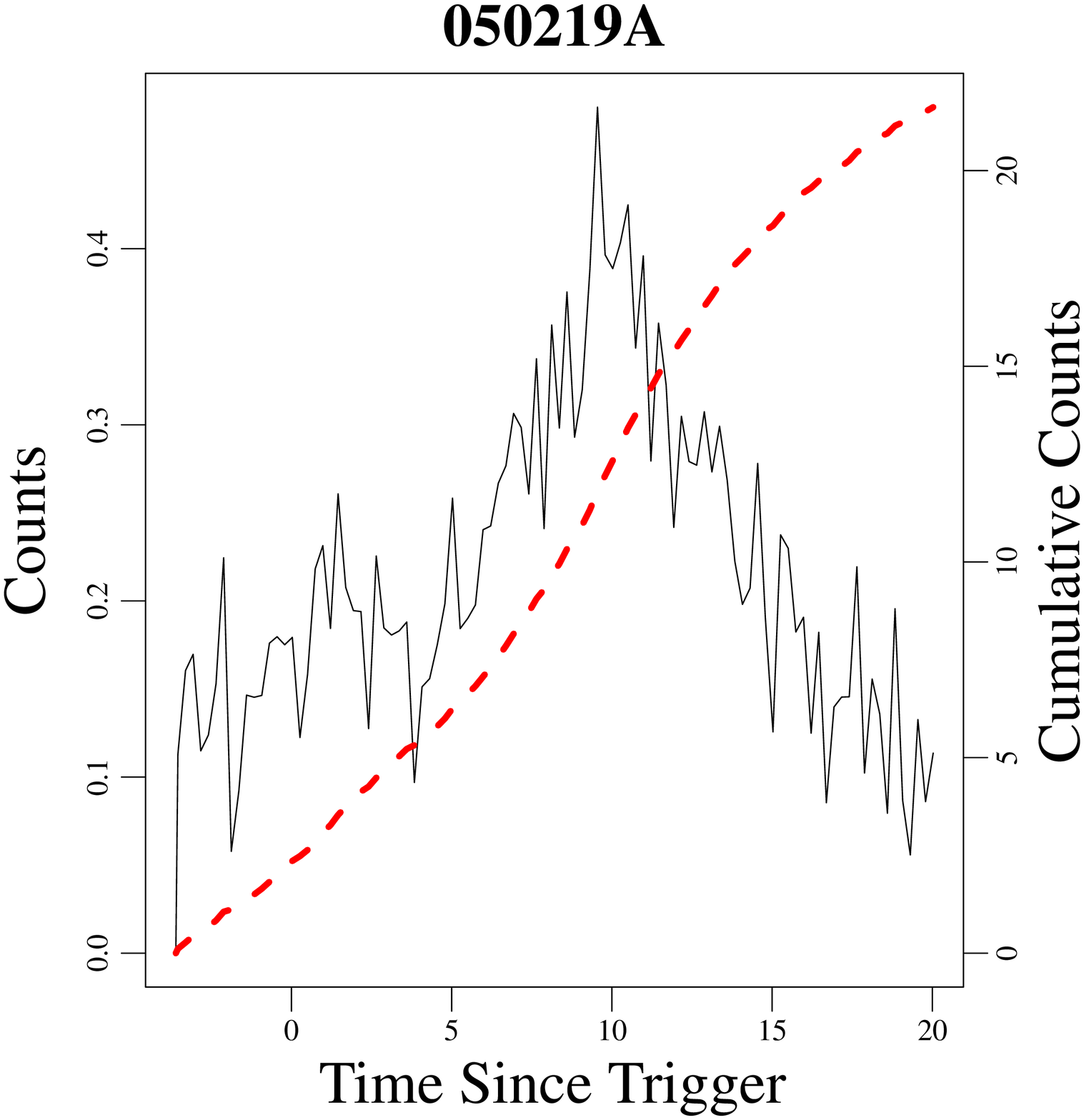}
 \end{center}
\end{minipage}
\begin{minipage}{0.25\hsize}
\begin{center}
    \FigureFile(40mm,40mm){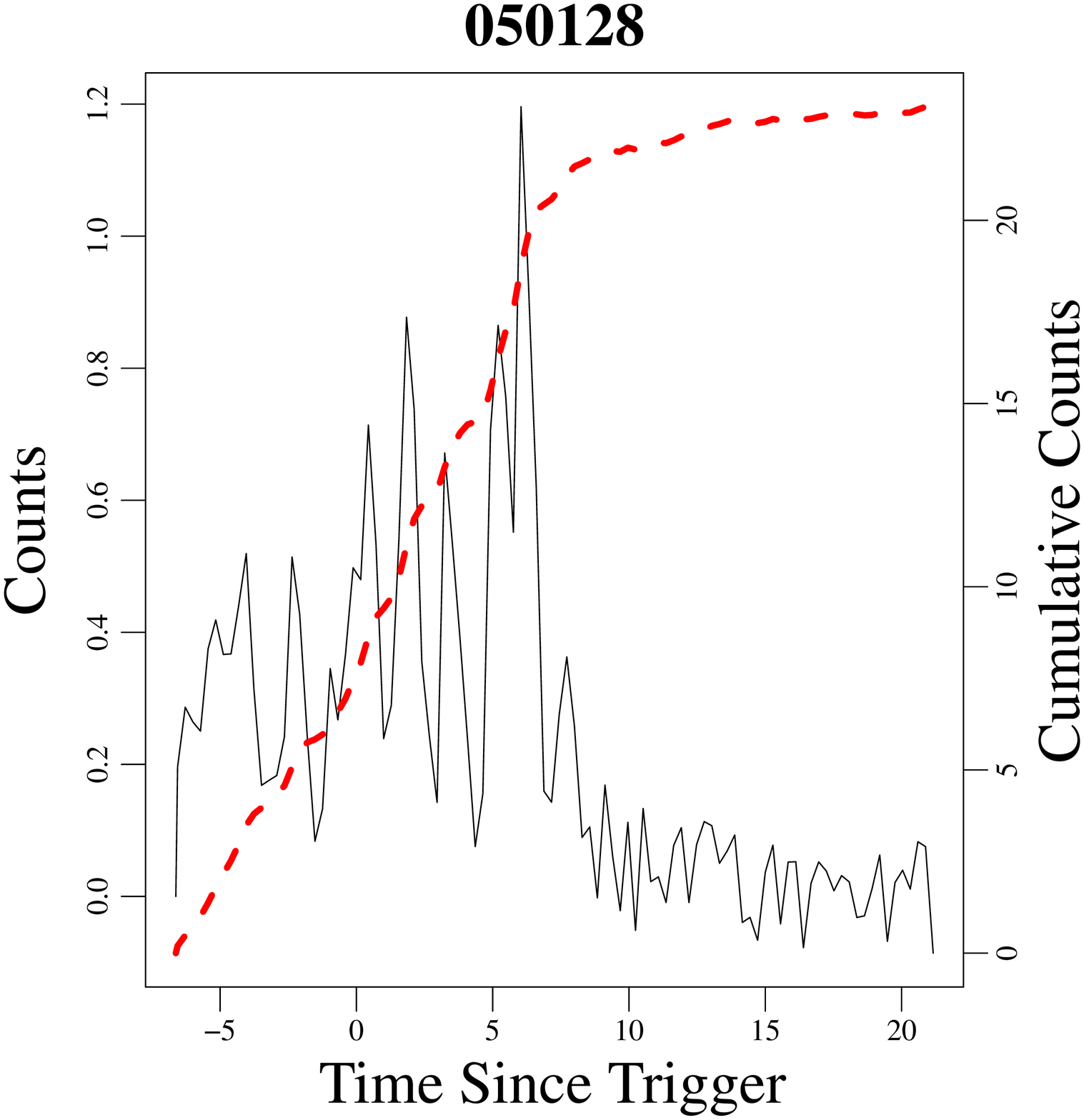}
\end{center}
\end{minipage}
\begin{minipage}{0.25\hsize}
\begin{center}
    \FigureFile(40mm,40mm){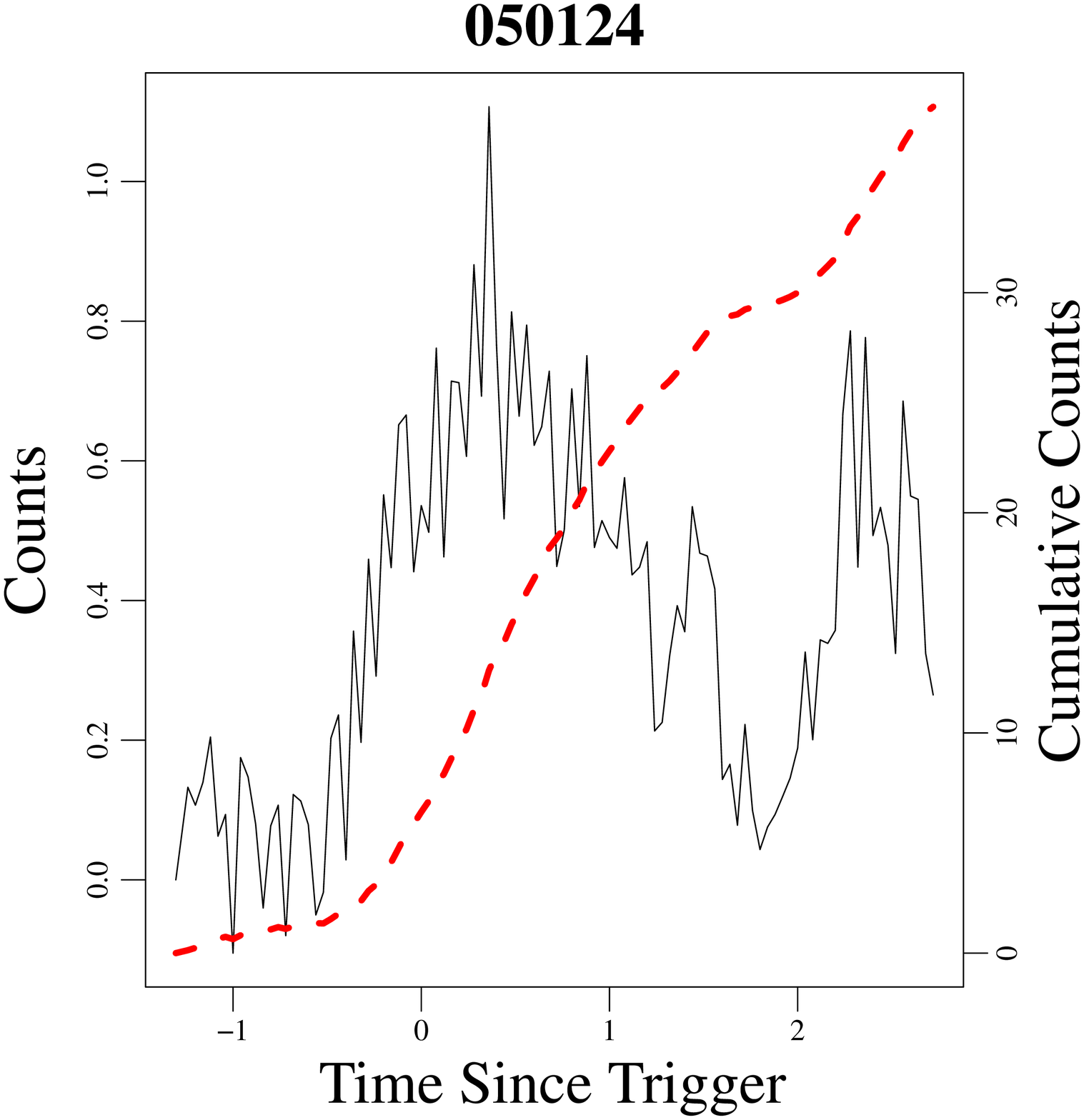}
 \end{center}
\end{minipage}\\
\begin{minipage}{0.25\hsize}
\begin{center}
    \FigureFile(40mm,40mm){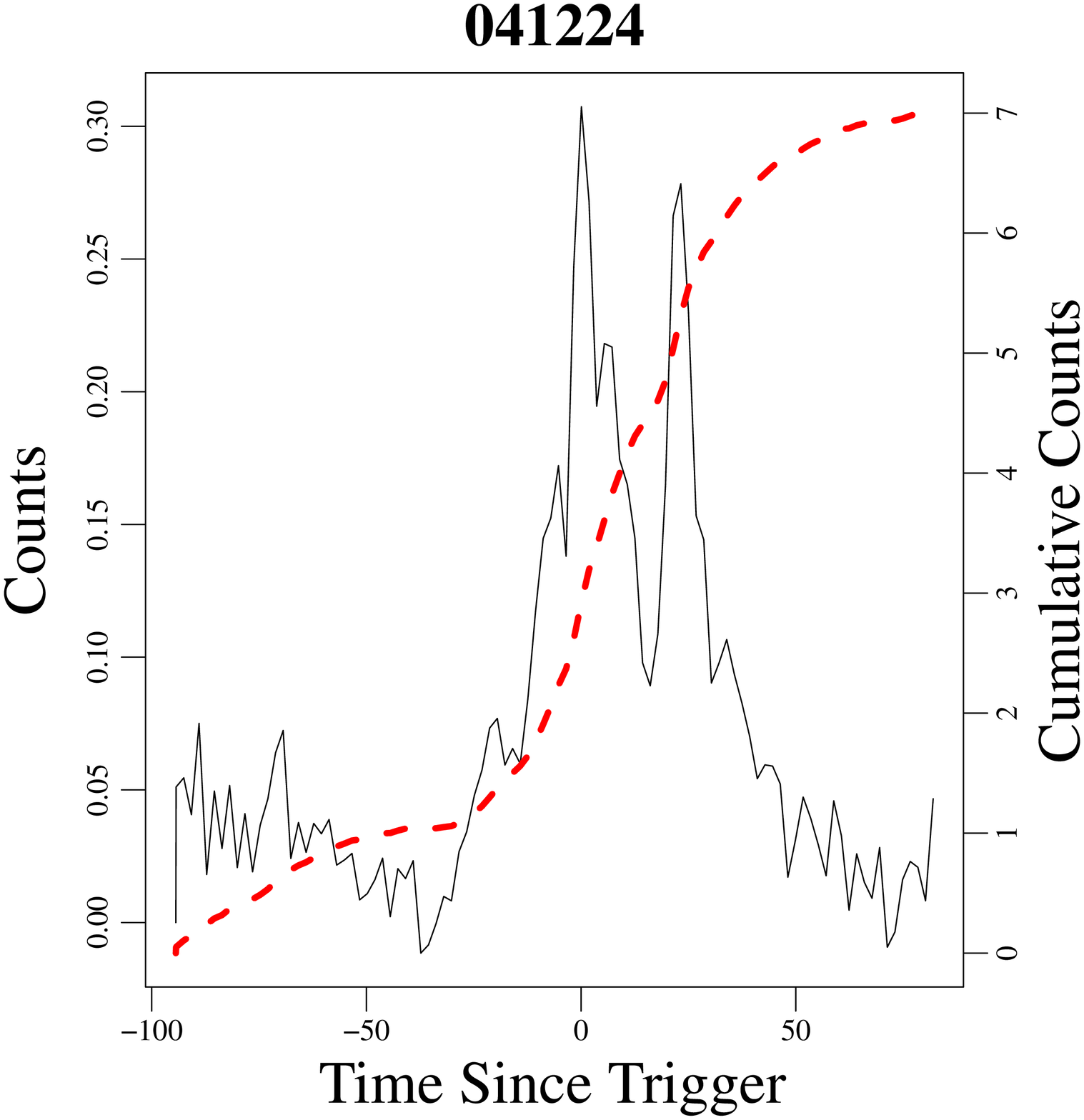}
\end{center}
\end{minipage}
\begin{minipage}{0.25\hsize}
\end{minipage}
\begin{minipage}{0.25\hsize}
\end{minipage}
\begin{minipage}{0.25\hsize}
\end{minipage}
\end{tabular}
   \caption{Light curves (black solid) and cumulative light curves (red doted) of Type I LGRBs.}\label{fig:A1}
\end{figure*}

\begin{figure*}[htb]
\begin{tabular}{cccc}
\begin{minipage}{0.25\hsize}
\begin{center}
    \FigureFile(40mm,40mm){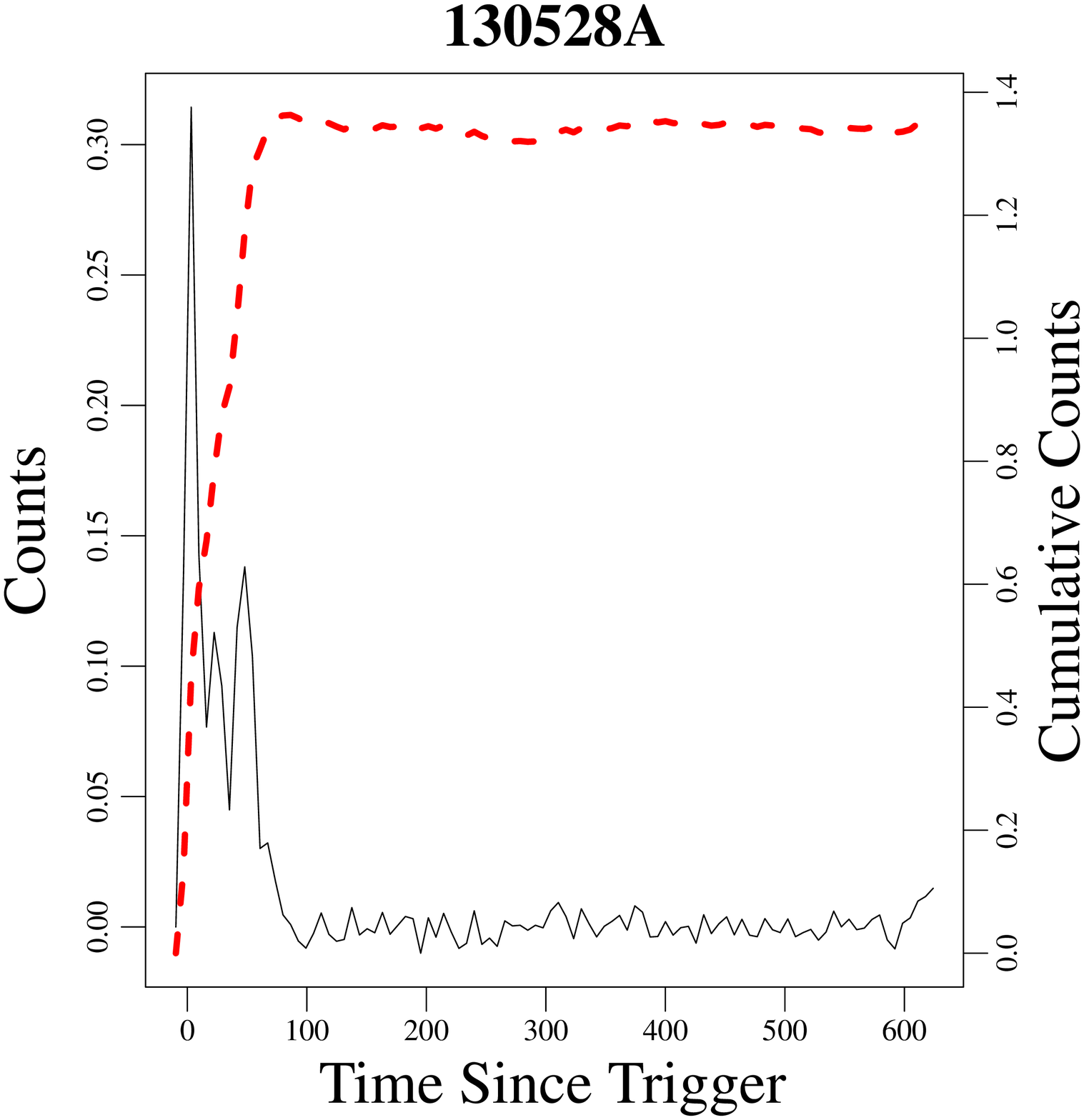}
\end{center}
\end{minipage}
\begin{minipage}{0.25\hsize}
\begin{center}
    \FigureFile(40mm,40mm){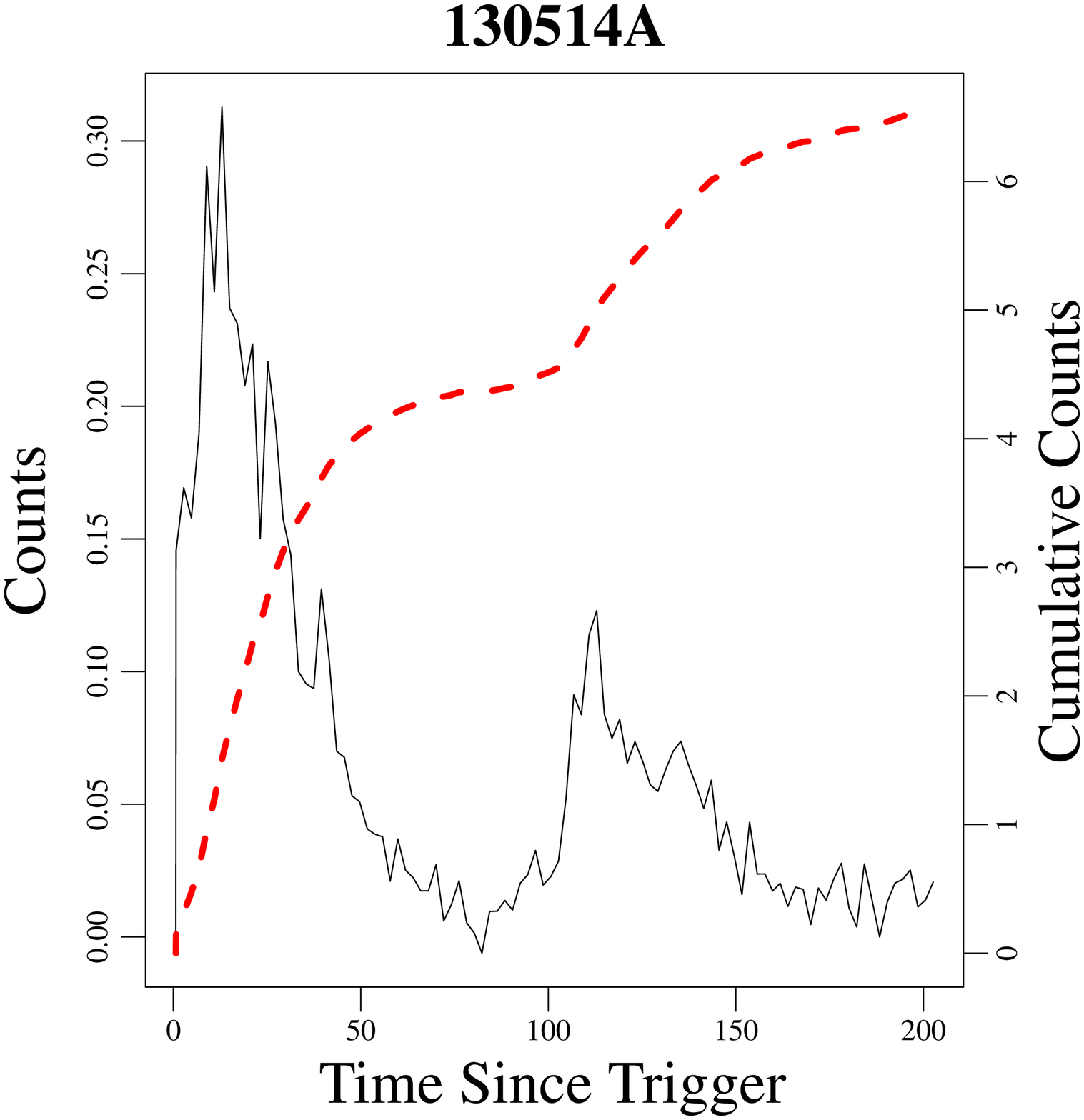}
 \end{center}
\end{minipage}
\begin{minipage}{0.25\hsize}
\begin{center}
    \FigureFile(40mm,40mm){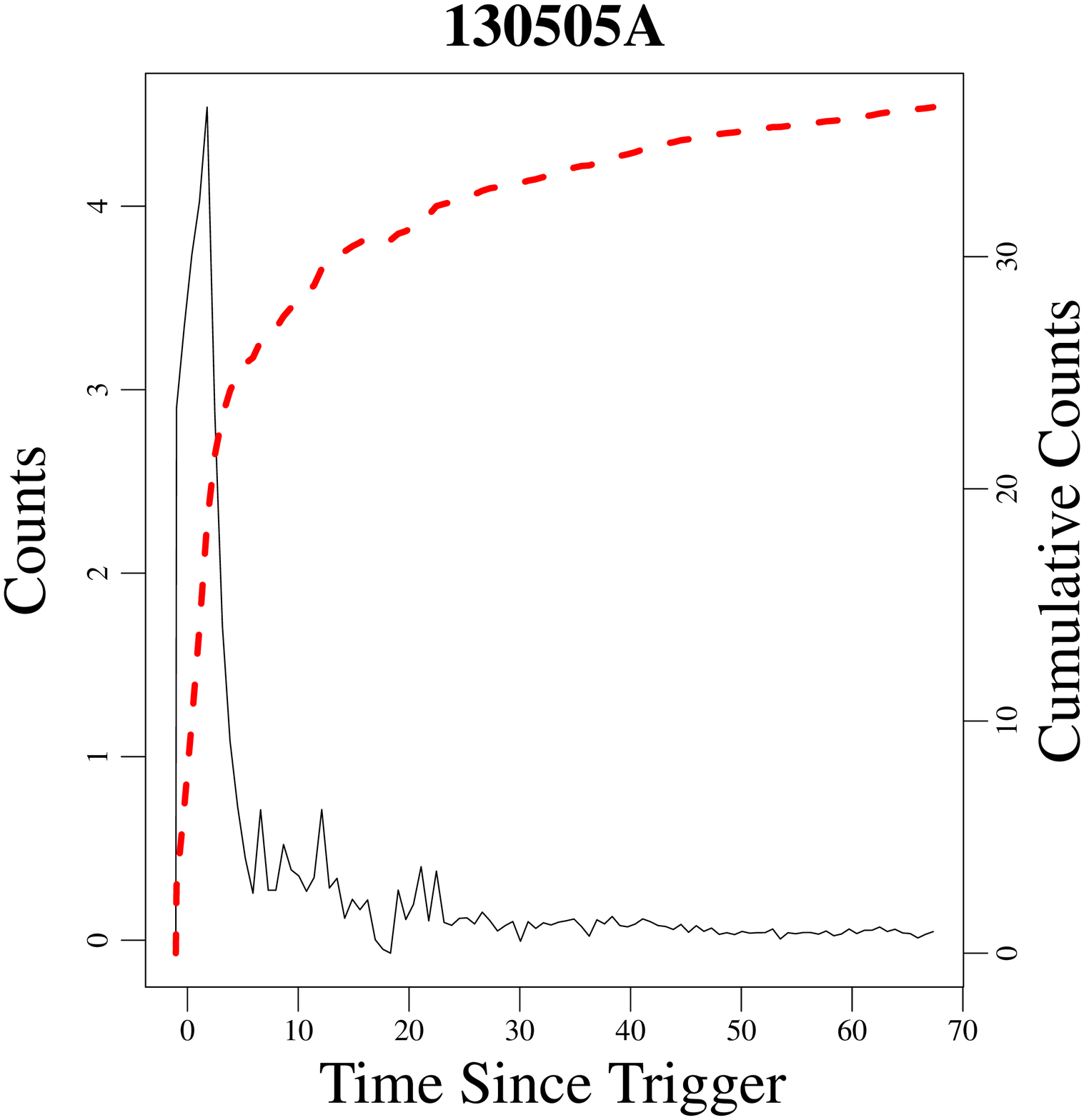}
\end{center}
\end{minipage}
\begin{minipage}{0.25\hsize}
\begin{center}
    \FigureFile(40mm,40mm){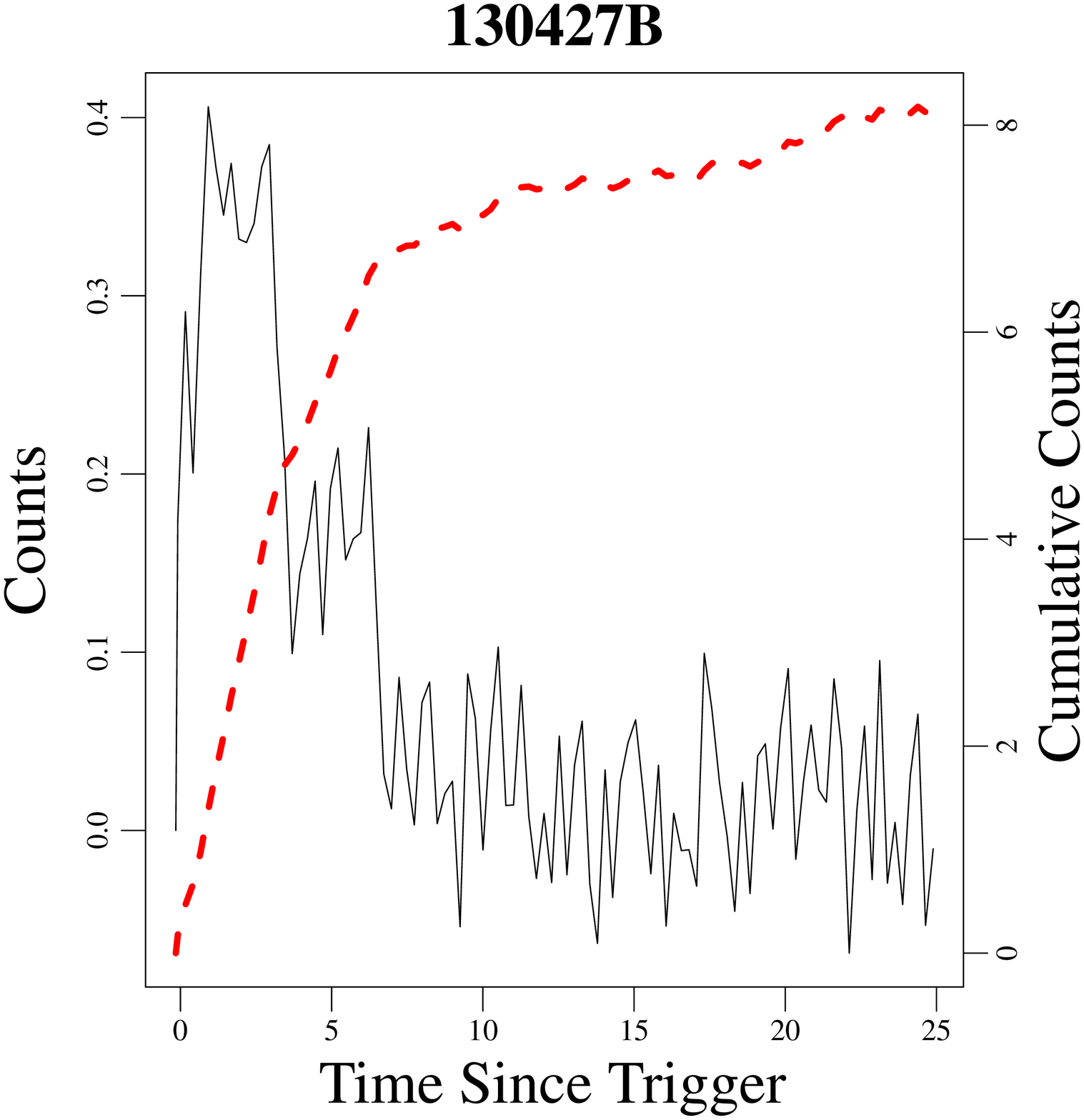}
 \end{center}
\end{minipage}\\
\begin{minipage}{0.25\hsize}
\begin{center}
    \FigureFile(40mm,40mm){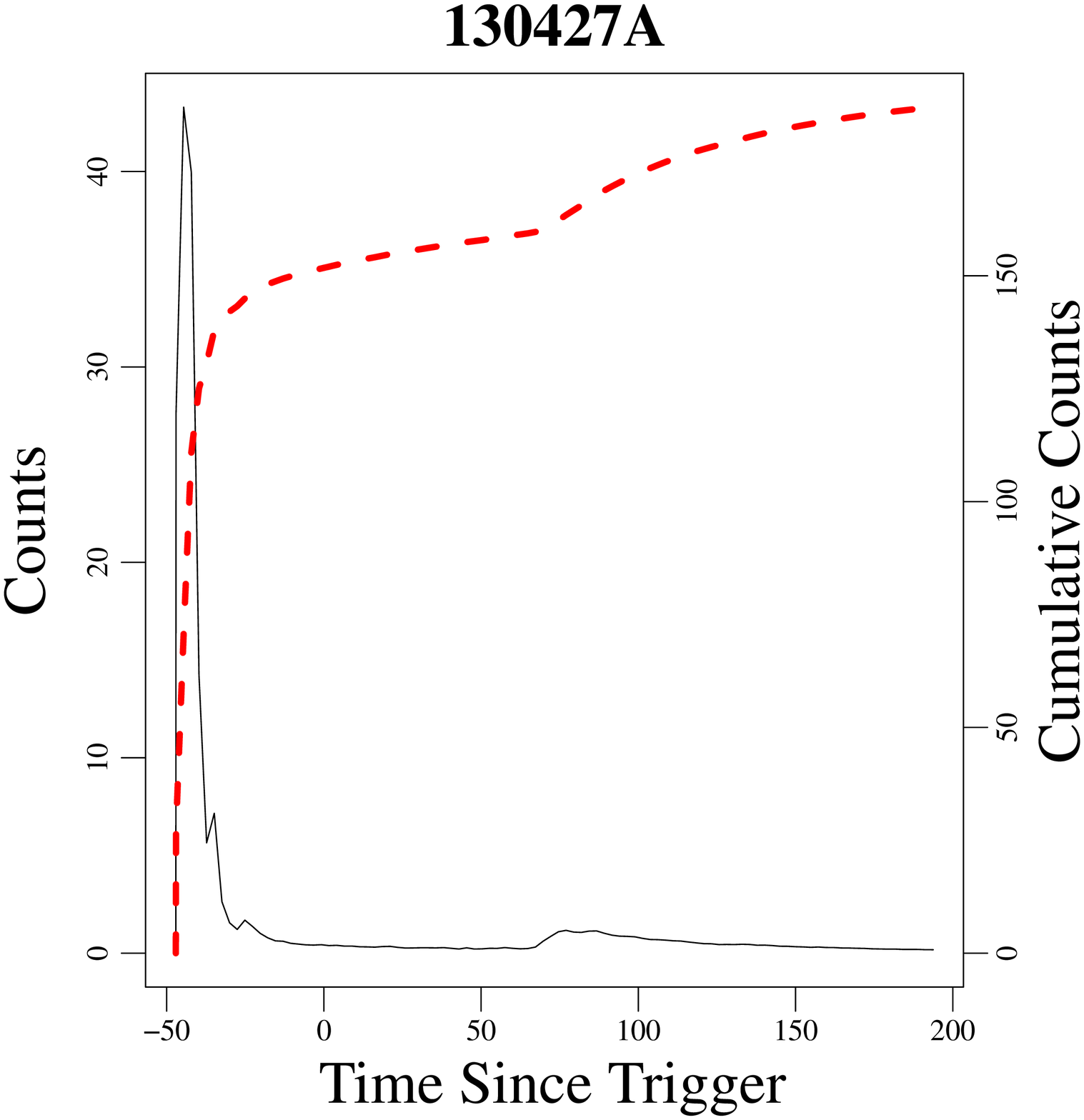}
\end{center}
\end{minipage}
\begin{minipage}{0.25\hsize}
\begin{center}
    \FigureFile(40mm,40mm){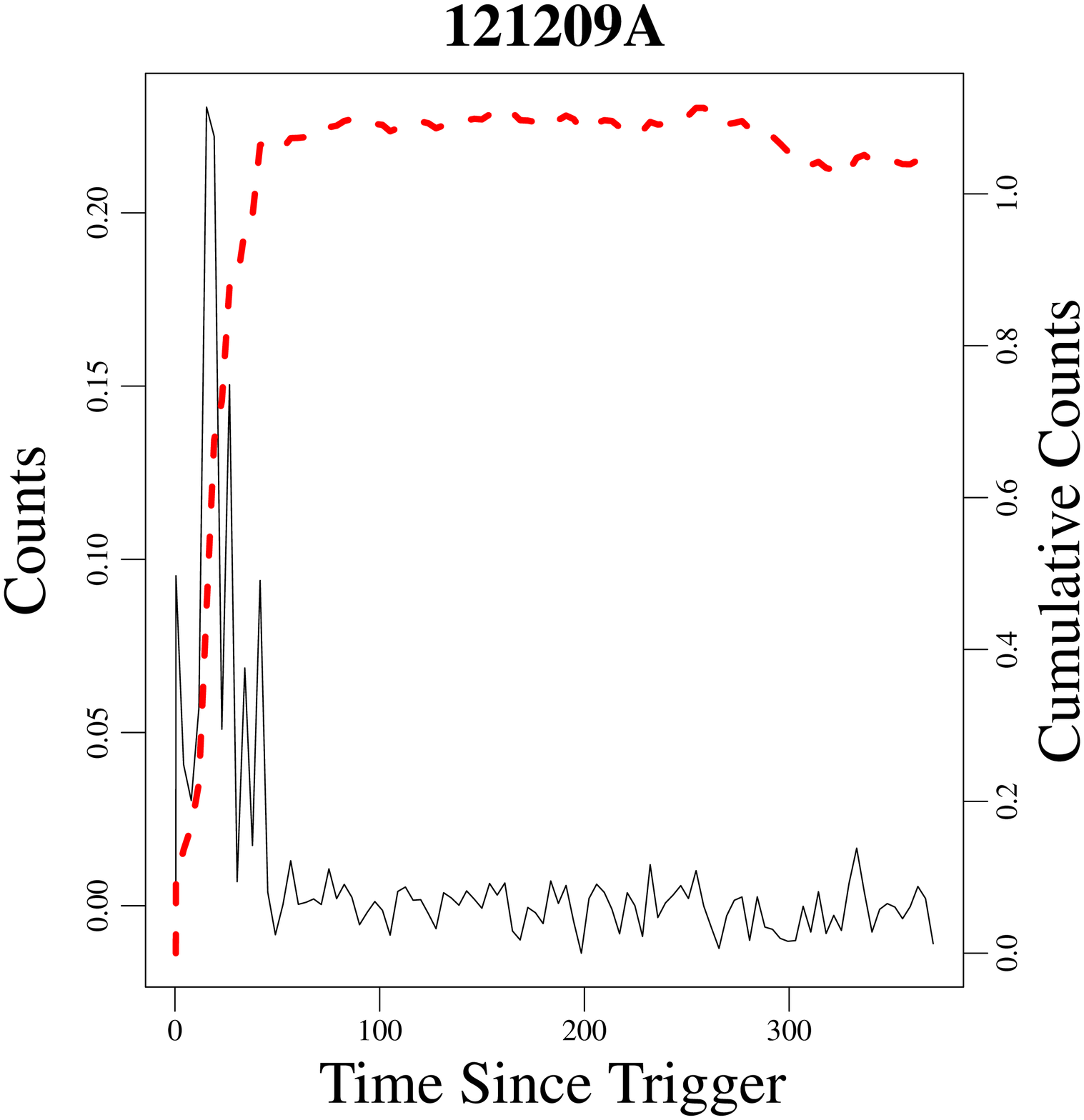}
 \end{center}
\end{minipage}
\begin{minipage}{0.25\hsize}
\begin{center}
    \FigureFile(40mm,40mm){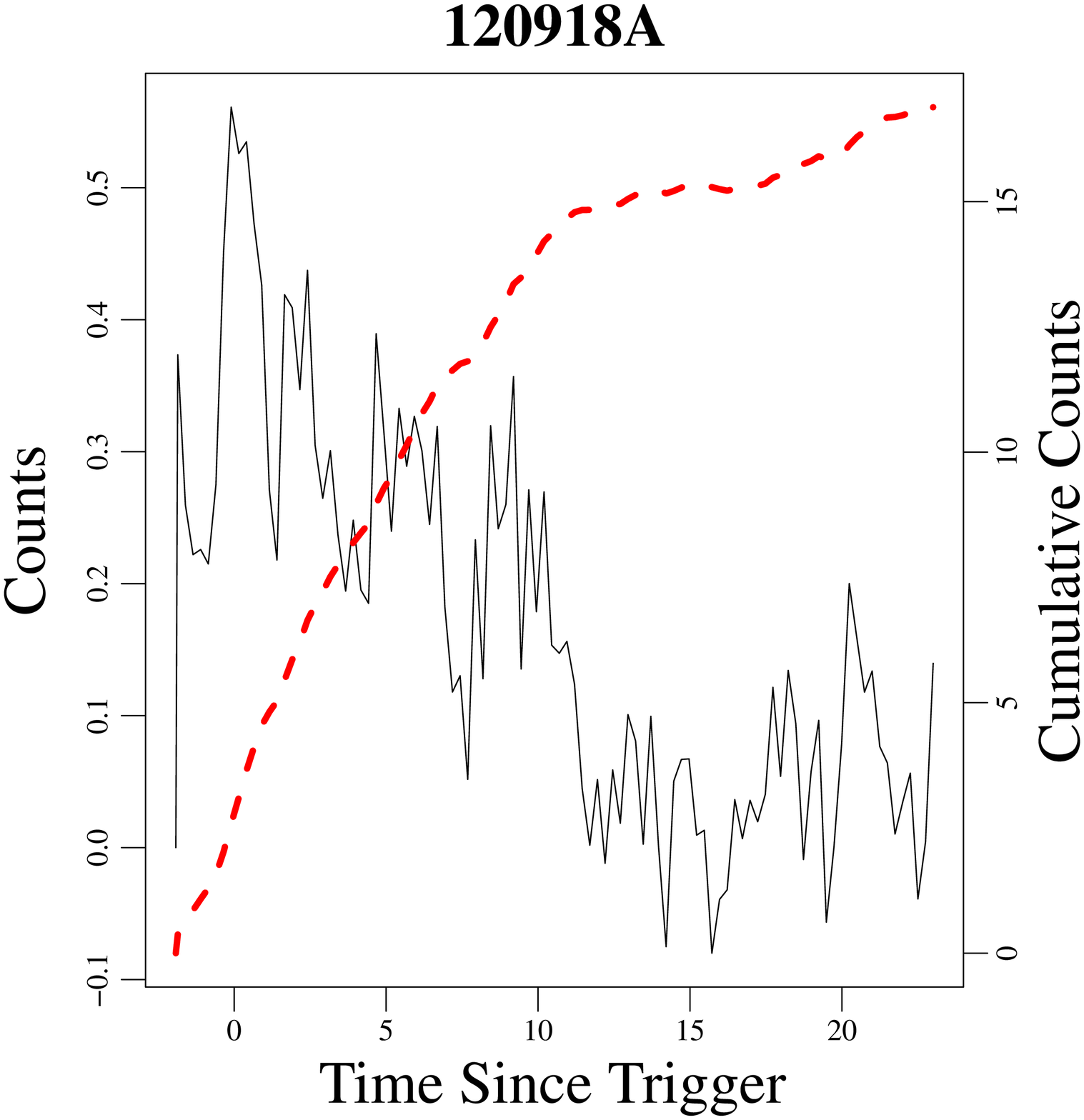}
\end{center}
\end{minipage}
\begin{minipage}{0.25\hsize}
\begin{center}
    \FigureFile(40mm,40mm){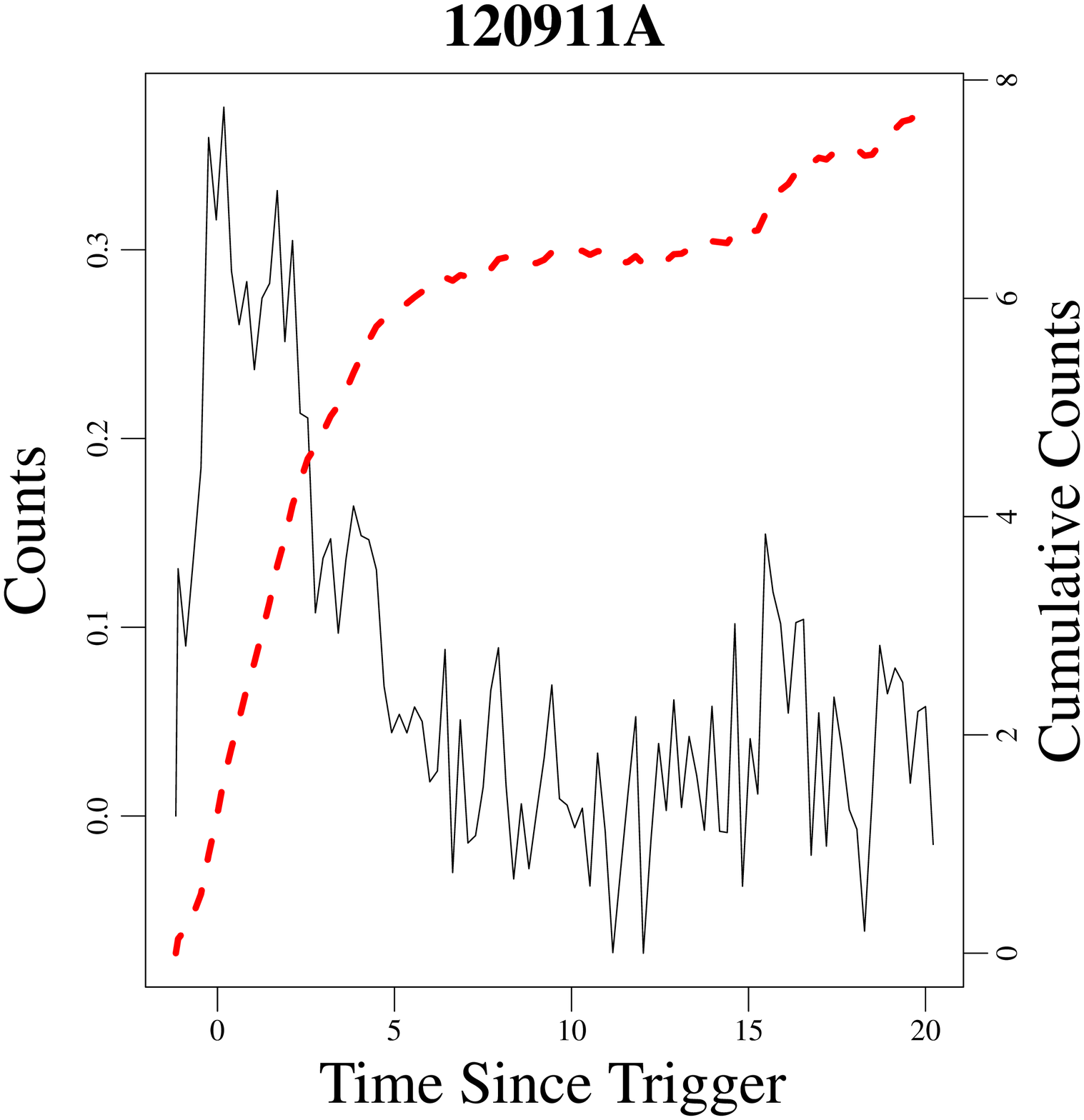}
 \end{center}
\end{minipage}\\
\begin{minipage}{0.25\hsize}
\begin{center}
    \FigureFile(40mm,40mm){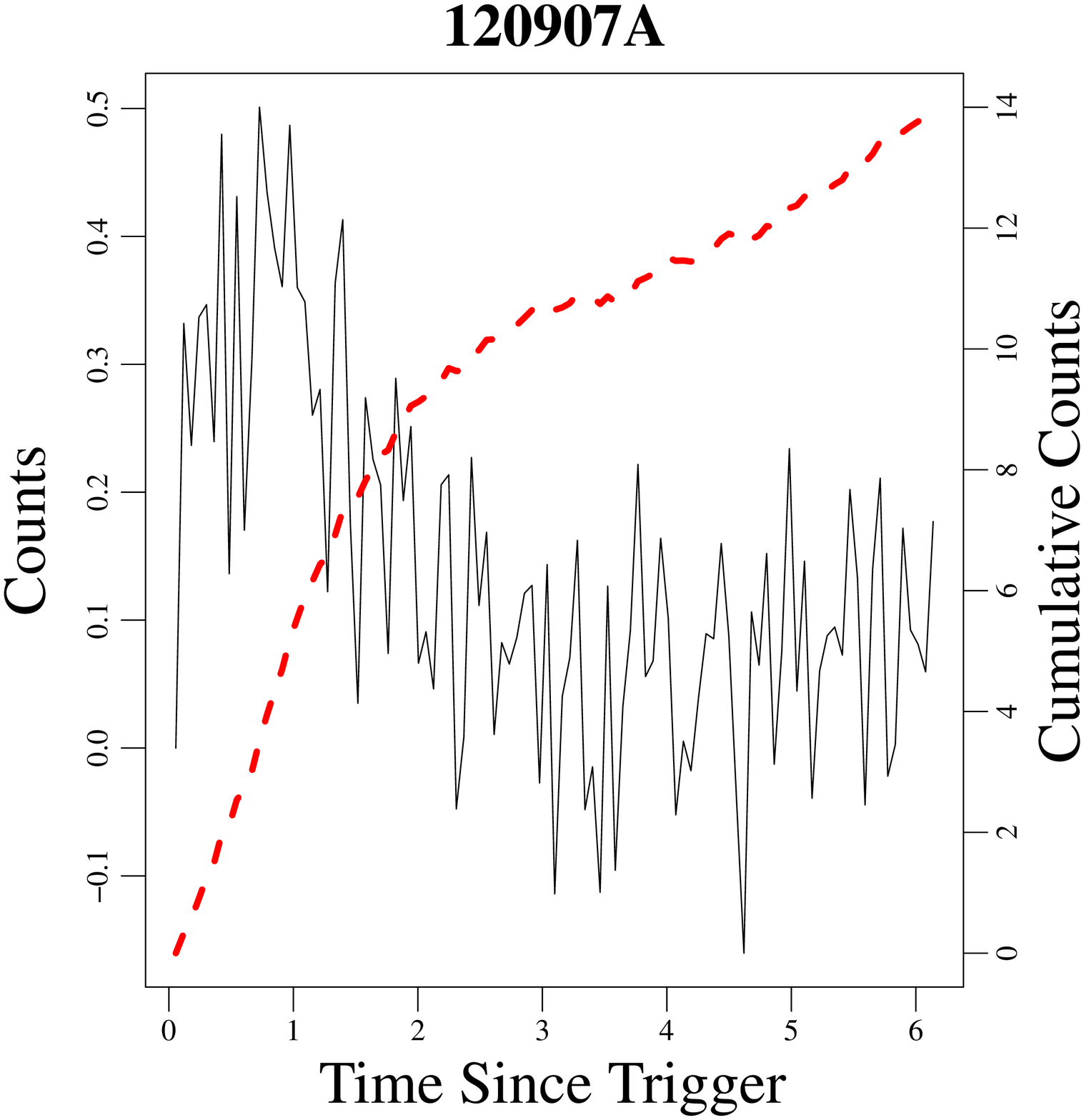}
\end{center}
\end{minipage}
\begin{minipage}{0.25\hsize}
\begin{center}
    \FigureFile(40mm,40mm){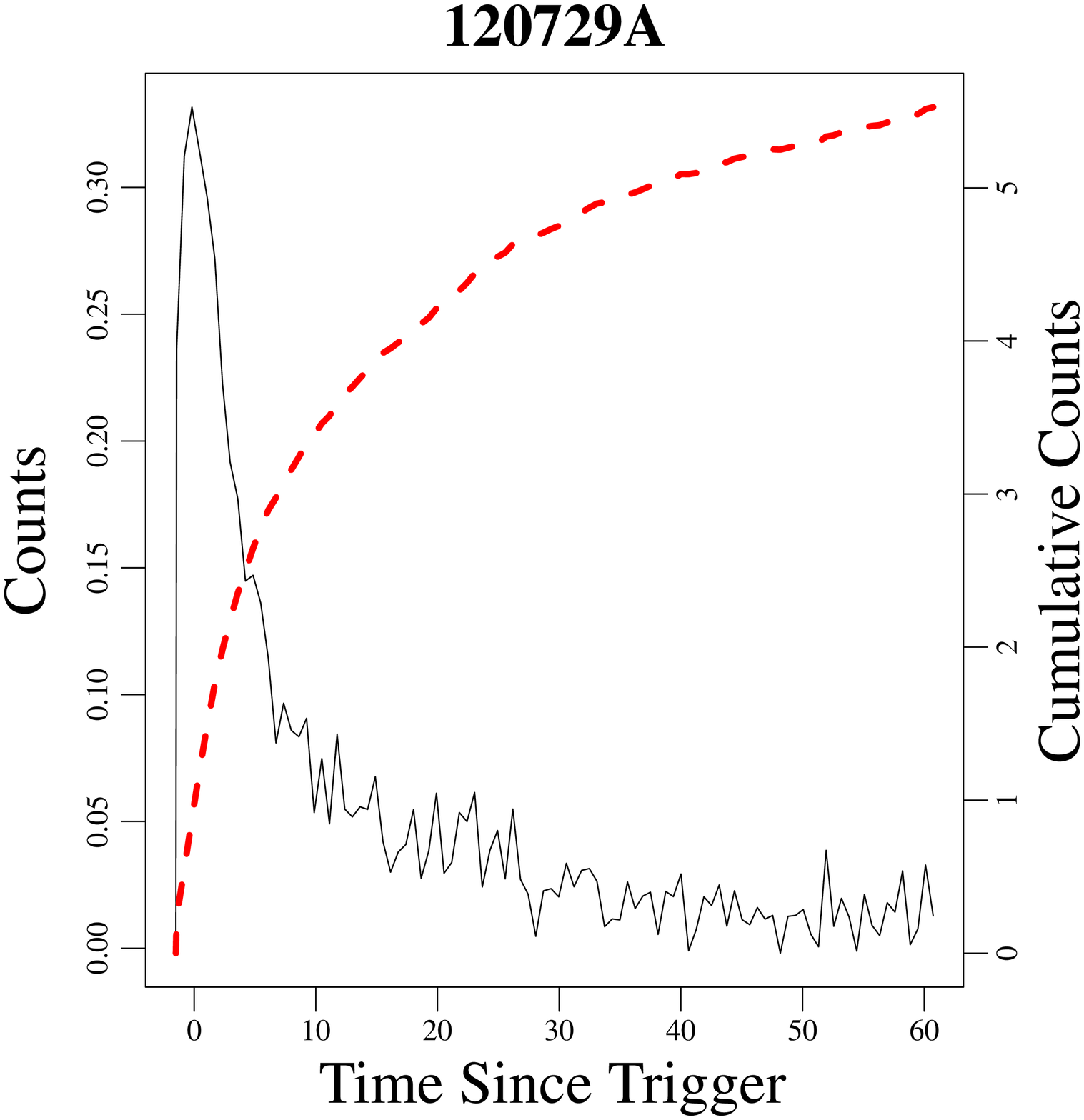}
 \end{center}
\end{minipage}
\begin{minipage}{0.25\hsize}
\begin{center}
    \FigureFile(40mm,40mm){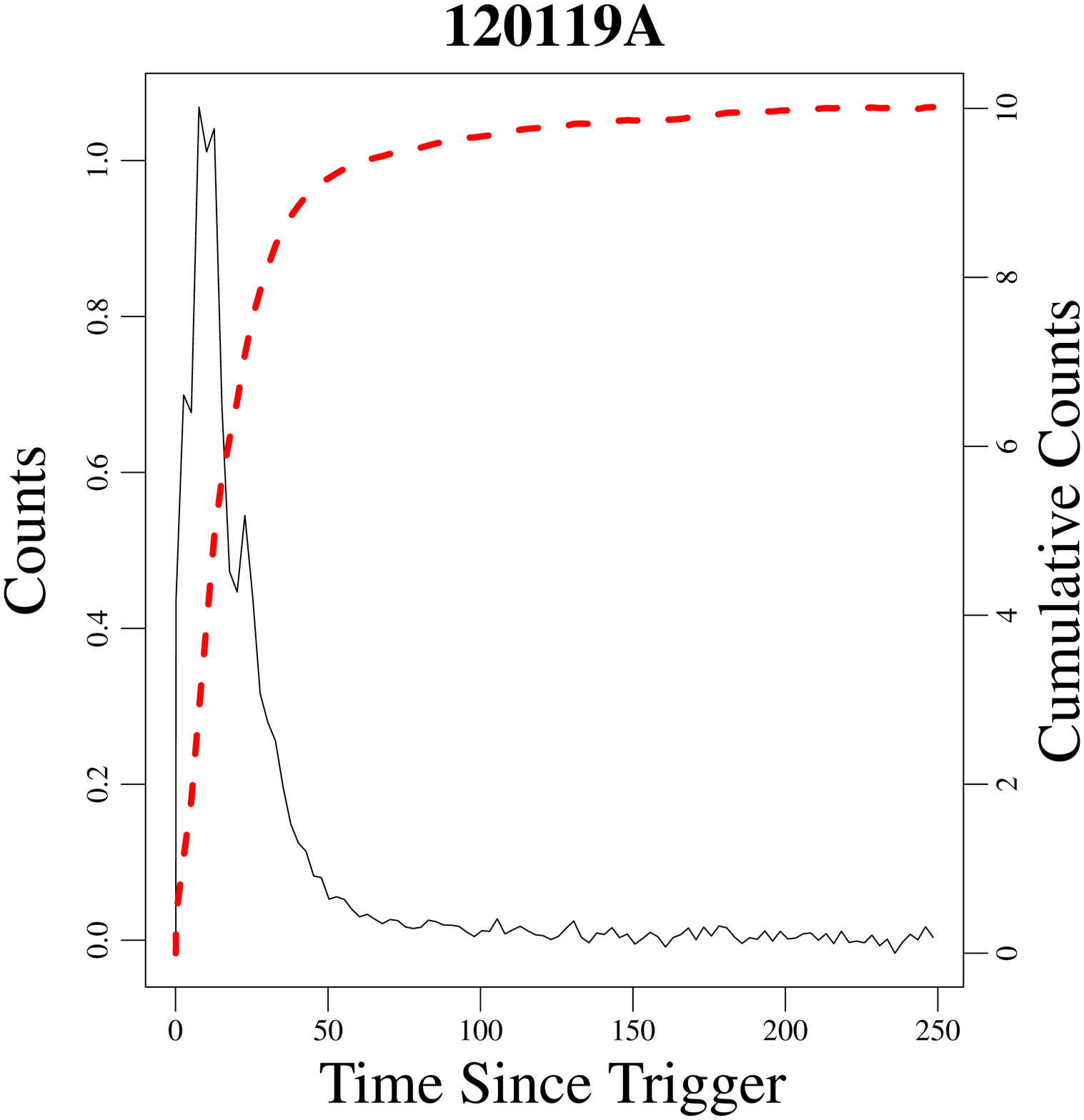}
\end{center}
\end{minipage}
\begin{minipage}{0.25\hsize}
\begin{center}
    \FigureFile(40mm,40mm){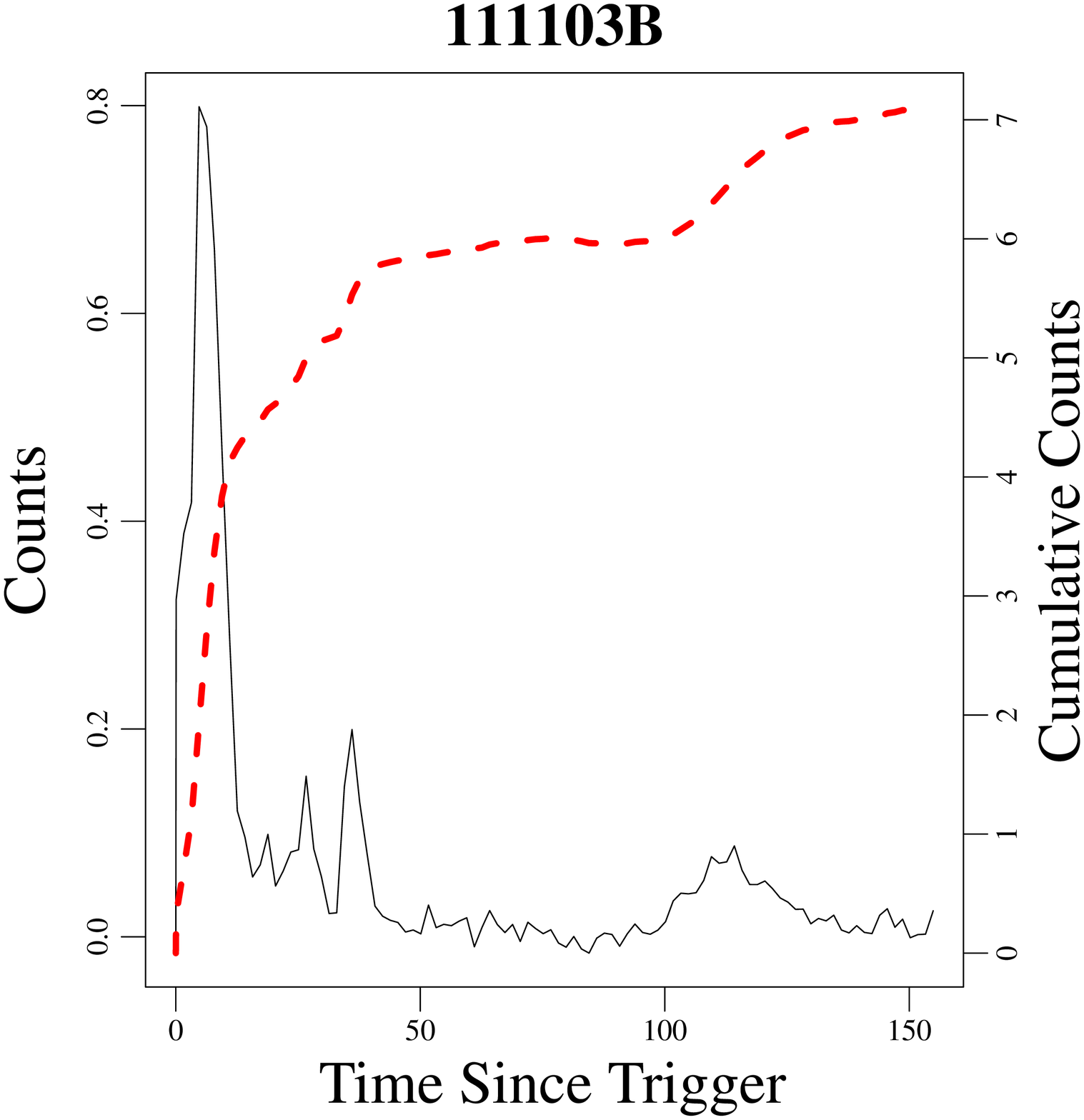}
 \end{center}
\end{minipage}\\
\begin{minipage}{0.25\hsize}
\begin{center}
    \FigureFile(40mm,40mm){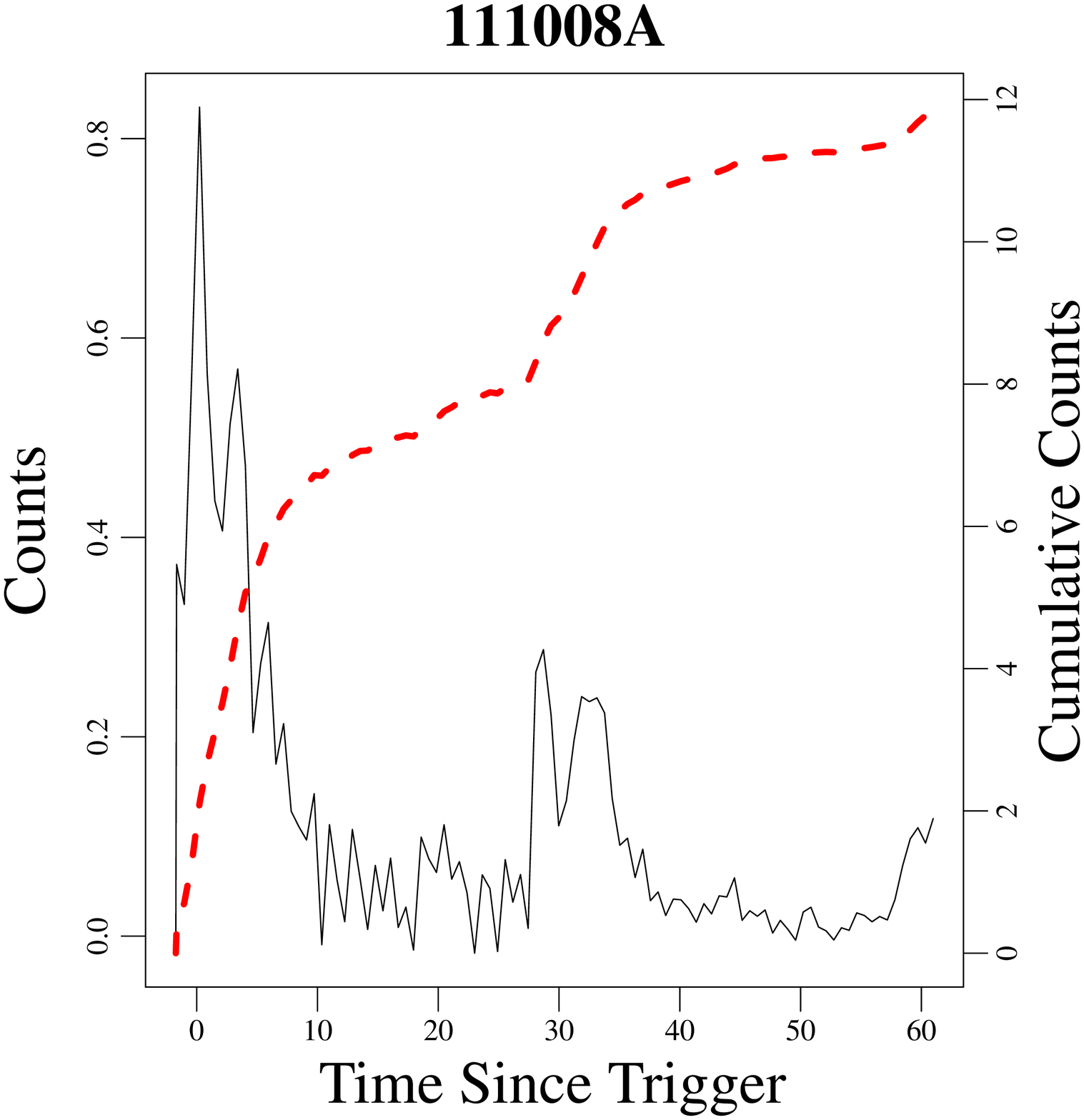}
\end{center}
\end{minipage}
\begin{minipage}{0.25\hsize}
\begin{center}
    \FigureFile(40mm,40mm){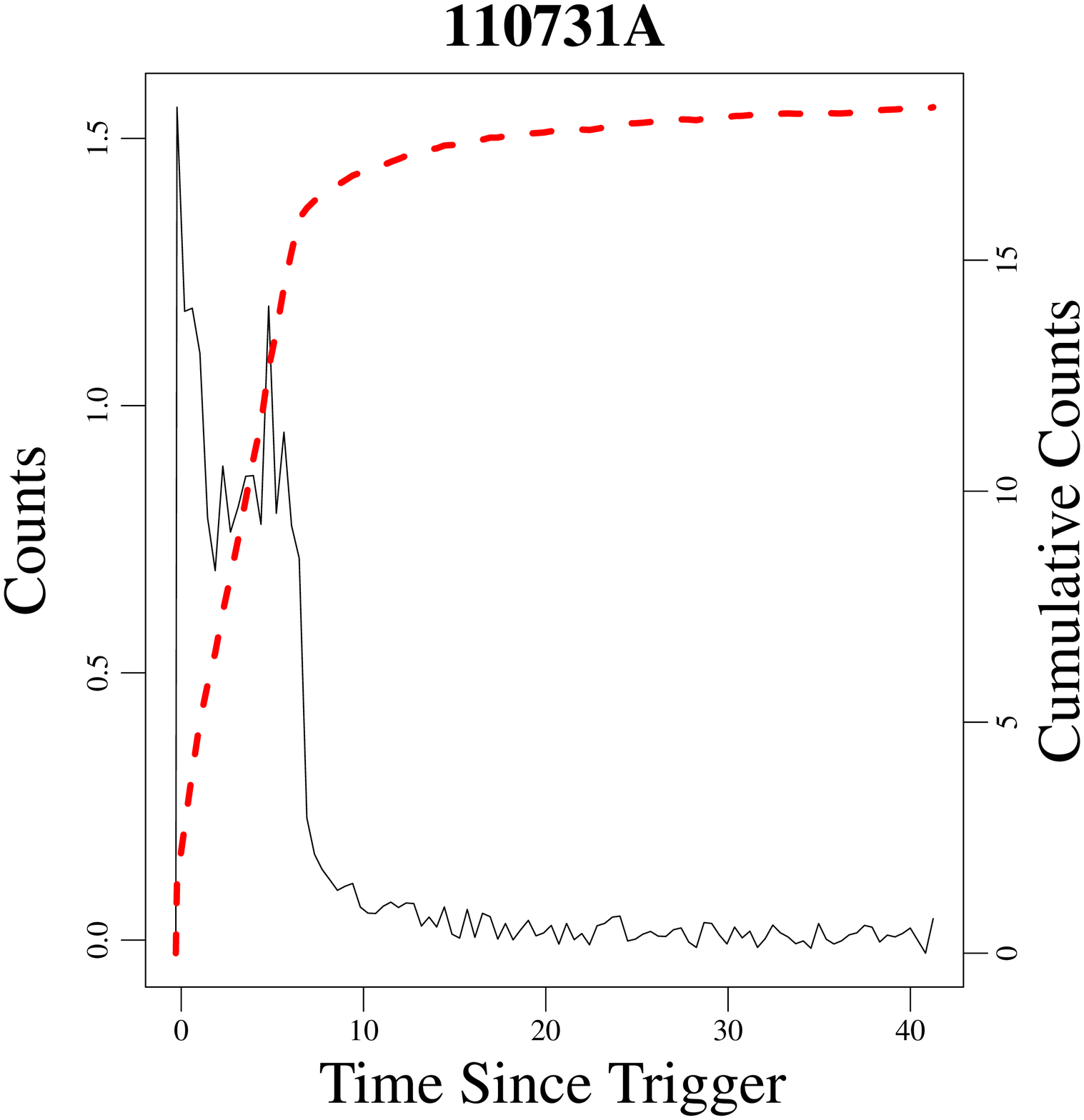}
 \end{center}
\end{minipage}
\begin{minipage}{0.25\hsize}
\begin{center}
    \FigureFile(40mm,40mm){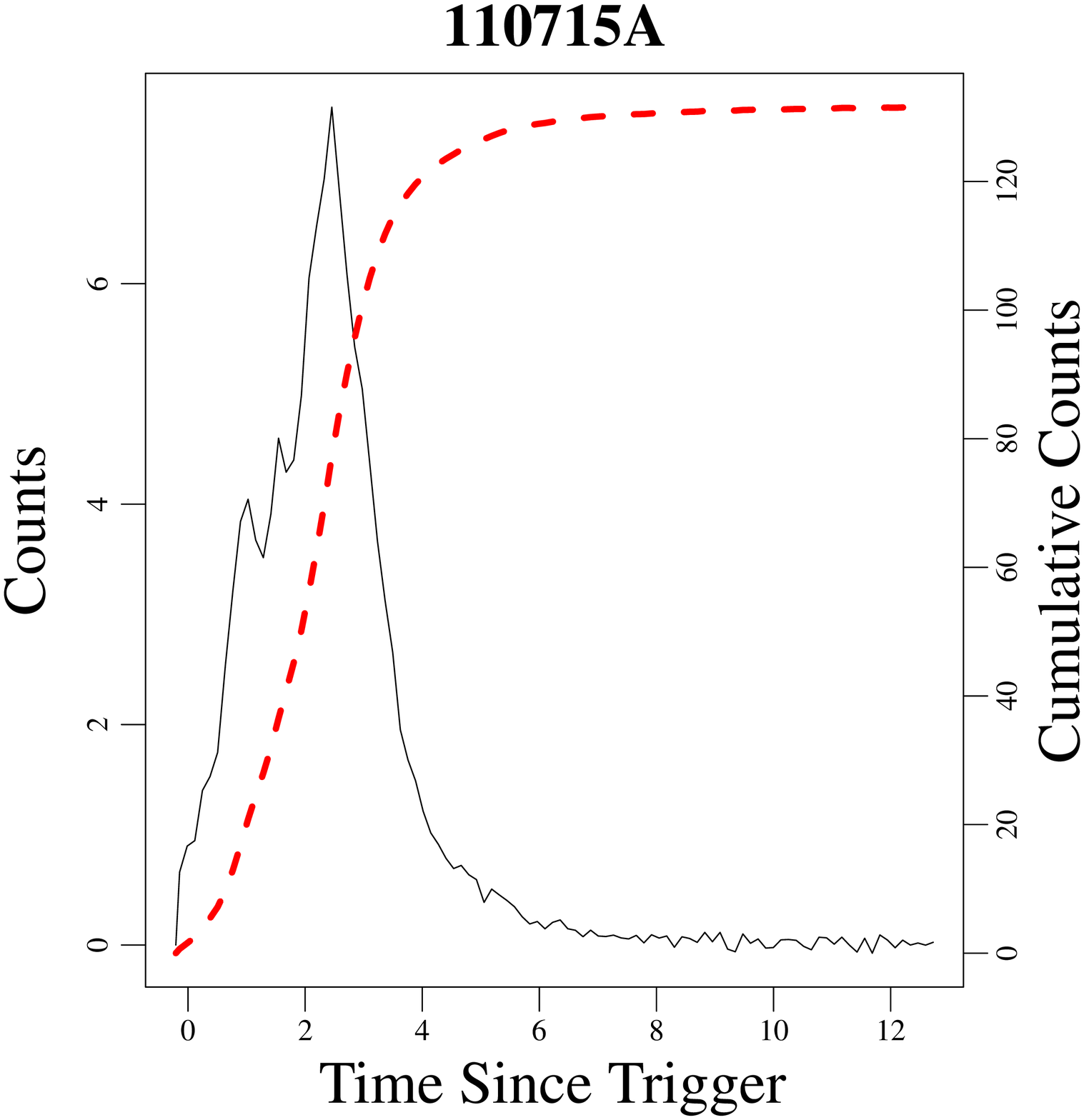}
\end{center}
\end{minipage}
\begin{minipage}{0.25\hsize}
\begin{center}
    \FigureFile(40mm,40mm){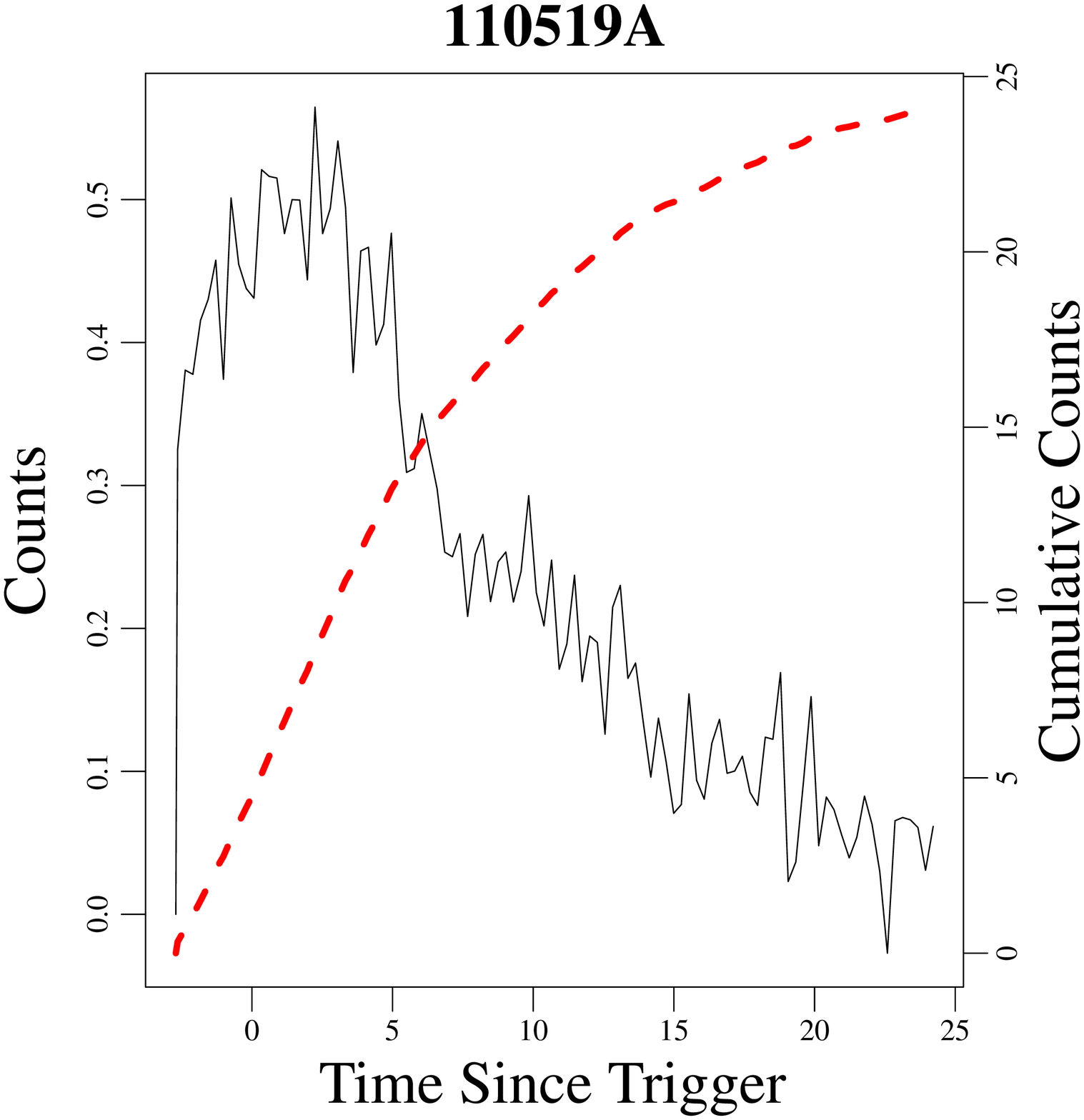}
 \end{center}
\end{minipage}\\
\begin{minipage}{0.25\hsize}
\begin{center}
    \FigureFile(40mm,40mm){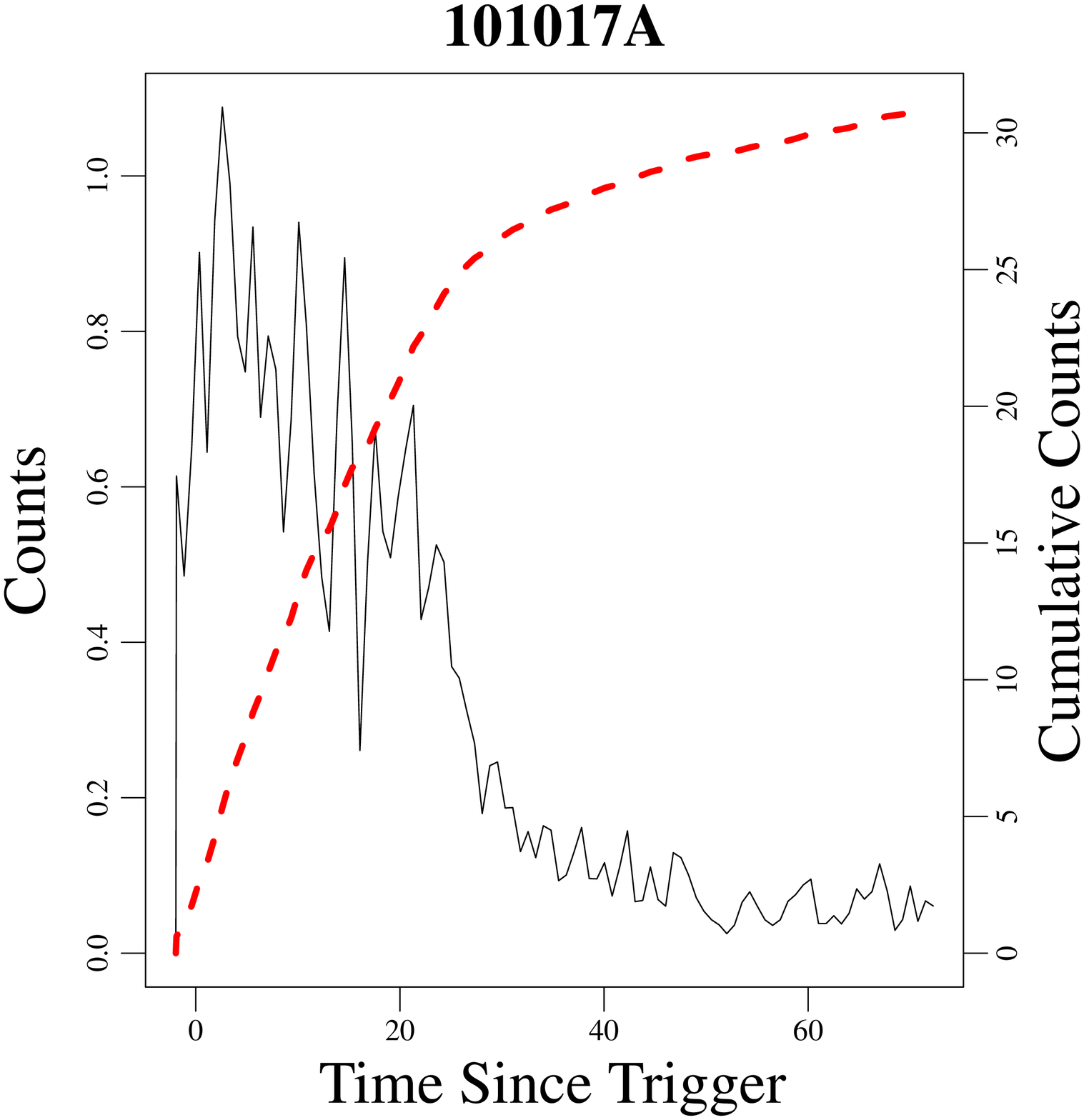}
\end{center}
\end{minipage}
\begin{minipage}{0.25\hsize}
\begin{center}
    \FigureFile(40mm,40mm){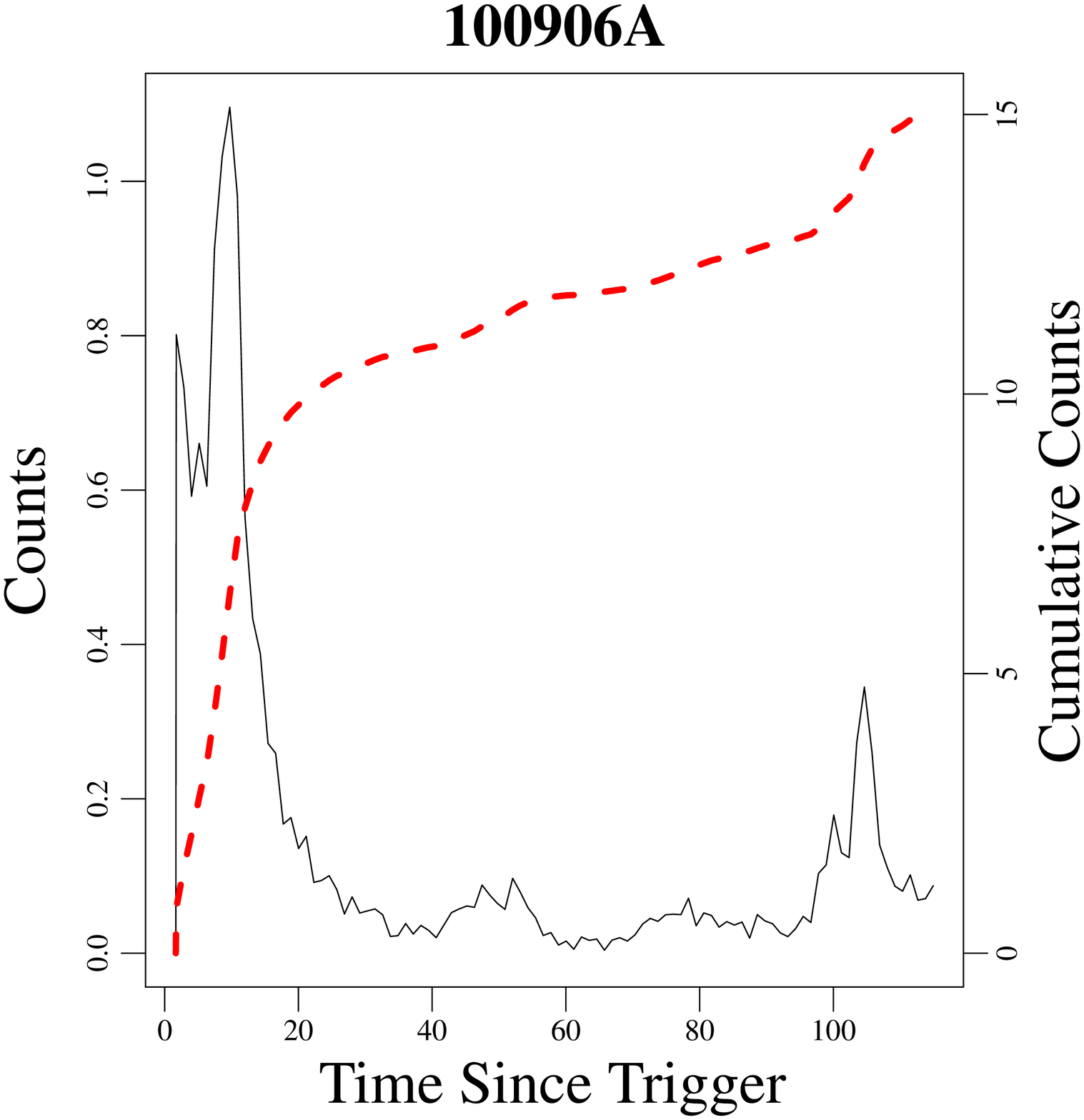}
 \end{center}
\end{minipage}
\begin{minipage}{0.25\hsize}
\begin{center}
    \FigureFile(40mm,40mm){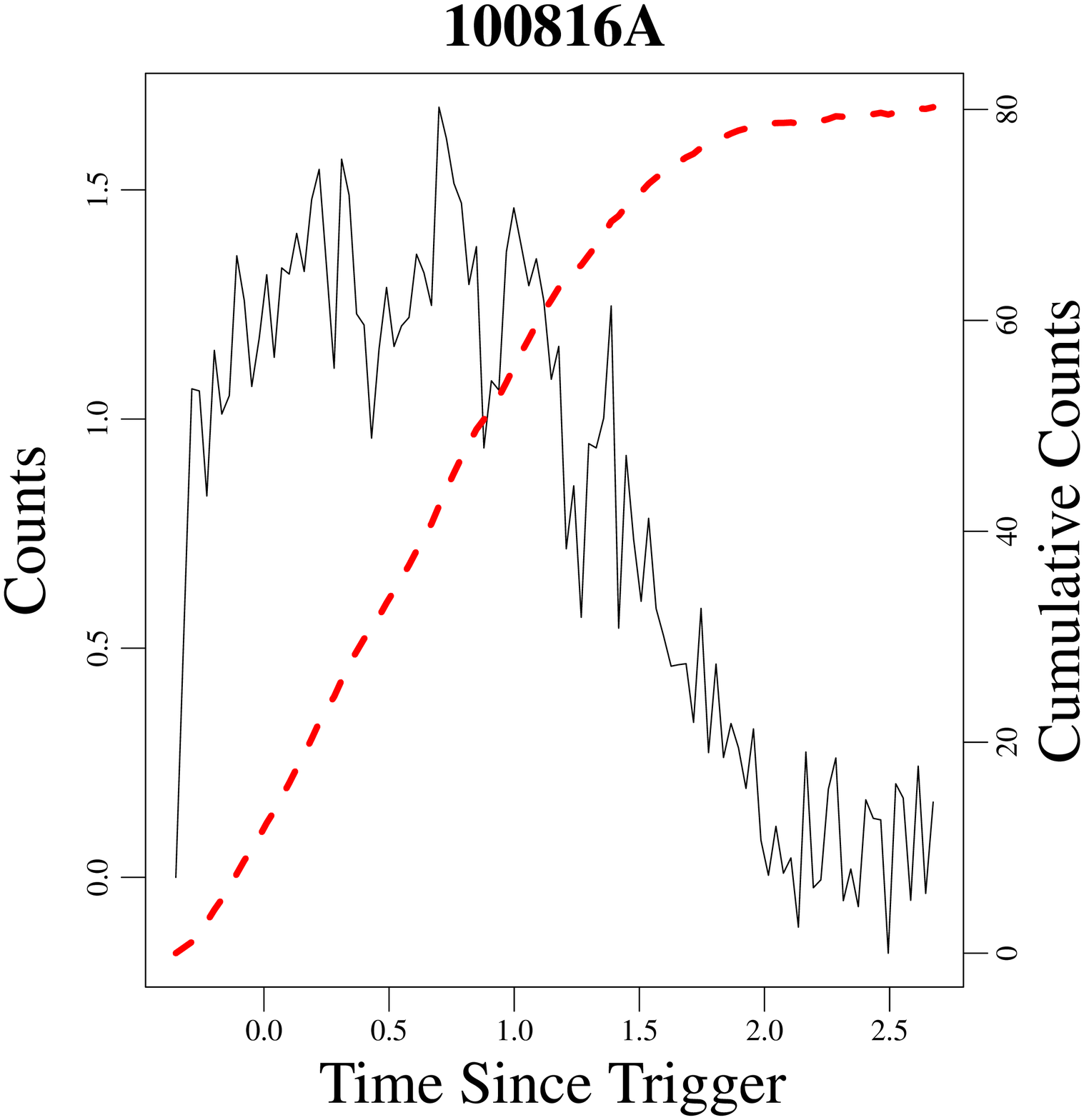}
\end{center}
\end{minipage}
\begin{minipage}{0.25\hsize}
\begin{center}
    \FigureFile(40mm,40mm){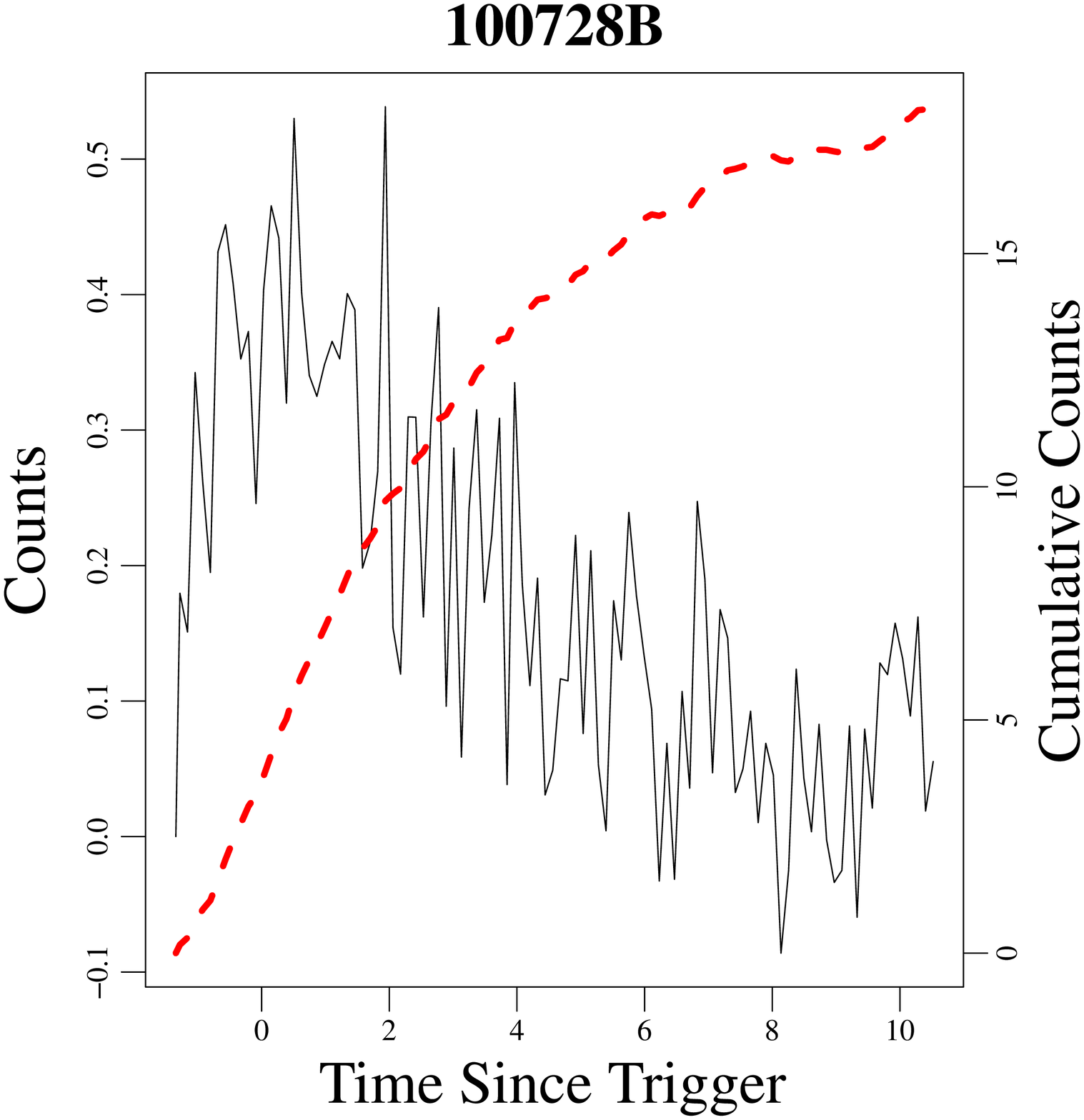}
 \end{center}
\end{minipage}\\
\begin{minipage}{0.25\hsize}
\begin{center}
    \FigureFile(40mm,40mm){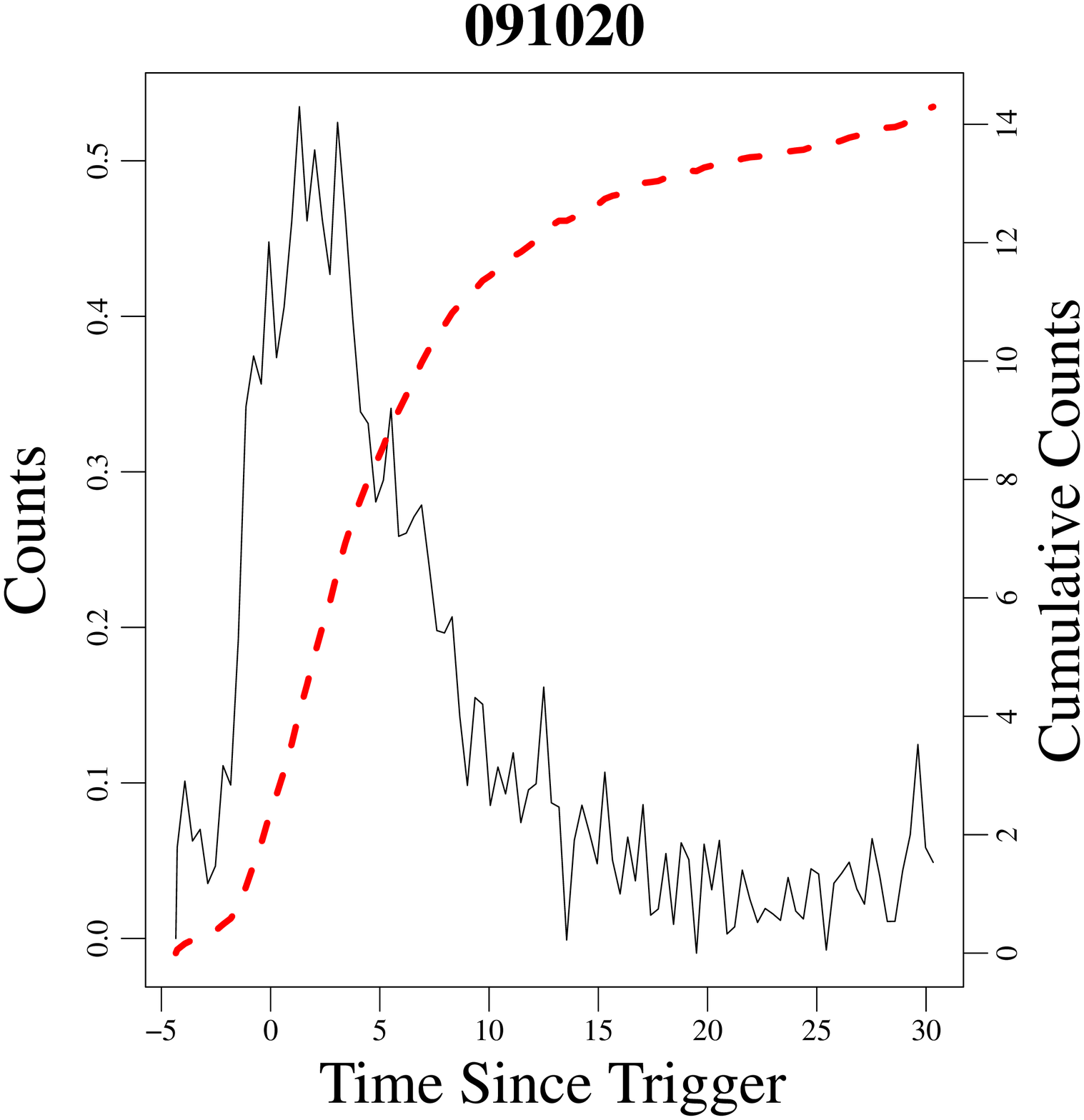}
\end{center}
\end{minipage}
\begin{minipage}{0.25\hsize}
\begin{center}
    \FigureFile(40mm,40mm){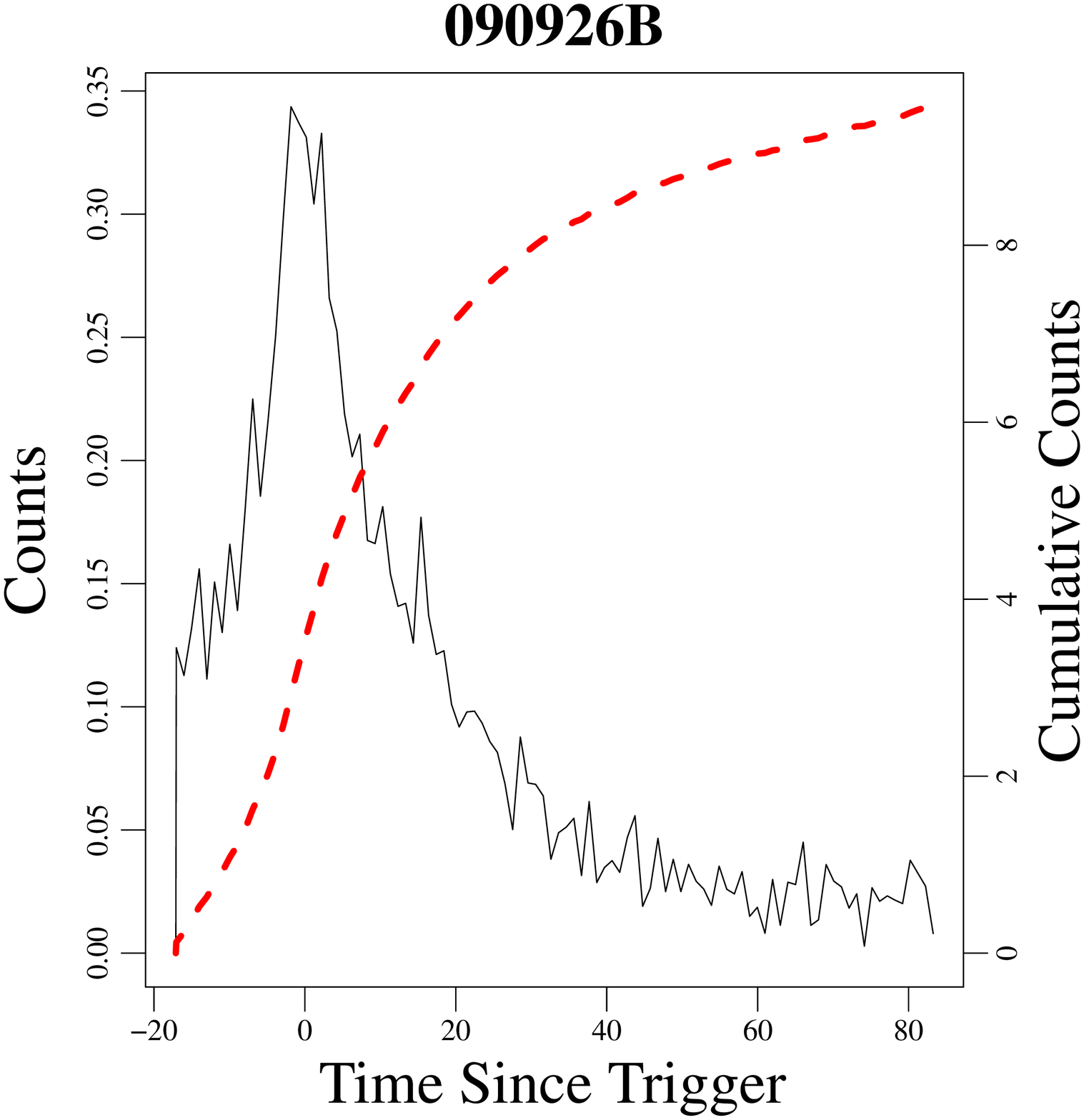}
 \end{center}
\end{minipage}
\begin{minipage}{0.25\hsize}
\begin{center}
    \FigureFile(40mm,40mm){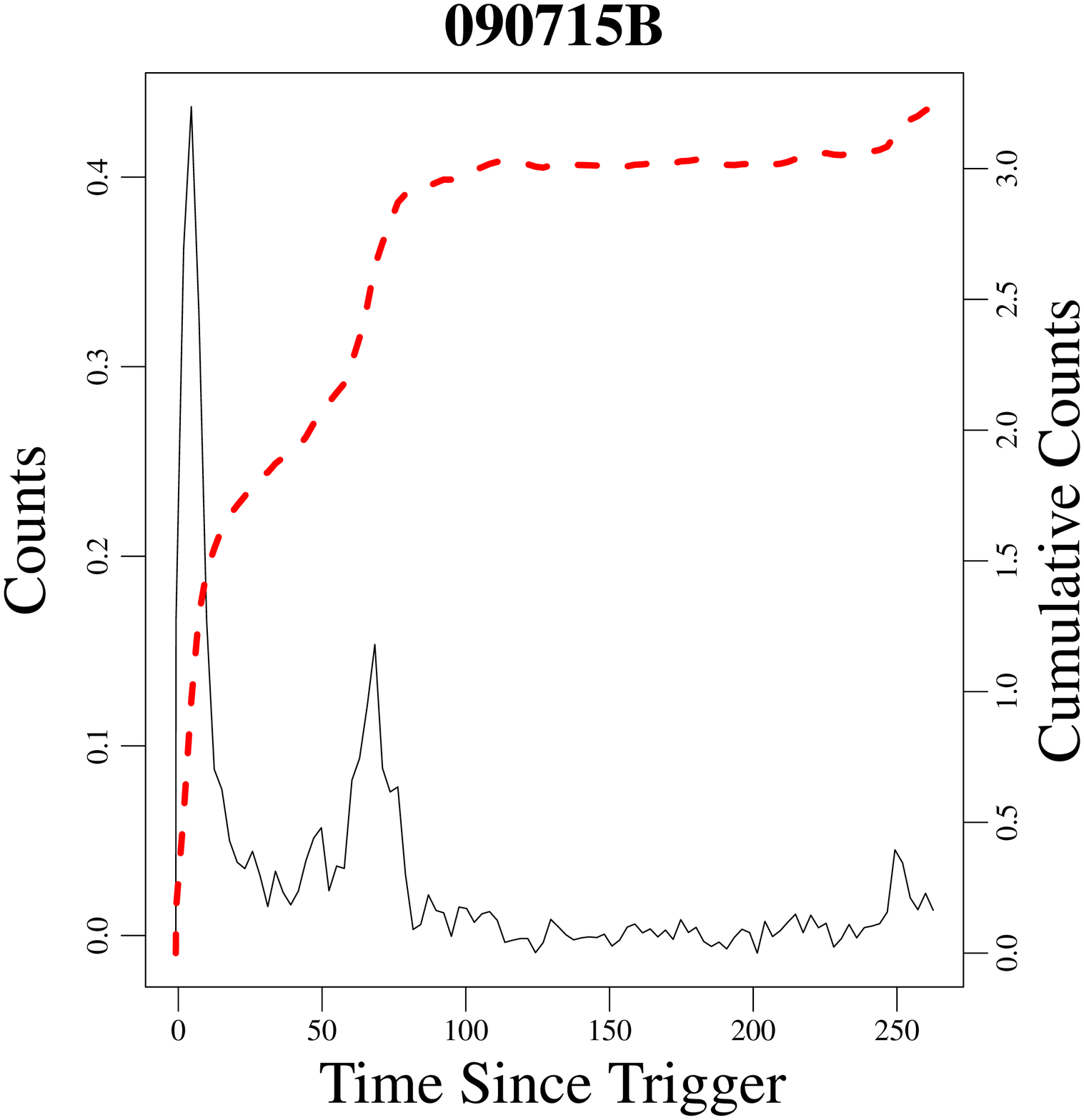}
\end{center}
\end{minipage}
\begin{minipage}{0.25\hsize}
\begin{center}
    \FigureFile(40mm,40mm){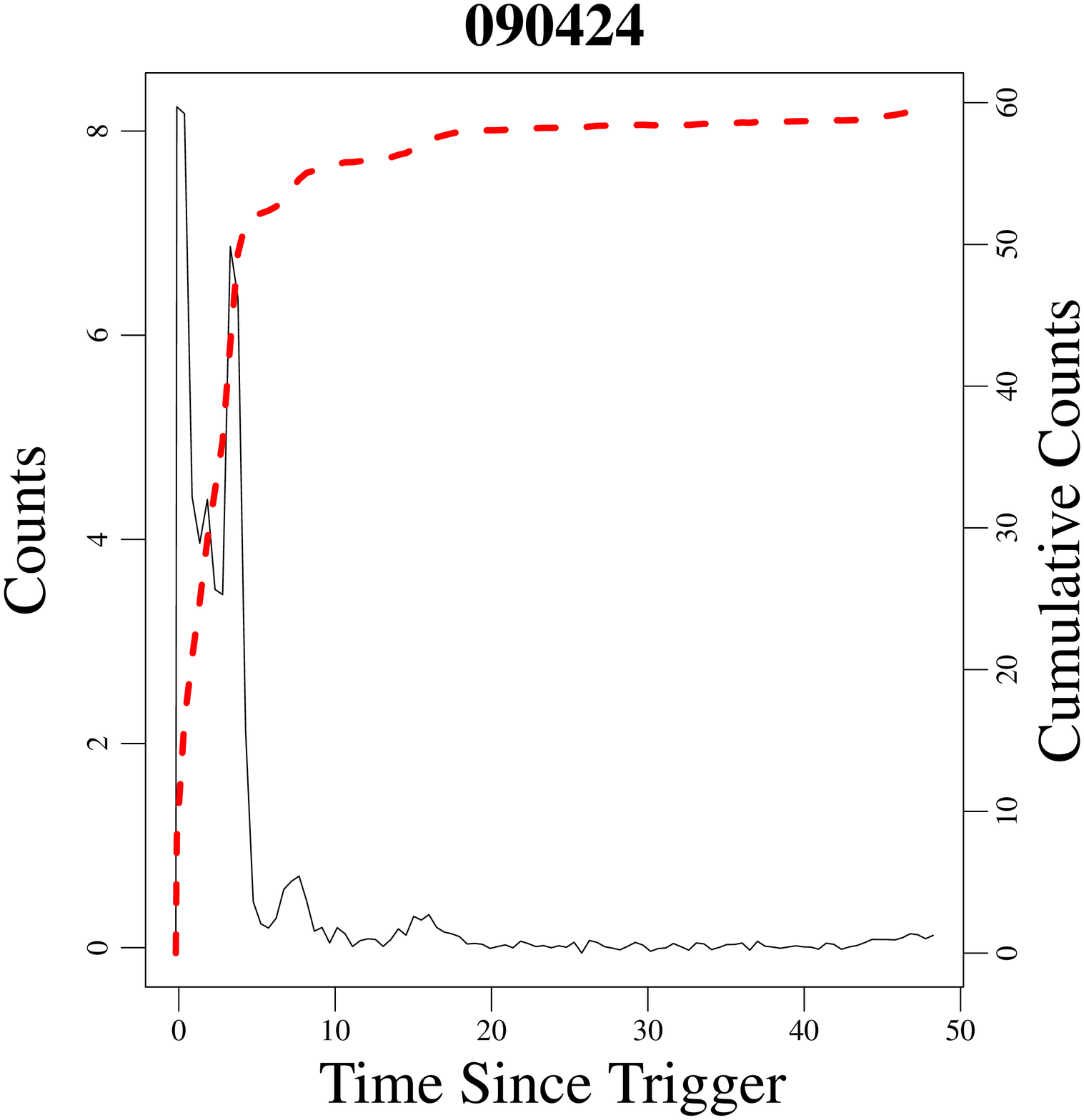}
 \end{center}
\end{minipage}\\
\end{tabular}
\caption{Light curves (black solid) and cumulative light curves (red doted) of Type II LGRBs.}\label{fig:A2-1}
\end{figure*}
%%%%%%%%%%%%%%%%%%%%%
\begin{figure*}[htb]
\begin{tabular}{cccc}
\begin{minipage}{0.25\hsize}
\begin{center}
    \FigureFile(40mm,40mm){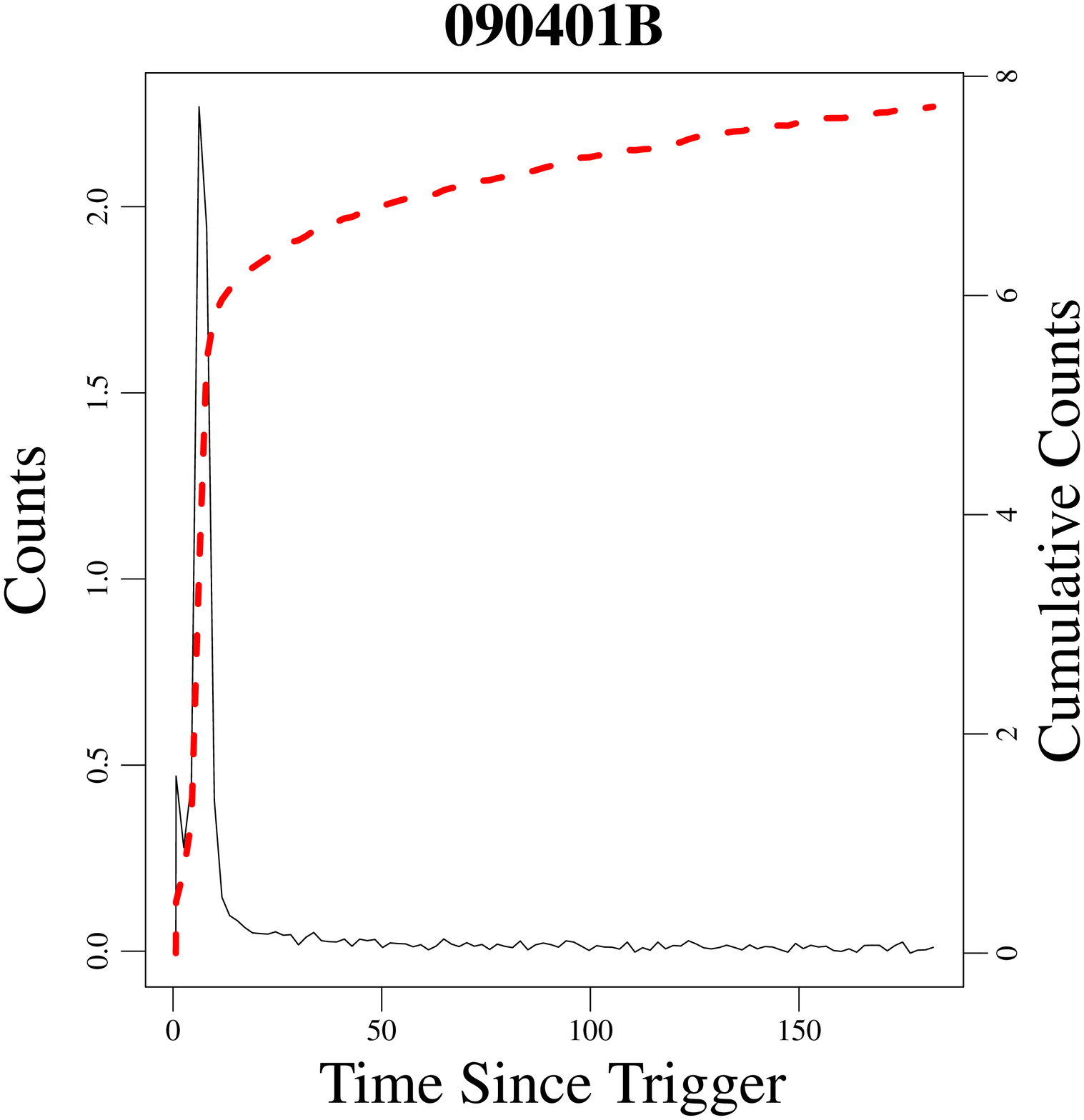}
\end{center}
\end{minipage}
\begin{minipage}{0.25\hsize}
\begin{center}
    \FigureFile(40mm,40mm){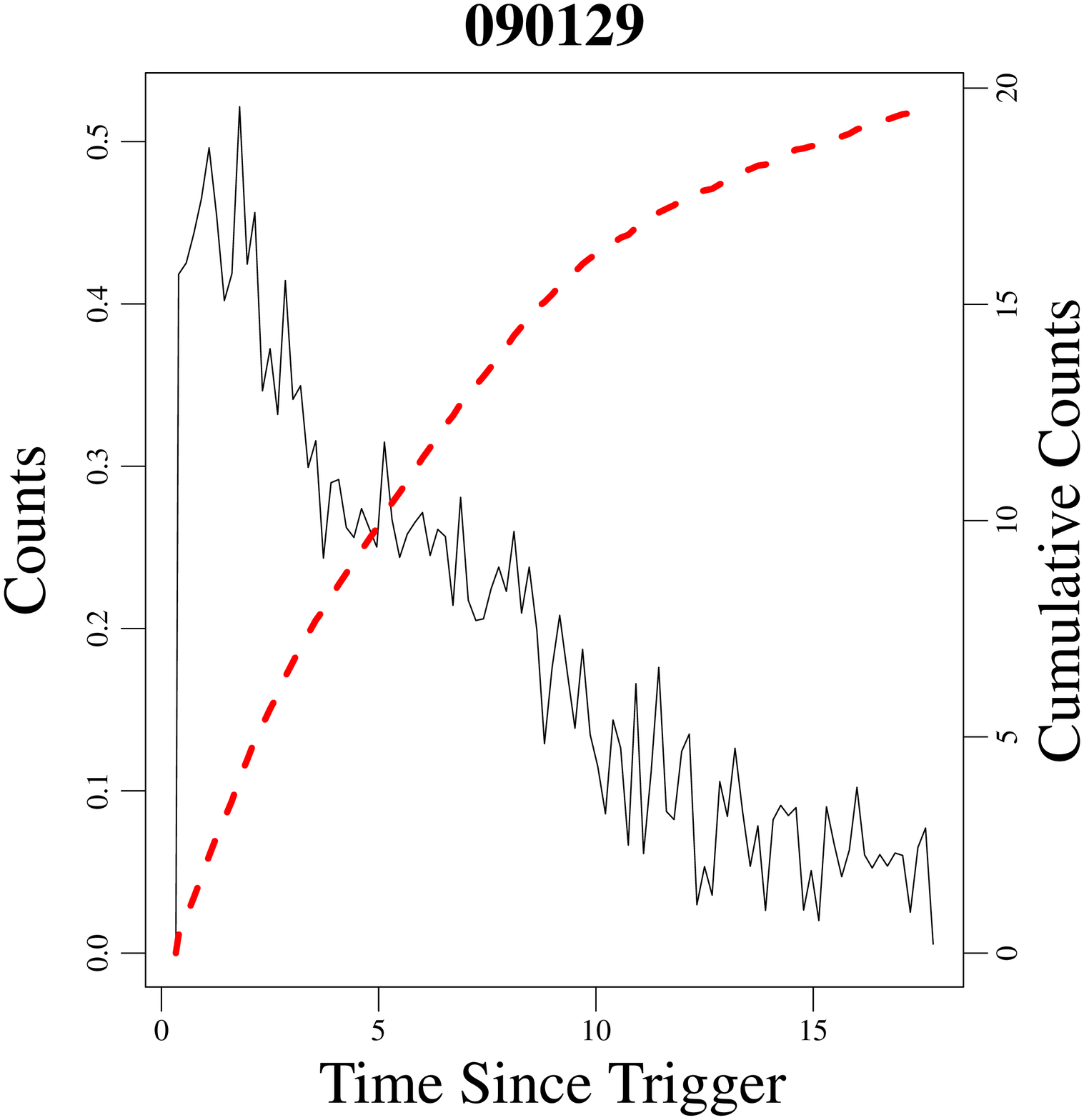}
 \end{center}
\end{minipage}
\begin{minipage}{0.25\hsize}
\begin{center}
    \FigureFile(40mm,40mm){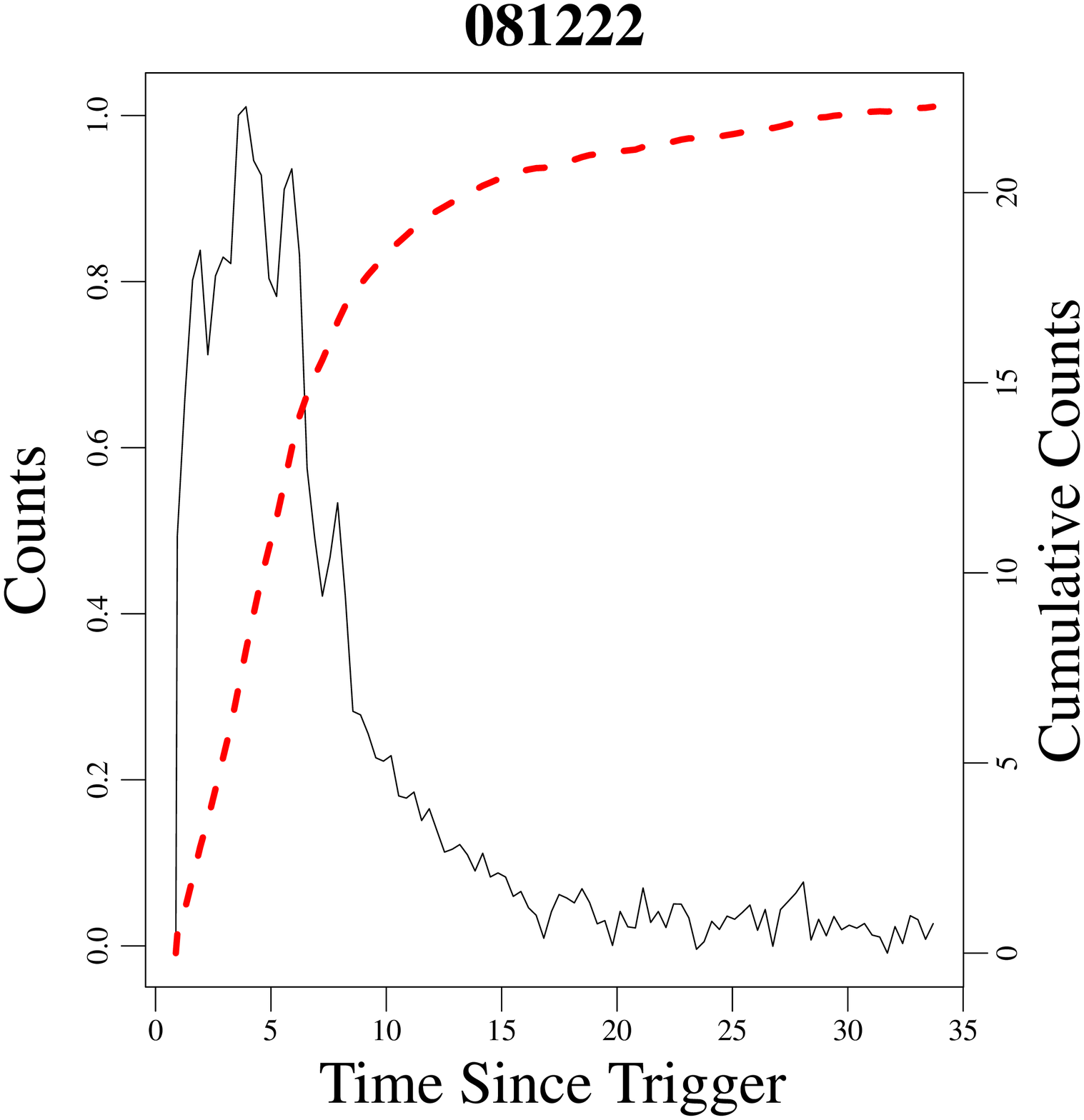}
\end{center}
\end{minipage}
\begin{minipage}{0.25\hsize}
\begin{center}
    \FigureFile(40mm,40mm){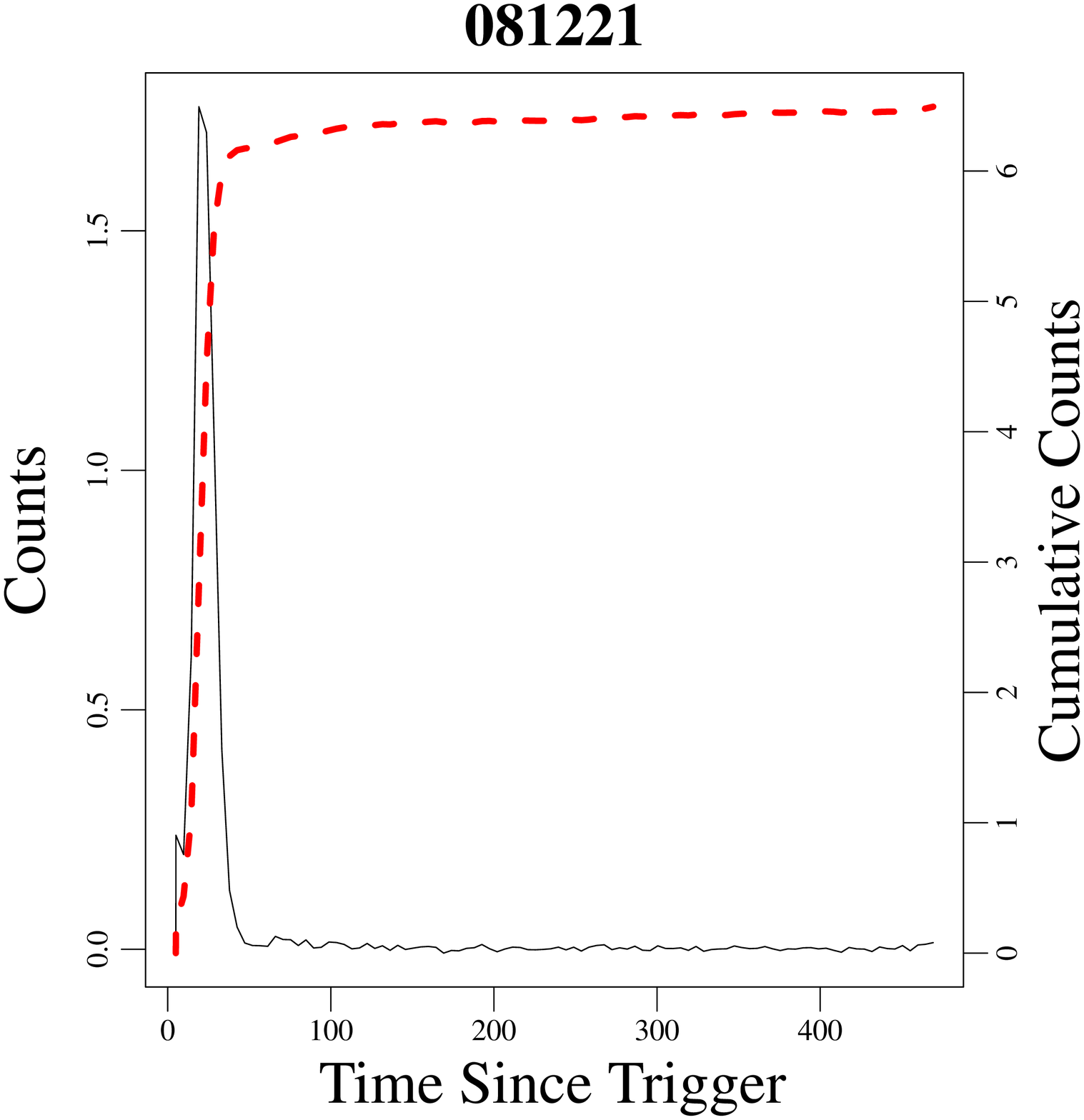}
 \end{center}
\end{minipage}\\
\begin{minipage}{0.25\hsize}
\begin{center}
    \FigureFile(40mm,40mm){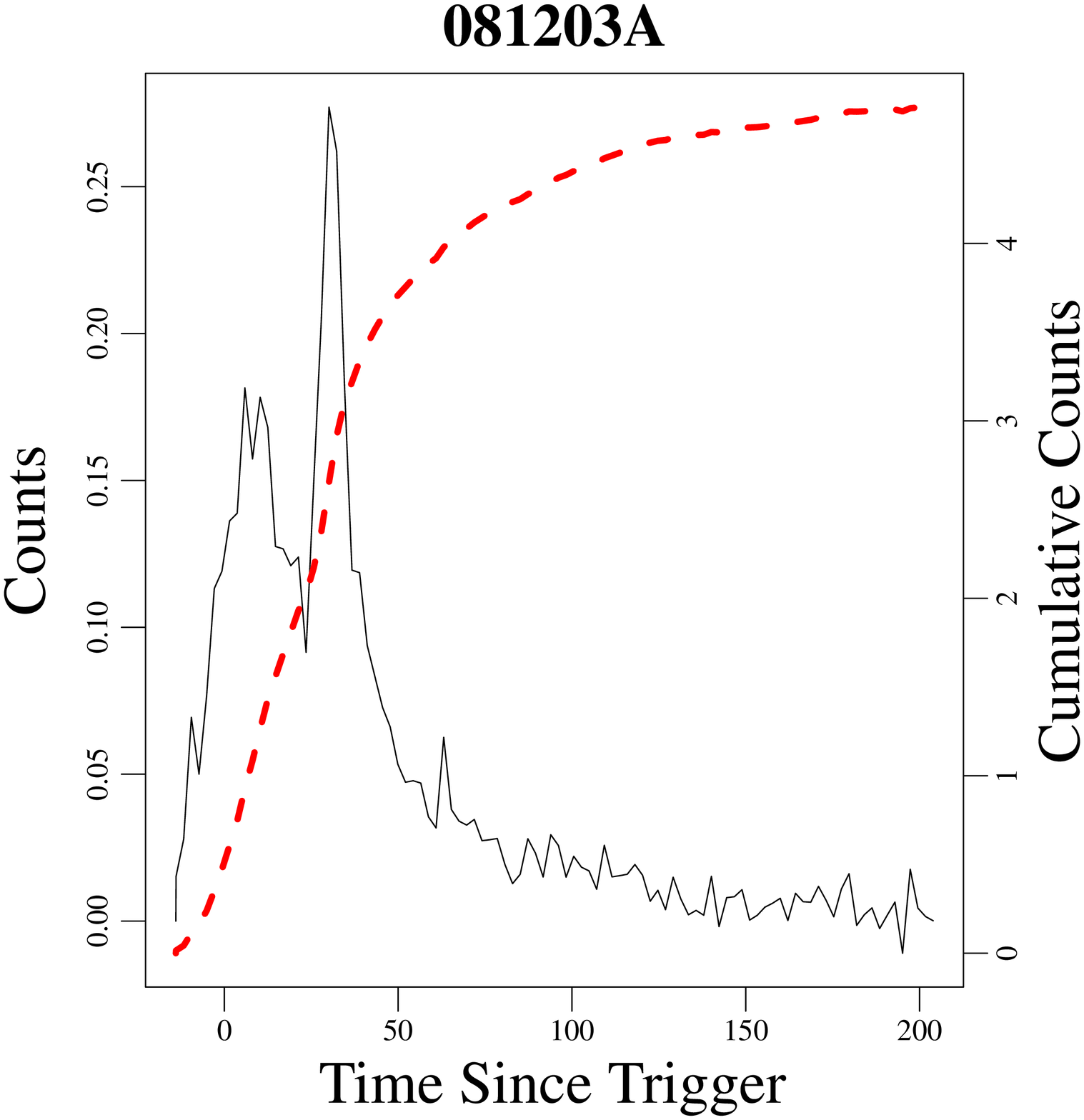}
\end{center}
\end{minipage}
\begin{minipage}{0.25\hsize}
\begin{center}
    \FigureFile(40mm,40mm){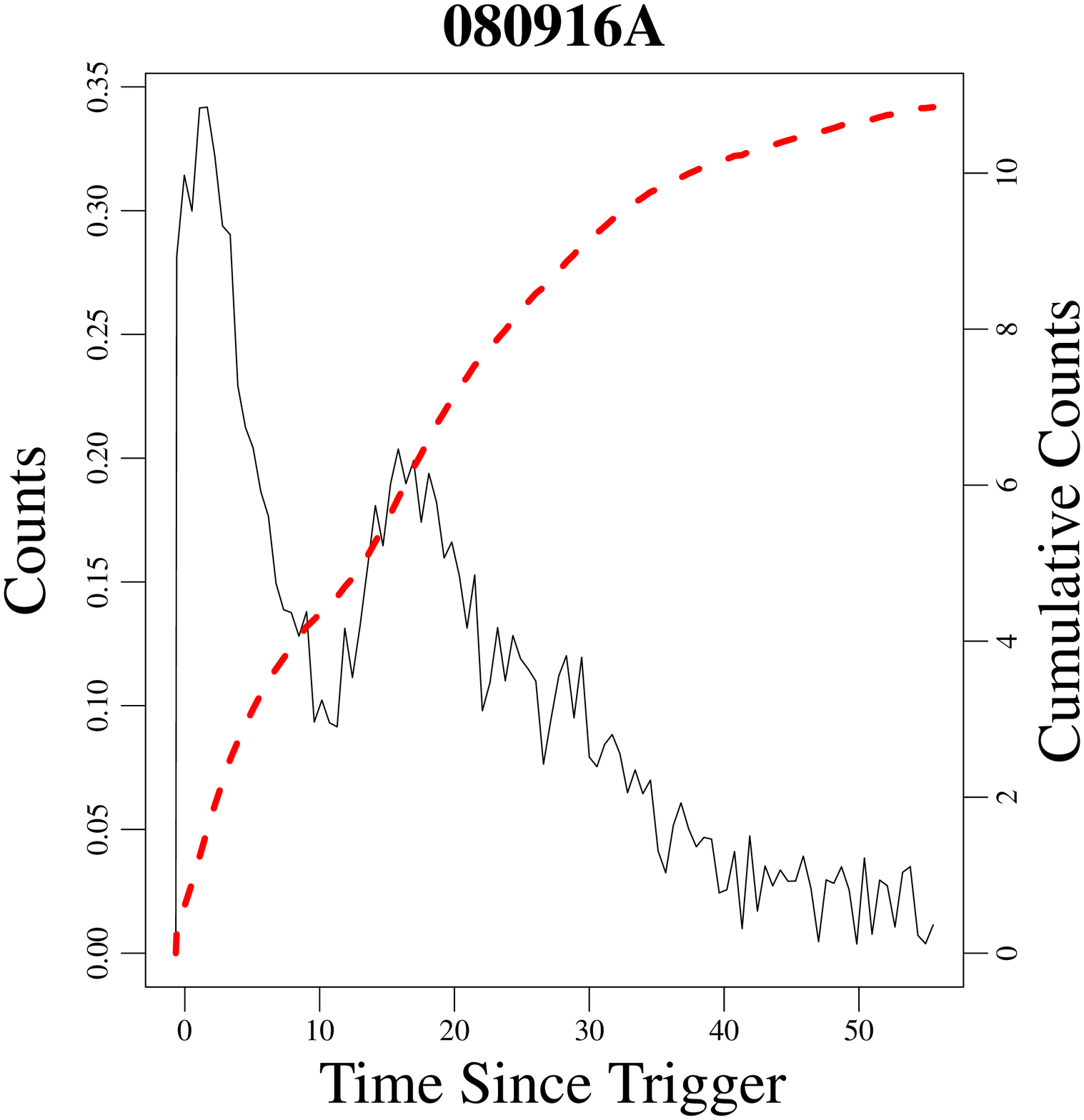}
 \end{center}
\end{minipage}
\begin{minipage}{0.25\hsize}
\begin{center}
    \FigureFile(40mm,40mm){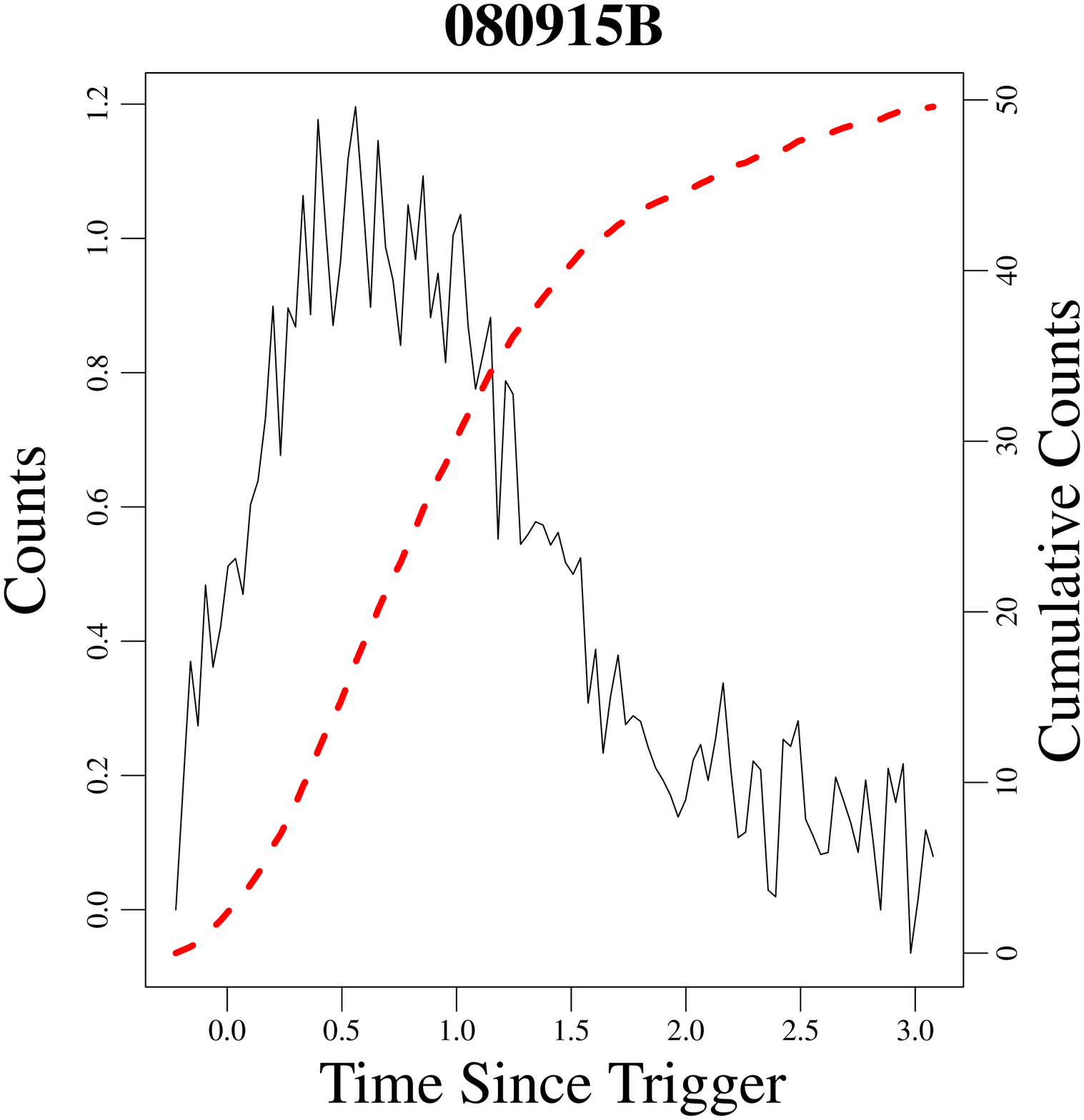}
\end{center}
\end{minipage}
\begin{minipage}{0.25\hsize}
\begin{center}
    \FigureFile(40mm,40mm){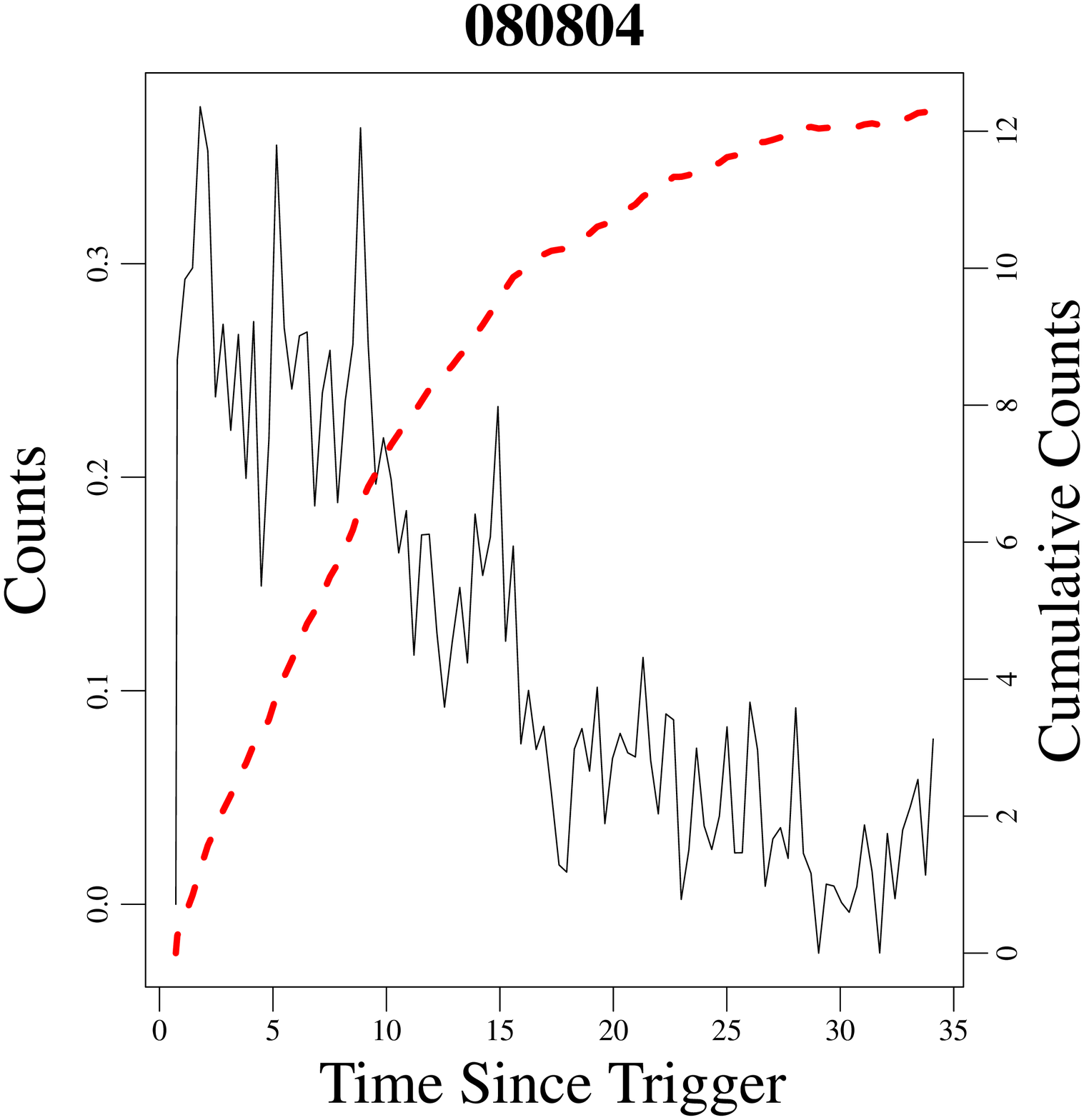}
 \end{center}
\end{minipage}\\
\begin{minipage}{0.25\hsize}
\begin{center}
    \FigureFile(40mm,40mm){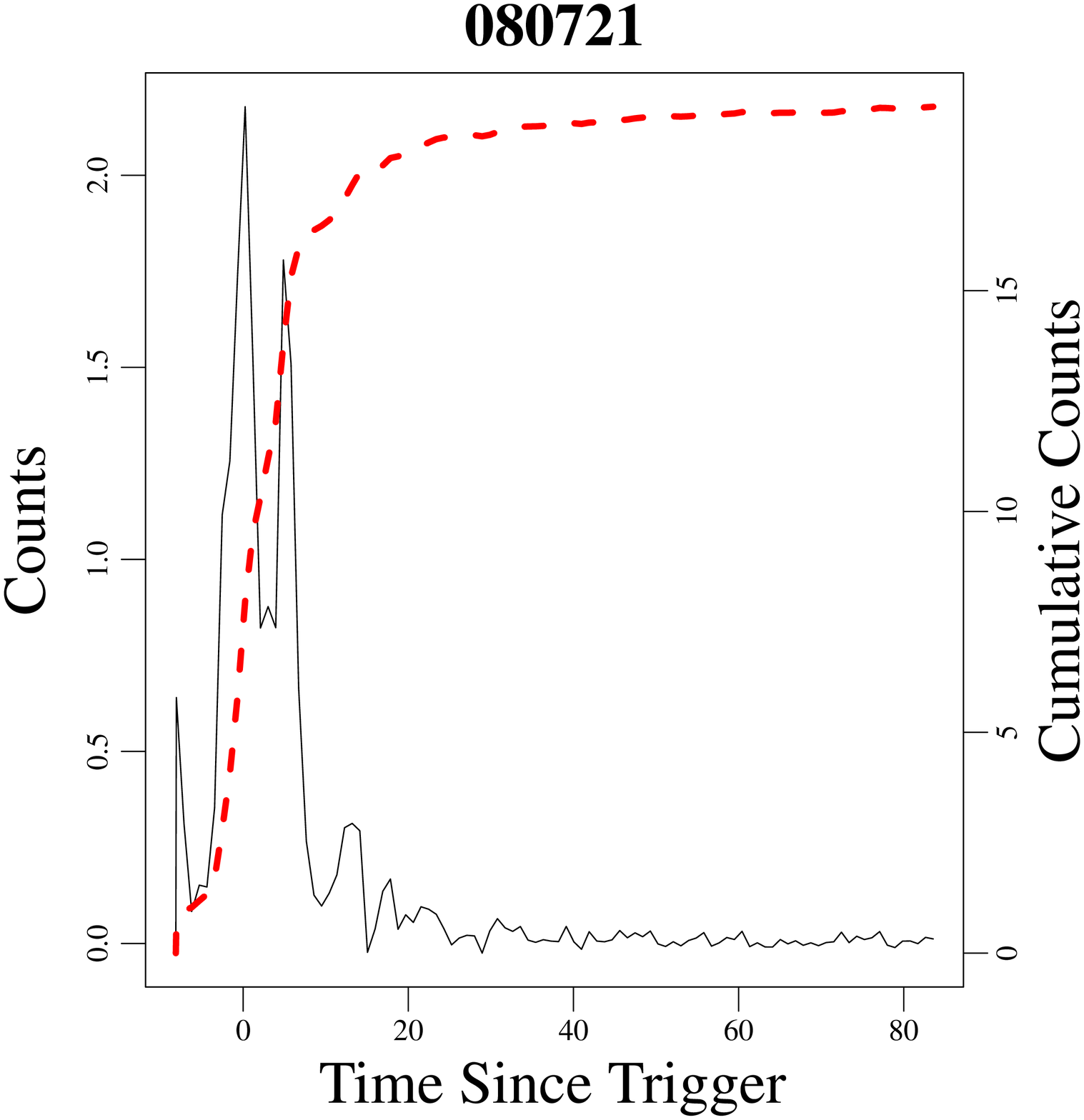}
\end{center}
\end{minipage}
\begin{minipage}{0.25\hsize}
\begin{center}
    \FigureFile(40mm,40mm){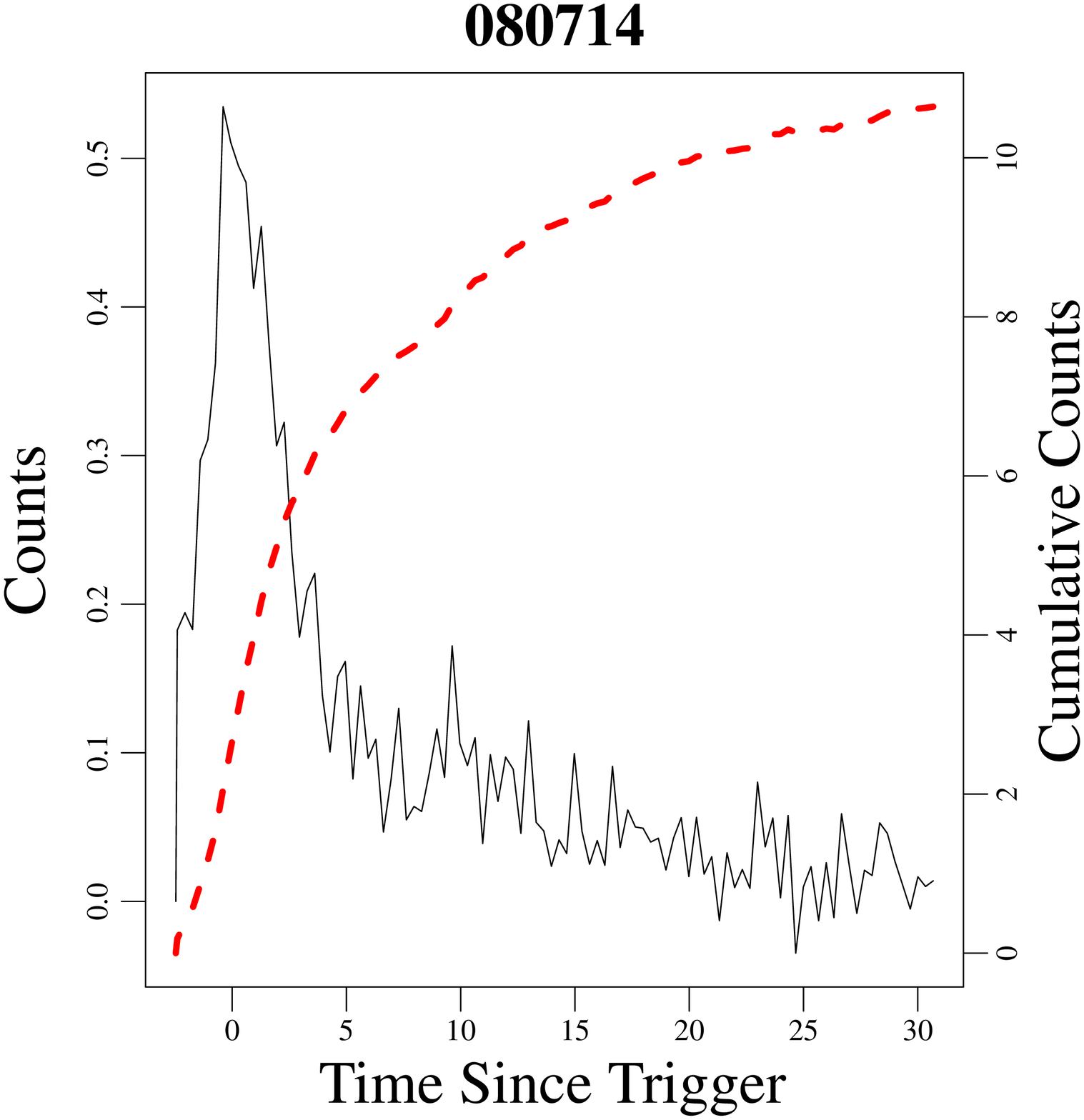}
 \end{center}
\end{minipage}
\begin{minipage}{0.25\hsize}
\begin{center}
    \FigureFile(40mm,40mm){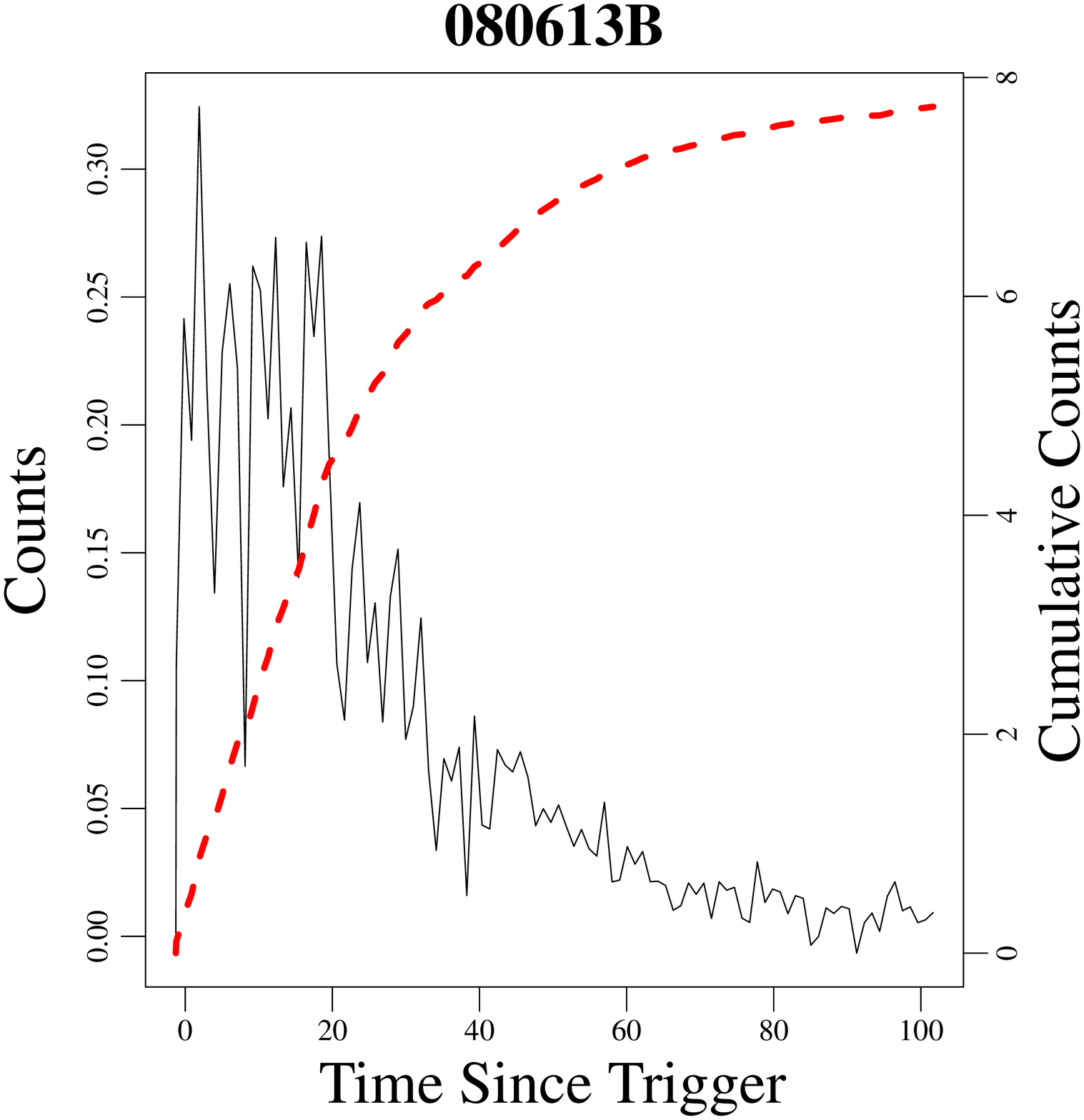}
\end{center}
\end{minipage}
\begin{minipage}{0.25\hsize}
\begin{center}
    \FigureFile(40mm,40mm){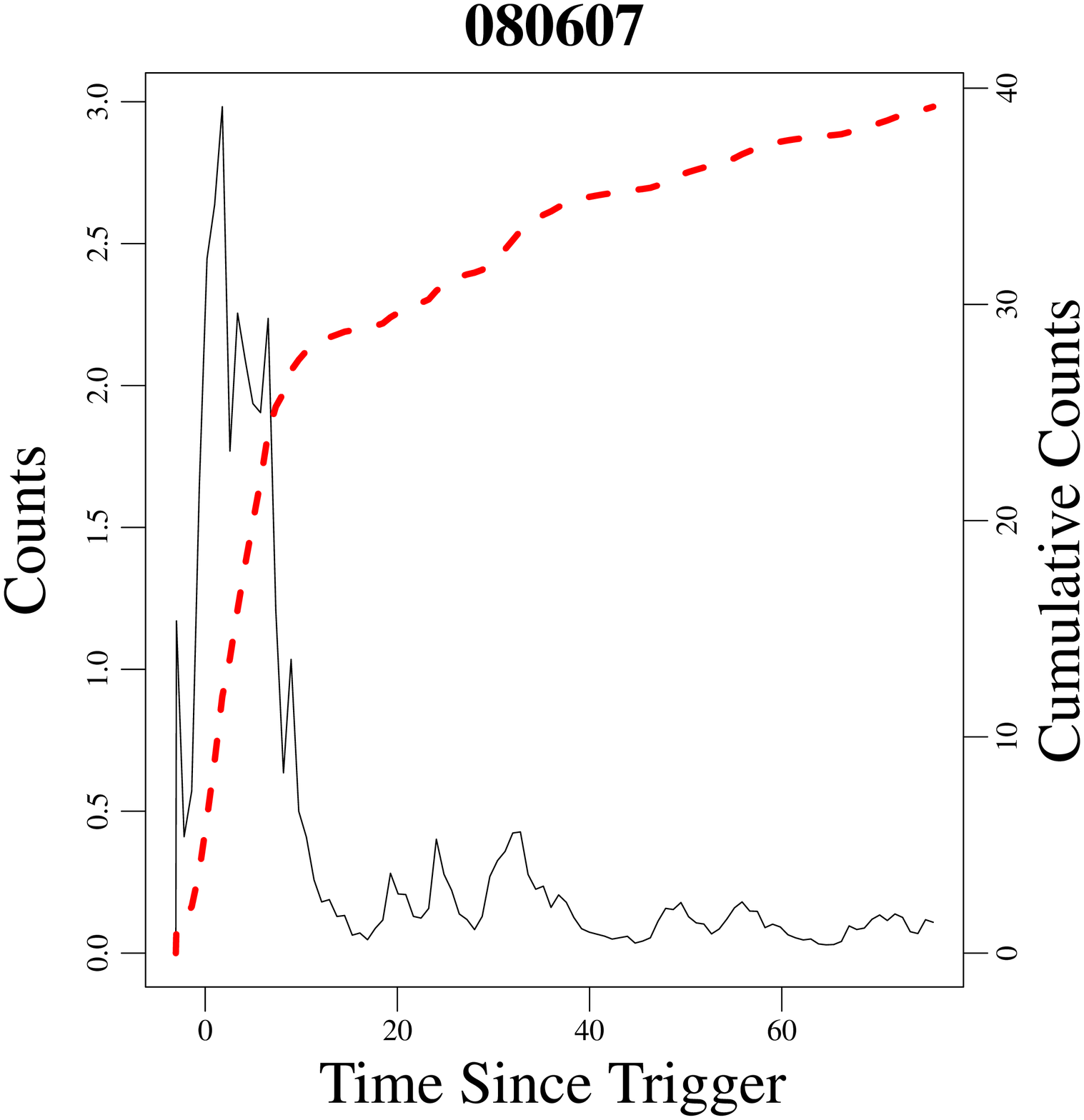}
 \end{center}
\end{minipage}\\
\begin{minipage}{0.25\hsize}
\begin{center}
    \FigureFile(40mm,40mm){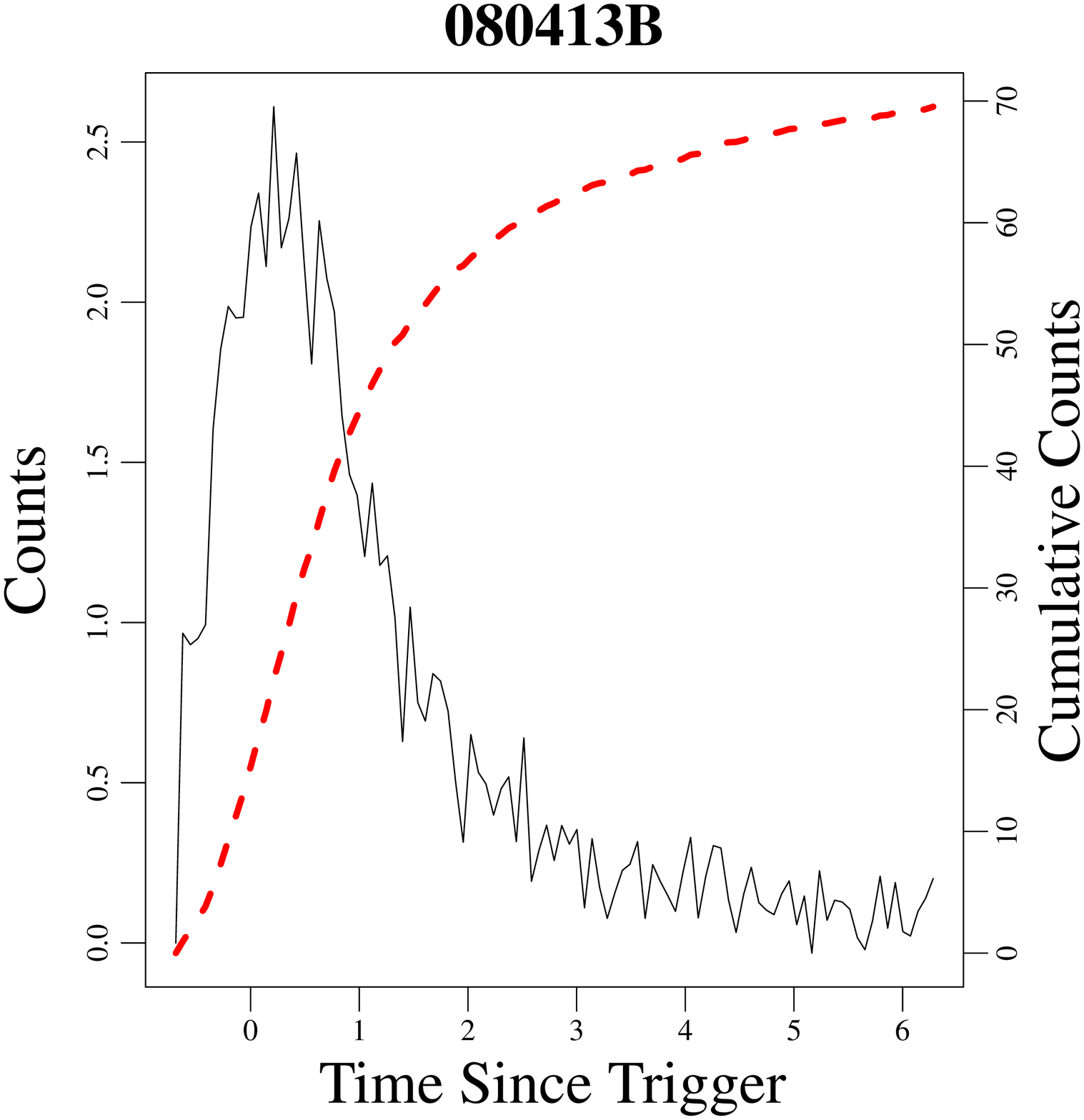}
\end{center}
\end{minipage}
\begin{minipage}{0.25\hsize}
\begin{center}
    \FigureFile(40mm,40mm){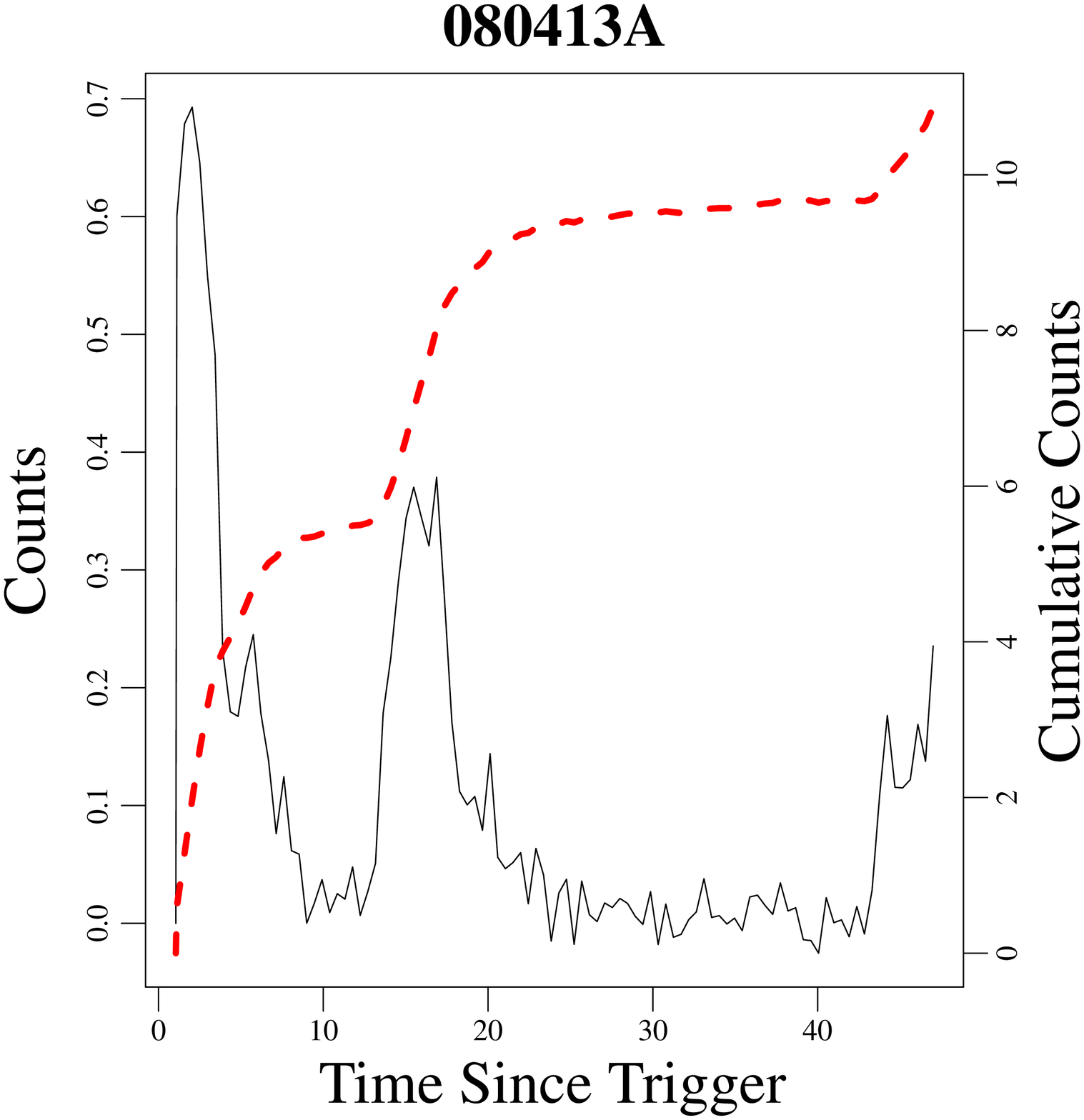}
 \end{center}
\end{minipage}
\begin{minipage}{0.25\hsize}
\begin{center}
    \FigureFile(40mm,40mm){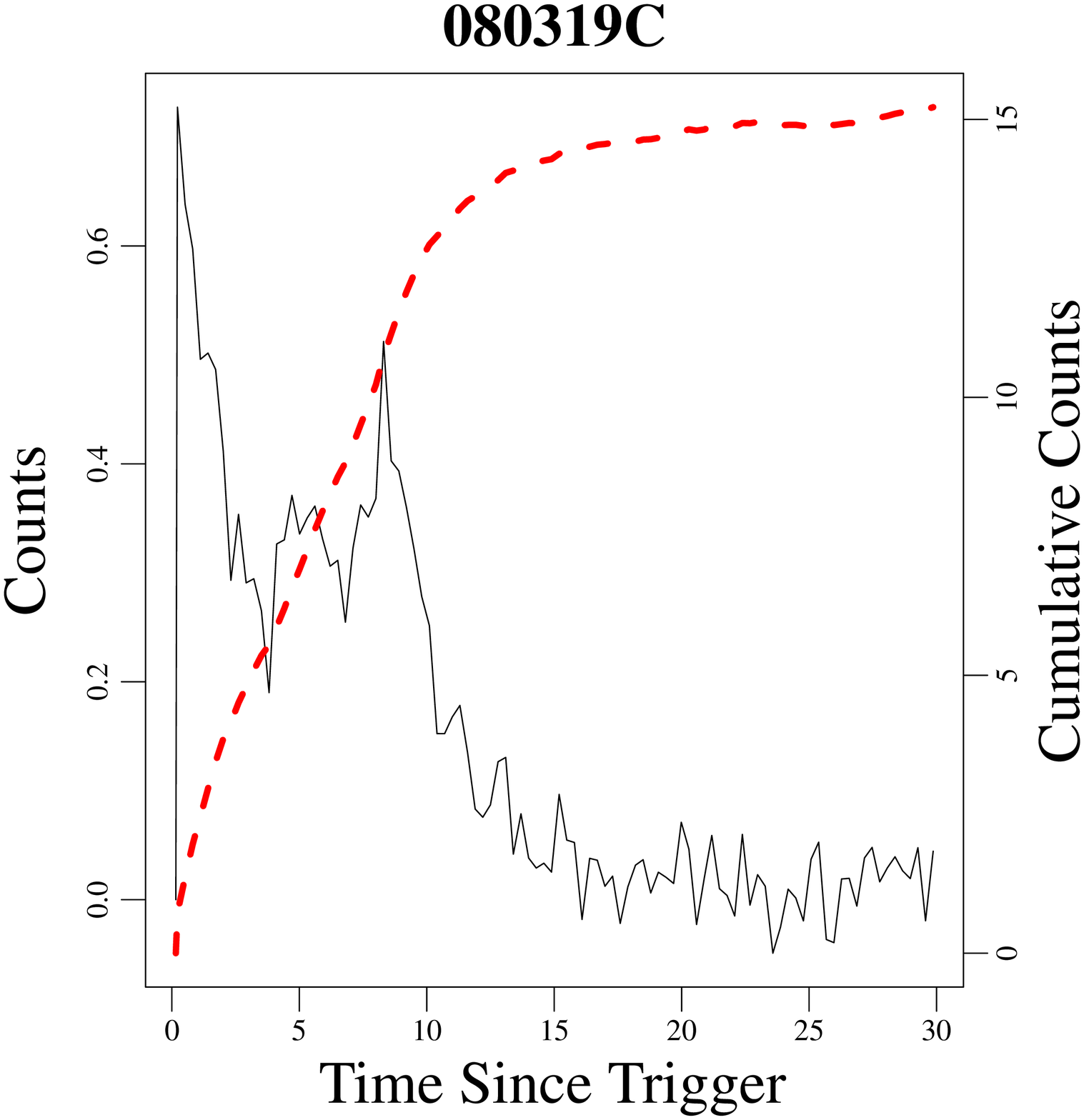}
\end{center}
\end{minipage}
\begin{minipage}{0.25\hsize}
\begin{center}
    \FigureFile(40mm,40mm){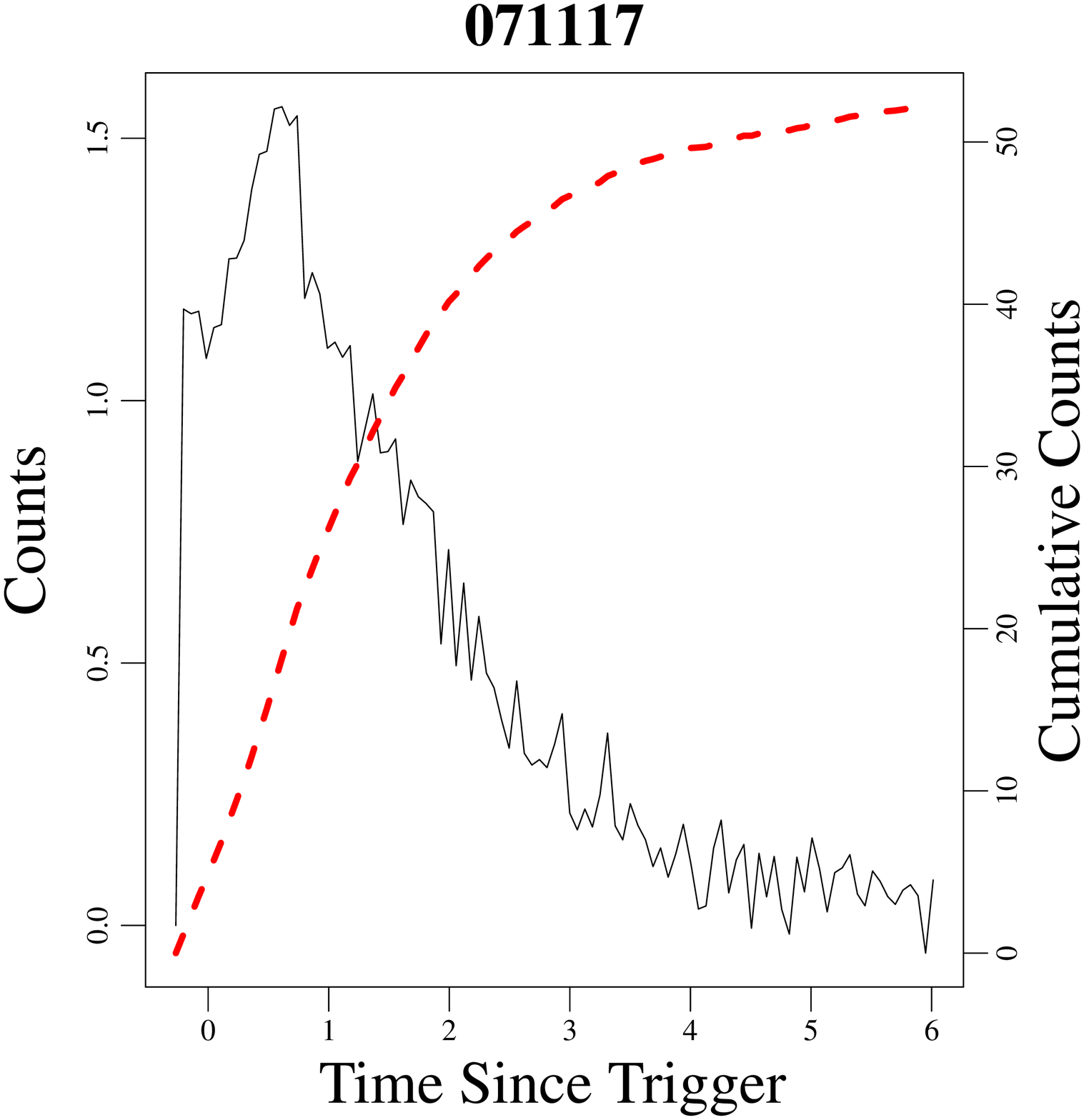}
 \end{center}
\end{minipage}\\
\begin{minipage}{0.25\hsize}
\begin{center}
    \FigureFile(40mm,40mm){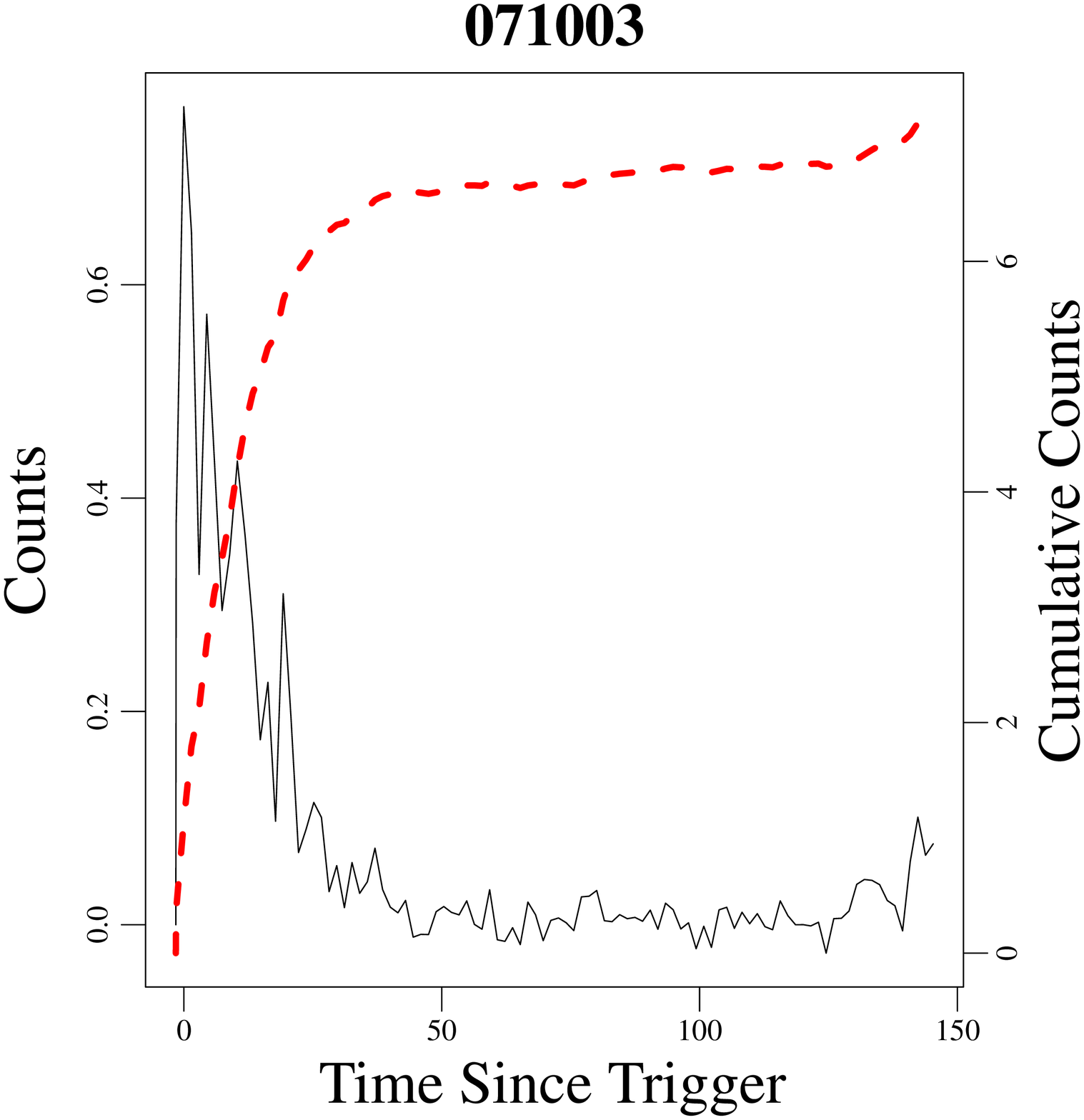}
\end{center}
\end{minipage}
\begin{minipage}{0.25\hsize}
\begin{center}
    \FigureFile(40mm,40mm){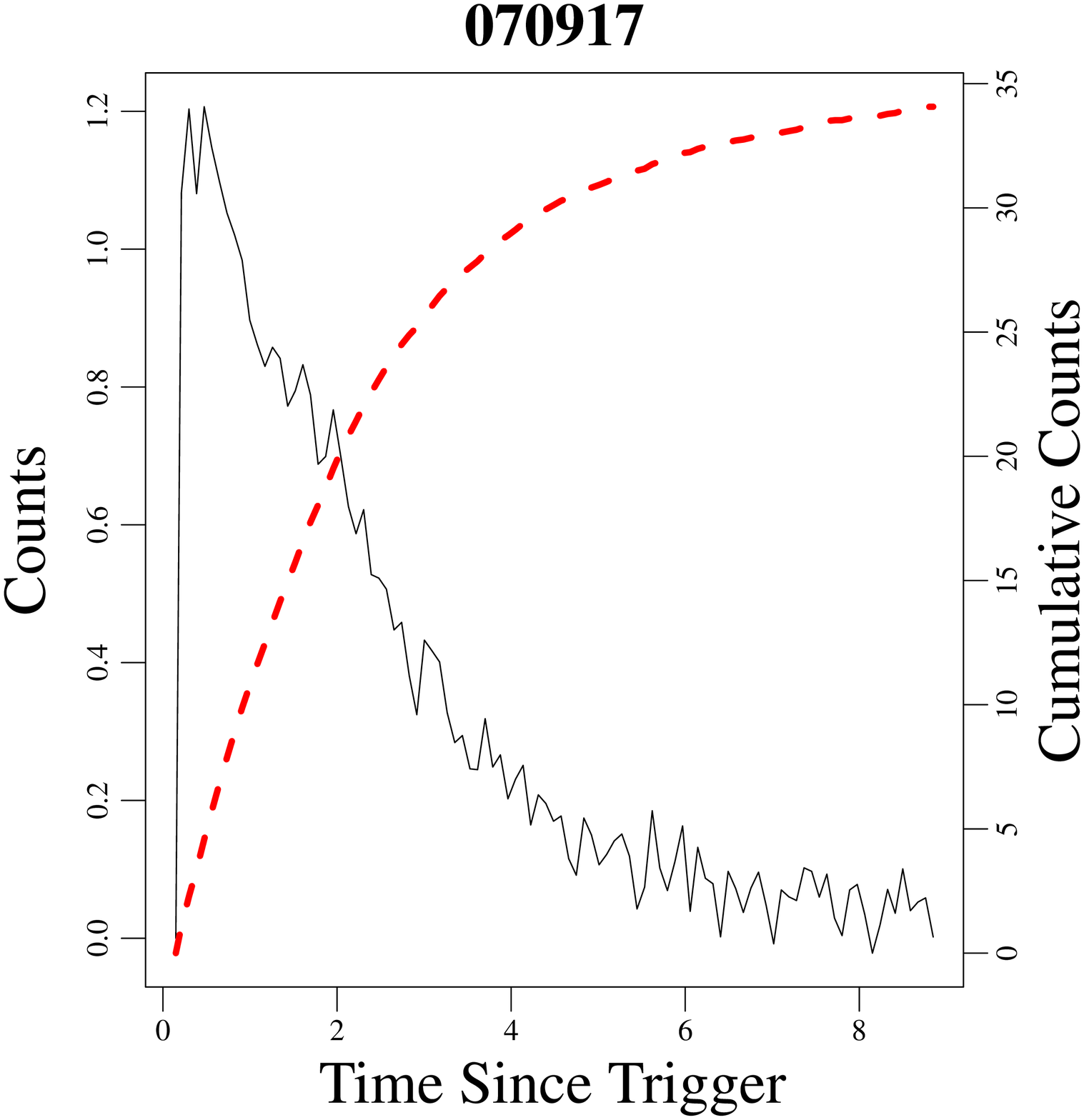}
 \end{center}
\end{minipage}
\begin{minipage}{0.25\hsize}
\begin{center}
    \FigureFile(40mm,40mm){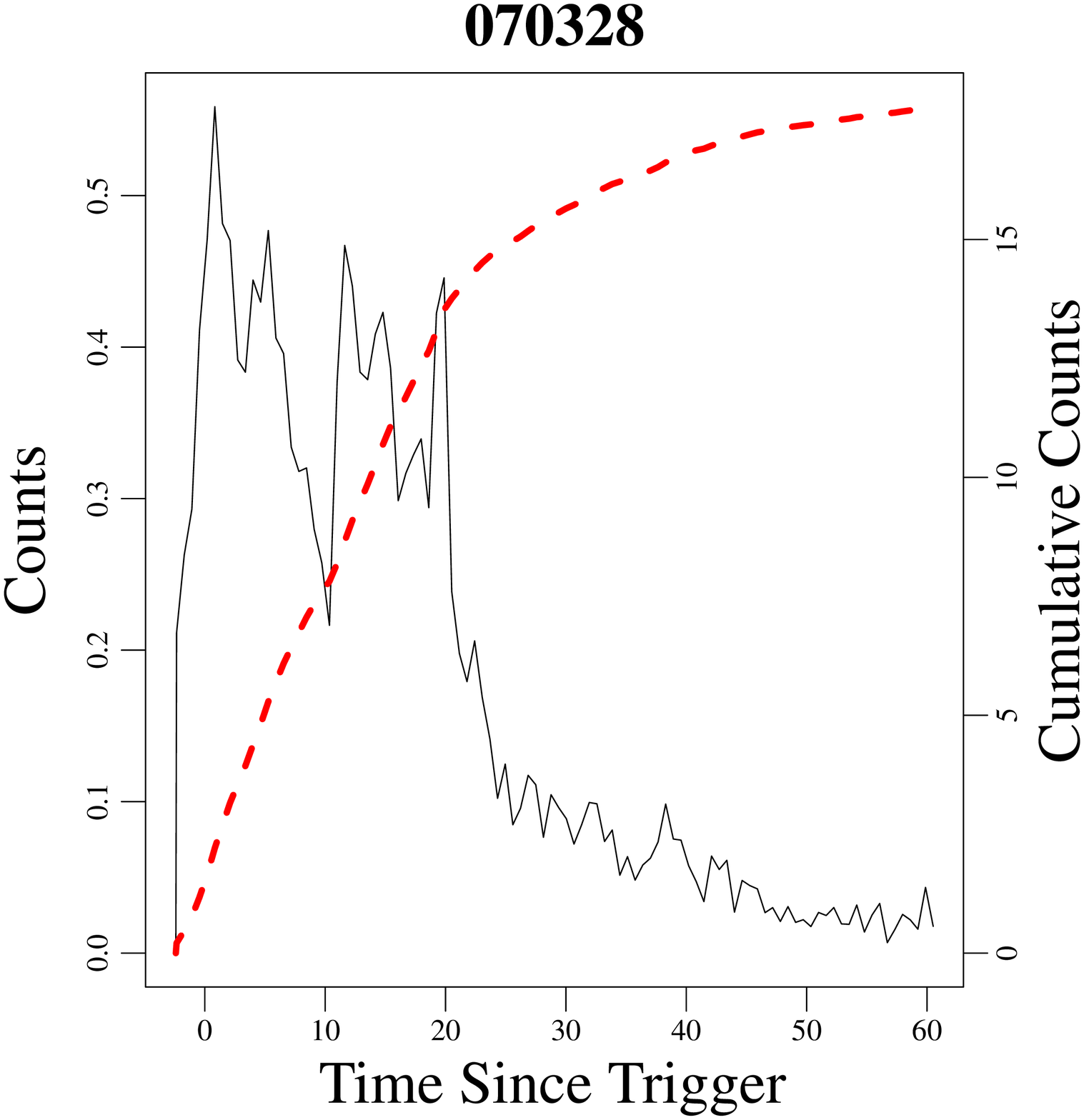}
\end{center}
\end{minipage}
\begin{minipage}{0.25\hsize}
\begin{center}
    \FigureFile(40mm,40mm){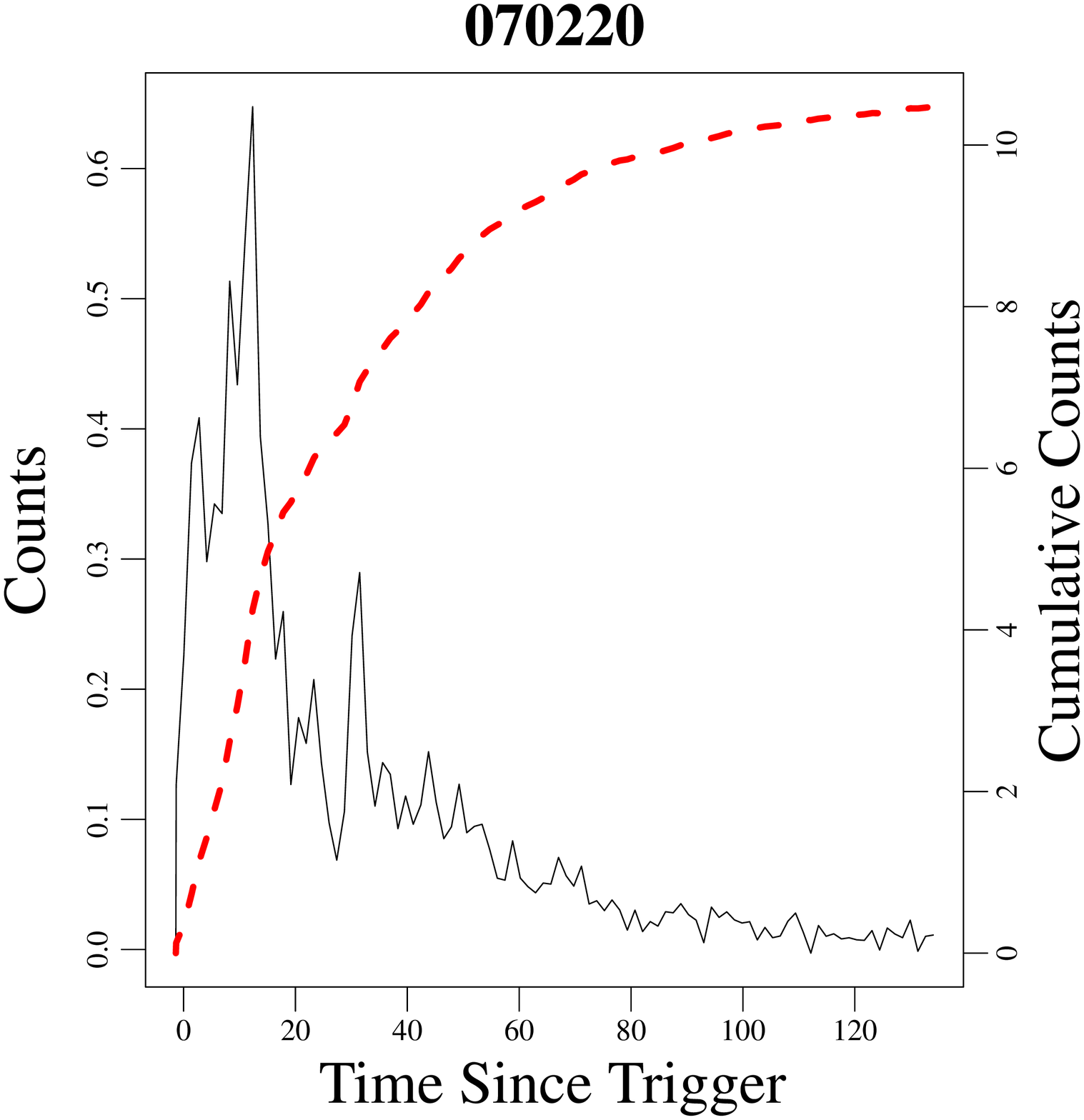}
 \end{center}
\end{minipage}\\
\begin{minipage}{0.25\hsize}
\begin{center}
    \FigureFile(40mm,40mm){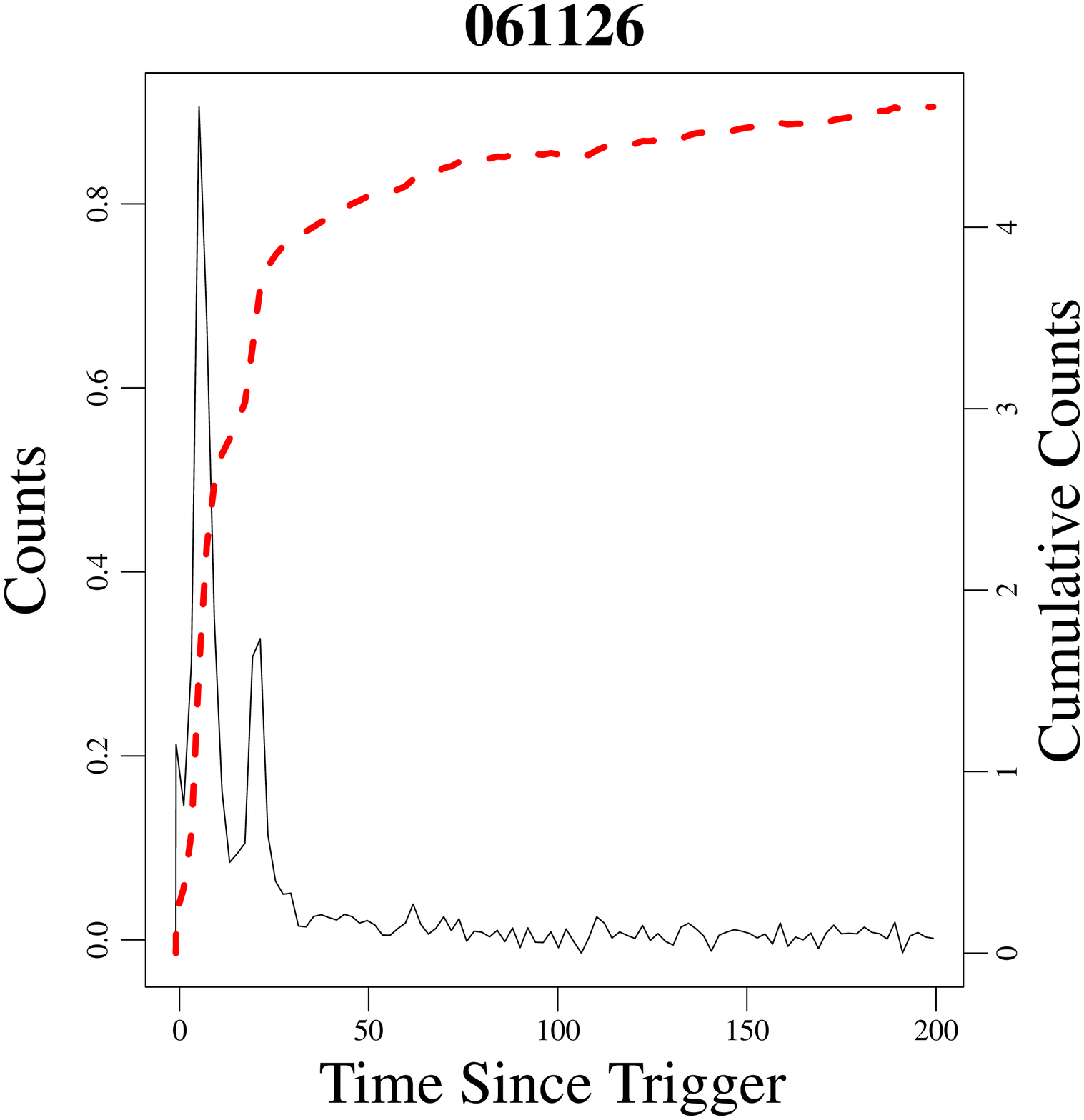}
\end{center}
\end{minipage}
\begin{minipage}{0.25\hsize}
\begin{center}
    \FigureFile(40mm,40mm){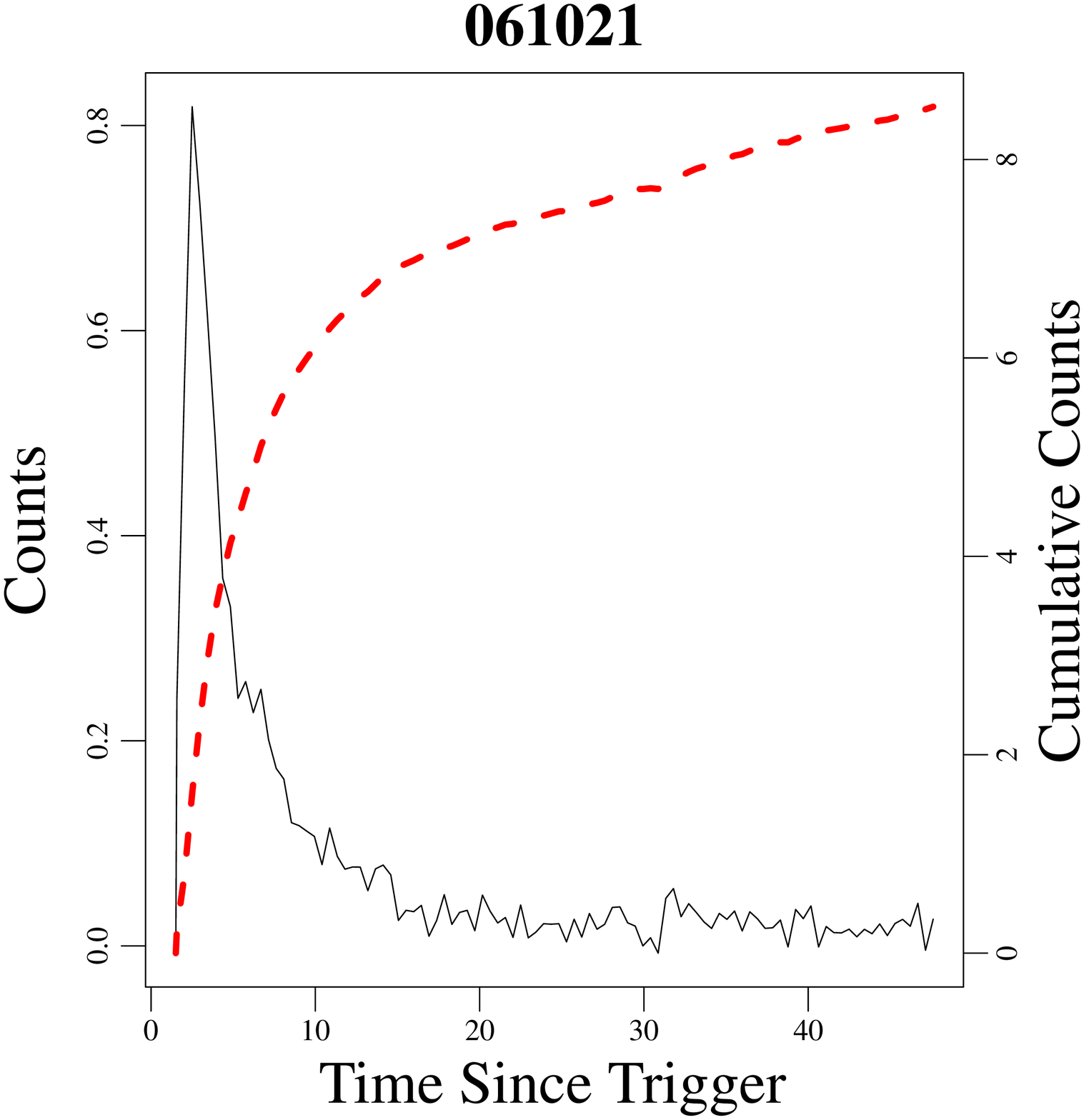}
 \end{center}
\end{minipage}
\begin{minipage}{0.25\hsize}
\begin{center}
    \FigureFile(40mm,40mm){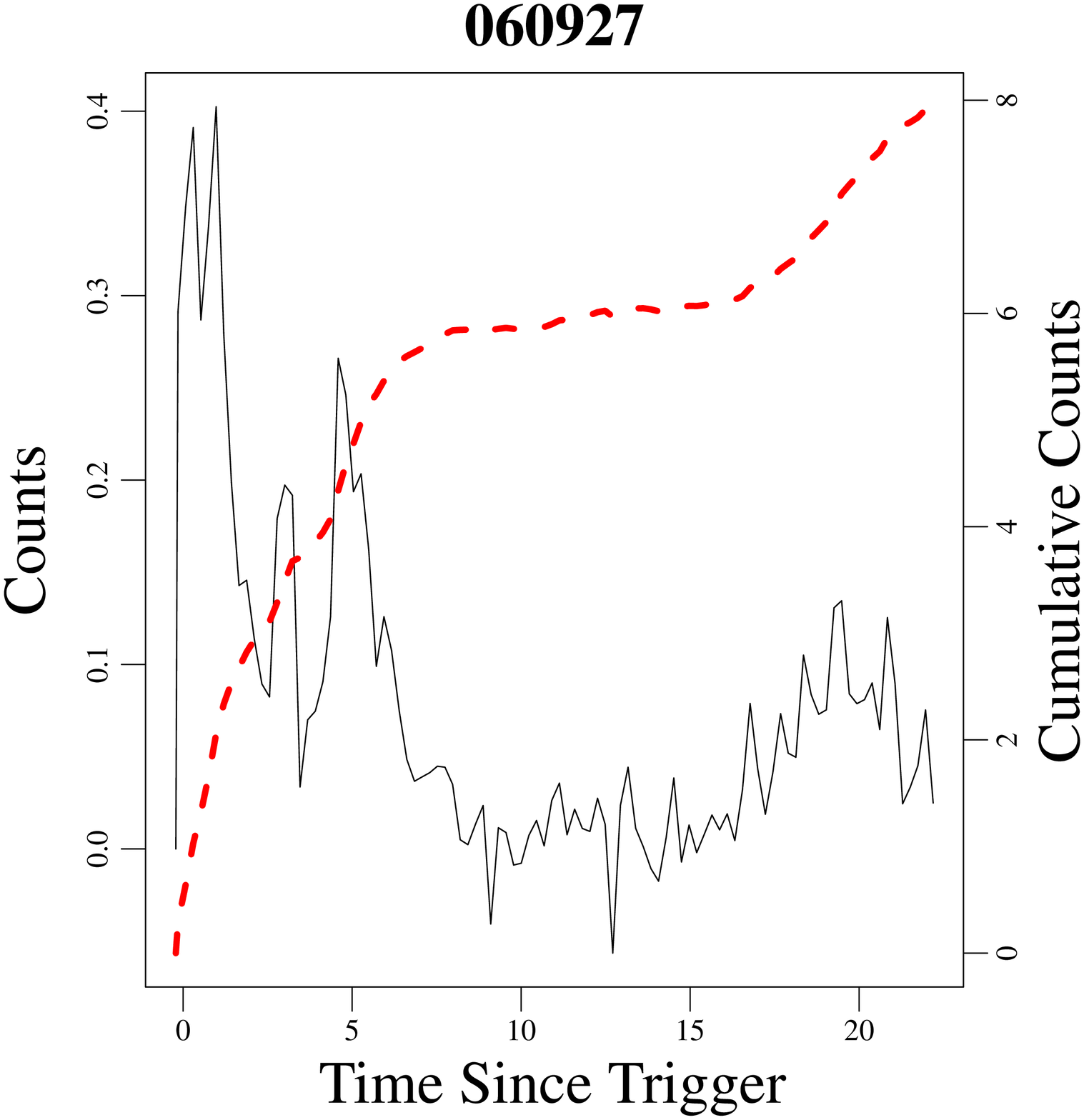}
\end{center}
\end{minipage}
\begin{minipage}{0.25\hsize}
\begin{center}
    \FigureFile(40mm,40mm){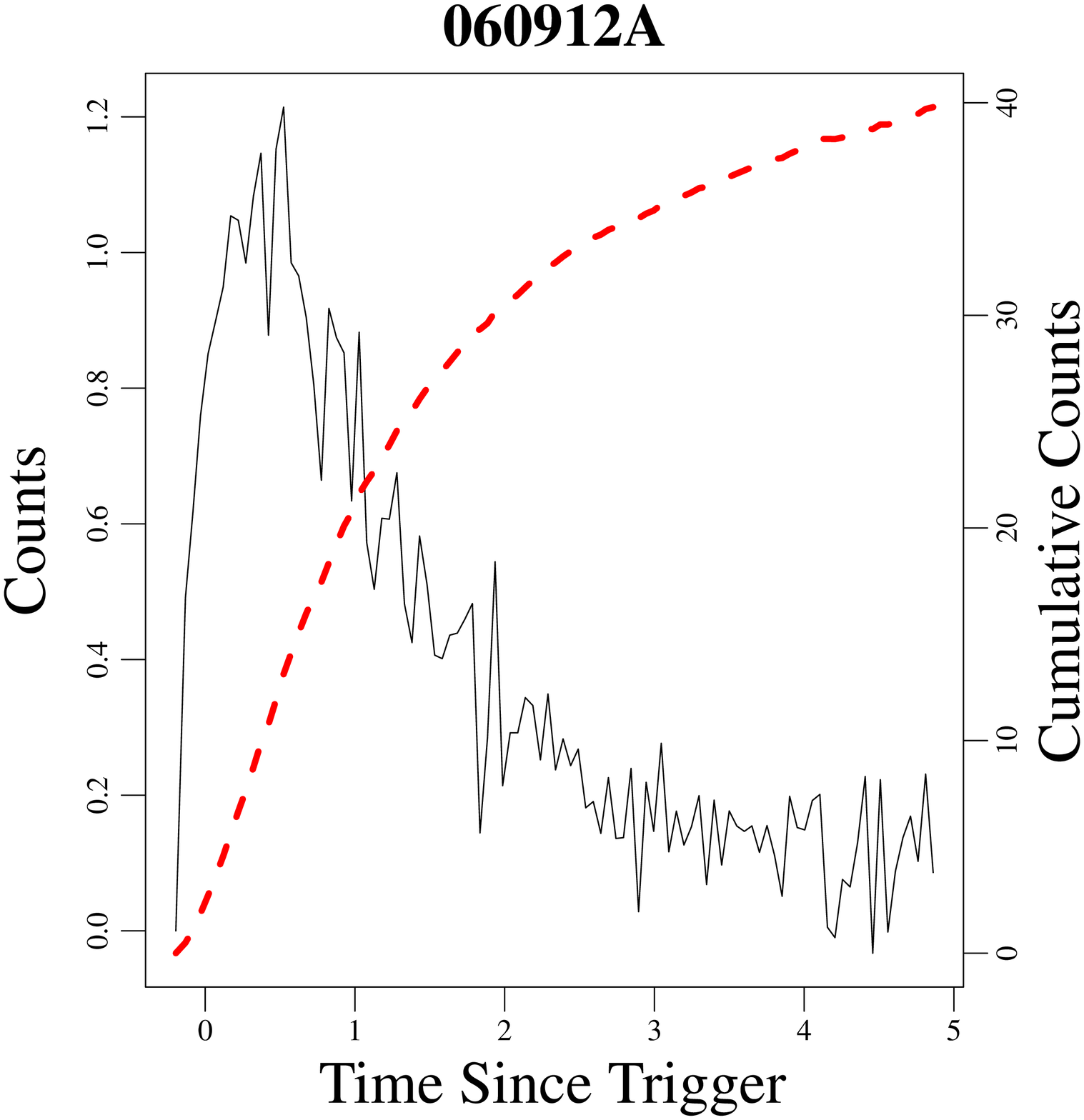}
 \end{center}
\end{minipage}\\
\end{tabular}
 \caption{Light curves (black solid) and cumulative light curves (red doted) of Type II LGRBs.}\label{fig:A2-2}
\end{figure*}
%%%%%%%%%%%%%%%%%%%%%
\begin{figure*}[htb]
\begin{tabular}{cccc}
\begin{minipage}{0.25\hsize}
\begin{center}
    \FigureFile(40mm,40mm){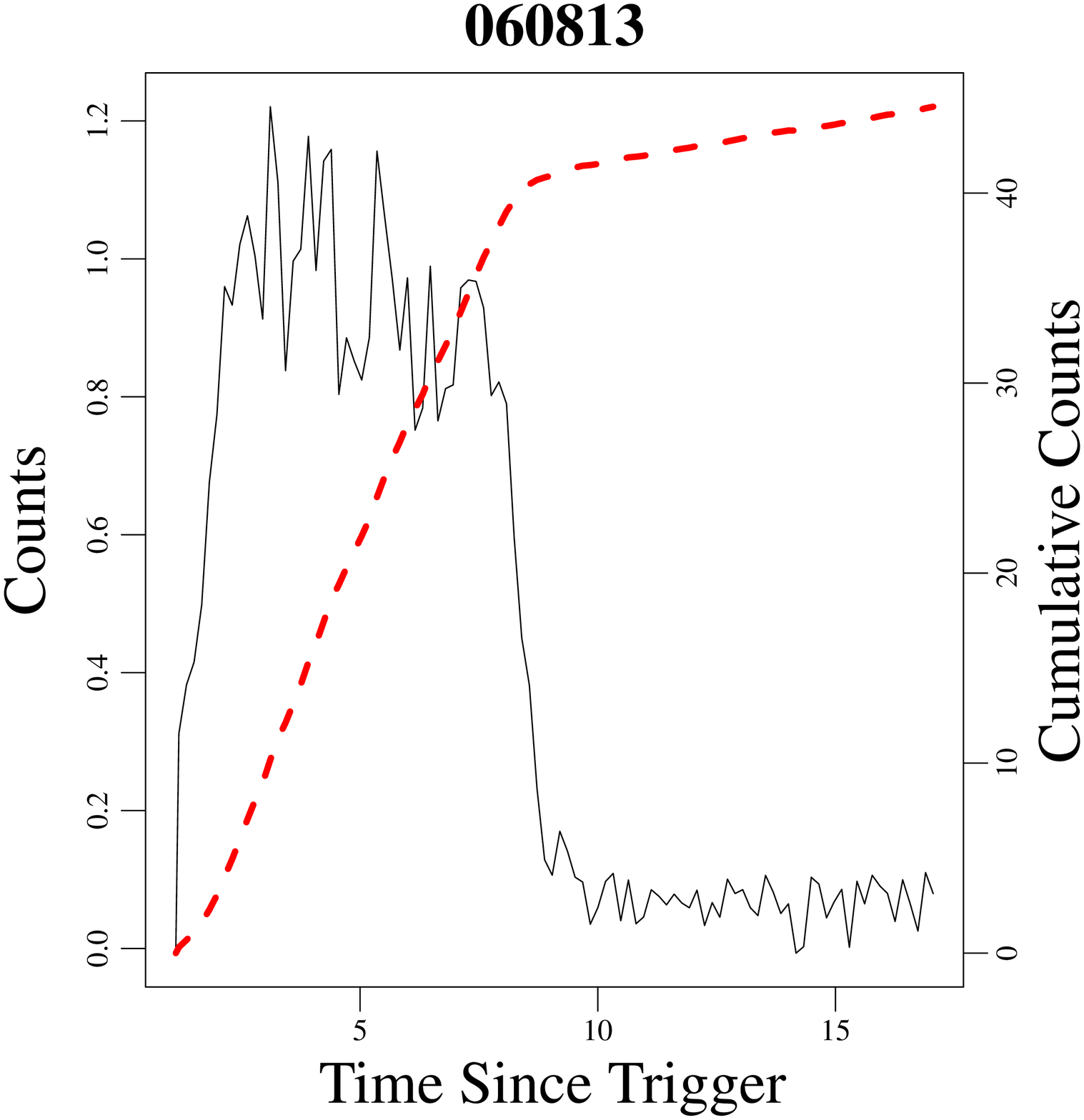}
\end{center}
\end{minipage}
\begin{minipage}{0.25\hsize}
\begin{center}
    \FigureFile(40mm,40mm){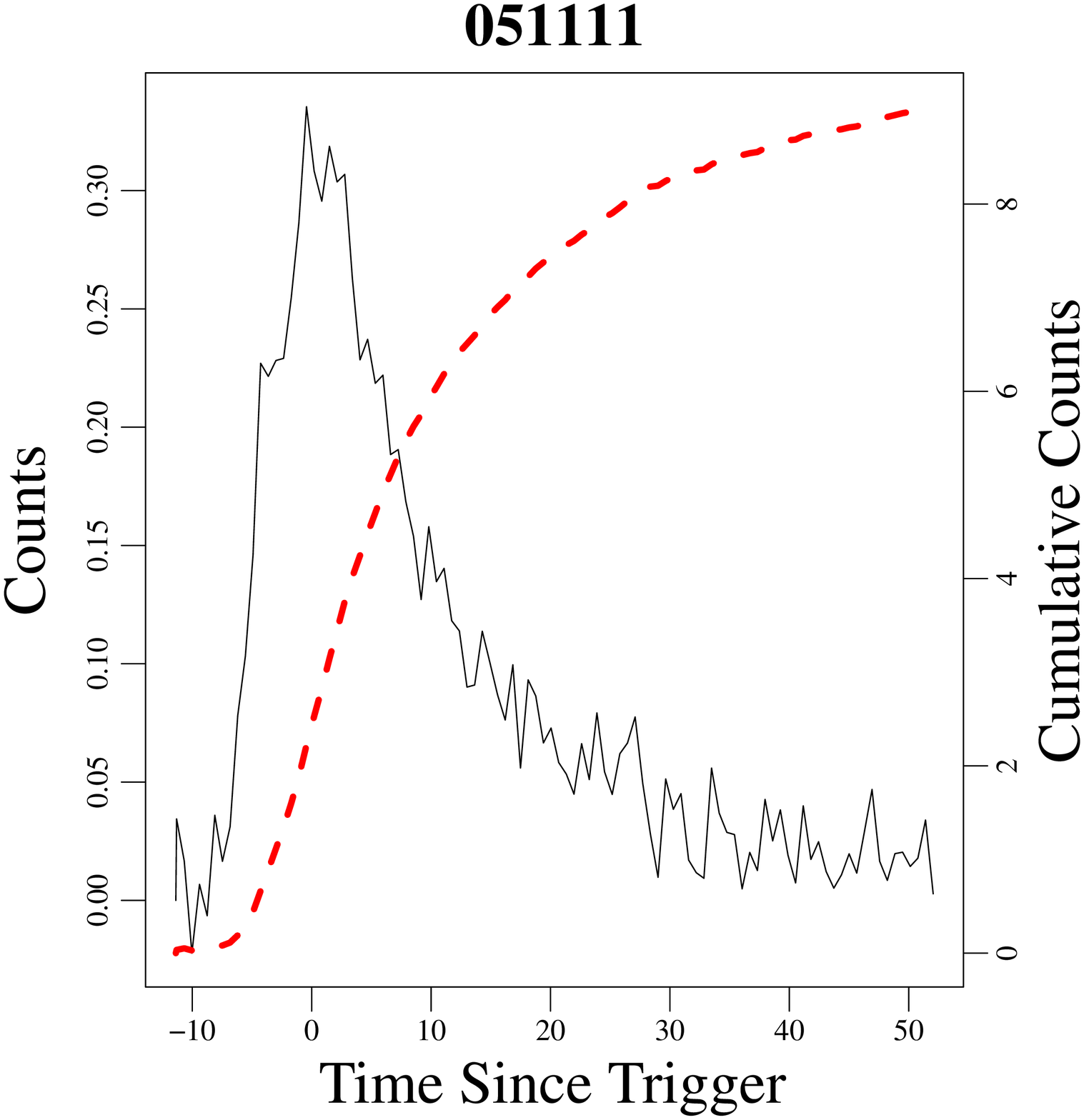}
 \end{center}
\end{minipage}
\begin{minipage}{0.25\hsize}
\begin{center}
    \FigureFile(40mm,40mm){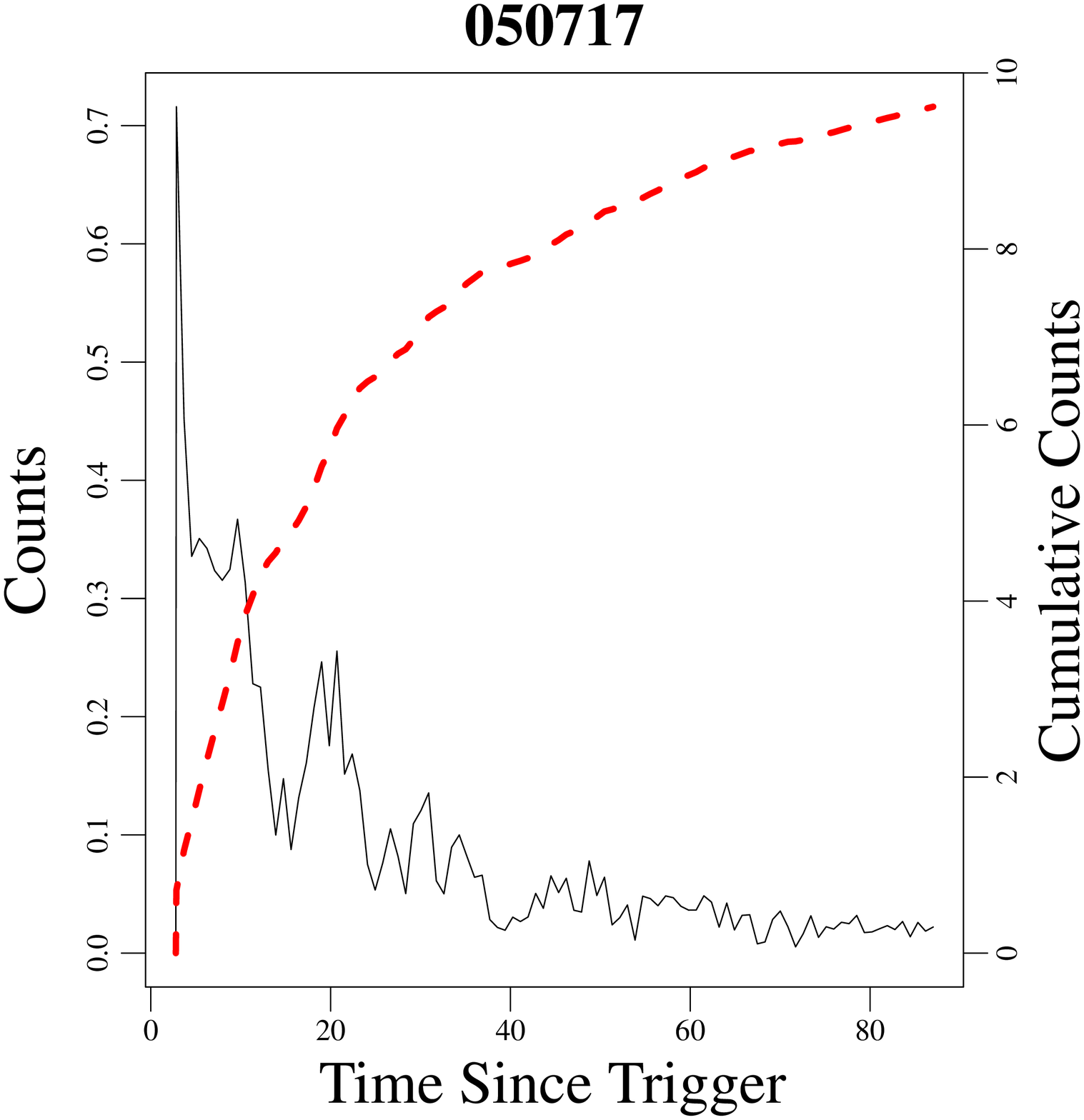}
\end{center}
\end{minipage}
\begin{minipage}{0.25\hsize}
\begin{center}
    \FigureFile(40mm,40mm){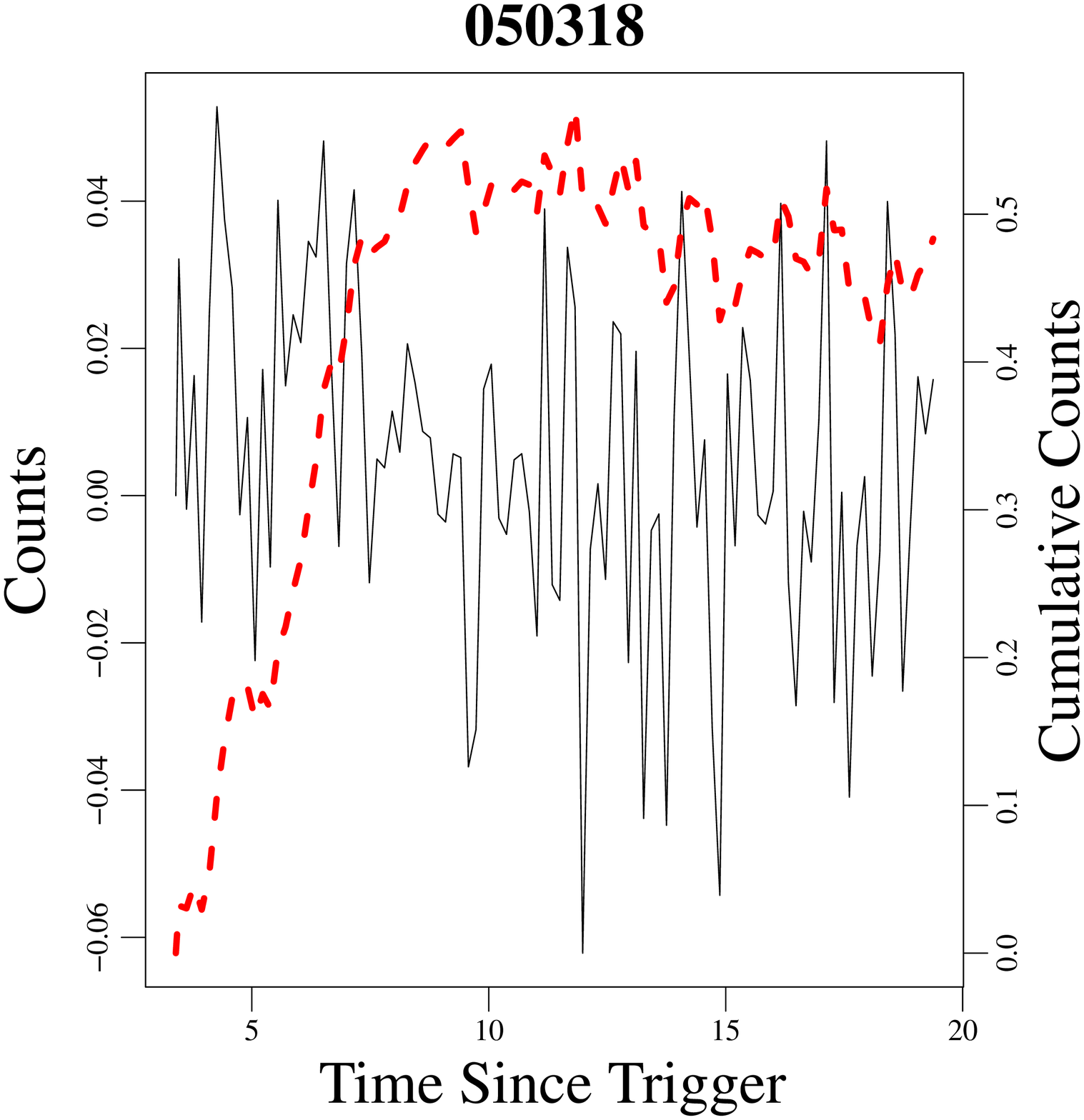}
 \end{center}
\end{minipage}\\
\begin{minipage}{0.25\hsize}
\begin{center}
    \FigureFile(40mm,40mm){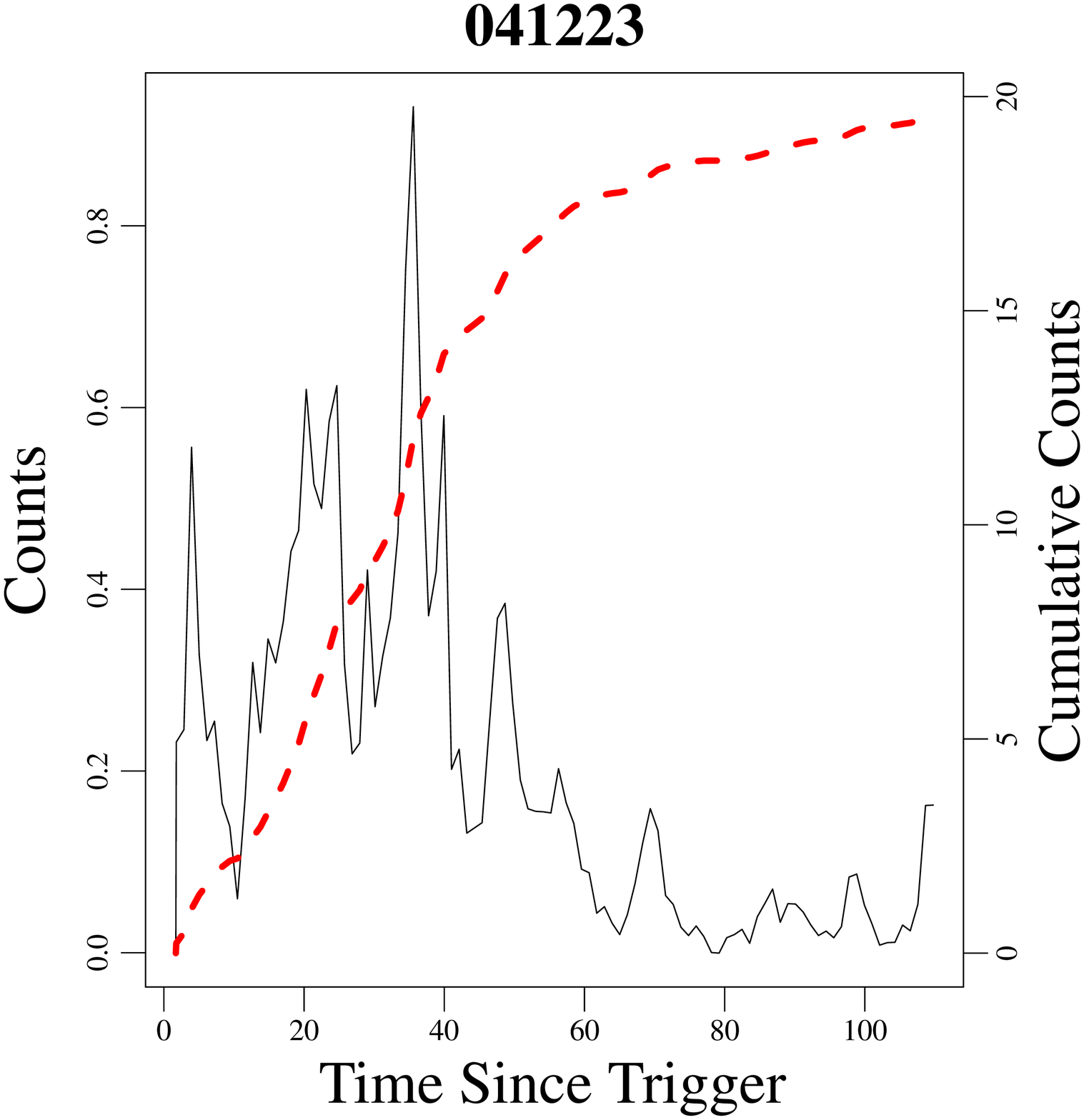}
\end{center}
\end{minipage}
\begin{minipage}{0.25\hsize}
\begin{center}
    \FigureFile(40mm,40mm){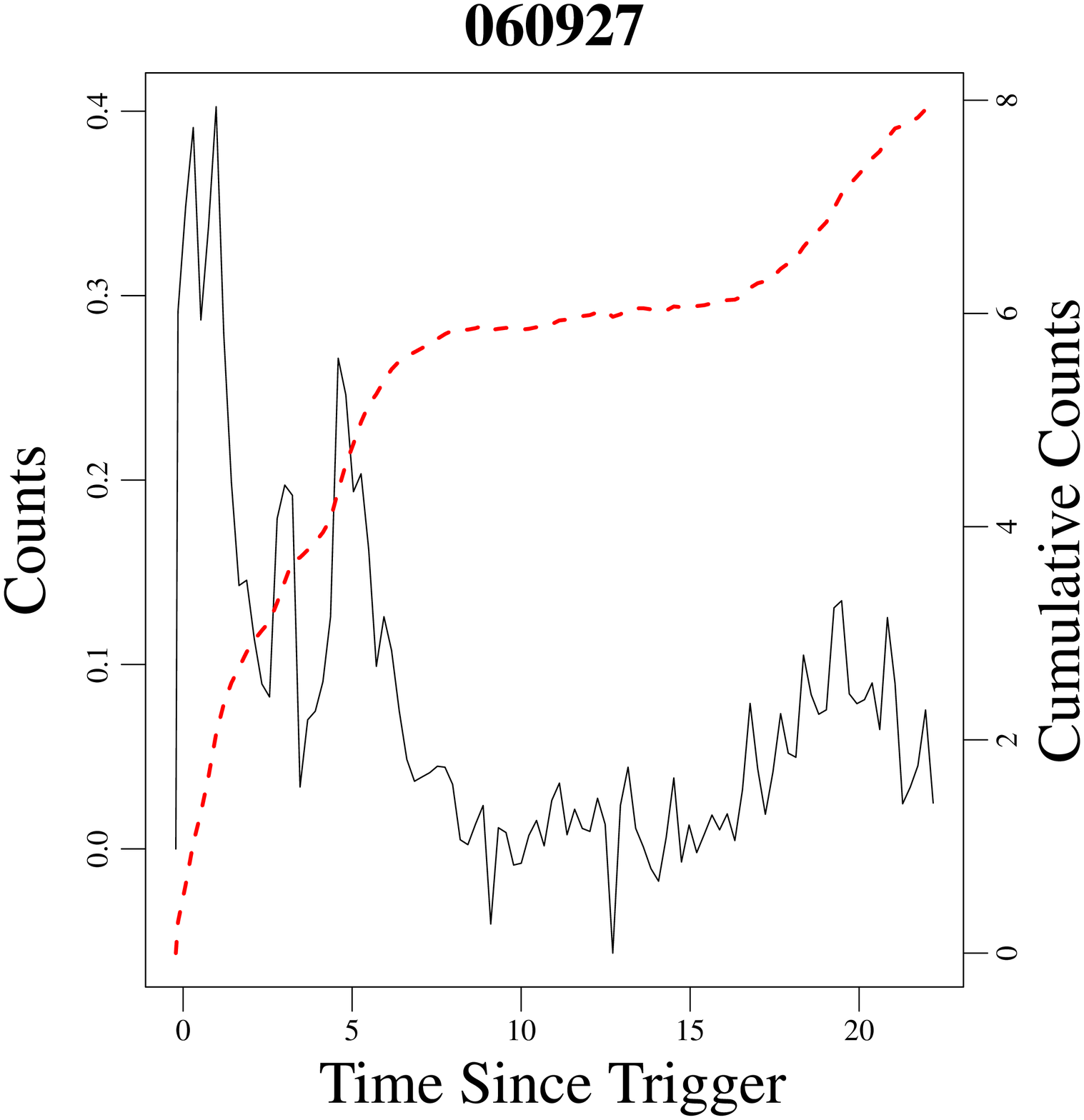}
 \end{center}
\end{minipage}
\begin{minipage}{0.25\hsize}
\begin{center}
    \FigureFile(40mm,40mm){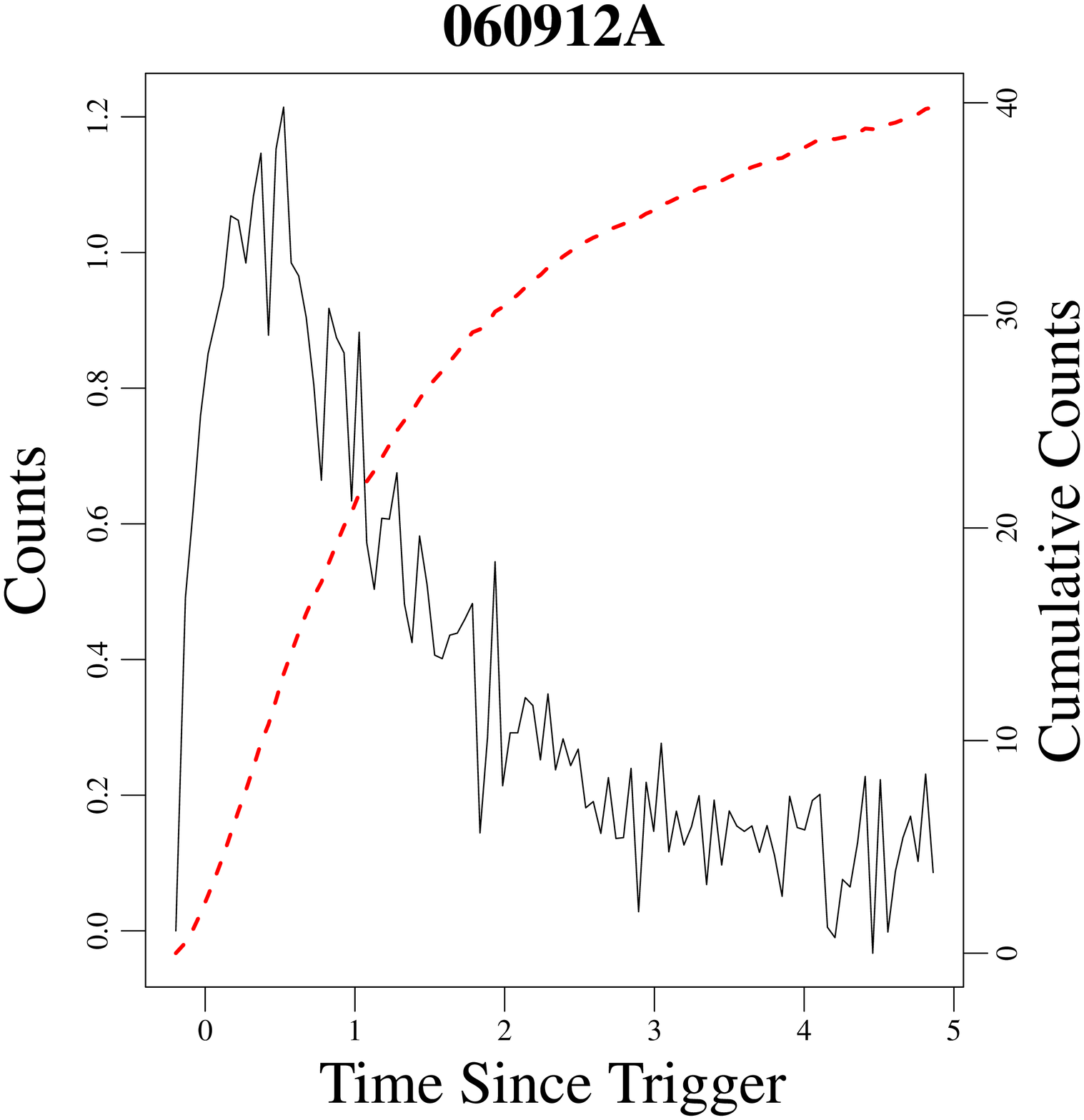}
\end{center}
\end{minipage}
\begin{minipage}{0.25\hsize}
\begin{center}
    \FigureFile(40mm,40mm){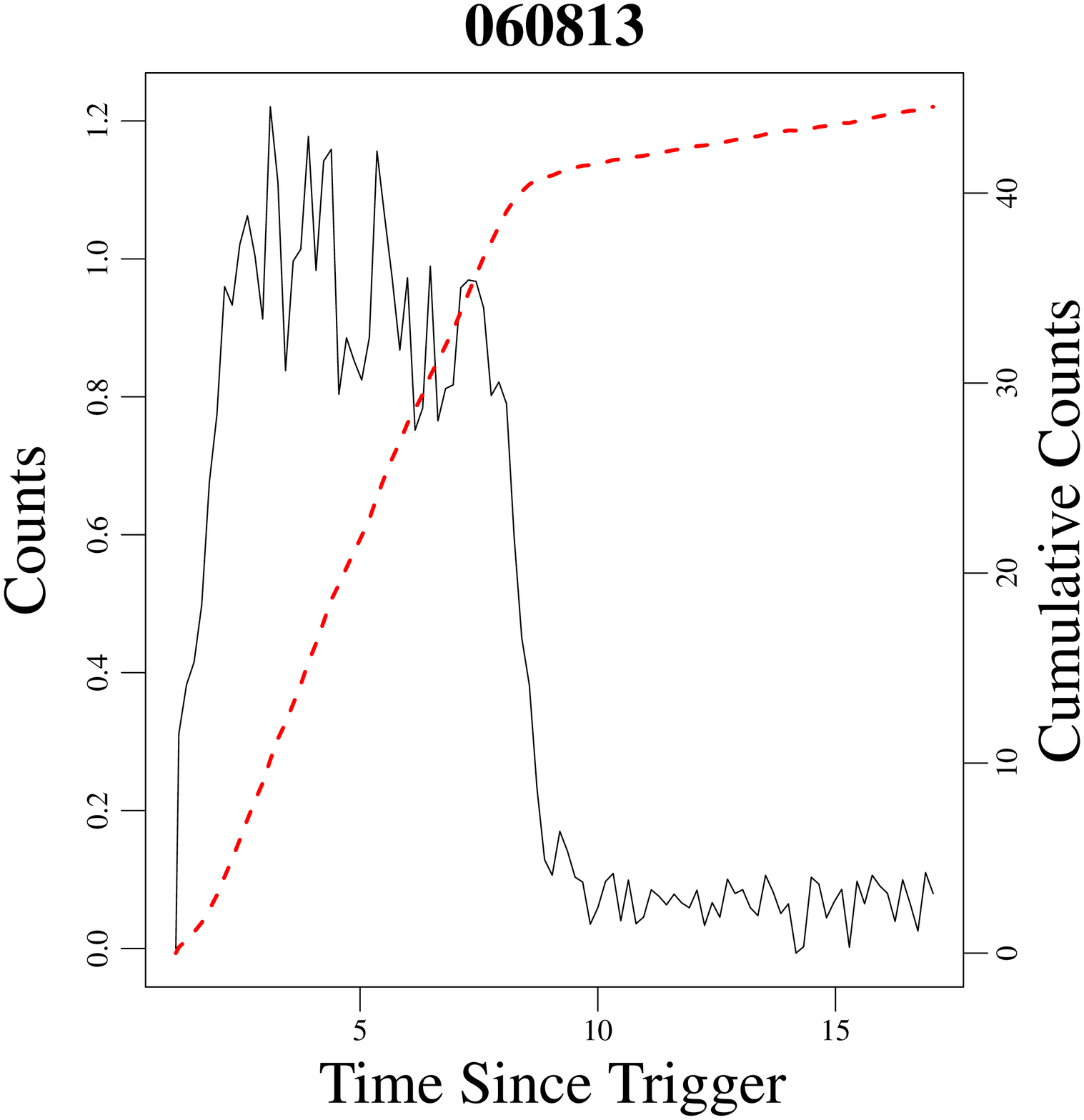}
 \end{center}
\end{minipage}\\
\begin{minipage}{0.25\hsize}
\begin{center}
    \FigureFile(40mm,40mm){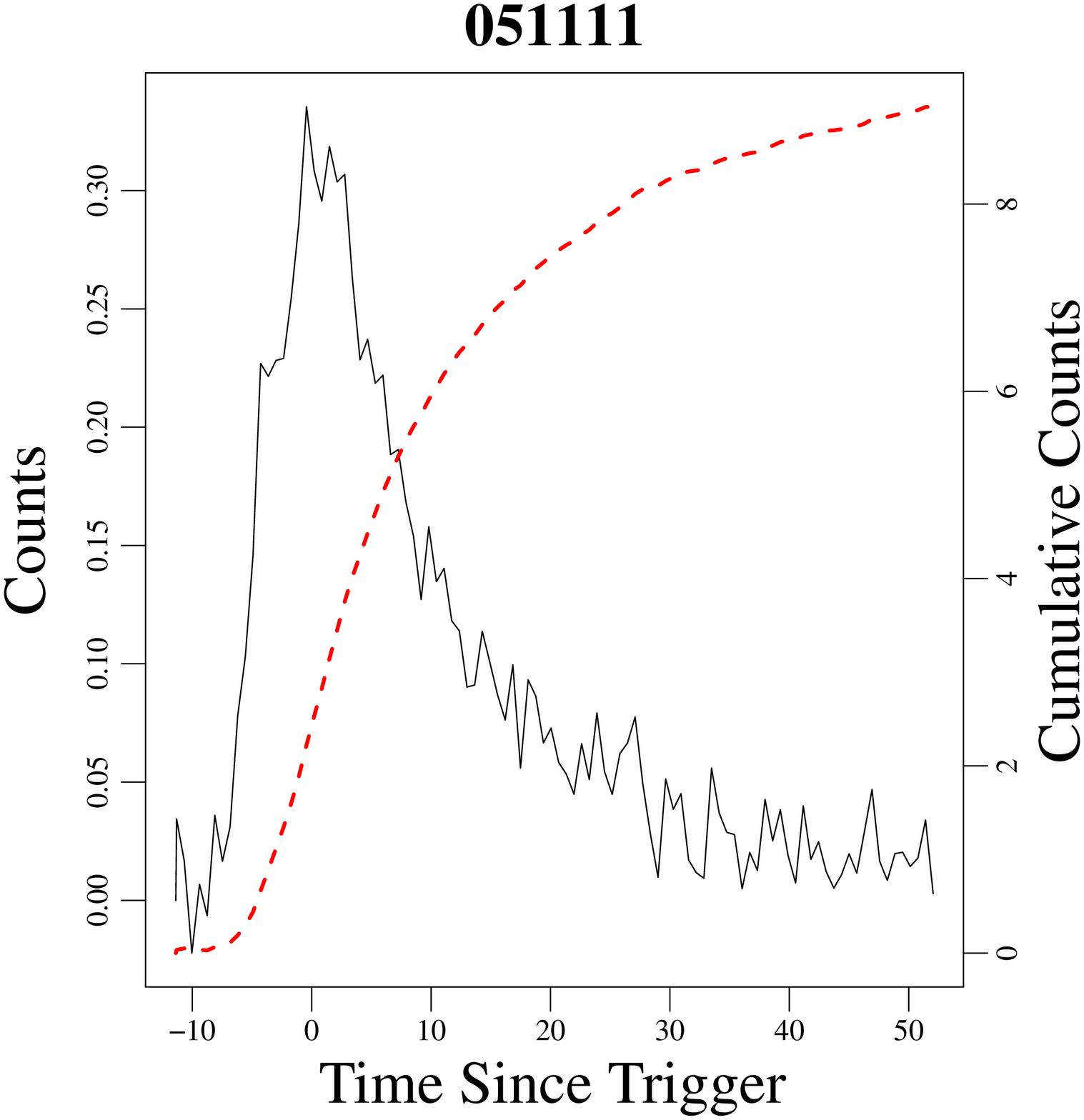}
\end{center}
\end{minipage}
\begin{minipage}{0.25\hsize}
\begin{center}
    \FigureFile(40mm,40mm){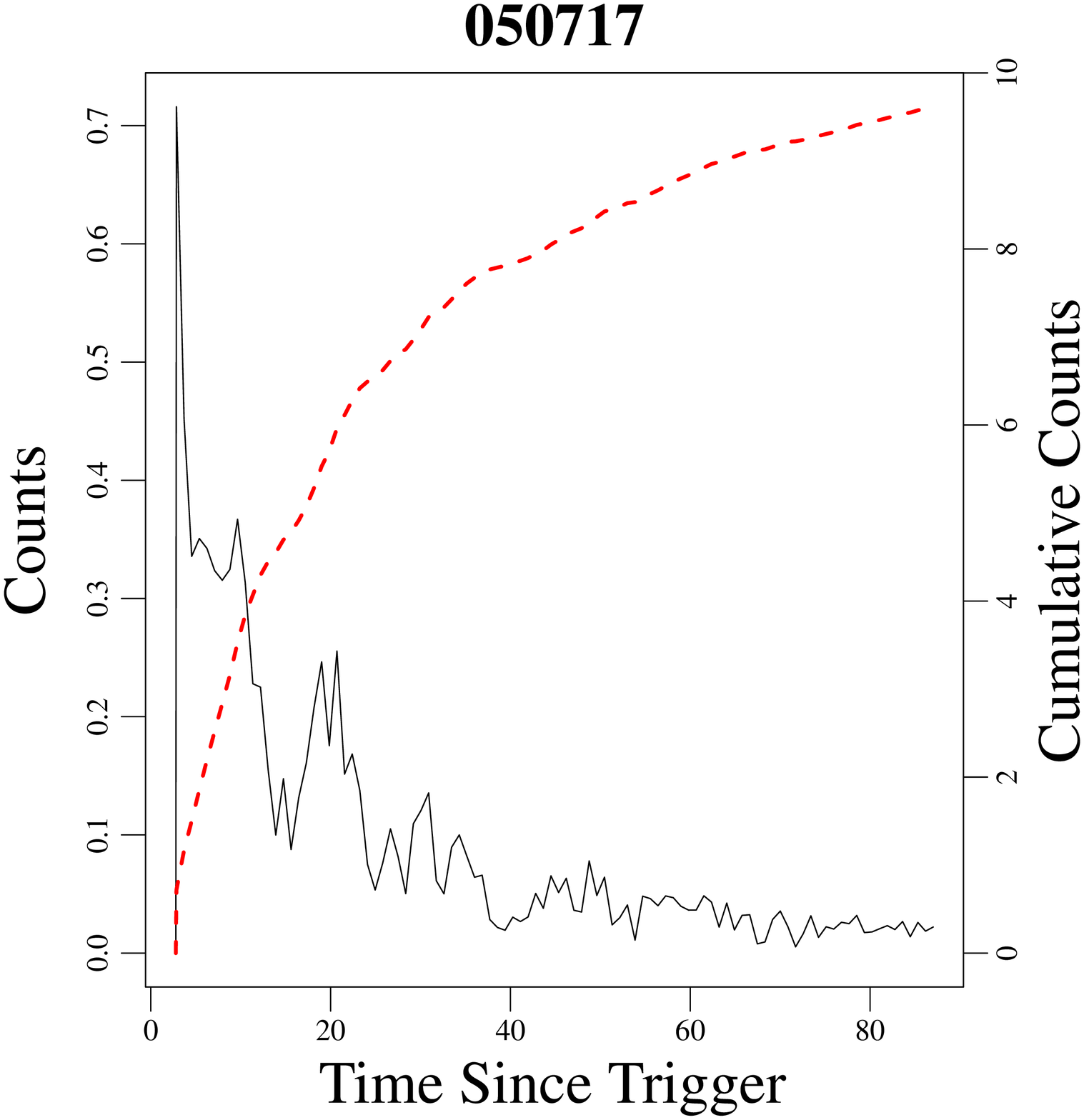}
 \end{center}
\end{minipage}
\begin{minipage}{0.25\hsize}
\begin{center}
    \FigureFile(40mm,40mm){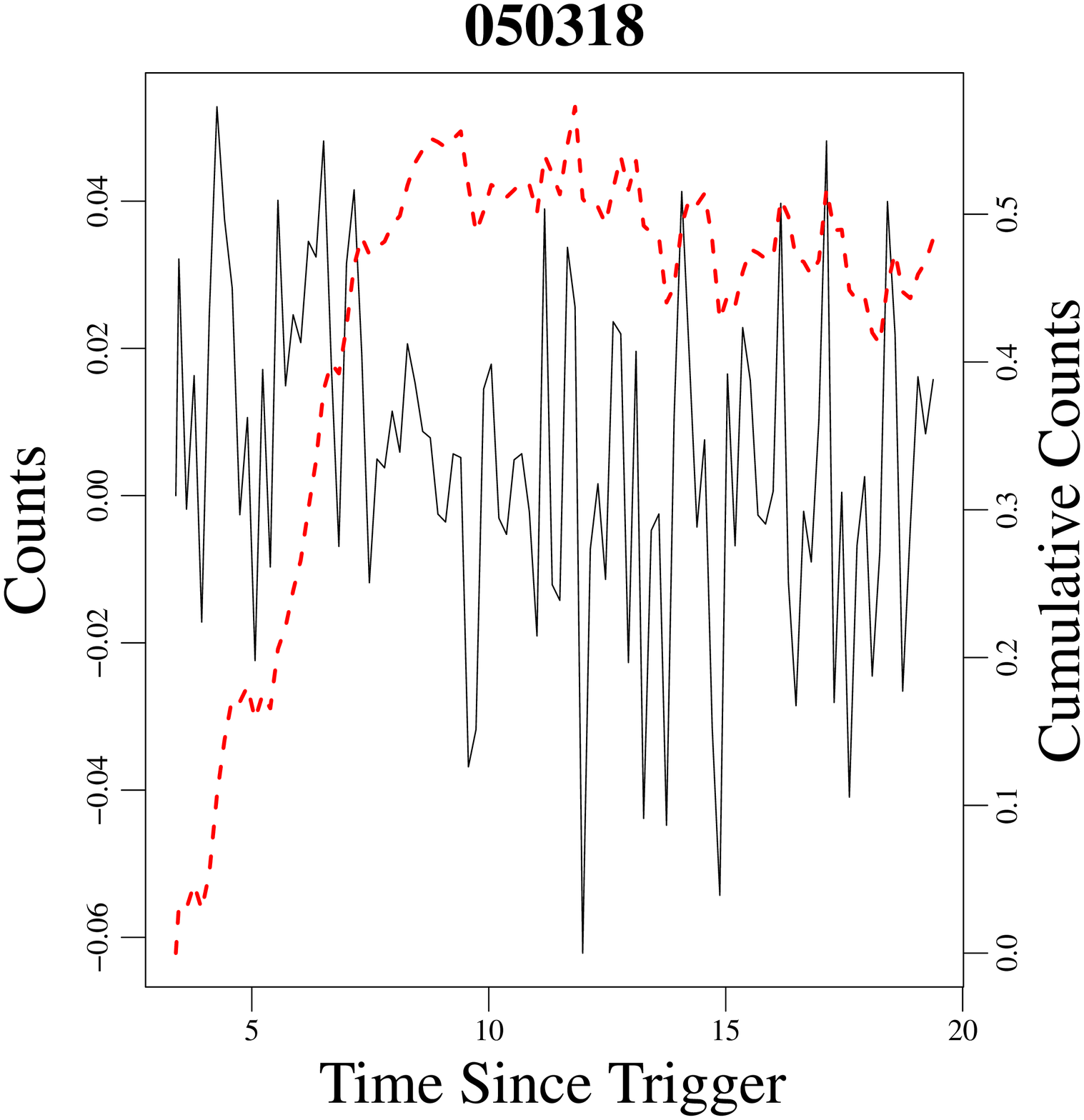}
\end{center}
\end{minipage}
\begin{minipage}{0.25\hsize}
\begin{center}
    \FigureFile(40mm,40mm){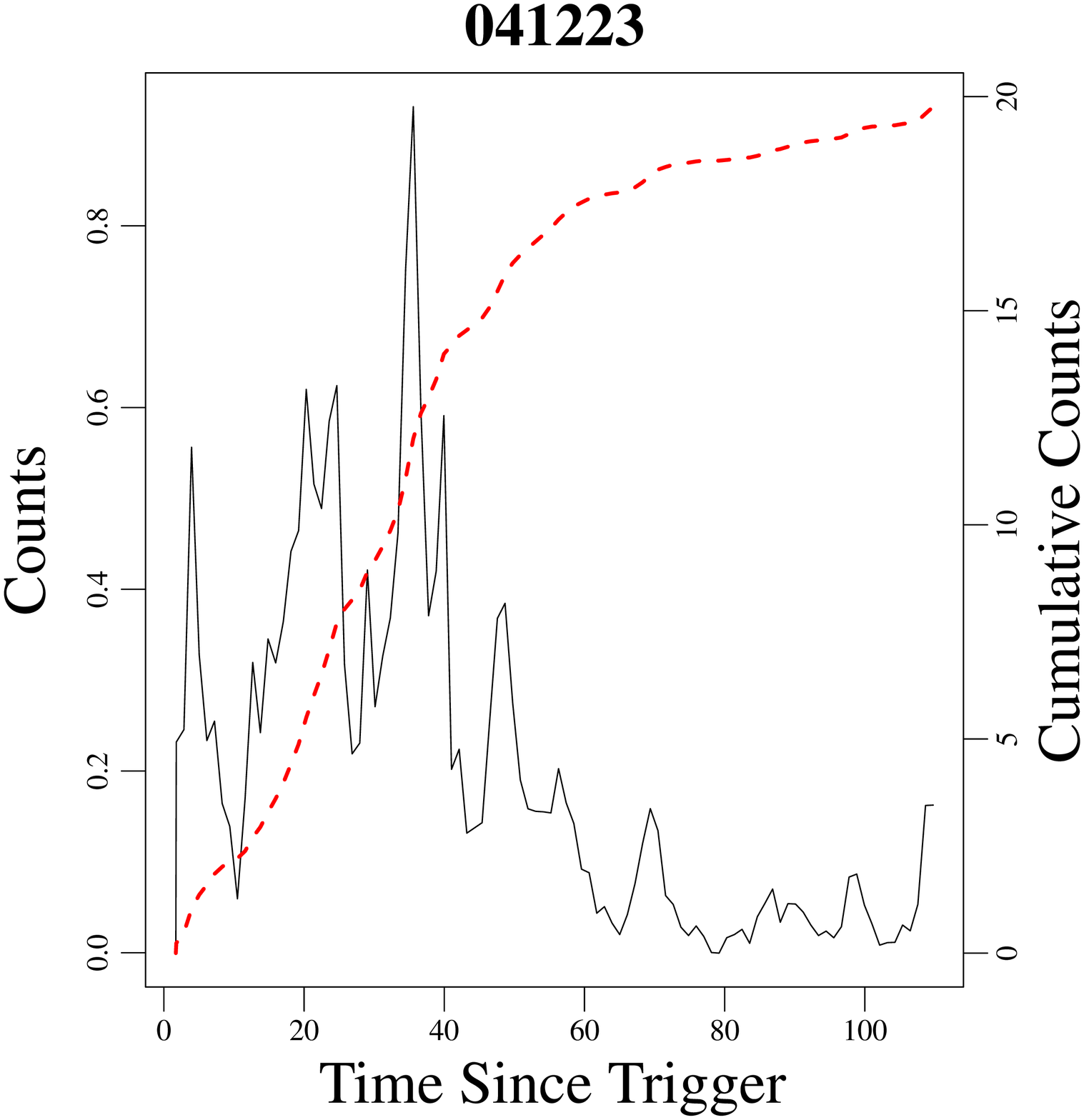}
 \end{center}
\end{minipage}\\
\end{tabular}
\caption{Light curves (black solid) and cumulative light curves (red doted) of Type II LGRBs.}\label{fig:A2-3}
\end{figure*}
%%%%%%%%%%%%%%%%%%%%%
\begin{figure*}[htb]
\begin{tabular}{cccc}
\begin{minipage}{0.25\hsize}
\begin{center}
    \FigureFile(40mm,40mm){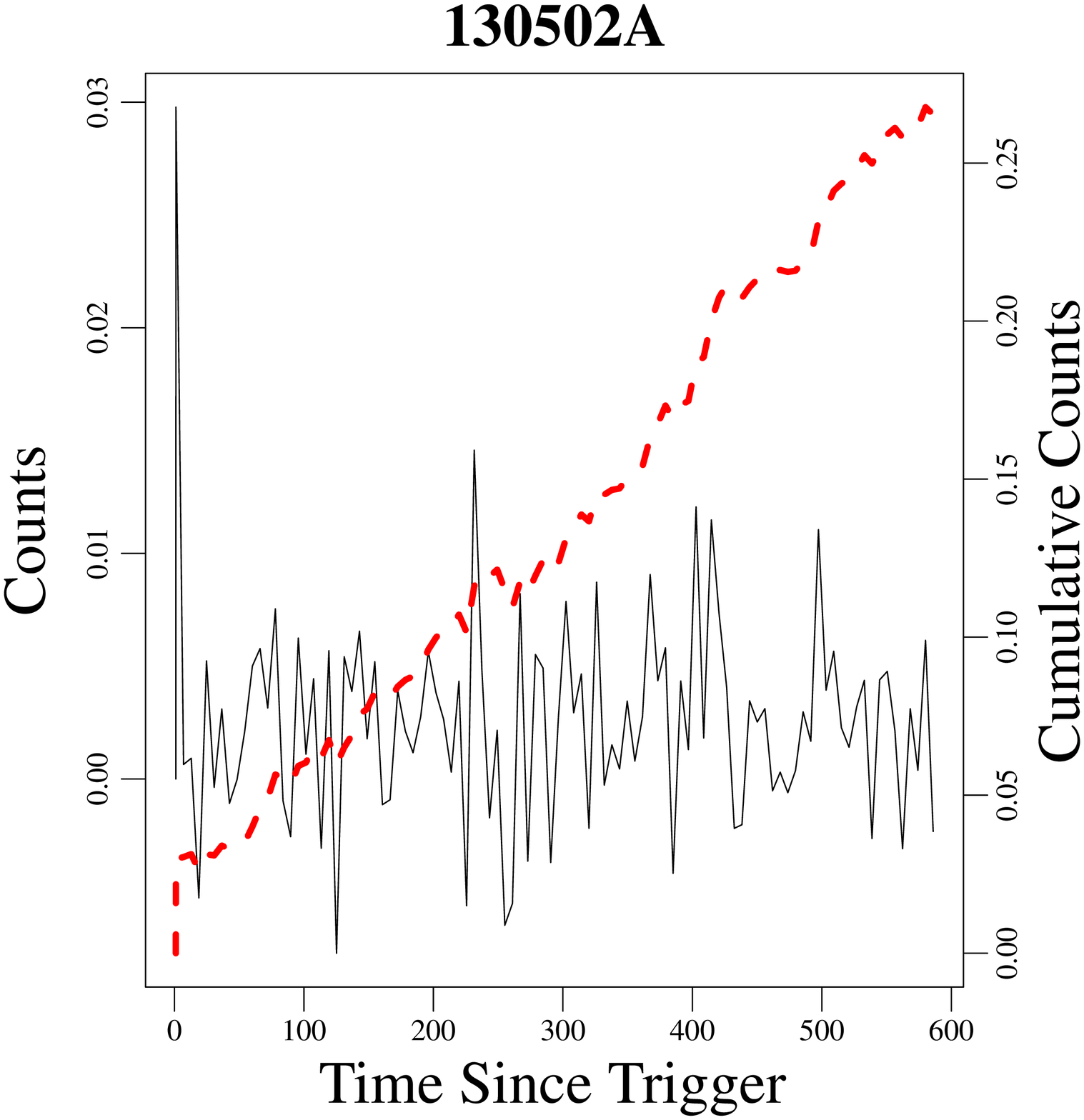}
\end{center}
\end{minipage}
\begin{minipage}{0.25\hsize}
\begin{center}
    \FigureFile(40mm,40mm){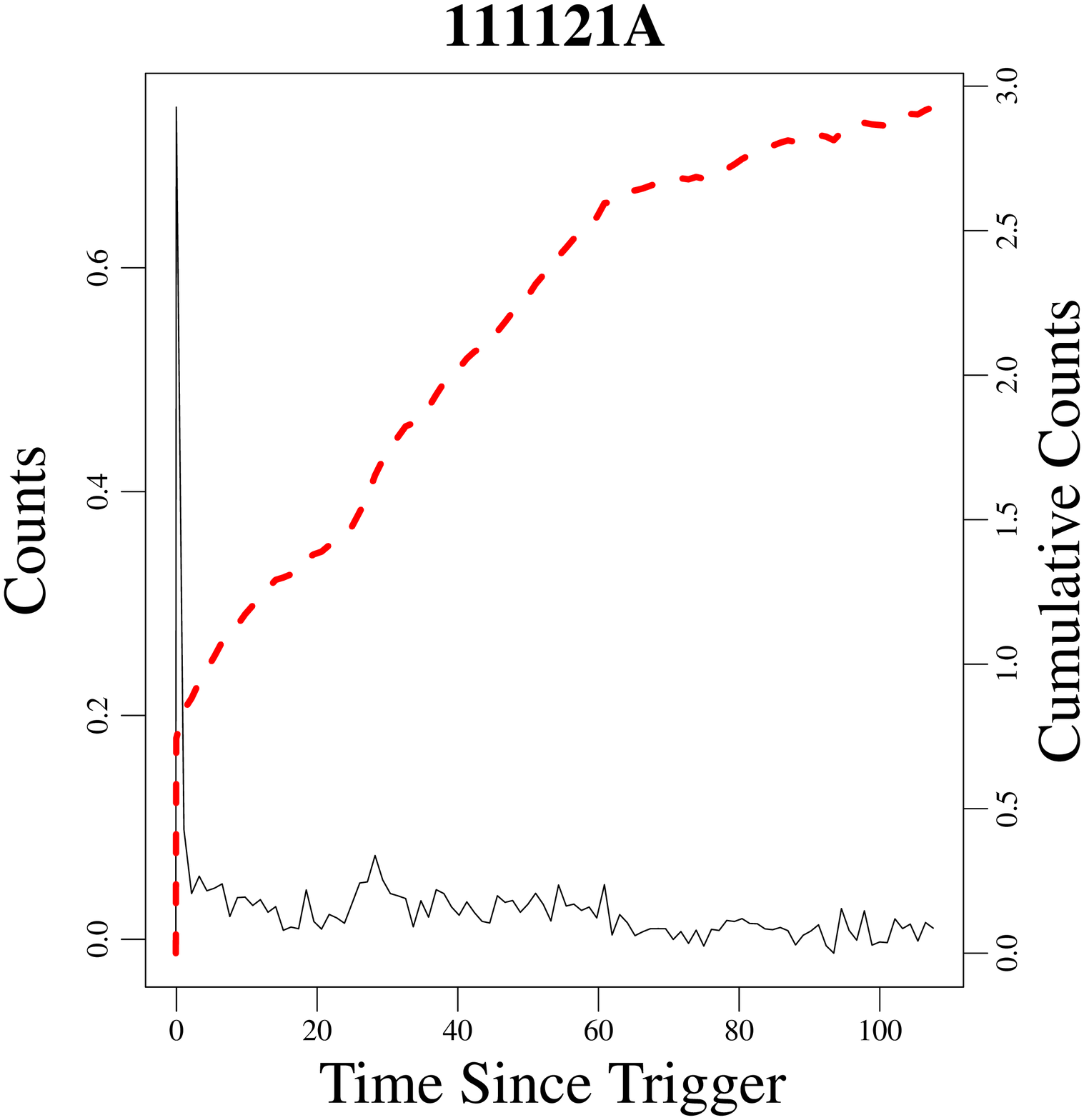}
 \end{center}
\end{minipage}
\begin{minipage}{0.25\hsize}
\begin{center}
    \FigureFile(40mm,40mm){./lc/AII-3.eps}
\end{center}
\end{minipage}
\begin{minipage}{0.25\hsize}
\begin{center}
    \FigureFile(40mm,40mm){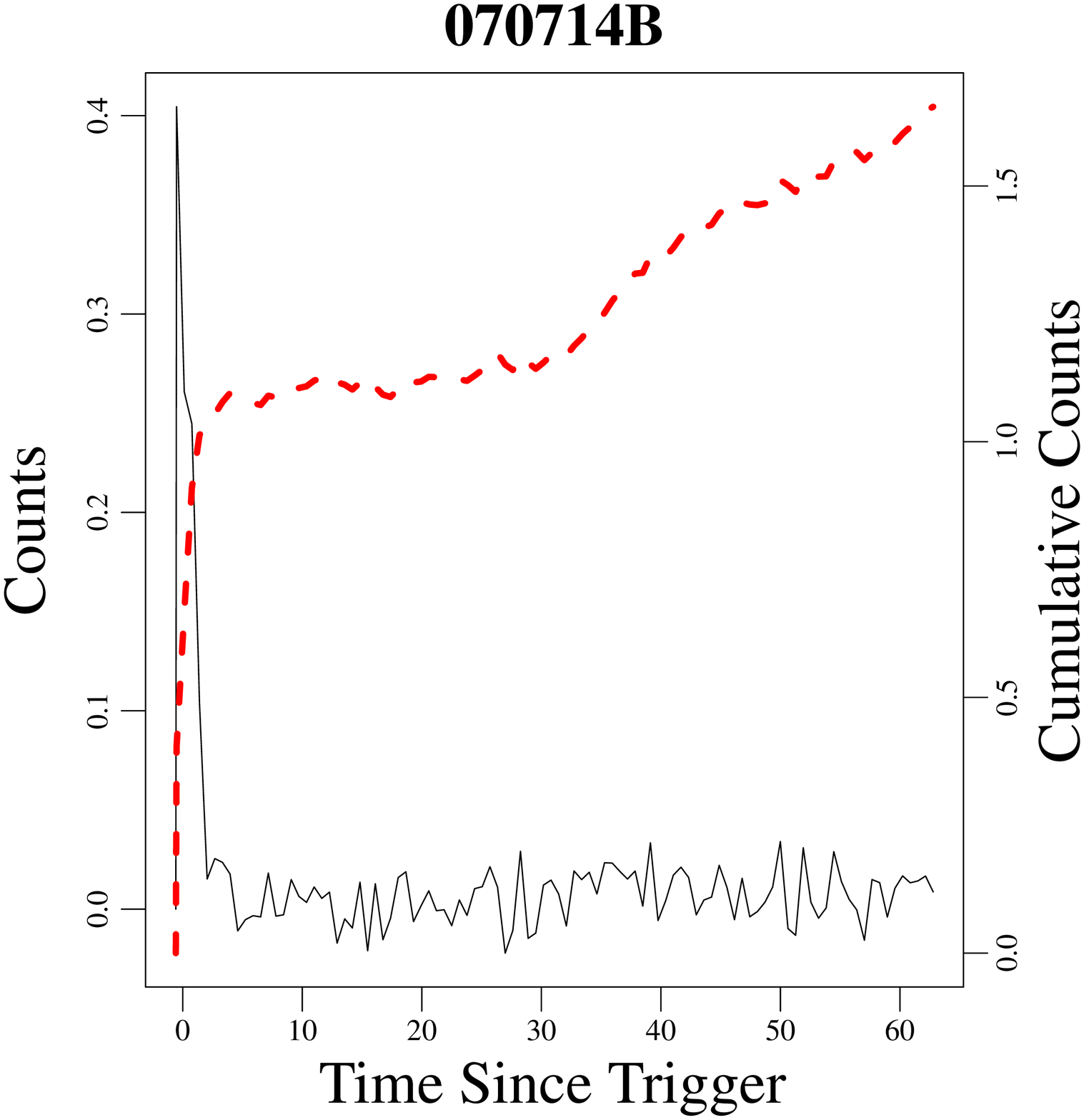}
 \end{center}
\end{minipage}\\
\begin{minipage}{0.25\hsize}
\begin{center}
    \FigureFile(40mm,40mm){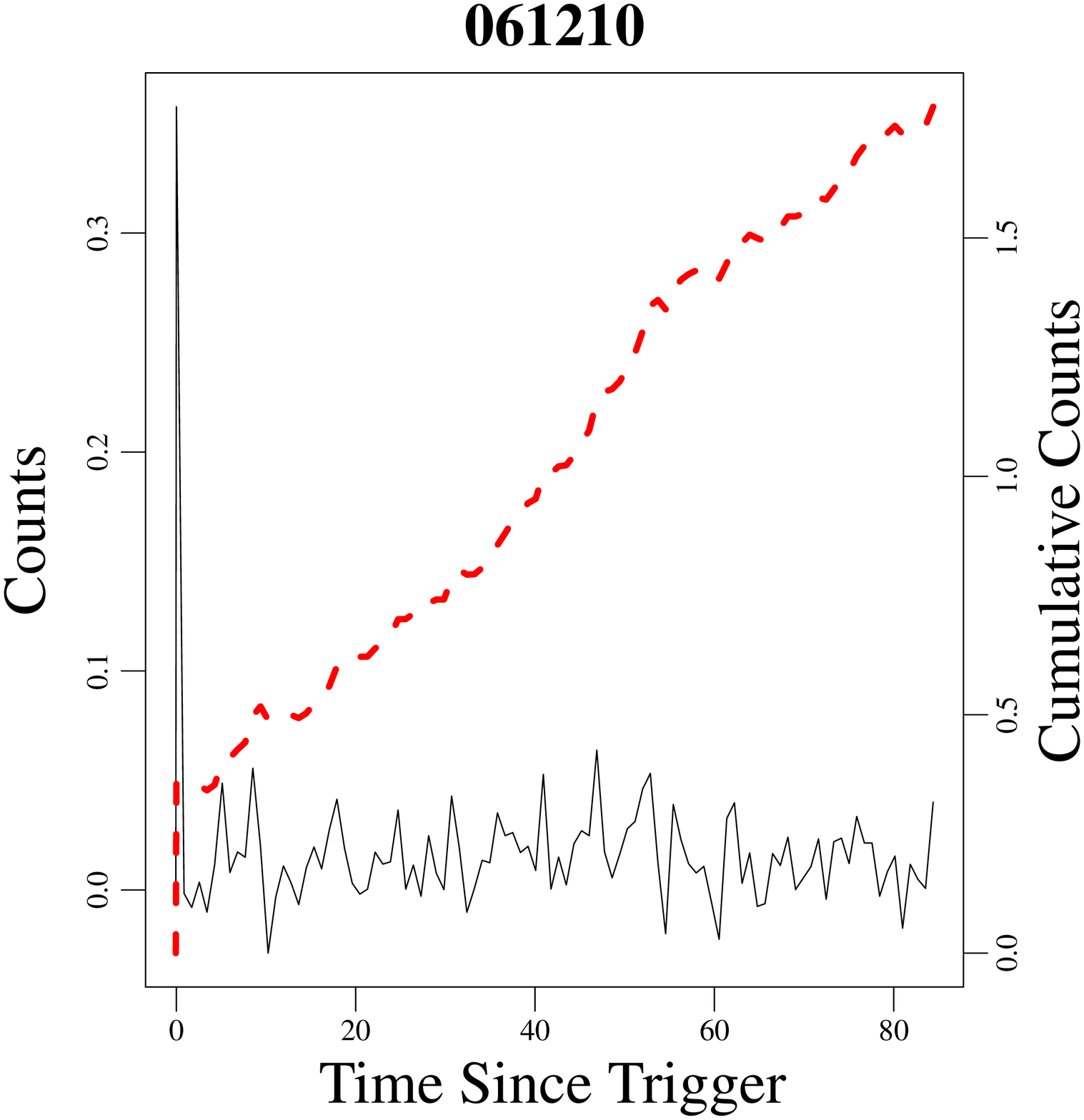}
\end{center}
\end{minipage}
\begin{minipage}{0.25\hsize}
\begin{center}
    \FigureFile(40mm,40mm){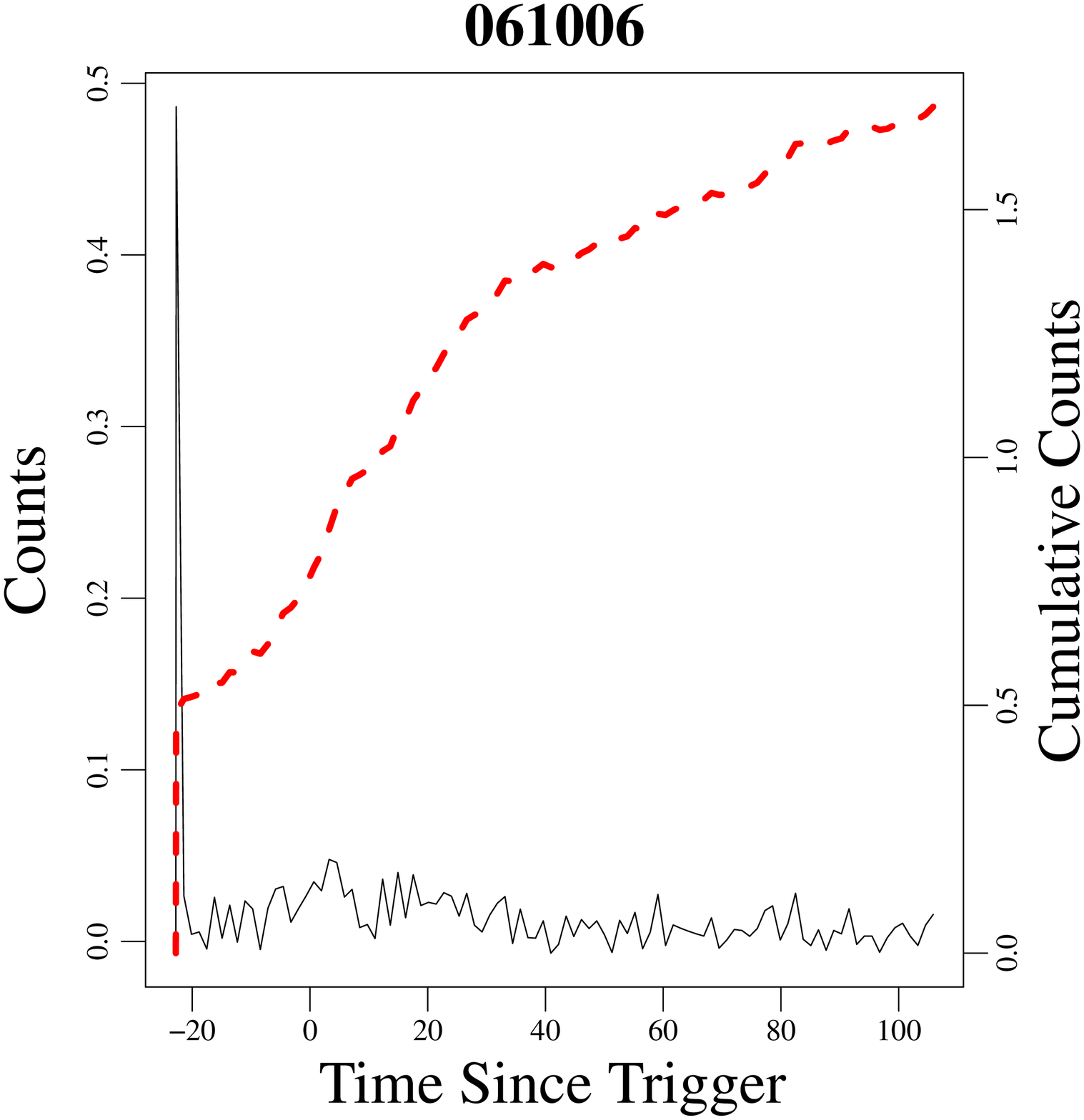}
 \end{center}
\end{minipage}
\begin{minipage}{0.25\hsize}
\begin{center}
    \FigureFile(40mm,40mm){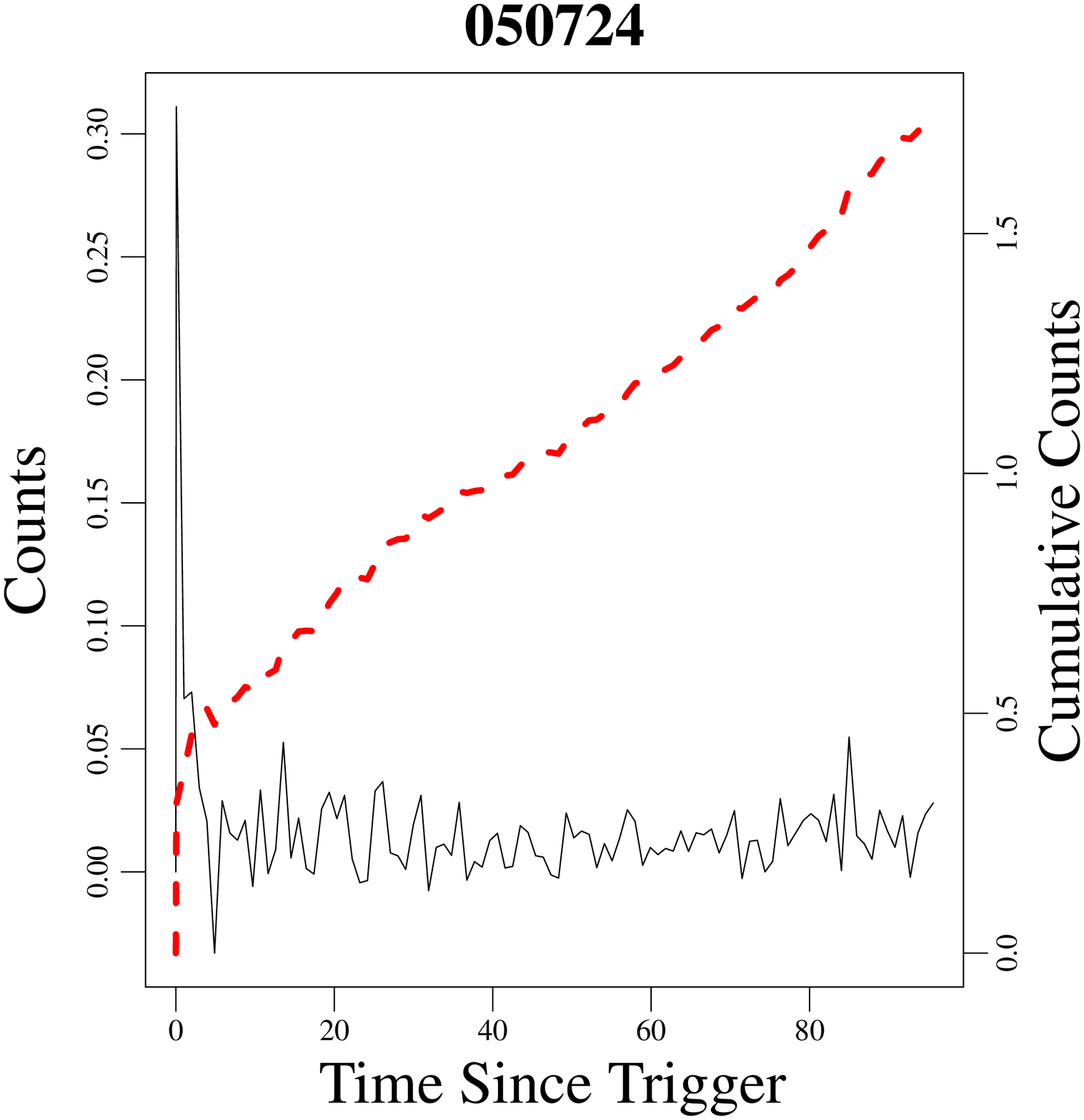}
\end{center}
\end{minipage}
\begin{minipage}{0.25\hsize}
\begin{center}
 \end{center}
\end{minipage}\\
\end{tabular}
\caption{Light curves (black solid) and cumulative light curves (red doted) of SGRBwEE.}\label{fig:A3}
\end{figure*}

%%%%%%%%%%%%%%%%%%%%%
%
% References
%
%\bibliographystyle{apj}
%\bibliography{tsutsui}
%\begin{thebibliography}{17}
%\expandafter\ifx\csname natexlab\endcsname\relax\def\natexlab#1{#1}\fi
%  
%\end{thebibliography}

\end{document}